\documentclass[11pt, a4paper, oneside]{Thesis} 

\graphicspath{{Pictures/}} 

\usepackage[square,sort,compress,numbers]{natbib}
\usepackage{cite}
\usepackage{mathtools}
\renewcommand{\thefootnote}{\arabic{footnote}}
\usepackage{pifont}
\usepackage{cancel}
\usepackage{perpage}
\usepackage{amsmath,latexsym} 
\usepackage{accents} 
\usepackage{float}
\usepackage{epigraph}
\title{\ttitle} 
\DeclareUnicodeCharacter{2212}{-}
\DeclareUnicodeCharacter{2212}{+}
\newcommand\blfootnote[1]{%
     \begingroup
     \renewcommand\thefootnote{}\footnote{#1}%
     \addtocounter{footnote}{-1}%
      \endgroup
    }
\begin{document}

\setstretch{1.3} 

\fancyhead{} 
\rhead{\thepage} 
\lhead{} 

%

\thesistitle{Astrophysical Objects in Modified Theories of Gravity}
\documenttype{THESIS}
\supervisor{\textbf{Prof. Pradyumn Kumar Sahoo}}
\supervisorposition{Professor}
\supervisorinstitute{BITS, Pilani, Hyderabad Campus}
\examiner{}
\degree{Ph.D. Research Scholar}

\coursecode{DOCTOR OF PHILOSOPHY}
\coursename{Thesis}
\authors{\textbf{SNEHA PRADHAN}}
\IDNumber{2022PHXF0026H}
\addresses{}
\subject{}
\keywords{}
\university{\texorpdfstring{\href{http://www.bits-pilani.ac.in/} 
                {Birla Institute of Technology and Science, Pilani}} 
                {Birla Institute of Technology and Science, Pilani}}
\UNIVERSITY{\texorpdfstring{\href{http://www.bits-pilani.ac.in/} 
                {\textbf{BIRLA INSTITUTE OF TECHNOLOGY AND SCIENCE, PILANI}}} 
                {\textbf{BIRLA INSTITUTE OF TECHNOLOGY AND SCIENCE, PILANI}}}



\department{\texorpdfstring{\href{http://www.bits-pilani.ac.in/pilani/Mathematics/Mathematics} 
                {Mathematics}} 
                {Mathematics}}
\DEPARTMENT{\texorpdfstring{\href{http://www.bits-pilani.ac.in/pilani/Mathematics/Mathematics} 
                {Mathematics}} 
                {Mathematics}}
\group{\texorpdfstring{\href{Research Group Web Site URL Here (include http://)}
                {Research Group Name}} 
                {Research Group Name}}
\GROUP{\texorpdfstring{\href{Research Group Web Site URL Here (include http://)}
                {RESEARCH GROUP NAME (IN BLOCK CAPITALS)}}
                {RESEARCH GROUP NAME (IN BLOCK CAPITALS)}}
\faculty{\texorpdfstring{\href{Faculty Web Site URL Here (include http://)}
                {Faculty Name}}
                {Faculty Name}}
\FACULTY{\texorpdfstring{\href{Faculty Web Site URL Here (include http://)}
                {FACULTY NAME (IN BLOCK CAPITALS)}}
                {FACULTY NAME (IN BLOCK CAPITALS)}}

\maketitle

\clearpage
\setstretch{1.3} 

\pagestyle{empty} 
\pagenumbering{gobble}

\addtocontents{toc}{\vspace{2em}} 
\frontmatter 
\Certificate
\Declaration
\begin{acknowledgements}

I express my sincere thanks and gratitude to my supervisor, \textbf{Prof. Pradyumn Kumar Sahoo}, Professor, Department of Mathematics, BITS, Pilani, Hyderabad Campus, for his unwavering support, guidance, and vast experience throughout my Ph.D. journey. I am deeply grateful for his endless patience and faith in me, which inspired me to develop a deeper understanding of the subject and fostered my aptitude for scientific research.

I sincerely thank the members of my Doctoral Advisory Committee (DAC), \textbf{Prof. Bivudutta Mishra} and \textbf{Prof. Sashideep Gutti}, for their valuable suggestions and constant encouragement, which significantly contributed to improve of my research work.

I am privileged to thank the Head of the Department, the DRC convener, all faculty members, and the staff of the Department of Mathematics for their support throughout this Ph.D. journey.


I gratefully acknowledge the National Board for Higher Mathematics for supporting my research through the NBHM Project Fellowship (File No. 02011/3/2022/-R \&D-II/5893) from 19.03.2022 to 18.03.2025. I also thank BITS, Pilani, Hyderabad Campus, for providing the necessary facilities and institute funding from 19.03.2025 onward.

 I also thank my friends, colleagues, and collaborators for their valuable suggestions and fruitful discussions throughout my Ph.D. journey. 
 
I extend my heartfelt gratitude to my M.Sc. guide, late \textbf{Prof. Tanuka Chattopadhaya}, who inspired me to love astrophysics.

I am grateful to my parents, \textbf{Mrs. Nandita Pradhan} and \textbf{Mr. Pravangshu Pradhan} and my \textbf{grandparents} for their unconditional love, support, and encouragement throughout my academic journey.

I am especially grateful to my best friend, \textbf{Surajit Bera}, whose understanding and unwavering support have brought light and clarity to my life. His presence helped me appreciate the beauty of simplicity.

Last but not least, I fondly remember \textbf{Dr. Animesh Bag}, whose guidance inspired me to dream bigger and pursue higher studies.

 



\vspace{0.5 cm}
\textbf{Sneha Pradhan},\\
ID: 2022PHXF0026H.
\end{acknowledgements}

\begin{abstract}
The primary objective of this work is to investigate compact astrophysical objects within the framework of modified theories of gravity. In particular, we aim to explore how deviations from general relativity influence the internal structure, stability of neutron stars, and strange stars. By combining theoretical modeling with observational data, this work aims to constrain model parameters and evaluate the viability of alternative gravity theories for ultra dense matter.\\
Chapter~\ref{Chapter1} presents the theoretical framework of this thesis, reviewing compact objects including white dwarfs, neutron stars (and strange stars), and black holes and their astrophysical relevance. It outlines the fundamentals of general relativity, relativistic stellar structure, and exterior spacetime solutions, emphasizing the motivations for modified gravity. The geometrical trinity of gravity through curvature, torsion, and nonmetricity is introduced. The chapter concludes with a summary of key equations of state and the essential physical and stability conditions for compact star models.\\
In chapter~\ref{Chapter2}, charged compact stars are studied in $f(Q)$ gravity using conformal motion and the MIT bag equation of state. Two models based on power-law and linear conformal factors are constructed and matched with the Bardeen exterior spacetime. Their physical viability are examined through energy conditions and equilibrium analysis. Furthermore, comparison with the Reissner–Nordström exterior highlights the robustness and physical preference of the Bardeen based configurations.\\
In chapter~\ref{Chapter3}, we study strange star models in modified 
$f(T)$ gravity using the gravitational decoupling approach via minimal geometric deformation. Exact deformed solutions are obtained by adopting the Buchdahl ansatz with a quadratic polytropic EoS. A comparative analysis between GR,  
$f(T)$ gravity, and $f(T)$+MGD highlights the role of deformation effects on stellar structure. Observational constraints confirm that the model supports massive and physically viable strange star configurations.\\
Chapter~\ref{Chapter4} focuses on strange stars in $f(T)$ gravity using the complete geometric deformation approach. Exact solutions are obtained under the vanishing complexity condition, allowing anisotropy and dark matter effects to be incorporated through the decoupling parameter $\alpha$. The models are tested with observations and stability verified via causality, Herrera’s cracking condition that demonstrates the viability of strange stars admixed with dark matter. \\
In Chapter~\ref{Chapter5}, neutron stars are studied in $f(Q)$ gravity through a Bayesian analysis, testing linear, logarithmic and exponential models while fixing the DDME2 equation of state to isolate gravitational effects. Confrontation with NICER mass–radius data and Bayes factor evaluation favors the exponential model, yielding radii and tidal deformabilities consistent with current constraints. Chapter~\ref{Chapter7} concludes the thesis and outlines future research directions.
\end{abstract} 

\Dedicatory{\bf \begin{LARGE}
Dedicated to
\end{LARGE} 
\\
\vspace{0.2cm}
\it Every existence in the universe that made me smile genuinely\\}


\lhead{\emph{Contents}} 
\tableofcontents 
\addtocontents{toc}{\vspace{1em}}
\addtocontents{toc}{\vspace{1em}}
\lhead{\emph{List of Tables}}
\listoftables 
\addtocontents{toc}{\vspace{1em}}
\lhead{\emph{List of Figures}}
\listoffigures 
\addtocontents{toc}{\vspace{1em}}



\lhead{\emph{List of Symbols and Abbreviations}}

\listofsymbols{ll}{
\textbf{Notation} & \textbf{Description} \\[0.6cm]
$\mathcal{M}$ & Manifold\\
$g_{\alpha\beta}$ & Metric tensor \\
$g$ & Determinant of $g_{\alpha\beta}$ \\
$\Gamma^{\alpha}{}_{\beta\gamma}$ & General affine connection \\
$\Big\{\begin{matrix}
\alpha \\[-4pt]
\beta\gamma
\end{matrix}\Big\}$ & Levi-Civita connection \\
$K^{\alpha}{}_{\beta\gamma}$ & Contorsion tensor \\
$L^{\alpha}{}_{\beta\gamma}$ & Disformation tensor \\
$\nabla_{\gamma}$ & Covariant derivative \\
${R}^{\alpha}{}_{\beta\gamma\sigma}$ & Riemann tensor \\
$R_{\mu\nu}$ & Ricci tensor \\
$R$ & Ricci scalar \\
$\mathcal{T}_{\mu\nu}$ & Stress-energy tensor \\
$P^\alpha{}_{\beta\gamma}$ & Non-metricity conjugate \\
$Q_{\alpha\beta\gamma}$ & Non-metricity tensor \\
$Q$ & Non-metricity scalar \\
$T$ & Torsion scalar \\[0.3cm]

GR & General Relativity \\
TEGR & Teleparallel Equivalent of General Relativity \\
STEGR & Symmetric Teleparallel Equivalent of General Relativity \\
EoS & Equation of State \\
MGD & Minimal Gravitational Decoupling \\
CGD & Complete Gravitational Decoupling \\
HR & Hertzsprung--Russell \\
CKVs & Conformal Killing Vectors \\
R--N & Reissner--Nordstr\"om \\
}

\addtocontents{toc}{\vspace{2em}}

%
%


\clearpage 





\mainmatter 

\pagestyle{fancy} 

\chapter{Introduction} 
\label{Chapter1}

\definecolor{maroon}{RGB}{128, 0, 0}
\lhead{\textcolor{maroon}{\textit{\textbf{Chapter 1:}}} \emph{\textcolor{maroon}{\textbf{Introduction}}}} 

\epigraph{\hspace{1cm}We start our fascinating journey }{at the end point of the star's evolution}

The Universe is filled with a diverse range of astrophysical objects, extending from small rocky planets and main-sequence stars to exotic high-energy systems such as neutron stars and black holes. These objects are formed and evolve through complex physical processes governed by the fundamental laws of gravitation, thermodynamics, and nuclear physics. Among these, gravity plays the most crucial role, as it dictates the large-scale structure and dynamics of the Universe. In the present thesis, the primary focus is on compact or dead stars, which represent the final evolutionary stages of stellar life. These objects not only illuminate the final fate of stars, but also serve as powerful laboratories for probing the limits of known physics and exploring potential extensions of Einstein’s theory of gravity \cite{Einstein1916}.

\section{Compact objects}
\begin{figure}[h!]
    \centering
    \includegraphics[height=9cm, width=10cm]{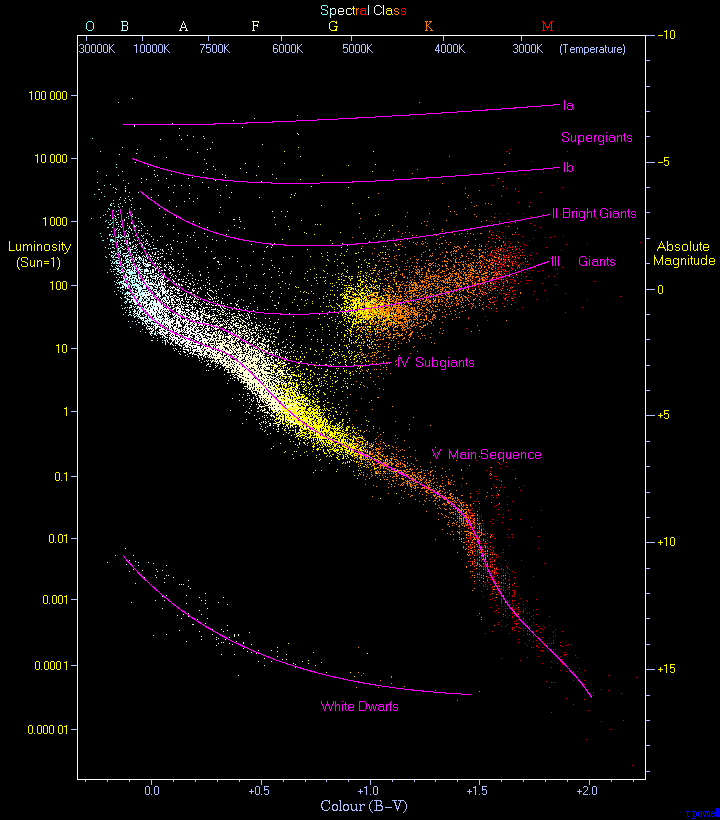}
    \caption[Hertzsprung--Russell diagram from]
{Hertzsprung--Russell (HR) diagram displaying stellar luminosity against spectral type for approximately four million stars within 5000 light-years of the Sun. The main sequence (diagonal band), red giant branch, horizontal branch, and white dwarf region are clearly visible, representing different stages of stellar evolution.\textit{Credit: ESA/Gaia/DPAC}~\cite{Gaia_HR2018}.}
    \label{fig:HRDiagram}
\end{figure}

Stars form within the vast interstellar clouds composed primarily of hydrogen and helium gas, with small amounts of cosmic dust. These clouds, known as nebulae, serve as stellar nurseries of the Universe where new stars originate. According to the Jeans instability criterion, when specific regions within these clouds experience an increase in gravitational attraction, the gas and dust begin to condense \cite{jeans1902}. As this condensation progresses, the region accumulates additional mass, which enhances its own gravitational field and leads to a subsequent collapse. The collapse raises both the temperature and pressure within the central region, resulting in the formation of a protostar \cite{shu1987}. The protostar continues to contract and heat up over thousands of years. When the central temperature and density reach a critical threshold, nuclear fusion begins, primarily converting hydrogen into helium. The energy generated through this process produces sufficient internal pressure to balance the inward gravitational pull, establishing a state of hydrostatic equilibrium. At this stage, the star attains stability and enters into the main-sequence phase, during which it continues to shine steadily for millions or even billions of years, depending on its mass. These stable stars are known as main-sequence stars, as represented in the HR diagram in Fig.-\ref{fig:HRDiagram}, which classifies stars according to their luminosity and surface temperature \cite{KDB1992}.

However, as the nuclear fuel in the stellar core is gradually exhausted, the gravitational force begins to dominate over the internal pressure support. The progressive conversion of hydrogen into helium during the main-sequence phase steadily increases the mean molecular weight of the core, leading to a gradual reduction of pressure support relative to gravity. As a result, the core undergoes a slow contraction throughout the main-sequence lifetime, releasing gravitational thermal energy. This contraction accelerates once the core hydrogen is fully depleted, eventually triggering hydrogen shell burning around the inert helium core. This contraction raises the core temperature, initiating further nuclear reactions that convert helium into heavier elements, such as carbon and oxygen. In 1939, \citet{Bethe1939} demonstrated that stellar core temperatures between \(10\) and \(30\) million Kelvin are optimal for thermonuclear reactions known as the carbon–nitrogen (C–N) cycle:  
\begin{eqnarray}
&&\hspace{0cm}\mathrm{C}^{12} + \mathrm{H}^{1} \rightarrow\ \mathrm{N}^{13} + \gamma, \hspace{2cm}
\mathrm{N}^{13} \rightarrow\ \mathrm{C}^{13} + \beta^{+} + \nu, \nonumber\\
&&\hspace{0cm}\mathrm{C}^{13} + \mathrm{H}^{1} \rightarrow\ \mathrm{N}^{14} + \gamma, \hspace{2cm}
\mathrm{N}^{14} + \mathrm{H}^{1} \rightarrow\ \mathrm{O}^{15} + \gamma, \nonumber\\
&&\hspace{0cm}\mathrm{O}^{15} \rightarrow\ \mathrm{N}^{15} + \beta^{+} + \nu, \hspace{2cm}
\mathrm{N}^{15} + \mathrm{H}^{1} \rightarrow\ \mathrm{C}^{12} + \mathrm{He}^{4} + \gamma.\nonumber
\end{eqnarray}
In this process, four hydrogen nuclei fuse into one helium nucleus, while carbon and nitrogen act merely as catalysts.  
Later, \citet{Gamow1938} proposed another sequence of reactions that does not require heavy elements, called the proton–proton (p–p) chain:  
\begin{align}
\mathrm{H}^{1} + \mathrm{H}^{1} &\rightarrow\ \mathrm{H}^{2} + \beta^{+} + \nu, \nonumber\\
\mathrm{H}^{2} + \mathrm{H}^{1} &\rightarrow\ \mathrm{He}^{3} + \gamma, \nonumber\\
\mathrm{He}^{3} + \mathrm{He}^{3} &\rightarrow\ \mathrm{He}^{4} + 2\,\mathrm{H}^{1}\nonumber.
\end{align}
Both mechanisms effectively convert four protons into one helium nucleus, releasing energy corresponding to the mass difference of \(0.0294\,\mathrm{amu}\) between the reactants and the product.  
This implies that a fraction of about $7 \times 10^{-3}$ of the hydrogen mass is converted into energy. Assuming a hydrogen mass fraction of $X = 0.7$, a star would therefore lose approximately $5 \times 10^{-3}$ of its total mass, producing an energy output of about $4.5 \times 10^{18}\,\mathrm{erg\,g^{-1}}$, which is sufficient to sustain the sun’s luminosity \cite{KDB1992}.
Generally, the p–p chain dominates in low-mass stars with relatively cooler cores, whereas the C–N cycle operates predominantly in massive stars with higher central temperatures. 

Consequently, the star evolves into a self-gravitating thermodynamically unstable system, becoming progressively hotter even as it radiates energy. This instability eventually leads the star to its final stage of evolution, forming a compact remnant,  either a white dwarf or neutron star (quark star), which is supported by quantum mechanical forces due to the Pauli exclusion principle, or a massive black hole based on the size of the progenitor stars. 

In the following subsections, we shall discuss in detail the properties and physical characteristics of various compact stellar objects, namely white dwarfs, neutron stars, strange quark stars, and the final collapse entity, i.e., black holes.

\subsection{White dwarf}

In the stellar life cycle, white dwarfs constitute one of the most common and well-understood forms of compact objects. These stars no longer burn nuclear fuel. Instead, they are slowly cooling as they radiate away their residual thermal energy.
The term \textit{white dwarf} was first proposed by Luyten \cite{Luyten1922a,Luyten1922b,Luyten1922c}. However, the first white dwarf, \textit{40~Eridani~B}, was first observed as a faint stellar companion by William~Herschel in 1783 \cite{Herschel1783}, and its nature was later confirmed spectroscopically by Adams \cite{Adams1914}. 
The existence of another white dwarf, Sirius~B, was first predicted by Friedrich~Bessel in~1844 from perturbations in Sirius’ motion and was subsequently confirmed observationally by Alvan~Graham~Clark on 1862 \cite{Bessel1844}. 
Following the development of Fermi-Dirac statistics \cite{Dirac1926,Fowler1926}, scientists proposed that stellar remnants could maintain hydrostatic equilibrium through electron degeneracy pressure counterbalancing gravitational collapse. 
Building on this idea, S. Chandrasekhar \cite{Chandrasekhar1931a,Chandrasekhar1931b} developed a relativistic model of a degenerate electron gas and derived the maximum possible mass of such an object, known as \textit{the Chandrasekhar limit} ($1.44\,M_{\odot}$), beyond which the electron degeneracy pressure cannot withstand gravity.

Low mass stars with initial mass up to about $10\,M_{\odot}$ eventually evolve into white dwarfs as their terminal stage. 
When hydrogen fusion in the core is exhausted, the star ignites helium burning through the triple–alpha process, synthesizing carbon and oxygen, and causing a sharp rise in luminosity.
To balance this enhanced energy flux, the outer layers expand, forming a \textit{core-halo} structure characterized by a cool, red surface, known as the \textit{red giant} phase. 
Once the core temperature becomes insufficient ($\sim10^{9}$\,K) to ignite carbon fusion, the star expels its envelope as a planetary nebula, leaving behind the dense, degenerate carbon-oxygen remnant core known as \textit{white dwarf}.

A white dwarf, which no longer generates energy through nuclear fusion, resists gravitational collapse through electron degeneracy pressure. As the star contracts, its density becomes extremely high, pushing electrons into a highly compressed state. According to the Pauli exclusion principle, each quantum state can accommodate at most two electrons. In the dense interior of a white dwarf, all available energy levels become fully occupied, rendering the star degenerate. In this state, gravity can no longer compress the star further as there is no available phase space for additional electrons, allowing the white dwarf to remain stable. When a carbon-oxygen white dwarf approaches the Chandrasekhar mass limit ($M_{\rm Ch} \simeq 1.44\,M_{\odot}$), the electron degeneracy pressure is no longer sufficient to counterbalance gravity. The resulting collapse triggers a \textit{Type Ia supernova} through carbon detonation. White dwarfs typically have masses in the range $0.5-1\,M_{\odot}$ and radii slightly larger than earth, but their densities reach $\rho \sim 10^{6}\,\mathrm{g\,cm^{-3}}$, roughly $2\times 10^5$ times the average density of earth. Observationally, they exhibit luminosities $L \sim 10^{-2}\,L_{\odot}$ and surface temperatures $T_s \sim 10^{4}\,\mathrm{K}$.

\subsection{Neutron star}
Neutron stars, the extremely compact remnants formed during the final stages of stellar evolution, serve as exceptional natural laboratories for investigating various extreme astrophysical phenomena. These include studying compact X-ray sources and radio pulsars within the context of intense gravitational fields. Moreover, neutron stars provide a unique opportunity to explore the characteristics and behavior of ultra-dense matter within their interiors, offering deep insights into the physics of matter under conditions that cannot be replicated on earth.

\hspace{1cm}The concept of neutron stars was first proposed in 1934 by European astronomers Walter Baade and Fritz Zwicky \cite{ns1}, just two years after Chadwick’s \cite{ns2} groundbreaking discovery of the neutron. They suggested that neutron stars originate as remnants of supernova explosions. About five years later, Oppenheimer and Volkoff \cite{ns3} developed a more realistic theoretical model for these stars. Due to their extreme compactness and very small size, direct detection of neutron stars proved challenging at the time. However, significant breakthroughs came with Iosif Shklovsky’s X-ray observations in 1967 \cite{ns4} and the discovery of radio pulsars by Jocelyn Bell and Antony Hewish in 1967 \cite{jcbell}, which provided an effective method for detecting and studying neutron stars. This discovery later earned the Nobel Prize in Physics in 1974. The underlying cause of supernova explosions has remained a mystery for a long time, although it was understood that the primary driving mechanism is the release of gravitational binding energy associated with the formation of neutron stars. Subsequently, Pacini \cite{pacini} proposed that a rapidly rotating neutron star with a strong magnetic field can emit electromagnetic radiation that may trigger a supernova event. The landmark detection of the gravitational wave event GW170817, resulting from the collision of neutron stars by LIGO and Virgo in August 2017 \cite{abott}, has helped to further our understanding of neutron stars and their role in astrophysical phenomena.

~~~~In the previous section, we discussed that main-sequence stars with relatively low masses generally evolve into white dwarfs. However, for more massive stars $(\geq 10\,M_\odot)$, the exhaustion of nuclear fuel results in an intense gravitational collapse, leading to the formation of a core that is both significantly more compact and hotter than a typical white dwarf. This gravitational compression drives further nuclear fusion among stellar constituents such as H, He, and C, ultimately resulting in the synthesis of Fe and the development of a core predominantly consisting of $^{56}$Fe. When the mass of the core exceeds the Chandrasekhar limit ~\cite{Chandrasekhar1931a}, the electron degeneracy pressure can no longer counterbalance the inward gravitational force, and the star undergoes further collapse to form a neutron star.
Neutrino emission plays a crucial role during core formation in massive stars, as evidenced by the decay process: $^{56}\mathrm{Ni} \rightarrow ^{56}\mathrm{Fe} + 2e^+ + 2\nu_e.$ 
If a white dwarf continues to compress and its density exceeds about $10^7$g/cm$^3$, the electrons gain very high kinetic energy. Under such extreme conditions, electrons can combine with protons to produce neutrons and neutrinos through the reaction: $e^- + p \rightarrow n + \nu_e$. Being highly non-interactive, neutrinos escape the core and carry away energy, depleting the electrons that previously supported the star against gravitational collapse. This process is also accompanied by endothermic photodisintegration of iron:
$\gamma + ^{56}\mathrm{Fe} \rightarrow 13\ ^{4}\mathrm{He} + 4n$. 
The interplay of these mechanisms accelerates core collapse once the Chandrasekhar mass limit is surpassed, achieving core densities on the order of $10^{14}$\,g/cm$^3$, which are comparable to those found in atomic nuclei. At this stage, the core behaves like a massive neutron-rich nucleus, exhibiting elastic properties under continued compression from gravity. Consequently, a strong shock wave is generated that blows away much of the star’s outer material in a supernova explosion as a nebula leaving behind a neutron star at the nebula’s center.

~~~~The pioneering study by Oppenheimer and Volkoff in 1939~\cite{ns3} demonstrated that neutron stars composed of non-interacting neutrons could have radii of about 10 \,km and masses close to $1\,M_\odot$. Later, this mass limit, known as the Tolman-Oppenheimer-Volkoff (TOV) mass limit, was established to be approximately $2.1\,M_\odot$~\cite{bombaci}. The neutron degeneracy pressure can stably support stars up to about $0.7\,M_\odot$, while beyond this, repulsive nuclear forces become the key support against gravity. If a neutron star accumulates mass above the TOV limit, it will inevitably collapse into a black hole.

\subsubsection{Strange quark star}
Although neutrons and protons were initially considered fundamental particles, this perspective was challenged following independent proposals by GellMann and Zweig in 1964 \cite{gellmann, zweig}. They demonstrated that all hadrons are, in fact, composed of quarks, introducing the concept of three primary quark flavors, up (u), down (d), and strange (s). Subsequently, Soviet physicists Ivanenko and Kurdgelaidze introduced the idea of compact stellar objects made of quark matter in 1965 \cite{ivanenko}. Later in 1970, Itoh presented a model for quark stars composed of three flavors, specifically up, down, and strange quarks, and explored the conditions for hydrostatic equilibrium in these ultra-dense bodies known as \textit{strange quark stars} \cite{itoh}.

It is widely recognized that $^{56}$Fe, with a binding energy per nucleon of 931 MeV about 8 MeV less than the nucleon mass, is considered the most stable configuration for conventional hadronic matter. However, according to the strange matter hypothesis, originally suggested by Bodmer \cite{bodmer}, Witten \cite{witten}, and Terazawa \cite{terazawa}, the true ground state of the strong interaction might actually be strange quark matter (SQM), rather than $^{56}$Fe. This hypothesis proposes that because strange quarks have additional available quantum states, the combination of up, down, and strange quarks (three-flavor quark matter) leads to a lower energy configuration than just up and down quarks (two-flavor matter). A neutron star is typically considered the final stage in the evolution of massive stars. If the gravitational force exceeds the neutron degeneracy pressure, the star may collapse further, with neutrons transitioning into an extremely dense quark matter phase. There is also an alternate scenario in which a quark star may form through the accumulation of external quark matter in a region with a high concentration of strange quarks. Further discussions regarding the properties and formation of strange quark star will be provided later in this chapter.

\subsection{Black hole}
A black hole represents the final stage in the evolution of a compact star when no form of quantum mechanical degeneracy pressure can withstand gravitational collapse. In the framework of general relativity (GR), a sufficiently massive and compact stellar object can wrap spacetime so extremely that it forms a black hole, a region where gravitational attraction is so powerful that nothing, not even light, can escape its grasp. The interface distinguishing the black hole's interior from the external universe is termed the event horizon. The notion of such an object, capable of trapping all electromagnetic radiation, was independently introduced by Michell \cite{michell} and Laplace \cite{laplace} in the late 18th century using Newtonian principles. They posited that a massive object of mass $M$ and radius $R$ would have an escape velocity greater than the speed of light if $R < R_{\text{lim}} = \frac{2GM}{c^2}$, where $G$ is the gravitational constant and $c$ is the speed of light, which implies that no radiation could escape and the object would appear perfectly black.

~~~~Following Einstein's discovery of GR, Schwarzschild \cite{schwarzschild} discovered the first exact solution for a non-rotating (static) black hole. However, the solution's true physical interpretation was largely overlooked for decades, until Finkelstein (1958) \cite{finkelstein} highlighted the event horizon concept inherent in the Schwarzschild metric. The solution for a rotating black hole was derived by Kerr (1963) \cite{kerr}, while Newman et al. (1965) \cite{newman} extended it to include electric charge, yielding the Kerr–Newman geometry. Landmark contributions by Penrose and Hawking \cite{penrose65,hawking_penrose} established that within GR, gravitational collapse of compact stellar remnants inevitably leads to a singularity. Penrose further formulated the cosmic censorship conjecture \cite{penrose02}, postulating that such singularities are always concealed behind an event horizon, thereby designating the black hole as the conclusive outcome of gravitational collapse.
\begin{figure}[h!]
    \centering
    \includegraphics[height=6cm, width=9cm]{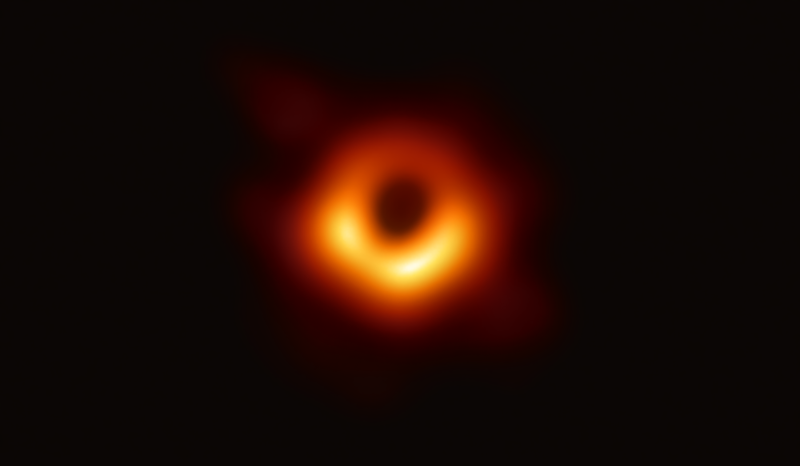}
    \caption{First detect image of the supermassive black hole located at the center of the M87 galaxy. (Image credit: EHT Collaboration, 2019).}
    \label{EHT}
\end{figure}

~~~~The X-ray source \textit{Cygnus X-1}, recognized as the first confirmed candidate for a stellar-mass black hole, was independently discovered by Bolton \cite{bolton72} and Webster \& Murdin \cite{webster72}. Subsequent astronomical studies provided further evidence for the existence of stellar-mass black holes in X-ray binary systems~\cite{remillard06} and supermassive black holes residing at the centers of galaxies~\cite{kormendy95}. The first-ever direct image of a black hole, captured by the Event Horizon Telescope Collaboration~\cite{eht19}, is shown in Fig.~\ref{EHT}. Although the imaged object corresponds to a supermassive black hole, it is important to note that black holes can exist across a broad range of masses, extending from stellar-mass to supermassive scales. According to GR, the mass of a stellar-mass black hole is expected to lie within the range of approximately $3$–$100\,M_{\odot}$, while observationally identified candidates in X-ray binary systems typically exhibit masses between $3$ and $20\,M_{\odot}$~\cite{casares14}. In the past two decades, advances in observational detection have greatly improved the study of astrophysical black holes. A major milestone was the first detection of gravitational waves by the LIGO experiment in 2015, corresponding to the binary black hole merger event GW150914~\cite{abott}, which revealed a $\sim60~M_{\odot}$ black hole formed from the merger of two $\sim30~M_{\odot}$ components.

\section{Beyond GR: Alternative gravity
theories}

In this section, we aim to construct the relativistic stellar structure model in GR. Furthermore, we present the motivation for moving beyond the conventional Einstein's framework of gravity embodied in GR. Drawing from both theoretical arguments and observational evidence, modified gravity theories are developed to address the limitations of GR, all while retaining its many well-established predictions, including phenomena like gravitational waves and black holes. Additionally, this section provides an introduction to alternative gravitational models that will be examined in the following sections of this thesis, such as $f(T)$ and $f(Q)$.

\subsection{Einstein  field equation}

In the framework of GR, Einstein’s field equations exhibit a profound level of complexity arising from their nonlinear structure and the intricate coupling between spacetime curvature and the distribution of matter and energy. Due to these nonlinearities, obtaining exact analytical solutions is feasible only under specific and idealized circumstances. One of the most notable solutions is the Schwarzschild metric, which describes the external gravitational field of a static, spherically symmetric mass. However, to explore the internal configuration of such a compact object, one must solve the Einstein field equations within the matter distribution. This problem is typically approached through numerical methods, leading to a system of coupled differential equations collectively known as the Oppenheimer–Volkoff equations, which govern the conditions for hydrostatic equilibrium in relativistic stellar structures \cite{ns3}.

~~~~The most general form of the spacetime interval in spherical coordinates, which applies to static and isotropic regions such as the exterior and interior of a non-rotating, isolated star, is given by \cite{Weinberg1972}
\begin{equation}
ds^2 = -A(r)\,dt^2 + B(r)\,dr^2 + W(r)\,r^2 \left( d\theta^2 + \sin^2\theta\, d\phi^2 \right) ,
\end{equation}
where, $A(r)$ and $B(r)$ are arbitrary functions of the radial coordinate $r$ that maintain spherical symmetry, and $W(r)$ is typically set to unity for simplicity. Thus, the metric reduces to \cite{Weinberg1972}
\begin{equation}\label{metric1st}
ds^2 = -e^{\nu(r)}\,dt^2 + e^{\lambda(r)}\,dr^2 + r^2 \left( d\theta^2 + \sin^2\theta\, d\phi^2 \right),
\end{equation}
where $\nu(r)$ and $\lambda(r)$ are arbitrary functions of the radial coordinate $r$, known as the \textit{metric potentials}. The exponential parametrization $g_{tt} = -e^{\nu(r)}$ and $g_{rr} = e^{\lambda(r)}$ is adopted because (i) it automatically preserves the Lorentzian signature of the metric, since $e^{\nu(r)}, e^{\lambda(r)} > 0$ for any real-valued $\nu, \lambda$, eliminating the need for explicit positivity constraints on the metric coefficients; (ii) it considerably simplifies the structure of the resulting field equations and curvature invariants, as derivatives naturally appear in terms of $\nu$ and $\lambda$ rather than logarithms of $A$ and $B$; and (iii) it provides a direct correspondence with the Schwarzschild exterior solution, for which $e^{\nu(r)} = e^{-\lambda(r)} = 1 - 2M/r$, facilitating the matching of interior and exterior geometries at the stellar surface. The line element $ds^2 = g_{\mu\nu}dx^\mu dx^\nu$ defines the metric tensor $g_{\mu\nu}$, which can be written as
\begin{equation}
g_{\mu\nu} =
\begin{pmatrix}
 -e^{\nu(r)} & 0 & 0 & 0 \\
  0 & e^{\lambda(r)} & 0 & 0 \\
  0 & 0 & r^2 & 0 \\
  0 & 0 & 0 & r^2 \sin^2\theta
\end{pmatrix} .
\end{equation}
This metric adopts the signature $(-+++)$, which is commonly used in the modern literature on GR.
The spacetime metric \( g_{\mu\nu} \) sets the geometric and causal structure of the manifold, while the motion of matter and light reveals how this structure is curved by mass and energy.

~~~~In the context of GR, the motion of free massive test particles provides direct insight into the geometric nature of spacetime. In analogy with special relativity, one postulates that in any spacetime $(\mathcal{M}, g_{\mu\nu})$, where $\mathcal{M}$ denotes the manifold, free particles follow timelike trajectories $x^{\mu}(\lambda)$, parametrized by $\lambda$, that extremize the proper time. This requirement naturally leads to the proper time functional
\begin{equation}
\tau = \int d\lambda \left| g_{\mu\nu} \frac{dx^{\mu}}{d\lambda} \frac{dx^{\nu}}{d\lambda} \right|^{1/2}.
\end{equation}

Varying this action with respect to the coordinates \( x^{\mu} \) yields the \textit{geodesic equation} \cite{Weinberg1972}
\begin{equation}
\frac{d^{2}x^{\alpha}}{d\lambda^{2}} 
+ \frac{1}{2} g^{\alpha\sigma} 
\left( \partial_{\mu} g_{\sigma\nu} 
+ \partial_{\nu} g_{\sigma\mu} 
- \partial_{\sigma} g_{\mu\nu} \right)
\frac{dx^{\mu}}{d\lambda} 
\frac{dx^{\nu}}{d\lambda} = 0,\label{christ}
\end{equation}
which governs the trajectories of free particles solely under the influence of spacetime geometry. This relation is fundamental: within Einstein’s theory, gravitation does not act as a conventional force but instead arises from the \textit{curvature} of spacetime. Observational phenomena such as the bending of starlight near massive objects and the gravitational redshift provide empirical support for this geometric viewpoint.
To describe the geometry of spacetime in GR, several tensorial quantities are defined, each of which has distinct mathematical and physical significance. \\ 
 \textbf{Levi-Civita Connection:}
    In GR, for the above geodesic equation, the Christoffel symbol is denoted as $\Gamma^{\alpha}{}_{\mu\nu}$ or
$\Big\{\!\begin{matrix}
\alpha \\[-4pt]
\mu\nu
\end{matrix}\!\Big\}$
and defined by \cite{Narlikar2002}    \begin{eqnarray}\label{christ}
    \Gamma^{\alpha}{}_{\mu\nu}=\Big\{\!\begin{matrix}
\alpha \\[-4pt]
\mu\nu
\end{matrix}\!\Big\} = \frac{1}{2} g^{\alpha\sigma} \left( \partial_{\mu} g_{\sigma\nu} + \partial_{\nu} g_{\sigma\mu} - \partial_{\sigma} g_{\mu\nu} \right).
    \end{eqnarray}
 It encodes how coordinate basis vectors change throughout curved spacetime, provides the coefficients for the covariant derivative, and prescribes how vectors are parallel transported across spacetime. Although not tensors, they are uniquely determined by the requirements of metric compatibility ($\nabla_{\alpha} g_{\mu\nu} = 0$) and zero torsion and play a central role in geodesic motion. For the static spherically symmetric metric (\ref{metric1st}), the non-vanishing Christoffel symbols are obtained as
\begin{eqnarray}
\Gamma^{t}{}_{t r} = \Gamma^{t}{}_{r t} = \frac{\nu'(r)}{2},\quad 
\Gamma^{r}{}_{t t} = \frac{\nu'(r)}{2}\,e^{\nu(r)-\lambda(r)},\quad \Gamma^{r}{}_{r r} = \frac{\lambda'(r)}{2}, \quad \Gamma^{r}{}_{\theta\theta} = -r\,e^{-\lambda(r)} \nonumber\\ 
\Gamma^{r}{}_{\phi\phi} = -r\,e^{-\lambda(r)}\sin^{2}\theta, \quad
\Gamma^{\theta}{}_{r\theta} = \Gamma^{\phi}{}_{r\phi} = \frac{1}{r}, \quad \Gamma^{\theta}{}_{\phi\phi} = -\sin\theta\cos\theta, \quad 
\Gamma^{\phi}{}_{\theta\phi} = \cot\theta.
\end{eqnarray}
These symbols are symmetric in their two lower indices:
$ \Gamma^{\alpha}_{~\mu \nu} = \Gamma^{\alpha}_{~\nu \mu}$. This symmetry reflects the torsion-free property of Levi-Civita connections. That essentially means that the connection does not twist as one moves along a manifold.
These Christoffel symbols serve as the essential connection between the geometry of spacetime and the geodesic equation, laying the groundwork for computing the Riemann curvature tensor in the subsequent analysis.

 \textbf{Riemannian curvature tensor:} Using the Levi-Civita connection, one can construct the most general measure of curvature, which is the \textit{Riemannian curvature tensor}. It is defined as
    \begin{eqnarray}
    R^{\alpha}_{\,\, \sigma\mu\nu} = \partial_{\mu} \Gamma^{\alpha}_{\ \nu\sigma} - \partial_{\nu} \Gamma^{\alpha}_{\ \mu\sigma} + \Gamma^{\alpha}_{\ \mu\lambda} \Gamma^{\lambda}_{\ \nu\sigma} - \Gamma^{\alpha}_{\ \nu\lambda} \Gamma^{\lambda}_{\ \mu\sigma}.
    \end{eqnarray}
    In particular, if one takes a closed loop in the manifold and transports a vector
at a point in the tangent space around the loop, the Riemann tensor measures the deviation of the vector in the tangent space from its original position. It encodes the intrinsic curvature of spacetime. The (Riemann) curvature tensor,
$R_{\alpha\sigma\mu\nu}$,
possesses the following symmetry properties
\begin{equation}
R_{\alpha\sigma\mu\nu}
= - R_{\sigma\alpha\mu\nu}
= - R_{\alpha\sigma\nu\mu},
\qquad
R_{\alpha\sigma\mu\nu}
= R_{\mu\nu\alpha\sigma}.
\end{equation}

Furthermore, the Riemann tensor satisfies the first Bianchi identity
\begin{equation}
R_{\alpha\sigma\mu\nu}
+ R_{\alpha\mu\nu\sigma}
+ R_{\alpha\nu\sigma\mu}
= 0.
\end{equation}

These symmetries encode fundamental geometric constraints of curvature on a pseudo-Riemannian manifold. \\
    \textbf{Ricci tensor:} By contracting the Riemann tensor on one upper and one lower index, we obtain the \textit{Ricci tensor}
    \begin{eqnarray}
    R_{\mu\nu} = R^{\lambda}_{\ \mu\lambda\nu}~.
    \end{eqnarray}
    The Ricci tensor succinctly measures how geodesic bundles, collections of nearby trajectories, and converge or diverge due to spacetime curvature. It quantifies local changes in spacetime volume caused by gravitational tidal effects, directly linking matter and energy content to geometric deformations. The nonzero components of the Ricci tensor $R_{\mu\nu}$ for the given metric are
\begin{eqnarray*}
   &&\hspace{0cm} R_{tt} = e^{\nu-\lambda}\Bigg[\frac{\nu''}{2}+\frac{{\nu'}^{2}}{4}
-\frac{\nu'\lambda'}{4}+\frac{\nu'}{r}\Bigg], \qquad
R_{rr} = -\Bigg[\frac{\nu''}{2}+\frac{{\nu'}^{2}}{4}
-\frac{\nu'\lambda'}{4}-\frac{\lambda'}{r}\Bigg],\\
 &&\hspace{0cm} R_{\theta\theta} = 1-e^{-\lambda}+\frac{r}{2}e^{-\lambda}\big(\lambda'-\nu'\big), \qquad
R_{\phi\phi} = \sin^{2}\theta\,R_{\theta\theta}.
\end{eqnarray*}
Where, $''$ denotes the second order derivative w.r.t the radial coordinate $r$.\\
\textbf{Ricci scalar:} A further contraction of the Ricci tensor with the metric yields the \textit{Ricci scalar}
    \begin{eqnarray}
    &&\hspace{0cm}R = g^{\mu\nu} R_{\mu\nu}\,=\; e^{-\lambda}\Bigg[-\nu''-\frac{{\nu'}^{2}}{2}+\frac{\nu'\lambda'}{2}
- \frac{2(\nu'-\lambda')}{r}\Bigg] \;+\; \frac{2\big(1-e^{-\lambda}\big)}{r^{2}}.
    \end{eqnarray}
   The Ricci scalar encapsulates the overall curvature at a point into a single invariant quantity. The Ricci scalar has the physical interpretation of measuring the deviation of the volume of a unit geodesic ball from the standard unit ball. It plays an essential role in the Einstein-Hilbert action and indicates overall geometric behavior, such as expansion or contraction of nearby geodesics.
    
    \textbf{Einstein tensor:} Combining the Ricci tensor and Ricci scalar, The \textit{Einstein tensor} is defined as
    \begin{eqnarray}
    G_{\mu\nu} = R_{\mu\nu} - \frac{1}{2} g_{\mu\nu} R.
    \end{eqnarray}
   It adheres to the symmetry property, i.e., $G_{\mu\nu}=G_{\nu\mu}$. This symmetry follows from the symmetry of the Ricci tensor $R_{\mu\nu}$ and the metric tensor $g_{\mu\nu}$. The non-vanishing components of the Einstein tensor for the given metric (\ref{metric1st}) are
\begin{align}\label{en1}
G^{t}{}_{t} &= e^{-\lambda}\Big(\frac{\lambda'}{r}-\frac{1}{r^{2}}\Big)+\frac{1}{r^{2}},
\\[6pt] \label{en2}
G^{r}{}_{r} &= e^{-\lambda}\Big(\frac{\nu'}{r}+\frac{1}{r^{2}}\Big)-\frac{1}{r^{2}},
\\[6pt] \label{en3}
G^{\theta}{}_{\theta} &= G^{\phi}{}_{\phi}
= e^{-\lambda}\Bigg[\frac{\nu''}{2}+\frac{{\nu'}^{2}}{4}-\frac{\nu'\lambda'}{4}
+\frac{\nu'-\lambda'}{2r}\Bigg].
\end{align}
 This makes it especially suitable for equating with the energy momentum tensor in the field equations. Following the discussion on the geometric structure of spacetime, now we will move towards the distribution of matter within the stellar interior.\\
 Another important curvature invariant frequently used in general relativity is the \textit{Kretschmann scalar}, defined as
\begin{equation*}\label{kretschmann}
    \mathcal{K} \equiv R_{\mu\nu\gamma\sigma} R^{\mu\nu\gamma\sigma},
\end{equation*}
Being a scalar quantity constructed from the full Riemann tensor, $\mathcal{K}$ is invariant under arbitrary coordinate transformations, which makes it a particularly useful diagnostic tool for distinguishing between \emph{coordinate singularities} and \emph{true (physical) singularities} of a given spacetime. A coordinate singularity arises merely from a poor choice of coordinates and can be removed by an appropriate coordinate transformation, whereas a true singularity corresponds to a genuine breakdown of the spacetime geometry, where curvature invariants such as $\mathcal{K}$ diverge.

    \textbf{Stress-energy momentum tensor:}  
\justifying
The matter content plays a crucial role in shaping the curvature of spacetime through its contribution to the energy–momentum tensor $\mathcal{T}_{\mu\nu}$. It plays a central role in GR, as it represents the distribution and flow of energy and momentum throughout spacetime, acting as the source of the gravitational field in Einstein's field equations. In GR, a perfect fluid is characterized by isotropic pressure, i.e., the pressure is the same in all directions. However, many realistic astrophysical objects exhibit anisotropic pressure, where the radial pressure differs from the tangential pressure. Such matter distributions are called anisotropic fluids. For an anisotropic matter distribution, $\mathcal{T}_{\mu\nu}$ is generally expressed as
\begin{equation}
   \mathcal{T}_{\mu\nu} = (\rho + p_t)u_{\mu}u_{\nu} + p_t g_{\mu\nu} + (p_r - p_t)v_{\mu}v_{\nu},
    \label{eq:stress_tensor}
\end{equation}
where $\rho$, $p_r$, and $p_t$ denote the energy density, radial pressure, and tangential pressure of the fluid distribution, respectively. The four-velocity of the fluid $u_{\mu}$ satisfies the normalization condition $u_{\mu}u^{\mu} = -1$, while $v_{\mu}$ is a unit spacelike four-vector in the radial direction satisfying $v_{\mu}v^{\mu} = 1$ and $u_{\mu}v^{\mu} = 0$.
Physically, $\mathcal{T}_{\mu\nu}$ encapsulates all information regarding the energy content, momentum flux, and internal stresses of the system. In the case of isotropic matter, where $p_r = p_t = p$, Eq.~\eqref{eq:stress_tensor} reduces to the familiar perfect fluid form \cite{Weinberg1972}
\begin{equation}
    \mathcal{T}_{\mu\nu} = (\rho + p)u_{\mu}u_{\nu} + p g_{\mu\nu}.
\end{equation}
However, for anisotropic configurations, the difference between the radial and tangential pressures ($p_r - p_t$) introduces additional degrees of freedom that enable a more realistic description of compact astrophysical objects such as neutron stars, gravastars, and strange quark stars. Such anisotropy may arise due to the presence of superfluid phases, strong magnetic fields, phase transitions, viscosity, or other microphysical effects within dense matter.

~~~~Mathematically, the stress-energy momentum tensor is symmetric ($\mathcal{T}_{\mu\nu} = \mathcal{T}_{\nu\mu}$) and satisfies the covariant conservation law $\nabla^{\mu} \mathcal{T}_{\mu\nu} = 0$, which ensures local conservation of energy and momentum. Its components directly influence the curvature of spacetime, thereby governing the dynamical structure and stability of self-gravitating systems. However, the non-vanishing components of the stress-energy momentum tensor for an anisotropic fluid are \cite{Weinberg1972}
\begin{eqnarray}\label{emt}
\mathcal{T}^{\mu}_{~\nu} =
\begin{pmatrix}
-\rho\, & 0 & 0 & 0 \\[6pt]
0 & p_r\,  & 0 & 0 \\[6pt]
0 & 0 & p_t\,  & 0 \\[6pt]
0 & 0 & 0 & p_t\, 
\end{pmatrix}.
\end{eqnarray}
 \textbf{Einstein's field equations:}  
\justifying
Finally, Einstein field equations encapsulate the fundamental relation between spacetime geometry and its content of matter-energy, and are given by \cite{Weinberg1972}
\begin{equation}\label{master}
    G_{\mu\nu} = 8\pi \, \mathcal{T}_{\mu\nu}.
\end{equation}
In the above equation, we have adopted geometrized units, in which the gravitational constant and the speed of light are set to unity, i.e., $G = c = 1$. This equation succinctly expresses the core principle of GR: \emph{matter tells spacetime how to curve, and curved spacetime tells matter how to move,} The proportionality constant $8\pi $ ensures consistency with Newtonian gravity in the weak-field limit, while the tensorial nature of the equation preserves general covariance, making the theory valid in all coordinate systems.

~~~~Using the non vanishing components of $G_{\mu\nu}$ from Eqs.~(\ref{en1}-\ref{en3}) and anisotropic fluid components $\mathcal{T}_{\mu\nu}$ from Eq.~(\ref{emt}) we can obtain the three independent field equations by using the master Eq.~(\ref{master}) as
\begin{align}
& e^{-\lambda}\!\left(\frac{\lambda'}{r}-\frac{1}{r^{2}}\right)+\frac{1}{r^{2}} \;=\; 8\pi\,\rho, 
\label{eq:fe_tt}\\[6pt]
& e^{-\lambda}\!\left(\frac{\nu'}{r}+\frac{1}{r^{2}}\right)-\frac{1}{r^{2}} \;=\; 8\pi\,p_{r},
\label{eq:fe_rr}\\[6pt]
& e^{-\lambda}\!\Bigg[\frac{\nu''}{2}+\frac{{\nu'}^{2}}{4}-\frac{\nu'\lambda'}{4}
+\frac{\nu'-\lambda'}{2r}\Bigg] \;=\; 8\pi\,p_{t}.
\label{eq:fe_thth}
\end{align}
These are known as \textit{Einstein's field equations}, which serve as the dynamical laws governing the coupling between the matter--energy content of the universe and the geometry of spacetime. The breadth of phenomena described by these equations can be systematically classified according to the relevant physical regime: in the \emph{weak-field, low-velocity limit}, they reduce to Newton's law of gravitation and account for planetary motion, the perihelion precession of Mercury, and the gravitational bending of light~\cite{will2014}; in the \emph{strong-field, stationary regime}, they admit exact solutions describing compact objects such as neutron stars, white dwarfs, and black holes (Schwarzschild, Reissner--Nordstr\"{o}m, Kerr, and Kerr--Newman spacetimes)~\cite{misner1973}.
\subsection{Relativistic stellar structure in GR}
Einstein's field equations~(\ref{eq:fe_tt})--(\ref{eq:fe_thth}) provide the three independent relations between the metric functions $\nu(r),\lambda(r)$ and the matter variables $\rho(r),p_r(r),p_t(r)$. It is convenient to extract the gravitational mass contained within radius $r$ by defining the Misner–Sharp mass function $m(r)$ through the metric component $e^{-\lambda(r)}$ \cite{Narlikar2002} as
\begin{equation}
e^{-\lambda(r)} = 1-\frac{2 m(r)}{r},
\qquad\Longrightarrow\qquad
m'(r)=4\pi r^{2}\rho(r),
\label{eq:mass_def_repeat}
\end{equation}
which follows directly from Eq.~(\ref{eq:fe_tt}). To relate the metric derivative $\nu'(r)$ to the matter content we can use Eq.~(\ref{eq:fe_rr}) to obtain
\begin{equation}
\frac{\nu'}{2} \;=\; \frac{\big[m(r)+4\pi r^{3}p_r(r)\big]}{r\big[r-2 m(r)\big]},
\label{eq:nuprime_repeat}
\end{equation}
thereby making explicit the contribution of pressure to the effective gravitational mass.
The remaining dynamical relation arises from local conservation of energy–momentum
\begin{equation}
\nabla_{\mu}\mathcal{T}^{\mu}{}_{\nu}=0,
\end{equation}
whose only nontrivial component for a static, spherically symmetric configuration is the radial one $\nabla_{\mu}\mathcal{T}^{\mu}{}_{\nu}=0$. Substituting the diagonal anisotropic form $\mathcal{T}^{\mu}{}_{\nu}=\mathrm{diag}(-\rho,p_r,p_t,p_t)$ and using Eq.~\eqref{eq:nuprime_repeat} yields the hydrostatic equilibrium condition
\begin{equation}
\frac{\mathrm{d}p_r}{\mathrm{d}r} + \frac{\nu'}{2}(\rho+p_r)-\frac{2}{r}(p_t-p_r)=0.
\label{eq:tov_intermediate_repeat}
\end{equation}
Finally, substituting \eqref{eq:nuprime_repeat} into \eqref{eq:tov_intermediate_repeat} gives the generalized TOV equation as
\begin{equation}
\frac{\mathrm{d}p_r}{\mathrm{d}r} \;=\; -\,\frac{(\rho + p_r)\,\big[m(r) + 4\pi r^{3} p_r\big]}{r\big[r-2 m(r)\big]} \;+\; \frac{2}{r}\,(p_t-p_r).
\label{eq:anisotropic_tov_repeat}
\end{equation}
Thus, starting from the metric ansatz (Eq.~\ref{metric1st}) and the Einstein field equations (\ref{eq:fe_tt})--(\ref{eq:fe_thth}), Eqs.~(\ref{eq:mass_def_repeat}) and (\ref{eq:anisotropic_tov_repeat}) form a closed system describing the interior of an anisotropic, static, spherically symmetric object. In the isotropic limit, Eq.~(\ref{eq:anisotropic_tov_repeat}) reduces to the standard TOV equations governing stellar hydrostatic equilibrium of relativistic stars given by
\begin{eqnarray}
&&\hspace{0cm}\frac{dm(r)}{dr} = 4 \pi r^2 \rho(r), \\
&&\hspace{0cm}\frac{dp(r)}{dr} = - \frac{[\rho(r) + p(r)] [m(r) + 4 \pi r^3 p(r)]}{r \,[r - 2 m(r)]}.
\end{eqnarray}
These equations, together with an appropriate equation of state (EoS) $p = p(\rho)$ and suitable boundary conditions, form the foundation for modeling the structure and stability of compact stars in GR. In practice, the gravitational mass $M$ and the radius $R$ of a star are obtained by numerically integrating the TOV equations outward from the center ($r = 0$) until the pressure vanishes at the stellar surface, i.e., $p(r = R) = 0$. The radius $R$ at which this condition is satisfied defines the stellar boundary, and the enclosed mass at that point, $M = m(R)$, corresponds to the gravitational mass of the star.

\subsection{Exterior spacetime: Schwarzschild solution}

The preceding discussion concerns the \textit{interior} of a static, spherically symmetric stellar configuration, where the matter distribution is nonzero, i.e. $\mathcal{T}_{\mu\nu} \neq 0$. To obtain a complete description of the stellar system, one must also determine the metric in the \textit{exterior region}, where no matter is present and hence the spacetime is vacuum, satisfying $\mathcal{T}_{\mu\nu}=0$. The metric in the exterior region must continuously match the interior solution at the stellar boundary $r=R$.
In the vacuum region, Einstein's field equation (\ref{eq:fe_tt})--(\ref{eq:fe_thth}) reduces to
\begin{eqnarray}
    R_{\mu\nu} - \frac{1}{2} R\, g_{\mu\nu} = 0.
\end{eqnarray}
Taking the trace of the above equation with $g^{\mu\nu}$ gives
\begin{eqnarray}
    R_{\mu\nu}g^{\mu\nu} - \frac{1}{2}R\, g_{\mu\nu} g^{\mu\nu} = 0
    \;\;\Rightarrow\;\; R = 0.
\end{eqnarray}
Thus, in a vacuum, the Ricci scalar curvature vanishes, and the field equations simplify to $R_{\mu\nu}=0$.
For the above metric, the $tt$-component of the Einstein equations yields
\begin{equation}
\frac{1}{r^2}\left(1 - e^{-\lambda}\right)
- \frac{e^{-\lambda}\lambda'}{r} = 0 \implies \frac{d}{dr}\left(r e^{-\lambda}\right) = 1.
\end{equation}

After integration of the above equation, we obtain the following result.
\begin{equation}
e^{-\lambda(r)} = 1 + \frac{C}{r},
\end{equation}
where $C$ is an integration constant. Identifying this constant with the gravitational mass of the object as $C = -2M$, we obtain the following
\begin{equation}
e^{-\lambda(r)} = 1 - \frac{2M}{r}.
\end{equation}

Next, the $rr$-component of the Einstein equations gives
\begin{equation}
\frac{1}{r^2}\left(1 - e^{-\lambda}\right)
+ \frac{e^{-\lambda}\nu'}{r} = 0.
\end{equation}
Substituting the expression for $e^{-\lambda}$ and solving for $\nu'(r)$, we find
\begin{equation}
\nu'(r) = \frac{2M}{r(r - 2M)}.
\end{equation}
Integrating the above equation leads to the following
\begin{equation}
\nu(r) = \ln\left(1 - \frac{2M}{r}\right) + C_1,
\end{equation}
where $C_1$ is another integration constant. The requirement of asymptotic flatness, namely $g_{tt} \to -1$, as $r\to \infty$ fixes $C_1 = 0$. 
Therefore, the metric function $\nu(r)$ becomes
\begin{equation}
e^{\nu(r)} = 1 - \frac{2M}{r}.
\end{equation}

Finally, substituting the metric functions $e^{\nu(r)}$ and $e^{\lambda(r)}$ into the line element, we obtain the Schwarzschild exterior solution
\begin{equation}\label{outer}
ds^2 = -\left(1 - \frac{2M}{r}\right)dt^2
+ \left(1 - \frac{2M}{r}\right)^{-1}dr^2
+ r^2\left(d\theta^2 + \sin^2\theta\, d\phi^2\right),
\end{equation}

where $r_s = 2M$ is known as the \textit{Schwarzschild radius}. This is the well-known \textit{Schwarzschild metric}, an exact solution to Einstein's field equations in vacuum. It represents the gravitational field in the exterior region of any static, spherically symmetric mass distribution, assuming zero charge, zero angular momentum, and a vanishing cosmological constant. Although the Schwarzschild metric was originally derived in the context of a non-rotating black hole, it is, in fact, applicable to the exterior spacetime of \emph{any} spherically symmetric and static compact object, such as a neutron star, white dwarf, or strange quark star, provided the symmetry conditions are satisfied. This universality makes it a cornerstone of both black hole physics and the study of compact stellar objects in GR. The solution was first obtained by Karl Schwarzschild in 1916 \cite{schwarzschild}. Thus, the Schwarzschild solution not only describes the spacetime geometry around a static spherical mass, but it also reveals profound physical features inherent to black holes. A closer examination of the metric components uncovers two key characteristics of this spacetime:
\begin{itemize}
    \item \textbf{Central singularity:} From Eq.~(\ref{outer}), it can be seen that at the center ($r = 0$), the curvature invariants (such as the Kretschmann scalar) diverge. This divergence signifies the presence of a \textit{true physical singularity}, where the energy density becomes infinite and the classical description of spacetime ceases to be valid.
    
    \item \textbf{Event horizon:} At $r = r_s = 2M$, the metric coefficient $g_{tt}$ vanishes while $g_{rr}$ diverges, marking the location of the \textit{event horizon}. This surface acts as a one-way causal boundary, beyond which no signal, matter, or information can escape the gravitational influence of the black hole.
\end{itemize}
Hence, in Einstein’s GR, the Schwarzschild metric provides a unique and exact description of the spacetime geometry exterior to a spherically symmetric mass. The interior and exterior solutions together ensure a complete and continuous relativistic model of a compact star.

\subsection{Limitation of general relativity} Although Einstein’s GR has been remarkably successful in describing gravitational phenomena on a wide range of scales, several theoretical and observational challenges indicate that GR may not provide a complete description of gravity in all regimes. These issues have motivated the exploration of extended theories of gravity, which aim to generalize Einstein’s framework while remaining consistent with known experimental results. The principal motivations are as follows: \begin{enumerate} \item \textbf{Cosmological constant and dark energy problems \cite{prob1}:} The observed late-time accelerated expansion of the universe cannot be fully explained within GR without introducing a cosmological constant or exotic dark energy, both of which lead to severe theoretical and fine-tuning issues known as the old and new cosmological constant problems. \item \textbf{Dark matter and large scale structure \cite{prob2}:} GR has trouble accounting for the observed dynamics of galaxies and galaxy clusters, which suggests the presence of dark matter. Although dark matter can be introduced as an unknown form of matter, modified gravity explores whether changes in the law of gravity at galactic and cosmological scales could explain these phenomena without invoking new particle species. \item \textbf{Singularities and theoretical incompleteness \cite{hawking_penrose}:} GR predicts singularities (such as those inside black holes and the Big Bang), where physical quantities become infinite and the classical laws of physics break down. The presence of singularities signals a breakdown of the equivalence principle and indicates that GR cannot describe extremely strong gravity regimes. \item \textbf{Testing gravity beyond the solar system \cite{will2014}:} Most experimental tests of GR probe gravity at terrestrial or solar system scales, but gravity may behave differently on much larger (cosmological) or much smaller (quantum, strong-field) scales. Modified gravity theories are motivated by the desire to extend gravity models to unexplored regimes and to test the foundations of GR with new astrophysical and cosmological observations, such as gravitational waves and compact stars. \item \textbf{Model-independent constraints and geometrical generalization \cite{prob5}:} GR is built upon the Riemannian geometry of spacetime, assuming metric compatibility and vanishing torsion and non-metricity. Modified gravity theories incorporate additional degrees of freedom such as scalar, vector, or tensor fields, and sometimes extra dimensions which allow them to circumvent certain no-go theorems and generate distinctive observational signatures. These extensions enrich the spectrum of possible gravitational phenomena and open new avenues for connecting theoretical predictions with empirical data. \end{enumerate} In view of the limitations of GR and the motivations for extending the gravitational framework, it becomes natural to explore how spacetime geometry itself can be generalized. The following section delves into this generalization by examining the fundamental geometrical attributes of spacetime, like curvature, torsion, and non-metricity that serve as the foundation for various modified theories of gravity.

\subsection{Geometrical trinity of gravity: Curvature, Torsion and Non-metricity}
 In Einstein’s formulation, gravitation is entirely described through the curvature of spacetime, encoded by the Levi-Civita connection, which assumes both vanishing torsion and metric compatibility. However, relaxing these geometric constraints gives rise to alternative formulations of gravity, where additional geometrical quantities namely torsion and non-metricity play the fundamental roles. To understand these extensions, it is essential to first examine the three core geometrical entities that can characterize spacetime: curvature, torsion, and non-metricity, respectively.
\subsubsection{Affine connection} An affine connection $\Gamma^{\alpha}_{~\mu \nu}$ in a manifold $\mathcal{M}$ defines how vectors are transported parallelly and their geodesic paths. It can exist independently of a metric and is essential for describing locally inertial observers and spacetime structure, especially in curved geometries beyond flat Minkowski space. 
To formally introduce the concept of an affine connection, consider the parallel transport of a vector field \( V^\mu \) from a point \( x^\mu \) to a neighboring point \( x^\mu + dx^\mu \). The infinitesimal change in the vector under this displacement is governed by the connection coefficients \( \Gamma^{\alpha}_{\ \mu\nu} \), such that
\begin{equation}
dV^{\alpha} = - \Gamma^{\alpha}_{\ \mu\nu}\, V^{\mu}\, dx^{\nu}.
\end{equation}
Here, \( \Gamma^{\alpha}_{\ \mu\nu} \) denotes the components of the affine connection defined on the manifold.
This connection naturally gives rise to the notion of a covariant derivative, which generalizes the standard partial derivative to curved manifolds. For a vector field \( V^\mu \), the covariant derivative is defined as
\begin{eqnarray}
\nabla_\alpha V^\mu = \partial_\alpha V^\mu + \Gamma^\mu_{~\alpha\nu} V^\nu,
\end{eqnarray}
with this definition extending simply to tensors of any rank. A distinctive feature of metric-affine geometry is that, in the presence of a metric, the general affine connection can be uniquely decomposed into three separate contributions. 
The components of the affine connection, \({\Gamma}^\alpha_{~\mu\nu}\), can therefore be represented as \cite{heisenberg2023}
\begin{eqnarray}\label{affine}
\Gamma^\alpha_{\mu\nu} = \Big\{\begin{matrix}
\alpha \\[-4pt]
\mu\nu
\end{matrix}\Big\}\implies R^\alpha_{~\mu\nu\delta } + K^\alpha_{~\mu\nu} + L^\alpha_{~\mu\nu}.
\end{eqnarray}
Building of the decomposition of a general affine connection into the Levi-Civita connection $\Big\{\begin{matrix}
\alpha \\[-4pt]
\beta\gamma
\end{matrix}\Big\}$ resulting in the Riemann
curvature tensor $R^{\alpha}_{\mu \nu \delta}$, the contortion tensor \( K^\alpha_{~\mu\nu} \), and the disformation tensor \( L^\alpha_{~\mu\nu} \), we can classify the resulting geometries by selectively setting these components to zero. This yields the following important scenarios:
\begin{itemize}
    \item \( R^\alpha_{~\mu\nu\delta} = 0,\; K^\alpha_{~~\beta\gamma} = 0,\; L^\alpha_{~~\beta\gamma} = 0 \): In the following case, all geometric contributions vanish, and the resulting space is Euclidean or Minkowski, depending on the metric signature.
    \item \( R^\alpha_{~\mu\nu\delta} \neq 0,\; K^\alpha_{~~\beta\gamma} = 0,\; L^\alpha_{~~\beta\gamma} = 0 \): Here, only curvature is present. This case corresponds to Riemannian geometry, which serves as the mathematical foundation of GR.
    \item \( R^\alpha_{~\mu\nu\delta} = 0,\; K^\alpha_{~~\beta\gamma} \neq 0,\; L^\alpha_{~~\beta\gamma} = 0 \): In the next case, the contortion \( K^\alpha_{\beta\gamma} \) is the only non-zero term, forming the geometric basis for TEGR (the Teleparallel Equivalent of General Relativity) and its generalizations.
    \item \( R^\alpha_{~\mu\nu\delta} = 0,\; K^\alpha_{~~\beta\gamma} = 0,\; L^\alpha_{~~\beta\gamma} \neq 0 \):
    In this geometry, only the disformation tensor \( L^\alpha_{~\mu\nu} \) is non-vanishing, providing the setting for STEGR (the symmetric teleparallel equivalent of GR) and related extensions.
\end{itemize}
This classification reveals how different choices for these geometric objects underpin different physical and mathematical frameworks. This formalism is referred to as \textit{geometrical trinity of gravity,}
but although the connection itself is not a tensor, it generates fundamental tensorial quantities: curvature, contortion, and non-metricity. With the metric that defines distances and angles and the connection that governs parallel transport, we are now equipped to introduce key geometrical tensors that capture essential information about the structure of spacetime.
\subsubsection{Curvature: The geometrical significance}
\begin{figure}[H]
    \centering   \includegraphics[width=0.5\linewidth]{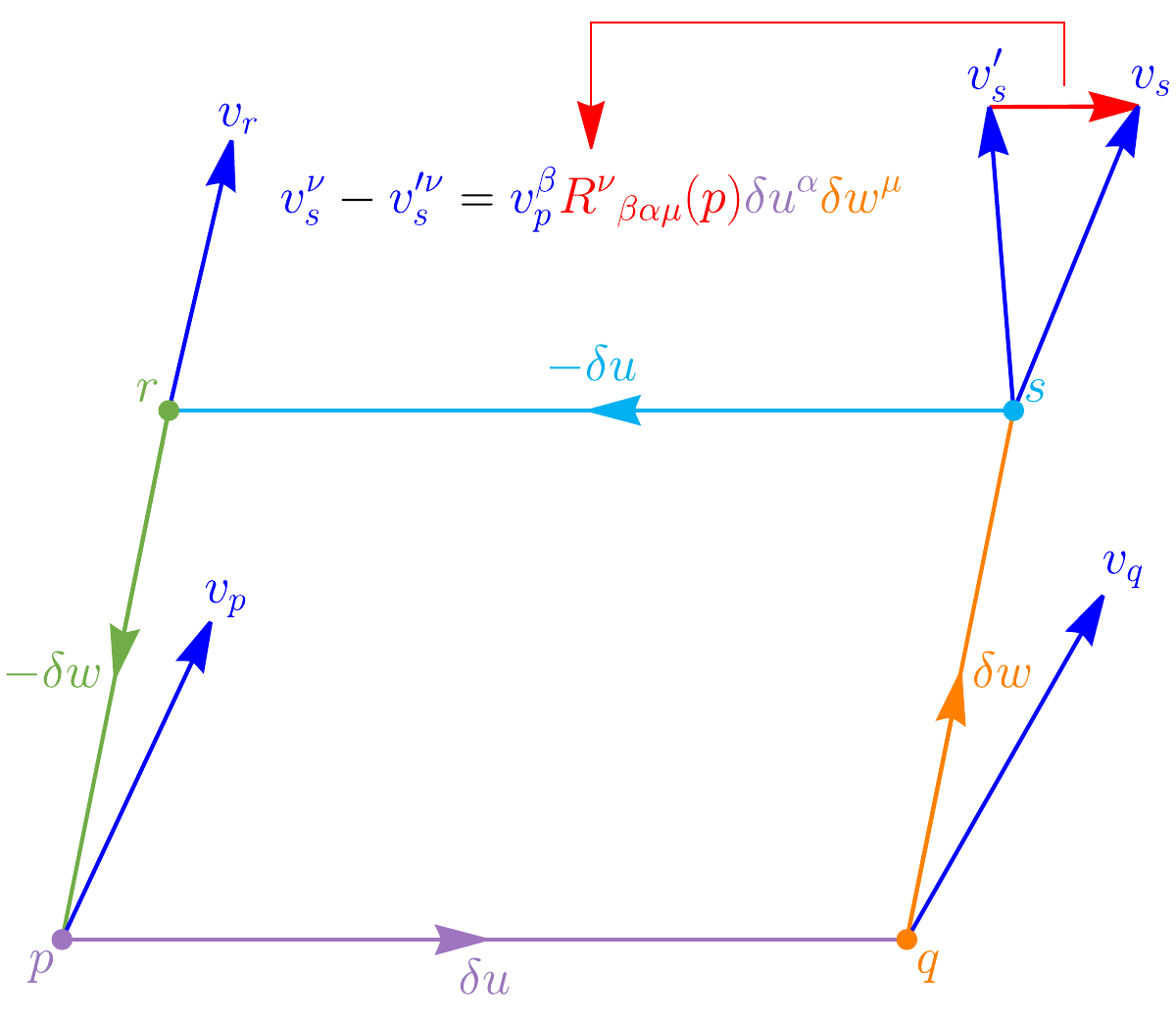}
    \caption{Moving a vector around a closed loop changes its direction, measured by $R^{\nu}_{~\beta\alpha\mu}$.}
    \label{curvature_fig}
\end{figure}

Curvature describes how the direction of a vector changes when it is parallel transported along a closed path in a given geometry. 
To visualize this, consider a vector $v_p$ defined at a point $p$ and successively transported along the infinitesimal loop $p \rightarrow q \rightarrow s \rightarrow r \rightarrow p$, as illustrated in Fig.~(\ref{curvature_fig}). 
In a flat geometry, such as Euclidean space, the vector returns to its original direction after completing the loop, implying that parallel transport around the closed path does not alter the vector’s orientation. 
However, in a curved geometry, the final vector fails to coincide with its initial direction, indicating that the space possesses non-vanishing curvature. 
The infinitesimal difference between the initial and final vectors, represented as $v_s' - v_s$, is proportional to the curvature tensor $R^{\nu}{}_{\beta\alpha \mu}$ \cite{heisenberg2023}
\begin{eqnarray}
    v_s' - v_s = v_p^{\beta} R^{\nu}{}_{\beta \alpha \mu}(p)\, \delta u^{\nu} \delta w^{\alpha}.
\end{eqnarray}
This tensor quantitatively measures the failure of parallel transport to preserve the direction of a vector when carried around a closed loop. 
Geometrically, it captures the intrinsic bending of the manifold and indicates how much the space deviates from being flat. 
In the framework of GR, the curvature tensor arises solely from the Levi-Civita connection, which is symmetric and metric-compatible. 
This ensures that torsion and non-metricity vanish, and curvature fully encodes the gravitational interaction. 
However, in more general geometric theories such as those incorporating torsion or non-metricity the total curvature may include contributions beyond the Levi-Civita connection, reflecting the richer geometric structure of spacetime.
\subsubsection{Torsion: The geometrical meaning}
\begin{figure}[H]
    \centering
    \includegraphics[width=01\linewidth]{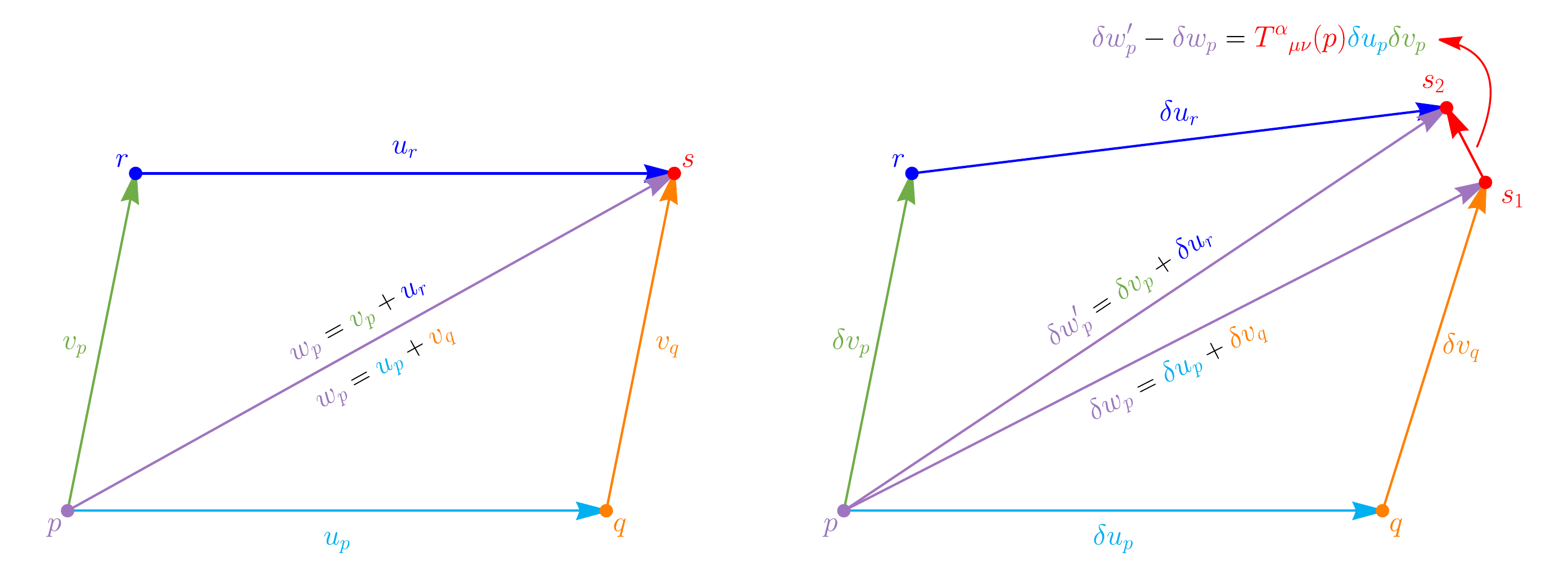}
    \caption{\textit{Left panel:~}In Euclidean geometry, parallel transport along two vectors forms a closed parallelogram.
\textit{Right panel:}~ In a space with torsion, the parallelogram does not close; the separation between the endpoints measures the torsion tensor $T^{\alpha}_{\ \mu\nu}$.}
    \label{torsion_fig}
\end{figure}

In Euclidean geometry, the addition of two vectors at a point \( p \) can be geometrically visualized as forming a parallelogram. 
Consider two infinitesimal vectors \( u_p \) and \( v_p \) at \( p \). 
If we move \( v_p \) along \( u_p \), or equivalently \( u_p \) along \( v_p \), we arrive at the same end point the tip of the parallelogram closes neatly, as illustrated in the left panel of Fig.~(\ref{torsion_fig}).
However, in a curved or more general space, this construction may not close. 
When we parallel transport one vector with another using a general connection, the order of transport becomes important: moving \( v_p \) along \( u_p \) may not coincide with moving \( u_p \) along \( v_p \). 
As a result, the two transported vectors end at different points, say \( s_1 \) and \( s_2 \), and the infinitesimal parallelogram fails to close (right panel of Fig.~(\ref{torsion_fig})). This ``failure to close'' is a direct manifestation of the \textit{torsion} of the connection. 
Mathematically, the torsion tensor \( T^{\alpha}{}_{\mu\nu} \) quantifies the difference between these two possible orders of infinitesimal displacement \cite{heisenberg2023}
\begin{equation}
    T^{\alpha}{}_{\mu\nu} = \Gamma^{\alpha}{}_{\mu\nu} - \Gamma^{\alpha}{}_{\nu\mu},
\end{equation}
which is simply the antisymmetric part of the general affine connection coefficients. If the torsion vanishes \( T^{\alpha}{}_{\mu\nu} = 0 \), the parallelogram closes and the geometry behaves like the Riemannian case of GR, governed by the symmetric Levi-Civita connection. On the other hand, when torsion is present, infinitesimal parallelograms do not close, indicating an intrinsic ``twist'' in the spacetime geometry. Therefore, in summary:
In the absence of torsion, parallel transport of two vectors \( u_p \) and \( v_p \) in different orders first \( u_p \) then \( v_p \), or vise versa, leads to the same final point, i.e. \( s_1 = s_2 \). However, when torsion is present, these two paths no longer coincide (\( s_1 \neq s_2 \)), and the resulting displacement between the endpoints is quantified by the torsion tensor \( T^{\alpha}{}_{\mu\nu} \).
Thus, The torsion tensor has an interesting geometrical interpretation: if one builds infinitesimal parallelograms in the manifold (by parallel transport), the failure of the parallelogram to close is proportional to the torsion tensor.
\subsubsection{Non-metricity: The geometrical interpretation} 
\begin{figure}[H]
    \centering
    \includegraphics[width=0.8\linewidth]{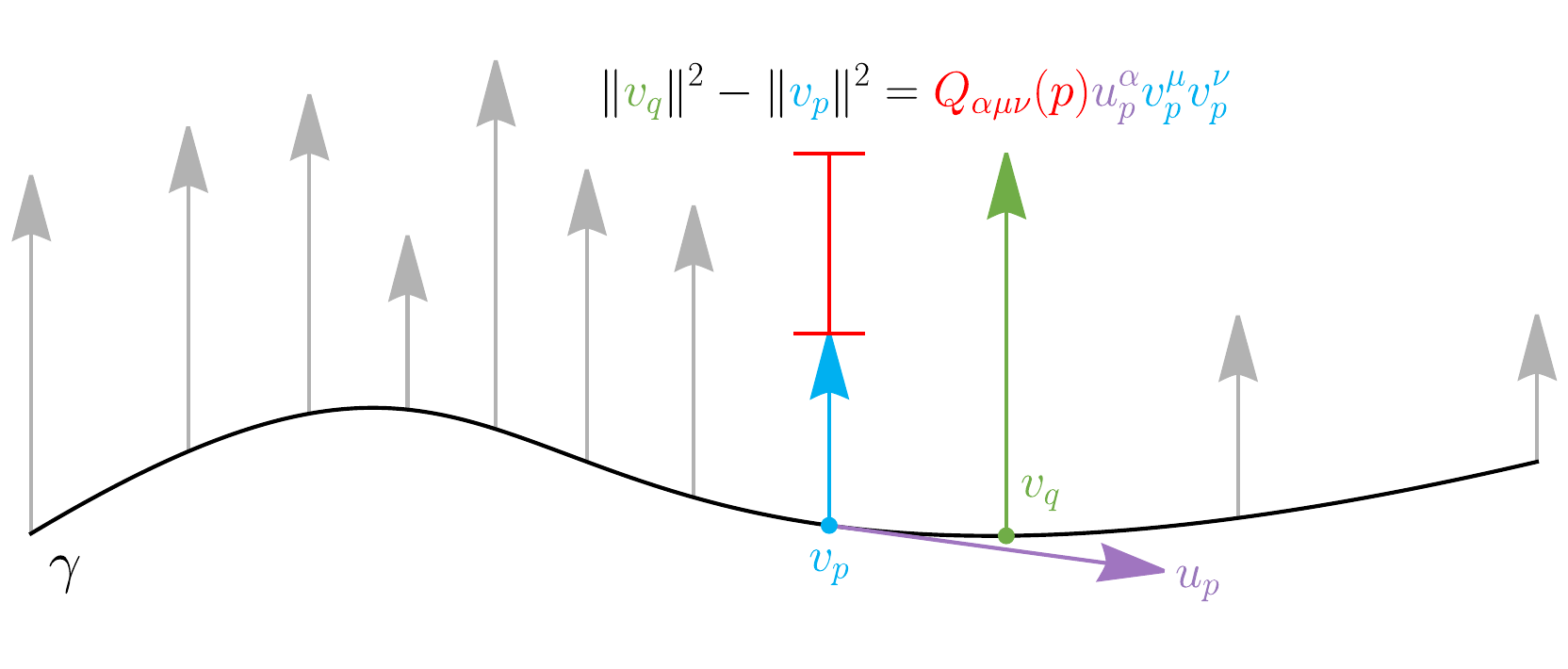}
    \caption{ 
Change in the length of a vector under parallel transport arises from non-metricity.}
\label{metricity_fig}
\end{figure}

In differential geometry, every vector has not only a direction but also a magnitude, which is measured by the metric tensor $g_{\mu\nu}$. In Euclidean geometry, when a vector is moved (parallel transported) along a path, its length and the angle between any two vectors remain unchanged. This is because the connection used there preserves the metric. In a general manifold, however, the connection may not preserve the metric that results, the length of a vector can change during parallel transport, a situation illustrated in Fig. (\ref{metricity_fig}).
Suppose a vector $v$ is being transported along a smooth curve $\gamma$, where $u$ is the tangent vector to that curve. In a \textit{metric-compatible} geometry, the covariant derivative of the metric tensor vanishes, i.e.,
$\nabla_\lambda g_{\mu\nu} = 0.$
This condition ensures that the inner product between any two vectors remains invariant during parallel transport. In simpler terms, it means that the length of a vector and the angle between vectors do not change as they move along the curve. This is precisely the case in GR. Thus, if one moves a vector around in spacetime according to GR, its magnitude remains constant, reflecting the preservation of the spacetime interval.

~~~~However, in a \textit{non-metric} geometry condition $\nabla_\lambda g_{\mu\nu} \neq 0$ holds. In this case, the connection allows the metric itself to vary from point to point under parallel transport. Physically, this means that as a vector is carried along a path from one point $p$ to another point $q$, its length or magnitude can change. From Fig. (\ref{metricity_fig}), we can note that the length changes between two nearby points $p$ and $q$, as measured by the non-metricity tensor \cite{heisenberg2023}
\begin{eqnarray}
    \|v_q\|^2 - \|v_p\|^2 = Q_{\alpha\mu\nu}(p)u_p^{\alpha}v_p^{\mu}v_p^{\nu},
\end{eqnarray}
where $u_p$ is the tangent vector of the curve $\gamma$ at the point $p$. The change arises because the metric used to measure distances and angles is no longer fixed during transport as it evolves along the curve. This deviation from metric compatibility marks a fundamental departure from GR and appears in generalized geometric frameworks such as those involving non-metricity, where the affine connection and the metric are treated as independent quantities.

\textbf{NOTE:}
\noindent\textbf{Key difference of curvature and non-metricity}

As far as we discussed, when a vector is parallel transported along a closed curve in a general geometric space, it may not return to its original state. 
The deviation between the initial and final vectors can manifest itself as a change in direction, in magnitude, or both. 
This direction change behavior is encoded in the curvature tensor $R^{\alpha}{}_{\beta\mu\nu}$, which can be decomposed into antisymmetric and symmetric parts with respect to the indices $\alpha$ and $\beta$:
\begin{eqnarray}
    R^{\alpha}{}_{\beta\mu\nu} = R^{\alpha}{}_{[\beta]\mu\nu} + R^{\alpha}{}_{(\beta)\mu\nu}.
\end{eqnarray}
The antisymmetric component $R^{\alpha}{}_{[\beta]\mu\nu}$ corresponds to the \textit{rotation} of the vector upon returning to its starting point and represents the familiar notion of curvature in Riemannian geometry. 
This is the only contribution present in GR, where the connection is metric-compatible and preserves the length of the vectors during transport.
\begin{figure}[H]
    \centering   \includegraphics[width=0.5\linewidth]{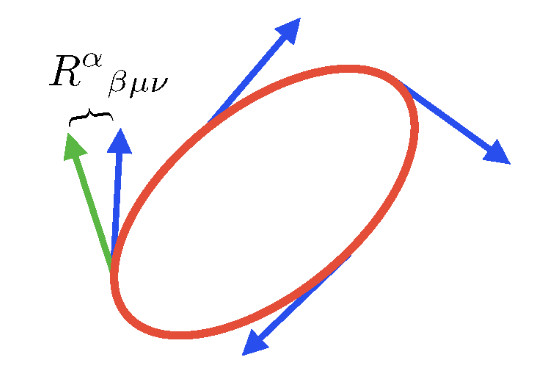}
    \caption{ The rotation of a vector transported along a closed curve given by $R^{\alpha}_{~~\beta \mu \nu}$.}
\label{rot_fig}
\end{figure}
In contrast, the symmetric part $R^{\alpha}{}_{(\beta)\mu\nu}$ arises in non-metric geometries and describes a change in the \textit{magnitude} (length) of the vector after parallel transport. 
Hence, in a general affine geometry, curvature can induce both a rotational and a stretching effect on transported vectors.
 Fig.~(\ref{rot_fig}) illustrates the rotation of a vector transported around a closed curve. In Riemannian geometry (as in GR), only the rotation component survives and is determined by the antisymmetric component $R^{\alpha}{}_{[\beta]\mu\nu}$ of the curvature tensor where we ignore the stretching effect. But in a more general non-metric geometry, an additional symmetric component $R^{\alpha}{}_{(\beta)\mu\nu}$ accounts for stretching or shrinking (length change) only appears when non-metricity is present.

\subsection{Modified theories of gravity: Representative examples}
Modified gravity theories include a distinct approach by modifying the geometric part of Einstein’s field equations rather than the matter (stress-energy) sector. In essence, these theories modify the underlying structure of gravity itself, leading to field equations that deviate from those of GR. 

\begin{figure}[H]
    \centering
    \includegraphics[width=1.1\linewidth]{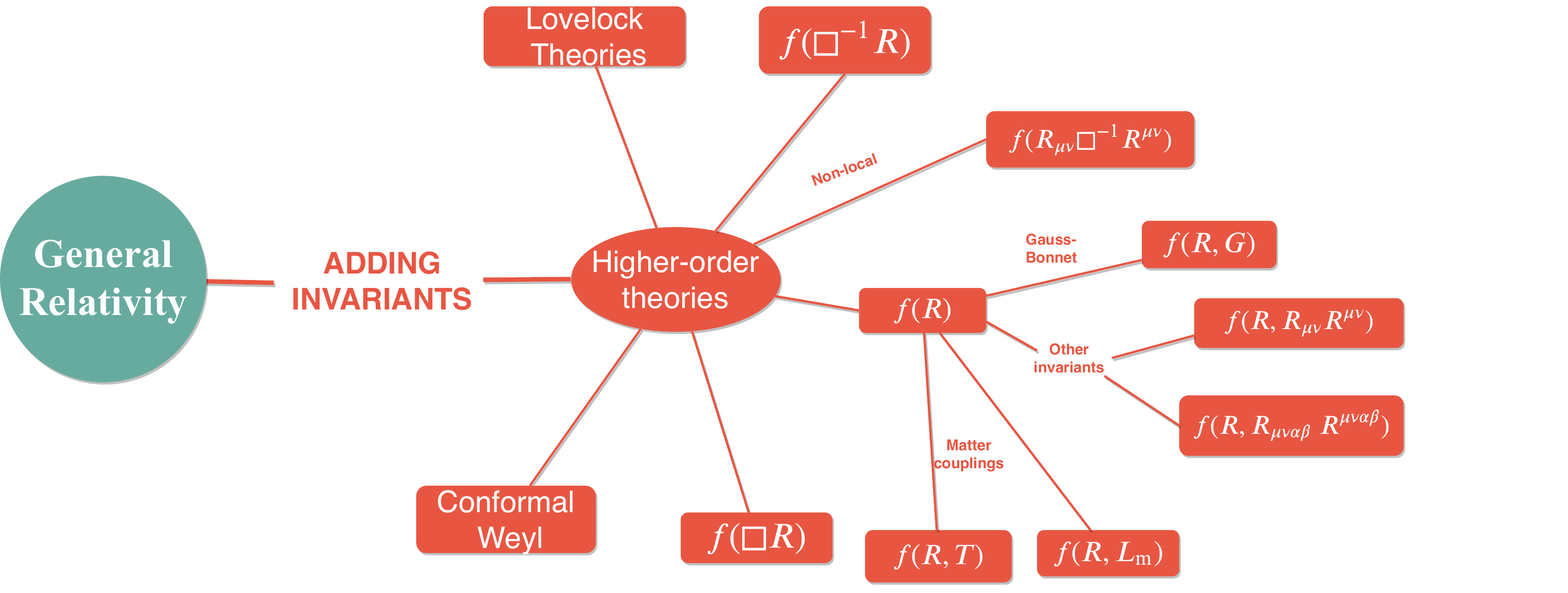}
    \caption{Schematic classification of higher-order gravity theories obtained by incorporating additional curvature invariants into the Lagrangian of GR.}
    \label{fig:higher_order_schematic}
\end{figure}

\textbf{Higher-order field equations:}
 Although Einstein's field equations contain derivatives only up to the second order, several modified gravity theories extend this idea by including higher-order derivatives. Although such extensions often introduce mathematical complications and may give rise to Ostrogradsky instability \cite{woodard2007} resulting in an unbounded Hamiltonian, certain models, such as $f(R)$ gravity \cite{sotiriou2010,defelice2010}, successfully circumvent this issue even though their field equations are of fourth order. 
A schematic representation of the various higher-order extensions of GR, obtained by introducing additional curvature invariants in the gravitational Lagrangian, is shown in Fig.~(\ref{fig:higher_order_schematic}).

\begin{figure}[H]
    \centering
    \includegraphics[width=0.8\linewidth]{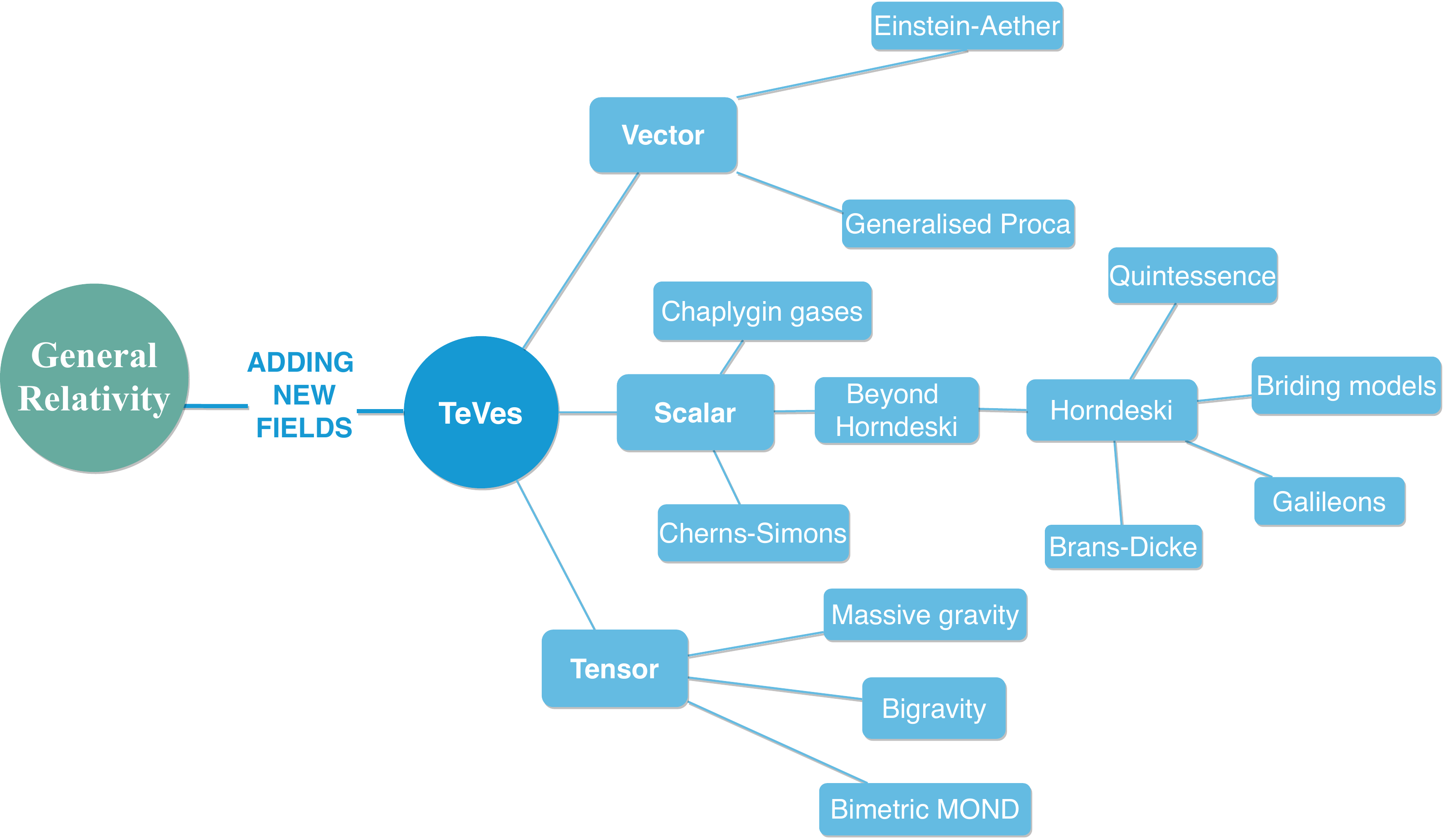}
    \caption{Schematic classification of higher-order gravity theories obtained by incorporating additional curvature invariants into the Lagrangian of GR.}
    \label{fig:TeVeS_schematic}
\end{figure}

\textbf{Tensor–Vector–Scalar extensions of GR:}
Another major class of modified gravity theories arises from introducing additional dynamical fields such as scalar, vector, or tensor fields into the framework of GR. These theories, collectively referred to as Tensor–Vector–Scalar (TeVeS) models \cite{bekenstein2004}, extend the gravitational interaction beyond the metric tensor by coupling it with extra degrees of freedom. The inclusion of new fields allows for richer phenomenology, offering explanations for cosmic acceleration, dark matter effects, and deviations from GR in strong-field regimes. Scalar-tensor theories, such as Brans–Dicke \cite{brans1961} and Horndeski gravity \cite{horndeski1974}, are among the most prominent examples. On the other hand, tensor extensions such as massive gravity and bi-gravity modify the spin-2 sector of gravity itself \cite{derham2011}. One may look at the schematic diagram in Fig.~(\ref{fig:TeVeS_schematic}).

\textbf{Changing geometry: Non-Riemannian theories of gravity}

\begin{figure}[H]
    \centering
    \includegraphics[width=\linewidth]{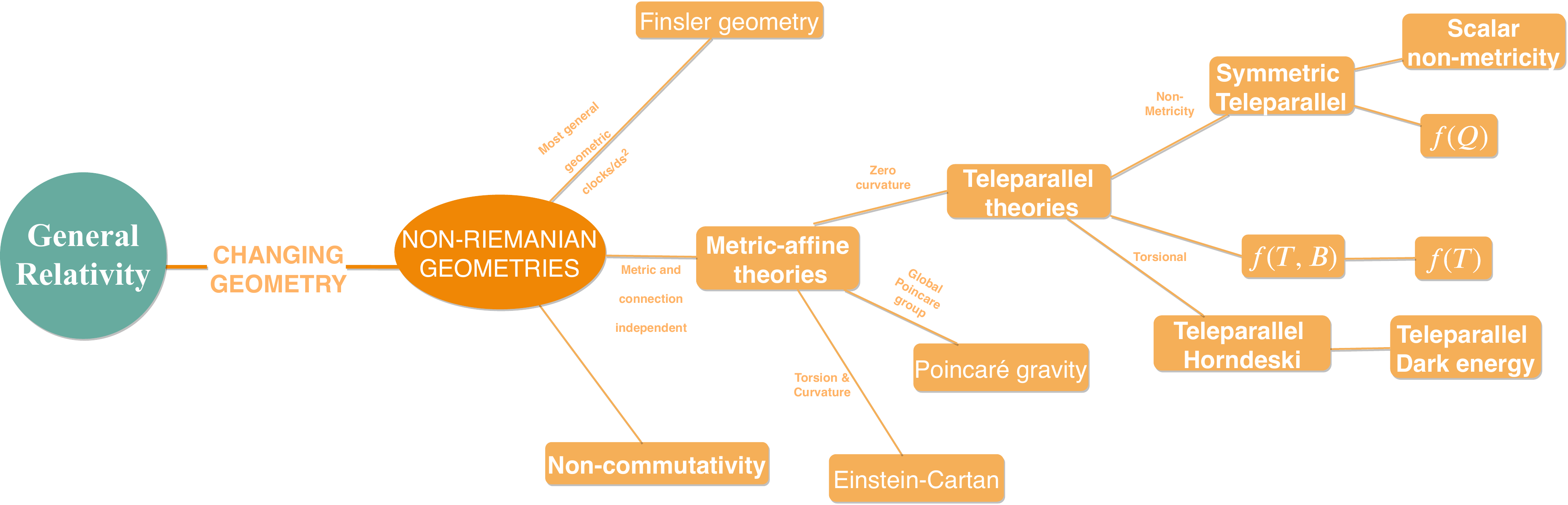}
    \caption{Schematic classification of higher-order gravity theories obtained by incorporating additional curvature invariants into the Lagrangian of GR.}
    \label{fig:nonriemannian_schematic}
\end{figure}

Another direction of modifying GR involves altering the underlying geometric structure of spacetime rather than adding new fields or higher-order invariants. These frameworks generalize Riemannian geometry by relaxing its constraints, such as metric compatibility or vanishing torsion. The resulting non-Riemannian geometries, including metric-affine \cite{hehl1995}, teleparallel \cite{aldrovandi2013}, and Finsler formulations offer alternative geometric interpretations of gravitation. For instance, teleparallel gravity describes gravity through torsion instead of curvature, while symmetric teleparallel gravity attributes it to non-metricity, giving rise to models like $f(T)$ and
$f(Q)$ gravity \cite{jimenez2018} respectively. Similarly, the Einstein-Cartan and Poincaré gauge theories incorporate both torsion and curvature as dynamical quantities, as can be seen in Fig.~(\ref{fig:nonriemannian_schematic}).

The following subsections present in detail the mathematical framework of some modified gravity models explored in this thesis.
\subsubsection{$f(T)$ gravity}
GR is a metric theory of gravitation formulated on a Riemannian manifold \( V_4 \), where the dynamics of spacetime are governed by the Einstein field equations. In this framework, the Levi-Civita connection is compatible with the metric and torsion free, which is used to construct the Riemann curvature tensor that encapsulates the gravitational effects. In contrast, the TEGR provides an alternative but dynamically equivalent formulation in which gravity is not attributed to spacetime curvature but rather to \textit{torsion}. This formulation is based on the idea of a globally flat spacetime, where curvature vanishes identically, and the gravitational interaction arises purely from the torsion of spacetime. One of the extensions of such gravity is $f(T)$ gravity. The possibility of such a construction was first realized by Weitzenböck \cite{weitzenbock1923}, who demonstrated that by choosing an appropriate connection distinct from the Levi-Civita one, it is possible to ensure global flatness while still describing gravitational interactions.
To understand this construction, let us consider a general affine connection that can be described into two fundamental components: a tetrad field and a spin connection. The general connection can thus be expressed as follows  
\begin{equation}
    \Gamma^{\lambda}_{\mu\nu} = e_b^{\lambda}\partial_\nu e^b_{\mu} + e_a^{\lambda}A^a_{b\nu}e^b_{\mu},
    \label{eq:tele1}
\end{equation}
where \( e^a_{\mu} \) are the tetrad fields, and \( A^a_{b\nu} \) denotes the spin connection. The spin connection can be related to the tetrad and affine connection as  
\begin{equation}
    A^a_{~~b\nu} = e^a_{\lambda}\partial_\nu e_b^{\lambda} + e^a_{\lambda}\Gamma^{\lambda}_{\mu\nu}e_b^{\mu}.
    \label{eq:tele2}
\end{equation}
Alternatively, using the covariant derivative associated with \( \Gamma^{\lambda}_{\mu\nu} \), this can be written as  
\begin{equation}
    A^a_{~~b\nu} = e^a_{\lambda}\nabla_\nu e_b^{\lambda}.
    \label{eq:tele3}
\end{equation}

In the teleparallel formulation, the \textit{Weitzenböck connection} is used, which characterizes a spacetime denoted \( W_4 \). This connection is defined as  
\begin{equation}
    \Gamma^{\lambda}_{\mu\nu} = e_b^{\lambda}\partial_\nu e^b_{\mu} = - e^b_{\mu}\partial_\nu e_b^{\lambda}.
    \label{eq:tele4}
\end{equation}
This particular choice corresponds to setting the spin connection to zero. A notable property of the Weitzenböck connection is that its associated Riemann curvature tensor vanishes identically  
\begin{equation}
    R^{\alpha}_{\ \mu\lambda\nu} = 0.
    \label{eq:tele5}
\end{equation}
Hence, the spacetime endowed with this connection is globally flat. Nevertheless, the connection still encodes gravitational effects through torsion rather than curvature.  
The Weitzenböck connection can be further decomposed into the sum of the Levi-Civita connection \( \Big\{\begin{matrix}
\alpha \\[-4pt]
\beta\gamma
\end{matrix}\Big\} \) and the contortion tensor \( K^{\lambda}_{\mu\nu} \), as  
\begin{equation}
    \Gamma^{\lambda}_{\mu\nu} = \Big\{\begin{matrix}
\alpha \\[-4pt]
\beta\gamma
\end{matrix}\Big\} + K^{\lambda}_{\mu\nu}.
    \label{eq:tele6}
\end{equation}
This relation serves as a bridge between GR and its teleparallel counterpart.
To understand the physical meaning of this connection, consider a vector \( V^a \) defined at a point \( p \) in the tangent space \( t_p \) of the manifold. Another vector \( V'^a \) defined at a nearby point \( p' \in t_{p'} \) is said to be parallel to \( V^a \) if it has identical tetrad components, i.e. \( V'^a = V^a \) or equivalently \( V'^{\mu} = V^{\mu} \). Under parallel transport, we require that  
\begin{equation}
    0 = dV^a = d(V^{\mu}e^a_{\mu}) = e^a_{\mu}dV^{\mu} + V^{\mu}\partial_\nu e^a_{\mu}dx^{\nu},
    \label{eq:tele7}
\end{equation}
which, after rearrangement and multiplication by the co-tetrad \( e_a^{\lambda} \), gives  
\begin{equation}
    dV^{\lambda} = - e_a^{\lambda}\partial_\nu e^a_{\alpha}V^{\alpha}dx^{\nu} = -\Gamma^{\lambda}_{\alpha\nu}V^{\alpha}dx^{\nu}.
    \label{eq:tele8}
\end{equation}
This expression is precisely the definition of the Weitzenböck connection. Therefore, the Weitzenböck connection characterizes a spacetime in which vectors remain parallel when transported along curves, a property known as \textit{distant parallelism} or \textit{teleparallelism}. The contortion tensor \( K^{\alpha}_{~~\mu\nu} \) in Eq.~(\ref{eq:tele6}) is defined by
\begin{equation}
    K^{\alpha}_{~~\mu\nu} = \frac{1}{2} T^{\alpha}_{~\mu\nu} + T_{(\mu\,~\nu)}^{~\,\alpha}
    = \frac{1}{2} g^{\alpha\lambda}\left( T_{\mu\lambda\nu} + T_{\nu\lambda\mu} + T_{\lambda\mu\nu} \right),
    \label{eq:contortion}
\end{equation}
where the torsion tensor \( T^{\alpha}_{\mu\nu} \) is defined as
\begin{equation}
    T^{\alpha}_{~~\mu\nu} = 2\,\Gamma^{\alpha}_{[\nu\mu]}.
    \label{eq:torsion}
\end{equation}
\noindent
In analogy to the role of the Einstein tensor in GR, one can construct a scalar quantity in teleparallel gravity from the torsion tensor, which will act as the gravitational Lagrangian. For this purpose, the \textit{superpotential} (or conjugate of the torsion tensor) is introduced and defined as
\begin{equation}
    S_{\alpha}^{~~\mu\nu} = \frac{1}{2} \left( K^{\mu\nu}_{~~\alpha} + \delta^{\mu}_{\alpha}\, T^{\beta\nu}_{~~\beta} 
    - \delta^{\nu}_{\alpha}\, T^{\beta\mu}_{~~\beta} \right).
    \label{eq:superpotential}
\end{equation}
This tensor plays an important role in defining the torsion scalar, which encapsulates the gravitational interaction in the teleparallel framework.
The \textit{torsion scalar} \( T \) is defined as the contraction of the superpotential with the torsion tensor
\begin{equation}
    T = T^{\alpha}_{~~\mu\nu} S_{\alpha}^{~~\mu\nu}.
    \label{eq:torsion_scalar}
\end{equation}
This scalar quantity serves as the teleparallel analog of the Ricci scalar \( R \) in GR. Unlike the curvature-based approach, teleparallel gravity uses torsion as the dynamical variable to describe gravitation.
The action for the TEGR can then be written as
\begin{equation}
    S = \frac{1}{2\kappa^2} \int d^4x\, e\, T + \int d^4x\, e\, \mathcal{L}_m,
    \label{eq:TEGR_action}
\end{equation}
where \( \kappa^2 = 8\pi  \), \( e = \det(e^a_{\mu}) = \sqrt{-g} \) is the determinant of the tetrad, \( g \) denotes the determinant of the metric tensor \( g_{\mu\nu} \) and \( \mathcal{L}_m \) represents the matter Lagrangian density. 
By varying the above action with respect to the tetrad field \( e^a_{\mu} \), the field equations of teleparallel gravity are obtained as \cite{aldrovandi2013} 
\begin{equation}
    e^{-1} \partial_{\nu}\left(e\, S_a^{~\mu\nu}\right)
    - e_a^{~\lambda} T^{\delta}_{~\nu\lambda} S_{\delta}^{~\mu\nu}
    + \frac{1}{4} e_a^{~\mu} T = \frac{\kappa^2}{2}\, e_a^{~\delta} \mathcal{T}_{\delta}^{~\mu},
    \label{eq:TEGR_field_eq}
\end{equation}
where \( \mathcal{T}_{\delta}^{~\mu} \) is the energy momentum tensor of the matter fields.
These equations are second-order differential equations, unlike the fourth-order ones found in curvature-based modified gravity theories, making teleparallel gravity computationally simpler in many contexts. It can also be shown that the torsion scalar \( T \) and the Ricci scalar \( R \) of the Levi-Civita connection are related through a total divergence term
\begin{equation}
    R = -T + 2\, e^{-1} \partial_{\mu}\left(e\, T^{\nu\mu}_{~~\nu}\right),
    \label{eq:R_T_relation}
\end{equation}
which establishes the dynamic equivalence between GR and its teleparallel counterpart (TEGR).
\noindent
In summary, teleparallel gravity provides an alternative geometrical interpretation of gravitation in which torsion, rather than curvature, mediates the gravitational interaction. While both GR and TEGR yield the same field equations and physical predictions, their geometric underpinnings differ fundamentally: GR attributes gravity to the curvature of spacetime, whereas TEGR describes it as a manifestation of spacetime torsion under the Weitzenböck connection. This alternative viewpoint also serves as the foundation for extensions such as \( f(T) \) gravity, which generalizes the torsion scalar in analogy to \( f(R) \) theories in curvature-based gravity.

\subsubsection{$f(Q)$ gravity}
Several extensions of the STEGR have been proposed to address unresolved issues in cosmology and astrophysics. These extensions modify the underlying framework either by introducing scalar fields or by considering functional generalizations of the non-metricity scalar. In the following subsections, we focus on the prominent example $f(Q)$ gravity theory, which aims to provide a deeper insight into observed astrophysical phenomena.
Before discussing these extensions, we briefly recall the essential features of STEGR. This gravity describes a class of non-Riemannian gravity theories in which the affine connection is chosen such that both the curvature and the torsion vanish identically. Consequently, gravitational interactions are encoded entirely in terms of the non-metricity of spacetime \cite{jimenez2019}. Mathematically, this is expressed as
\begin{equation}
   R^{\alpha}_{\ \mu\nu\delta}
= \partial_{\nu}\Gamma^{\alpha}_{\ \mu\delta}
- \partial_{\delta}\Gamma^{\alpha}_{\ \mu\nu}
+ \Gamma^{\beta}_{\ \mu\delta}\,\Gamma^{\alpha}_{\ \beta\nu}
- \Gamma^{\beta}_{\ \mu\nu}\,\Gamma^{\alpha}_{\ \beta\delta}
= 0 . ,\,\,\,\,
\text{and}
  \,\,\,\,  T^{\alpha}_{~\mu\nu} = 0 .
\end{equation}

Therefore, in STG the connection reduces to $\Gamma^{\alpha}_{\mu\nu} = \left\{^{\,\alpha}_{\mu\nu}\right\} + L^{\alpha}_{~~\mu\nu} .$
The disformation tensor $L^{\alpha}_{~~\mu\nu}$ quantifies the non-metricity
\begin{equation}
    L^{\alpha}_{~~\mu\nu} = \frac{1}{2}g^{\alpha\lambda}\left( Q_{\lambda\mu\nu}- Q_{\mu\lambda\nu} - Q_{\nu\lambda\mu}  \right)
    = \frac{1}{2} Q^{\alpha}_{\mu\nu} - Q^{\alpha}_{(\mu\nu)} ,
\end{equation}
This quantity $L^{\alpha}_{\mu\nu}$ is symmetric in the second and third indices, i.e. $L^{\alpha}_{~\mu\nu}=L^{\alpha}_{~\nu\mu}$, in other words, it can be represented as $L^{\alpha}_{[\mu\nu]} = 0$. In the above expression, the non-metricity tensor $Q_{\alpha\mu\nu}$ is defined by
\begin{equation}
    Q_{\alpha\mu\nu} = \nabla_\alpha g_{\mu\nu}
    = \partial_\alpha g_{\mu\nu} - \Gamma^\beta_{\alpha\mu}g_{\beta\nu} - \Gamma^\beta_{\alpha\nu}g_{\mu\beta} .
\end{equation}
The non-metricity tensor $Q_{\alpha\mu\nu}$ is symmetric in the last two indices, $Q_{\alpha[\mu\nu]} = 0 $, i.e., $Q_{\alpha \mu\nu}=Q_{\alpha \nu\mu}$,
and it furthermore satisfies the Bianchi identity, $\nabla_{[\alpha}Q_{\beta]\mu\nu} = 0$.
Furthermore, the \textit{non-metricity scalar} is defined as
\begin{eqnarray}
   &&\hspace{0cm} Q = -\frac{1}{4} Q_{\alpha\beta\mu} Q^{\alpha\beta\mu}
      +\frac{1}{2} Q_{\alpha\beta\mu} Q^{\beta\mu\alpha}
      +\frac{1}{4} Q_{\alpha} Q^{\alpha}
      -\frac{1}{2} \tilde{Q}^{\alpha} Q_{\alpha} ,\nonumber\\
      &&\hspace{1cm}\equiv g^{\mu\nu} \left( L^{\alpha}{}_{\alpha\beta} L^{\beta}{}_{\mu\nu} - L^{\alpha}{}_{\beta\mu} L^{\beta}{}_{\nu\alpha} \right),
\end{eqnarray}
where the trace of the non-metricity is given by
\begin{equation}
    Q_{\alpha} = g^{\mu\nu} Q_{\alpha\mu\nu} = Q_{\alpha~\mu}^{~\mu}
\quad \text{and} \quad
\tilde{Q}_{\alpha} = g^{\mu\nu} Q_{\mu\nu\alpha} = Q^{\mu}{}_{\mu\alpha}.
\end{equation}

Accordingly, contractions of the disformation tensor become
\begin{equation}
    L^{\lambda}{}_{\alpha\lambda} = -\frac{1}{2}Q_{\alpha} ,\qquad
    L^{\alpha\lambda}{}_{\lambda} = \frac{1}{2}Q^{\alpha} - \tilde{Q}^{\alpha} .
\end{equation}

Furthermore, we can define the superpotential or non-metricity conjugate tensor as
\begin{equation}
    P^{\alpha}{}_{\mu\nu} = \frac{1}{2} \frac{\partial Q}{\partial Q_{\alpha}{}^{\mu\nu}} .
\end{equation}
Due to the symmetry property of $Q_{\alpha\mu\nu}$, $P^{\alpha}{}_{\mu\nu} = P^{\alpha}{}_{\nu\mu}$.
The explicit form of the superpotential is
\begin{eqnarray}
   &&\hspace{0cm} P^{\alpha}{}_{\mu\nu} = \frac{1}{4} \left[ -Q^{\alpha}{}_{\mu\nu} + 2 Q_{(\mu\nu)}^{~\,\,\,\,\alpha}
    + Q^{\alpha} g_{\mu\nu}
    - \tilde{Q}^{\alpha} g_{\mu\nu}
    - \delta^{\alpha}_{(\mu}Q_{\nu)} \right] ,\nonumber\\
   &&\hspace{1cm}  = -\frac{1}{2} L^{\alpha}{}_{\mu\nu}
        + \frac{1}{4}(Q^{\alpha}-\tilde{Q}^{\alpha})g_{\mu\nu}
        - \frac{1}{4}\delta^{\alpha}_{(\mu} Q_{\nu)} .
\end{eqnarray}
or, equivalently, it can be expressed in terms of metric, disformation tensor, and non-metricity tensor as
\begin{equation}
P^{\alpha}_{~\mu\nu} = 
-\frac{1}{2} L^{\alpha}_{~\mu\nu}
+ \frac{1}{4} g_{\mu\nu}
\left(
g^{\alpha\beta} g^{\delta q} Q_{\beta q \delta}
- g^{\alpha\gamma} g^{\lambda\eta} Q_{\eta\gamma\lambda}
\right)
-\frac{1}{8}
\left(
\delta^{\alpha}_{\mu} g^{\xi\zeta} Q_{\nu\xi\zeta}
+ \delta^{\alpha}_{\nu} g^{zt} Q_{\mu zt}
\right).
\end{equation}

Finally, the non-metricity scalar can be compactly rewritten as
\begin{equation}
    Q = Q_{\alpha\mu\nu} P^{\alpha\mu\nu} .
\end{equation}
Next, incorporating Lagrange multipliers, the action for \( f(Q) \) gravity can be formulated as 
\begin{equation}
S = \int \sqrt{-g} \, d^4x 
\left[ 
\frac{1}{2 \kappa^2} f(Q) + \lambda_{\alpha}^{~\beta\mu} R^{\alpha}_{~\beta\mu} + \lambda_{\alpha}^{~\mu\nu} T^{\alpha}_{~\mu\nu} + \mathcal{L}_m
\right],
\end{equation}
where\( f(Q) \) is an arbitrary function of the non-metricity scalar \( Q \), 
\( \lambda_{\alpha}^{~\beta\mu} \) are the Lagrange multipliers, and \( \mathcal{L}_m \) represents the matter Lagrangian density. 
Varying the above action with respect to the metric inverse \( g^{\mu\nu} \) yields the modified field equations \cite{jimenez2018} 
\begin{equation}\label{fQfield}
-\kappa^2 \mathcal{T}_{\mu\nu} = 
\frac{2}{\sqrt{-g}} \nabla_{\alpha} \left( \sqrt{-g} f_Q P^{\alpha}_{~\mu\nu} \right)
+ \frac{1}{2} g_{\mu\nu} f 
+ f_Q \left( P_{\mu\alpha\beta} Q_{\nu}^{~\alpha\beta} - 2 Q_{\alpha\beta\mu} P^{\alpha\beta}_{~~\nu} \right),
\end{equation}
where \( f_Q = \partial f / \partial Q \), and the energy–momentum tensor is defined conventionally as  
\begin{equation}
\mathcal{T}_{\mu\nu} = -\frac{2}{\sqrt{-g}} \frac{\delta (\sqrt{-g} \mathcal{L}_m)}{\delta g^{\mu\nu}}.
\end{equation}

Similarly, variation of the action with respect to the affine connection gives  
\begin{equation}\label{rm}
\nabla_{\delta} \lambda_{\alpha}^{~\mu\delta} + \lambda_{\alpha}^{~\mu\nu}
= \sqrt{-g} f_Q P^{\mu\nu}_{~~\alpha} + H_{\alpha}^{~\mu\nu},
\end{equation}
where the hypermomentum tensor density is expressed as  
\begin{equation}
H_{\alpha}^{~\mu\nu} = -\frac{1}{2} \frac{\delta \mathcal{L}_m}{\delta \Gamma^{\alpha}_{~\mu\nu}}.
\end{equation}

Using the anti-symmetry properties of the Lagrange multiplier coefficients in \( \mu \) and \( \nu \), the above equation Eq.~(\ref{rm}) simplifies to  
\begin{equation}
\nabla_{\mu} \nabla_{\nu} \left( \sqrt{-g} f_Q P^{\mu\nu}_{~~\alpha} + H_{\alpha}^{~\mu\nu} \right) = 0.
\end{equation}
In the absence of hypermomentum (\( \nabla_{\mu} \nabla_{\nu} H_{\alpha}^{~\mu\nu} = 0 \)), this reduces to  
\begin{equation}
\nabla_{\mu} \nabla_{\nu} \left( \sqrt{-g} f_Q P^{\mu\nu}_{~~\alpha} \right) = 0.
\end{equation}
In a spacetime devoid of curvature and torsion, the affine connection takes the general form  
\begin{equation}
\Gamma^{\alpha}_{~\mu\nu} = \frac{\partial x^{\alpha}}{\partial \xi^{\lambda}} \, \partial_{\mu} \partial_{\nu} \xi^{\lambda},
\end{equation}

where $\xi^{\lambda} \equiv \xi^{\lambda}(x)$ are four arbitrary scalar fields that define an alternative coordinate system in which the affine connection vanishes identically. They encode the residual gauge freedom of the symmetric teleparallel framework once the conditions of vanishing curvature and torsion are imposed.
Now, we can make a special coordinate choice, the so-called coincident gauge, where the affine connection \( \Gamma^{\alpha}_{~\mu\nu} = 0 \), simplifies the theory considerably and reduces the non-metricity tensor to  
\begin{equation}
Q_{\alpha\mu\nu} = \partial_{\alpha} g_{\mu\nu}.
\end{equation}
This choice simplifies the computations by making the metric the only fundamental variable. However, it breaks the diffeomorphism invariance of the action except in the case of STGR. Alternatively, a covariant approach to \( f(Q) \) gravity may be used, in which the affine connection is treated explicitly even in the absence of curvature and torsion, thus preserving general covariance.
However, it is possible to rewrite the metric field Eq.~(\ref{fQfield}) as \cite{heisenberg2023}
\begin{equation}
f'(Q)\, G_{\mu\nu}
- \frac{1}{2}\bigl(f(Q) - f'(Q)\, Q\bigr) g_{\mu\nu}
+ 2 f''(Q)\, P^{\alpha}{}_{\mu\nu}\, \partial_{\alpha} Q
= \kappa^2 \, \mathcal{T}_{\mu\nu}.
\label{eq:fQ_field_equations}
\end{equation}

In this form, it is evident that \( f''(Q) = 0 \), (with \( f'(Q) = \text{constant} \)), 
reproduces the Einstein field equations with a cosmological constant. 

\section{Discussion on various equations of state}

The EoS plays a fundamental role in determining the internal composition, stability, and global structure of compact stars. In relativistic stellar structure, the EoS provides the closure relation between the pressure $p$ and the energy density $\rho$ (or equivalently, the baryon number density $n_B$). Depending on the physical motivation and the level of microscopic detail incorporated, EoS models may be broadly classified into two categories: (i) analytical or geometrical EoS, and (ii) realistic, microphysics-based, tabulated EoS. In this section, we summarize both classes and their relevance in compact-star modeling, particularly within modified gravity frameworks.
\subsection{Analytical equations of state}
Geometrical or analytical EoS refers to simplified relations between $p$ and $\rho$ that are not derived from nuclear many-body theory but are instead chosen for their mathematical tractability or phenomenological insight. These EoS are especially useful for studying the qualitative effects of modified gravity, anisotropy, electric charge, exotic matter, or geometric corrections on the stellar structure.

\begin{enumerate}
\item \textbf{Quadratic polytropic EoS:}
The quadratic polytropic EoS is expressed as \cite{chandrasekhar1939}
\begin{eqnarray}\label{eos1st}
    p_r = \gamma\, \rho^{1+\frac{1}{n}} + \beta\,\rho + \chi,
\end{eqnarray}
where $\gamma$, $\beta$, and $\chi$ are constant parameters and $n$ represents the polytropic index. This EoS introduces a combination of nonlinear and linear density contributions, enabling a unified description of different phases of dense matter. The parameter $\gamma$ controls the nonlinear part of the pressure, while $\beta$ and $\chi$ determine the linear and constant components, respectively. Due to its high nonlinearity, a polytropic index $n = 1$ is adopted for analytical simplification, reducing Eq.~(\ref{eos1st}) to the quadratic form $p_r = \gamma\rho^2 + \beta\rho + \chi$. In this representation, the term $\gamma\rho^2$ is typically associated with a Bose–Einstein condensate like a neutron liquid, capturing strong interaction effects at supranuclear densities, whereas the linear contribution $\beta\rho + \chi$ naturally connects to the phenomenology of quark matter, forming a bridge between baryonic and quark phases. Thus, this quadratic polytropic EoS provides a versatile framework for modeling hybrid compact stars that exhibit both hadronic and quark-matter characteristics.

\begin{enumerate}
    \item \textit{\textbf{MIT Bag EoS}:}  
The quadratic polytropic EoS reproduces the MIT Bag model when the parameters take the specific values $\gamma = 0$, $\beta = 1/3$, and $\chi = -4B_g/3$, where $B_g$ is the bag constant \citep{farhi1984}. Under this assignment, Eq.~(\ref{eos1st}) reduces to the well-known MIT bag relation:
\begin{equation}
     p_r = \frac{1}{3}(\rho - 4B_g).
 \end{equation}
The above EoS describes non-interacting quarks confined within a vacuum pressure. In this context, the linear term represents the kinetic pressure of quarks, while the constant term accounts for the vacuum energy required for confinement. The absence of the nonlinear term ($\gamma = 0$) highlights the purely quark-based nature of the MIT model. This EoS successfully captures the essential behavior of strange quark matter and is extensively used in modeling strange quark stars.

\item \textit{\textbf{Linear EoS}:}  
The linear portion of the quadratic polytropic EoS. represented by $p_r = \beta\rho + \chi$, also functions as an independent model known as the linear EoS. It is the generalized form of the MIT Bag EoS where the constant terms $\beta$ and $\chi$ play a crucial role in characteristic baryonic matter. It is frequently used to describe simple barotropic fluids in phenomenological studies. When $\beta$ is treated as the squared sound speed and $\chi$ acts as an offset density contribution, the linear EoS becomes a convenient analytical tool for exploring compact objects in modified gravity frameworks. In hybrid configurations, this linear term captures the contribution of quark matter even when nonlinear interactions (encoded through $\gamma$) are present. 

\end{enumerate}

\item \textbf{Van der Waals type EoS:} Motivated by the thermodynamics of non-ideal fluids, the Van der Waals equation incorporates both short-range repulsive and long-range attractive interactions \cite{hendi2016}. Its compact-star formulation is given by
\begin{equation}
    \left(p + \frac{a}{\rho^2}\right)(\rho - b) = c_s^2 \rho ,
\end{equation}
where $c_s^2=\frac{\partial p}{\partial \rho}$ is the speed of sound and the constants $a$ and $b$ carry clear physical interpretations: $b$ denotes an \textit{excluded volume} correction, accounting for the finite size of the matter constituents and producing a repulsive effect at high densities.  where, $a$ measures the strength of an effective \textit{attractive interaction}, leading to a reduction in pressure through the term $a/\rho^2$.
In relativistic astrophysics, Van der Waals-type EoS have been applied to model matter with significant self-interactions, including interacting dark fluid stars, anisotropic interiors, and exotic compact objects. Such an EoS exhibits physically important qualitative behavior: The pressure is stiffened at high densities because of repulsive interactions, while it is softened in intermediate density regimes as a result of attractive interactions. It describes the non-linear pressure-density relations that mimic realistic nuclear interactions.
Thus, the Van der Waals EoS serves as a useful phenomenological framework for studying microphysical interactions within geometrically modified gravity theories.

\item \textbf{Chaplygin gas EoS:} The (generalized) Chaplygin gas EoS is given by \cite{kamenshchik2001,bilic2002},
\begin{equation}
    p = -\frac{A}{\rho^\beta},
\end{equation}
where $A>0$ and $0 \le \beta \le 1$. This model is distinguished by its intrinsically negative pressure, which plays a crucial role in determining the dynamical behavior of the fluid. At extremely high densities, the negative pressure term becomes negligible, causing the fluid to behave similarly to dust or weakly interacting matter. However, as the density decreases, the magnitude of the negative pressure increases, generating a repulsive, dark-energy-like effect. Owing to this dual nature mimicking ordinary matter at high densities and dark energy at lower densities, the Chaplygin gas has been explored in the context of compact objects to model dark fluid interiors, exotic stars with repulsive cores, and configurations where internal negative pressure counterbalances gravitational collapse. The parameter $\beta$ governs the degree of generalization: $\beta = 1$ corresponds to the original Chaplygin gas, $0 < \beta < 1$ produces the generalized form with adjustable stiffness, and $\beta = 0$ reduces the EoS to a constant-pressure behavior similar to vacuum energy. This versatile EoS therefore provides a useful phenomenological framework for capturing exotic matter characteristics within modified gravity or anisotropic stellar models.

\end{enumerate}

\subsection{Realistic equations of state}

Realistic EoS constitute the most reliable description of matter inside neutron stars and other compact objects, as they are constructed from state-of-the-art nuclear many-body theories and are constrained by both laboratory experiments and astrophysical observations. These EoS incorporate microphysical interactions at sub-nuclear, nuclear, and supra-nuclear densities using approaches such as Relativistic Mean-Field (RMF) theory, Skyrme-type energy density functionals, variational methods, Chiral Effective Field Theory (ChEFT), and many-body perturbation techniques. Modern observational constraints, particularly the existence of $\sim 2 M_{\odot}$ pulsars and tidal deformability bounds from the gravitational-wave events GW170817 \cite{abott} and GW190814 \cite{abott2} play a key role in shaping viable models. Since realistic EoS do not usually permit simple analytic expressions, they are represented in tabulated form, providing numerical relations among pressure, energy density, baryon density, and other thermodynamic quantities.

\begin{enumerate}
    \item \textbf{RMF-based EoS:}  
    RMF theories model dense matter by treating nucleons as Dirac particles interacting through meson exchanges, most commonly the scalar $\sigma$, vector $\omega$, and isovector $\rho$ mesons. Well-established RMF parameterizations including TM1, NL3, GM1, GM3, DD-ME2, DD2, and the recently developed DD-MEX \cite{lattimer2016,shen1998} successfully reproduce nuclear saturation properties and heavy-ion phenomenology. In particular, density-dependent couplings in modern RMF sets significantly improve consistency with empirical data. Many of these EoS predict maximum neutron star masses above $2M_{\odot}$, making them compatible with the most massive observed pulsars.

    \item \textbf{Skyrme-Type and BSk Family:}  
    The Brussels-Skyrme (BSk) series (BSk19, BSk20, BSk21) \cite{goriely2010} represents one of the most comprehensive sets of unified crust core EoS models. Constructed using generalized Skyrme energy density functionals fitted to over 2000 measured nuclear masses and constrained by neutron-matter calculations, these models offer varying degrees of stiffness: BSk19 is comparatively soft, BSk20 has intermediate stiffness, and BSk21 is sufficiently stiff to support $2M_{\odot}$ neutron stars. Their unified treatment ensures thermodynamic consistency across the outer crust, inner crust, and core.

    \item \textbf{Chiral EFT-based EoS:}  
    At low to moderate densities ($\lesssim 2\rho_{\rm sat}$), ChEFT provides one of the most robust frameworks for describing nuclear interactions based on QCD symmetries. EoS developed by Hebeler et al. and Drischler et al. \cite{hebeler2013,drischler2021} use ChEFT constraints at low densities and match them to high-density extensions informed by phenomenological or RMF models. These EoS form a key ingredient in uncertainty quantification for neutron star radii and tidal deformabilities.

    \item \textbf{APR, SLy, FPS and related models:}  
    The APR model (Akmal-Pandharipande-Ravenhall) \cite{akmal1998} remains one of the most widely used realistic EoS, incorporating two-body and three-body forces through variational chain summation techniques. Other important models include the SLy4/SLy9 EoS, derived from Skyrme interactions tailored for neutron-rich matter \cite{douchin2001}, and the soft FPS model, historically significant in early neutron star structure calculations. These EoS provide valuable benchmarks for stellar modeling across different density regimes.

    \item \textbf{Hyperonic, hybrid and crossover EoS:}  
    At even higher densities, additional degrees of freedom may emerge, such as hyperons ($\Lambda$, $\Sigma$, $\Xi$), deconfined quarks, or mixed hadron-quark phases. Hyperonic EoS like GM1Y and DD2Y incorporate strange baryons, whereas hybrid and quark-matter models such as ALF2 and ALF4 include hadron-quark transitions. More recent QCD-inspired crossover models \cite{baym2018} aim to describe a smooth transition between hadronic and quark matter. These EoS are essential for exploring exotic phases and for interpreting multi-messenger observations from neutron star mergers.
\end{enumerate}
\noindent 
Realistic EoS are the cornerstone for connecting theoretical stellar models with modern astrophysical data from NICER, radio pulsar timing, and gravitational-wave measurements. In the context of modified gravity theories, such EoS play a crucial role, as they allow deviations from GR to be distinguished from uncertainties in the microphysics of dense matter. Their use therefore enhances the reliability and credibility of theoretical predictions.

\section{Physical properties of compact star}
In order to assess the physical viability of any compact star model, it is essential to examine a set of fundamental properties that arise from both gravitational theory and stellar matter characteristics. These physical features provide consistency checks on the internal structure, stability, and observational compatibility of the star. In the following subsections, we will briefly discuss these physical characteristics to establish the reliability of the proposed stellar model.

\subsection{Energy conditions}
The concept of energy conditions generalizes the notion that the energy density cannot be negative, extending it to the full stress-energy tensor. Permitting arbitrary positive and negative regions in energy density would destabilize the vacuum, allowing it to split into zones with positive and negative energy, which would lead to an ever-increasing ``\textit{energy density gap}''. To prevent such runaway instabilities, energy conditions impose mathematical constraints derived from the Raychaudhuri equations \cite{raychaudhuri}, governing the focusing of geodesic congruences (timelike, spacelike, or null). These conditions translate into key inequalities among thermodynamic quantities such as the energy density ($\rho$) and the principal pressures ($p_r$ and $p_t$), thus distinguishing gravitational attraction in strong-field scenarios. They provide essential criteria for validating whether a matter distribution is physically reasonable and for determining if the presence of exotic matter is required to describe a self-gravitating system.

There are four primary ``pointwise'' energy conditions encountered in relativistic physics: the Null Energy Condition (NEC), the Weak Energy Condition (WEC), the Strong Energy Condition (SEC), and the Dominant Energy Condition (DEC). A further criterion, less frequently encountered, is the Trace Energy Condition (TEC), now mainly of historical interest. The logical interrelations between these energy conditions are depicted schematically in Fig.~\ref{f1100}, where an arrow from one condition to another signifies a logical implication.

\begin{center}
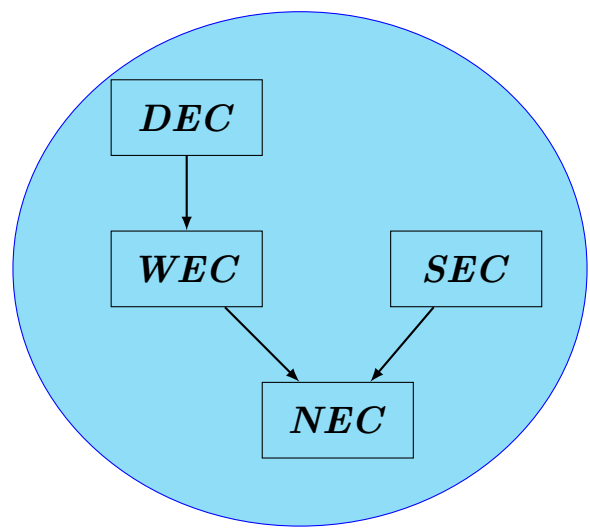

\begin{tikzpicture}
\draw[blue,fill= bluE] (7,0) ellipse (3.8 and 3.4);
\node[rectangle,draw = black, minimum width = 2cm, minimum height = 1cm] (dec) at (5.5,2) {\Large \textit{\textbf{DEC}}};
\node[rectangle,draw = black, minimum width = 2cm, minimum height = 1cm] (wec) at (5.5,0) {\Large \textit{\textbf{WEC}}};
\node[rectangle,draw = black, minimum width = 2cm, minimum height = 1cm] (nec) at (7.5,-2) {\Large \textit{\textbf{NEC}}};
\node[rectangle,draw = black, minimum width = 2cm, minimum height = 1cm] (sec) at (9.2,0) {\Large \textit{\textbf{SEC}}};
\draw[-latex, thick] (dec) to (wec);
\draw[-latex, thick] (wec) to (nec);
\draw[-latex, thick] (sec) to (nec);
\end{tikzpicture}
\captionof{figure}{A schematic diagram among the relation of energy condition: An arrow from A to B in the illustration indicates that A implies B.}
\label{f1100}
\end{center}

The \textit{NEC} requires the energy density measured by any null observer to be non-negative. It serves as a foundational requirement upon which both the WEC and the SEC are built. Formally, for any null vector $l^{m}$,
$\mathcal{T}_{mn} \, l^{m} l^{n} \geq 0,$ which translates to the effective inequalities 
\begin{eqnarray}
    \rho + p_r \geq 0, \qquad \rho + p_t \geq 0.
\end{eqnarray}
The \textit{WEC} stipulates that the energy density measured by any timelike observer cannot be negative, prohibiting the presence of locally negative energy. Mathematically, for every timelike vector $t^{m}$ $\mathcal{T}_{mn} \, t^{m} t^{n} \geq 0.$, This leads to the effective inequalities
\begin{eqnarray}
    \rho \geq 0, \qquad \rho + p_r \geq 0, \qquad \rho + p_t \geq 0,
\end{eqnarray}
Thus, a positive local mass-energy density is ensured.

The \textit{SEC} asserts that gravity remains attractive, enforcing the focus of timelike geodesics. This requirement implies that matter must tend to cluster gravitationally. However, observational data indicate that the SEC was violated during certain epochs in cosmic history, particularly between galaxy formation and the present, suggesting the influence of dark energy or some form of exotic matter. The SEC is expressed as
\begin{eqnarray}
    \left( \mathcal{T}_{mn} - \frac{1}{2} \mathcal{T} g_{mn} \right) t^{m} t^{n} \geq 0 \implies
    \rho + p_r \geq 0, \qquad \rho + p_t \geq 0, \qquad \rho + p_r + 2p_t \geq 0.
\end{eqnarray}
The \textit{DEC} ensures that the energy flux never exceeds the speed of light and that the energy propagates along causal, physically meaningful trajectories. In mathematical terms, for any pair of co-oriented timelike vectors $t^{m}$ and $\eta^{n}$, $\mathcal{T}_{mn} \, t^{m} \eta^{n} \geq 0$. This condition results in the following effective inequalities
\begin{eqnarray}
    \rho \geq 0, \qquad 
    \rho \pm p_r \geq 0, \qquad
    \rho \pm p_t \geq 0,
\end{eqnarray}
ensuring the causal propagation of energy and matter.

\subsection{Compactness parameter (CP)}
The compactness parameter is a dimensionless quantity that measures the strength of relativistic gravitational effects in a self-gravitating object. It is defined as the ratio of the gravitational mass $M$ to the radius $R$ \cite{lattimer2007}
\begin{equation}
\mathcal{C} = \frac{M}{R} = \frac{R_s}{2R},
\end{equation}
where $R_s = 2M$ is the Schwarzschild radius. For neutron stars, typical compactness values lie in the range
$0.1 \lesssim \mathcal{C} \lesssim 0.3$,
with the precise value depending on the EoS and the stellar-mass. The compactness parameter naturally appears in the TOV equation governing hydrostatic equilibrium in relativistic stars, where it controls the relative importance of relativistic corrections arising from pressure and spacetime curvature. As $\mathcal{C}$ increases, gravitational fields become stronger, requiring steeper pressure gradients to counterbalance gravity. Consequently, highly compact configurations demand a stiffer equation of state to remain stable against collapse.

\subsection{Buchdahl's limit}

A fundamental upper bound on the compactness of relativistic stars was established by Buchdahl in 1959 \cite{buchdahl1959} for static, spherically symmetric configurations composed of an isotropic perfect fluid within GR. Under the assumption that the energy density is non-increasing outward and that the pressure remains finite and positive throughout the stellar interior, Buchdahl showed that the compactness must satisfy 
\begin{equation}
\frac{2M}{R} < \frac{8}{9},
\quad \text{or, equivalently,}~~~~ R > \frac{9}{8} R_s .
\end{equation}
This result follows from the integration of the TOV equation and represents a purely relativistic constraint independent of the detailed microphysics of the EoS. The Buchdahl limit demonstrates that no stable, isotropic fluid configuration can exist arbitrarily close to its Schwarzschild radius without undergoing gravitational collapse and turning into a black hole. As this bound approaches, central pressures diverge, signaling the breakdown of hydrostatic equilibrium. Therefore, Buchdahl's limit provides a crucial theoretical benchmark for assessing the physical viability of compact star models.

\subsection{Gravitational redshift and stability}

An important observable consequence of strong gravitational fields in compact stars is the gravitational redshift, which measures the frequency shift of photons emitted from the stellar surface and received by a distant observer. For a static, spherically symmetric spacetime described by the exterior Schwarzschild metric, the gravitational redshift is given by \cite{wald1984}
\begin{equation}
z = \left( 1 - \frac{2M}{R} \right)^{-1/2} - 1 ,
\end{equation}
where $M$ and $R$ denote the gravitational mass and radius of the star, respectively. This expression follows directly from the temporal component of the Schwarzschild metric, $g_{tt} = -(1 - 2M/R)$, by comparing the proper time experienced by an emitter at the stellar surface to that measured by an observer at spatial infinity. The gravitational redshift therefore provides a direct probe of the compactness parameter and the strength of relativistic effects at the stellar surface.
 Invoking Buchdahl's compactness limit for isotropic fluid spheres $2M/R < 8/9$, one obtains an absolute upper bound on the surface gravitational redshift of any stable compact object
\begin{equation}
z < \left(1 - \frac{8}{9} \right)^{-1/2} - 1 = 2 .
\end{equation}
This result demonstrates that within the framework of GR, no stable, non-black hole object composed of isotropic matter can exhibit a surface redshift exceeding $z = 2$. As the Buchdahl limit is approached, the redshift grows rapidly, reflecting the increasing curvature of spacetime and the divergence of central pressure, signaling the onset of gravitational instability. Consequently, any astrophysical object with an observed surface redshift $z > 2$ would violate the standard assumptions of general relativistic stellar stability and would instead correspond to a black hole, for which the redshift formally diverges as the radius approaches the Schwarzschild radius, $R = 2M$ \cite{shapiro1983}.\\
For typical astrophysical objects, the magnitude of the gravitational redshift varies significantly: for the Sun, relativistic effects are extremely weak, yielding $z \sim 2 \times 10^{-6}$; for neutron stars, strong gravity leads to appreciable redshifts in the range $z \sim 0.2$–$0.5$, depending on the EoS and stellar compactness; while for black holes the surface redshift diverges, $z \rightarrow \infty$, reflecting the presence of an event horizon.

\subsection{Speed of sound and causality criterion}

The propagation of small perturbations in neutron star matter is characterized by the adiabatic speed of sound, which plays a fundamental role in determining the stiffness, causality, and stability of the EoS. As discussed in the previous section, within the relativistic perfect fluid framework, the energy momentum tensor is given by
\begin{equation}
\mathcal{T}^{\mu\nu} = (\rho + p) u^\mu u^\nu + p g^{\mu\nu}.
\end{equation}
 The fluid dynamics are governed by the local conservation of energy momentum, $\nabla_\mu \mathcal{T}^{\mu\nu} = 0$,
together with baryon number conservation,
$\nabla_\mu (n u^\mu) = 0$,
where $n$ is the baryon number density.
To derive the mathematical expression for the speed of sound, small linear perturbations are introduced around a static equilibrium configuration such as \cite{haensel2007}
\begin{equation}
\rho = \rho_0 + \delta \rho, \qquad
p = p_0 + \delta p, \qquad
u^\mu = u_0^\mu + \delta u^\mu .
\end{equation}
For cold neutron star interiors, the perturbations are effectively adiabatic, implying that the pressure perturbation is related to the energy density perturbation through
\begin{equation}
\delta p = \left( \frac{\partial p}{\partial \rho} \right)_s \delta \rho ,
\end{equation}
where the derivative is taken at constant entropy per baryon. Linearization of conservation equations and the combination of the resulting relations lead to a relativistic wave equation for density perturbations with the dispersion relation $\omega^2 = c_s^2 k^2$. The adiabatic speed of sound is therefore identified as \cite{rezzolla2013}
\begin{equation}
\boxed{
c_s^2 = \left( \frac{\partial p}{\partial \rho} \right)_s
}
\end{equation}
in the local rest frame of the fluid. Alternatively, when the EoS is expressed in terms of the baryon number density, the speed of sound can be written as
$c_s^2 = \frac{dp/dn}{d\rho/dn}$,
a form particularly useful for numerical calculations of neutron stars.
For \emph{anisotropic} compact stars, the speed of sound becomes direction-dependent. One therefore defines two independent sound speeds: the \textit{radial sound speed},  $c_{sr}^{2} \equiv \left(\frac{dp_r}{d\rho}\right)$,
which governs the propagation of perturbations along the radial direction, and the \textit{tangential sound speed},
    $c_{st}^{2} \equiv \left(\frac{dp_t}{d\rho}\right)$,
which characterizes propagation along the directions orthogonal to the radial one. Both quantities must individually satisfy the causality requirement $0 \leq c_{sr}^{2},\, c_{st}^{2} \leq 1$ throughout the stellar interior.
\paragraph{Herrera's cracking condition.}
For anisotropic stellar configurations, Herrera~\cite{herrera1992} introduced an additional local stability criterion widely known as the \emph{cracking condition} which examines the response of the fluid to local perturbations of the energy density. According to this criterion, a stable anisotropic configuration must satisfy
   $ -1 \leq c_{st}^{2} - c_{sr}^{2} \leq 0$,
throughout the stellar interior. The condition $c_{st}^{2} - c_{sr}^{2} \leq 0$ (no-cracking) ensures that no instability develops within the stellar matter, while the lower bound preserves causality. Regions where $c_{st}^{2} - c_{sr}^{2} > 0$ are said to exhibit \emph{cracking}, indicating the onset of local instability. 
From a physical perspective, the speed of sound quantifies the response of ultra-dense matter to local compressional perturbations and thus serves as a direct measure of the stiffness of the EoS. Larger values of $c_s^2$ correspond to a more rapid increase of pressure with energy density, enabling matter to sustain stronger gravitational fields and support more massive neutron stars.
\paragraph{Causality Criterion:}Beyond its role in characterizing the stiffness of the EoS, the adiabatic speed of sound is subject to a fundamental causality constraint, which requires that no physical perturbation propagates faster than the speed of light. In relativistic units ($G=c=1$), this condition imposes the bound \cite{rezzolla2013}
\begin{equation}
0 \leq c_s^2 \leq 1 ,
\end{equation}
throughout the stellar interior. This requirement provides a stringent upper limit on the admissible stiffness of dense matter, as an excessively rapid increase of pressure with energy density would imply superluminal signal propagation and hence violate relativistic causality. The causality condition becomes particularly restrictive at supranuclear densities, where large sound speeds are often required to support massive neutron stars against gravitational collapse. Consequently, the simultaneous enforcement of hydrostatic equilibrium, stability, and causality severely constrains the allowed EoS and the resulting mass-radius configurations of neutron stars. In the context of modified theories of gravity, although the definition of the speed of sound and the local causality bound on the sound speed remain unchanged, deviations from GR can alter the global stellar structure equations, thereby modifying the interplay between causality and macroscopic properties such as compactness and maximum mass.

\subsection{Adiabatic index and stability of neutron stars}

Based on the definition of the adiabatic speed of sound,
$c_s^2 = \left( \frac{\partial p}{\partial \rho} \right)_s$,
a closely related quantity of fundamental importance for stellar stability is the adiabatic index $\Gamma$, which provides a dimensionless measure of the stiffness of the EoS. It is defined as \cite{rezzolla2013}
\begin{equation}
\Gamma = \left( \frac{\partial \ln p}{\partial \ln n} \right)_s
= \frac{n}{p} \left( \frac{\partial p}{\partial n} \right)_s ,
\end{equation}
where $n$ denotes the baryon number density. For cold, catalyzed neutron star matter, the first law of thermodynamics yields the identity,
\begin{equation}
d\rho = \frac{\rho + p}{n}\, dn ,
\end{equation}
which allows the adiabatic index to be expressed directly in terms of the speed of sound as
\begin{equation}
\Gamma = \frac{\rho + p}{p}\, c_s^2.
\end{equation}
This relation establishes an explicit connection between the microscopic nature of dense matter and compressional perturbations, encoded in $c_s^2$, and the macroscopic stability properties of relativistic stars.
The adiabatic index plays a central role in determining the dynamical stability of neutron stars against radial perturbations. In the Newtonian limit, Chandrasekhar showed that a spherically symmetric self-gravitating fluid becomes unstable for $\Gamma < 4/3$, indicating insufficient pressure support against collapse \cite{chandrasekhar1964,shapiro1983}. In GR, relativistic corrections arising from strong gravity and pressure contributions tighten this condition to $\Gamma > 4/3 + \Delta_{\mathrm{GR}}$, where $\Delta_{\mathrm{GR}}$ depends on stellar compactness $\mathcal{C} = 2M/R$ and pressure profile \cite{shapiro1983}. These effects are closely related to the Buchdahl bound $2M/R < 8/9$, beyond which there is no stable isotropic configuration \cite{buchdahl1959}. As this limit approaches, enhanced relativistic effects require larger values of $\Gamma$ to maintain stability, particularly in the stellar core \cite{astashenok2013,doneva2013}. In modified gravity theories, although the formal definition of $\Gamma$ remains unchanged, modifications to field equations shift the critical stability threshold.

\section{Motivation to study compact stars in modified gravity}
From an astrophysical perspective, neutron stars and other compact objects are among the most powerful natural laboratories for testing modified theories of gravity. They allow exploration of the strong-field, high-curvature, and high-density regimes, where the limitations of GR become apparent. 
\begin{enumerate}
    \item \textbf{Strong-field regime and curvature effects:}
Neutron stars possess extremely high central densities (of order $10^{18} ~\text{kg}/\text{m}^3$) and intense gravitational fields where relativistic corrections dominate. These conditions deviate significantly from the weak-field regime where GR has been tested with the highest precision \cite{will2014,psaltis2008}. Modified gravity models introduce corrections to Einstein’s equations that become significant at these curvature scales, leading to observable deviations in mass–radius relations, compactness, or surface redshift.
\item \textbf{Equation of state dependence and degeneracy breaking:}
Determining the EoS for dense nuclear matter encounters degeneracy issues in GR-based models because different EoS choices can mimic similar observable properties \cite{lattimer2007}. Modified gravity theories alter the TOV equations, modifying the relation between pressure, density, and gravitational potential. This allows for disentangling gravitational effects from micro-physical EoS uncertainties, providing an independent handle to test gravity models against pulsar and gravitational-wave data.
\item \textbf{Maximum mass and stability constraints:}
Observations of massive pulsars, such as the PSR J0740+6620 or the PSR J0348+0432 exceeding two solar masses \cite{demorest2010,antoniadis2013}, challenge some EoS when interpreted strictly under GR. Modified gravity models, especially 
the frameworks $f(R)$ \cite{astashenok2013}, $f(T)$ \cite{aldrovandi2013}, or $f(R,\mathcal{T})$ \cite{harko2011}, can yield different upper limits for stellar-masses and radii through additional curvature or matter-coupling terms. These differences can reconcile observed high masses and radii without the assumption of exotic matter.
\item \textbf{Gravitational waves and binary mergers:}
Neutron star mergers, observed through events such as GW170817 \cite{abott}, offer constraints on the tidal deformability parameter, which encodes the star’s response to external fields. Modified gravity can alter the tidal Love numbers through changes in spacetime curvature, influencing the waveform phase evolution. These deviations can be compared with precise gravitational-wave data to test alternative gravity models on strong-field, dynamical scales.
\item \textbf{Surface redshift and compactness tests:}
Spectroscopic measurements of surface atomic lines and gravitational redshifts from neutron stars allow relativistic predictions to be tested. Under modified gravity, the redshift factor depends on the model parameters, such as coupling constants in 
$f(R,G,L_m)$ \cite{bertolami2007},
$f(R,\mathcal{T})$ gravity, providing direct observable consequences. Hence, accurate measurements of surface emissions and periodic oscillations can constrain deviations from GR predictions.
\end{enumerate}
In summary, neutron stars act as cosmic laboratories for probing high-energy gravitational physics. Their mass-radius behavior, cooling evolution, and tidal responses under modified gravity theories provide critical, testable deviations from GR that are inaccessible through solar-system or cosmological observations.
In this chapter, we explored torsion and nonmetricity-based extensions of GR, with particular emphasis on $f(T)$ and $f(Q)$ gravity, where gravitational interactions are reformulated in terms of torsion and nonmetricity rather than curvature. This discussion establishes the theoretical foundation for the analytical and numerical investigations presented in the subsequent chapters, which aim to assess the capability of $f(Q)$ gravity to describe neutron star structure in the strong-field regime.

\chapter{\fontsize{14}{16}\selectfont
A Comprehensive Study of Massive Compact Star Admitting Conformal Motion Under Bardeen Geometry} 

\label{Chapter2} 

\definecolor{maroon}{RGB}{128, 0, 0}
\lhead{\textcolor{maroon}{\textit{\textbf{Chapter 2:}}} \emph{\textcolor{maroon}{A Comprehensive Study of Massive Compact Star Admitting Conformal Motion Under Bardeen Geometry}}} 

\blfootnote{*The work in this chapter is covered by the following publication:\\
\textit{A Comprehensive Study of Massive Compact Star Admitting Conformal Motion Under Bardeen Geometry}, Nuclear Physics B, \textbf{1002}, 116523 (2024).}



In this chapter, we investigate the existence of charged compact star configurations within the framework of 
$f(Q)$ gravity, utilizing the conformal motion approach and incorporating the MIT bag EoS to describe the relation between pressure and energy density inside the compact star. The brief summary of this work is as follows.
\begin{itemize}
    \item Two distinct stellar models are constructed by considering power-law and linear forms of the conformal factor, allowing a comparative analysis between the models.
    \item To find the parameter values, the interior spacetime metric is smoothly matched with the Bardeen spacetime at the stellar boundary through appropriate junction conditions.

    \item The physical properties of several observed compact stars, namely PSR J1614–2230, PSR J1903+327, Vela X-1, Cen X-3 and SMC X-1, are examined to test the astrophysical viability of the models.

    \item A detailed comparative study of the two models is carried out by analyzing the behavior of energy density, pressures, equilibrium conditions, and the adiabatic index.

    \item A brief analysis is also performed by considering Reissner–Nordström (R-N) spacetime as an alternative exterior geometry for the matching condition.

    \item It is found that, in comparison to the R-N case, the Bardeen spacetime with additional asymptotic corrections provides more viable and physically appealing solutions.
    
    \item The overall analysis confirms that the resulting charged compact star models in the presence of conformal motion within $f(Q)$ gravity are physically acceptable, stable, and realistic.
    
\end{itemize}

\section{Introduction}

The study of compact stars has attracted considerable attention due to their extreme density and strong gravitational fields. Compact objects such as white dwarfs, neutron stars, and quark stars possess small radii and large masses, making them excellent laboratories for testing gravitational theories beyond GR. Earlier, stellar models were based on isotropic matter distributions; however, Ruderman \cite{RR} first emphasized the importance of pressure anisotropy in compact stars. Since then, anisotropic stellar configurations, including charged compact stars, have been extensively explored using various EoS \cite{RB,SK,KN}.

Conformal symmetries play a crucial role in simplifying Einstein’s field equations and in establishing a deeper connection between spacetime geometry and matter content. Several studies have shown that compact stars admitting Conformal Killing Vectors (CKVs) can lead to physically viable stellar models \cite{LH1,LH2,LH3,LH4,MFS,MK,ME}. In particular, charged strange quark star solutions admitting conformal motion have been successfully constructed within GR and alternative gravity theories.

Motivated by the limitations of GR in explaining dark matter and dark energy phenomena, various modified gravity theories such as $f(R)$, $f(G)$, $f(R,T)$, and $f(Q)$ have been proposed \cite{1,2,8,9,11}. Among them, symmetric teleparallel gravity, where gravity is described through non-metricity, has gained increasing attention \cite{jimenez2018}. This theory has been successfully applied to both cosmological and astrophysical scenarios \cite{OS1,OS2,SP,SM1}.

This chapter is organized as follows: In section \ref{secII}, we briefly discuss the mathematical formalism of $f(Q)$ gravity and derive the field equation. Next, by introducing the conformal motion technique, we develop two models in our study in section \ref{secIII}. Furthermore, we have examined various physically valid characteristics and stability of realistic stars, such as PSR J$1614-2230$, PSR
J$1903+327$, Vela X$-1$, Cen X$-3$, and SMC X$-1$ in section \ref{seciv1} and section \ref{secv2}. A brief comparative study has been done between the R-N spacetime and the Bardeen spacetime as an outer structure in section \ref{vi}. Finally, we conclude in section \ref{vii}.  

\section{Field equations in $f(Q)$ gravity}\label{secII}

To describe a charged compact star in the framework of $f(Q)$ gravity, we consider the contribution of both matter and electromagnetic fields. The electromagnetic energy momentum tensor is given by \cite{Jackson_ClassicalED}
\begin{equation}
\varepsilon_{ij}
=F_{ik}F_{j}^{\ k}-\frac{1}{4}g_{ij}F_{kl}F^{kl},
\end{equation}
where the electromagnetic field tensor is defined in terms of the four-potential $\mathcal{A}_i$ as
\begin{equation}
F_{ij}=\mathcal{A}_{i,j}-\mathcal{A}_{j,i}.
\end{equation}

The Maxwell equations in curved spacetime are expressed as
\begin{equation}\label{maxwell}
    F_{ij,k} + F_{ki,j} + F_{jk,i} = 0,
    \qquad
    \left(\sqrt{-g}\,F^{ij}\right)_{,j} = \frac{1}{2}\sqrt{-g}\,J^{i},
\end{equation}
where $J^{i}$ denotes the four-current density. For a static, spherically symmetric charged fluid distribution, only the temporal component $J^{0}$ is non-vanishing, leading to a purely radial electric field. Consequently, the only non-zero component of the electromagnetic field tensor is $F_{01}=-F_{10}$. Solving the Maxwell equations under these assumptions, the electric field takes the form of
\begin{equation}
E(r)=\frac{q(r)}{r^2},
\end{equation}
where $q(r)$ represents the total electric charge enclosed within radius $r$, and the charge density is given by
\begin{equation}
\sigma(r)=\frac{e^{-\lambda/2}}{4\pi r^2}\left(r^2E\right)'.
\end{equation}
Now, assuming the static, spherically symmetric spacetime metric \ref{metric1st},
 the non-metricity scalar in $f(Q)$ gravity is obtained as \cite{wang2022}
\begin{equation}
Q=-\frac{2e^{-\lambda(r)}\left(r\nu'(r)+1\right)}{r^2}.
\end{equation}
Using these mathematical components in Eq. \ref{fQfield}, the modified field equations of $f(Q)$ gravity for a charged isotropic fluid configuration can be written as
\begin{align}
\rho^{\text{eff}}+\frac{q^2}{r^4}
&=\frac{f(Q)}{2}
-f_Q\!\left[Q+e^{-\lambda}\!\left(\frac{\lambda'}{r}-\frac{1}{r}\right)
+\frac{1}{r^2}\right], \\
p^{\text{eff}}-\frac{q^2}{r^4}
&=-\frac{f(Q)}{2}
+f_Q\!\left(Q+\frac{1}{r^2}\right), \\
p^{\text{eff}}+\frac{q^2}{r^4}
&=-\frac{f(Q)}{2}
+f_Q\!\left[\frac{Q}{2}
-e^{-\lambda}\!\left\{
\frac{\nu''}{2}
+\left(\frac{\nu'}{4}+\frac{1}{2r}\right)
(\nu'-\lambda')
\right\}\right].
\end{align}

These equations govern the interior structure of charged compact stars in $f(Q)$ gravity and form the basis for constructing physically viable stellar models in the following sections.

\section{Conformal motion treatment}\label{secIII}

CKVs play an important role in simplifying the field equations by imposing symmetry on the spacetime geometry. A spacetime admitting conformal motion satisfies \cite{LH1}
\begin{equation}\label{conf}
\mathcal{L}_{\xi} g_{ij}=\Phi\, g_{ij},
\end{equation}
where $\mathcal{L}_{\xi}$ denotes the Lie derivative along the vector field $\xi$, and $\Phi$ is the conformal factor. If $\Phi=0$, Eq.~(\ref{conf}) reduces to a Killing vector, and a constant $\Phi$ corresponds to a homothetic vector.
For the static and spherically symmetric metric given in Eq.~(\ref{metric1st}), the conformal Killing equations lead to
\begin{equation}
\xi^{1}\nu'=\Phi, \qquad
\xi^{1}\lambda'+2\,\partial_{r}\xi^{1}=\Phi, \qquad
\xi^{4}=K, \quad ~\xi^1 =\frac{\Phi r}{2}.
\end{equation}
which admits the simultaneous solution \cite{LH2}
\begin{equation}\label{cnf}
e^{\nu}=H^{2}r^{2}, \qquad
e^{\lambda}=\left(\frac{I}{\Phi}\right)^{2}, \qquad
\xi^{i}=K\delta^{i}_{4}+\frac{r\Phi}{2}\delta^{i}_{1},
\end{equation}
where $H$, $I$, and $K$ are arbitrary constants. The symbols $\delta^{i}_{\,1}$ and $\delta^{i}_{\,4}$ are Kronecker deltas, which select the radial ($i = 1$) and temporal ($i = 4$) components of the CKV $\xi^{i}$, respectively, so that the only non-vanishing components are $\xi^{1} = r\Phi/2$ and $\xi^{4} = K$. In this work, we construct compact star models by choosing two different functional forms of the conformal factor $\Phi(r)$.

\subsection{Model-I}
In our first proposed model, under the assumption that $\Phi$ depends solely on the radial coordinate $r$, we have assumed the functional form of $\Phi(r)$ as follows
\begin{eqnarray}\label{phi1}
   \Phi(r)=I\sqrt{\psi(r)}.
\end{eqnarray}
The above power-law form of the conformal factor is well established in the context of GR and is motivated by the work \cite{MFS}. Following the discussion in the introduction, we model the interior quark matter using the MIT Bag model EoS, given by
\begin{equation}
p=\frac{1}{3}(\rho-4B_g),
\end{equation}
where $B_g$ is the bag constant \cite{farhi1984}.
Now, using Eqs.~(\ref{cnf},\ref{phi1}) and manipulating the field equations, we get the solution of $\psi(r)$ as
\begin{eqnarray}
   \psi(r)= \frac{1}{3m}(m+B_g r^2)-\frac{nr^2}{6m}+\frac{C}{r^2}.
\end{eqnarray}
Where $C$ is the integration constant. We have considered the linear model of $f(Q)=m\, Q+n$ which is motivated by some other studies given in the above references. Here, $m,n$ are the model parameters.
This study focuses on the case where $C$ is not equal to zero and uses Bardeen geometry as the external spacetime framework to examine compact stars. Therefore, we obtain the subsequent explicit and precise solution that accurately describes the internal geometrical and physical structure of a strange quark star.
\begin{eqnarray}
e^{\nu(r)} &=& H^{2} r^{2},~~~ E^2 = -\frac{m \big(r^2-12 C\big)}{6 r^4},~~~~
\qquad
\rho^{\text{eff}} = \frac{2B_g r^{4}-m(r^{2}+6C)}{2r^{4}}, \nonumber \\[0.3cm]
e^{\lambda(r)} &=& \frac{6 m r^{2}}{-r^{4}(n-2 B_g)+2 m (r^{2}+3C)},
~~~~\qquad
p^{\text{eff}} = -\frac{6 B_g r^{4}+m(r^{2}+6C)}{6r^{4}}.
\end{eqnarray}

Furthermore, the physical parameters exhibit a central singularity because of the employing of conformal symmetries. Indeed, this formalism cannot overcome the core singularity in the physical parameters. Nevertheless, the solutions of a core-envelope-type model can be explored to represent the envelope portion of a star. Now, we will consider appropriate boundary conditions to ensure compatibility between the solutions of the interior spacetime.

\subsubsection{Boundary and matching condition}
Now, one of the most important parts is determining the values of the constants. For that, we usually match the interior geometry with the outer spacetime. In this study, we have matched the interior spacetime with the exterior spacetime of Bardeen, which is given by \cite{Bardeen,ABG}
\begin{eqnarray}
    d s^{2} &=& -L(r) d t^{2}+L(r)^{-1} d r^{2}+r^{2} (d \theta^{2}+ \sin ^{2} \theta d\phi^{2}),~~~
\end{eqnarray}
where $L(r)=1-\frac{2 M r^{2}}{\left(\mathcal{Q}^{2}+r^{2}\right)^{\frac{3}{2}}}$. Here, $M$ is the star's total mass, and $\mathcal{Q}$ is the total charge surroundings of the outer region of the star. By applying a binomial expansion, one can obtain the expression of $L(r)=1-\frac{2 M}{r}+\frac{3 M \mathcal{Q}^{2}}{r^{3}}+O\left(\frac{1}{r^{5}}\right)$. Here, in this expression of $L(r)$, the presence of the fraction term $\frac{1}{r^3}$ distinguishes the Bardeen geometry from the usual R-N spacetime configuration. We shall ignore the term $O\left(\frac{1}{r^{5}}\right)$ and its subsequent quantity because of its modest value. By applying the continuity equation, we have matched the exterior and interior spacetime metric potentials at the boundary.
\begin{eqnarray}
    1-\frac{2 M}{r_{b}}+\frac{3 M \mathcal{Q}^{2}}{r_{b}{ }^{3}} &=& H^2\,r_{b}^2,\\
    \left(1-\frac{2 M}{r_{b}}+\frac{3 M \mathcal{Q}^{2}}{r_{b}{ }^{3}}\right)^{-1} &=& \frac{6 m r_b^2}{2 m \big(r_b^2+3 C-r_b^4 (n-2 B_g )\big)}.\nonumber
\end{eqnarray}
By imposing the above matching conditions, where $r_b$ denotes the stellar radius, we have derived the values of the following constants,
\begin{eqnarray}
    H &=&\pm \frac{\sqrt{5 m M-2 M n r_b^2+4 B_g  M r_b^2+4 M-2 r_b}}{\sqrt{12 m M r_b^2-2 r_b^3}},\\
    C &=& \frac{r_b^2 \left(-3 m^2 M-12 m M+4 m r_b+n r_b^3-2 B_g  r_b^3\right)}{6 m (r_b-6 m M)}.
\end{eqnarray}

\subsection{Model-II }

Here, we have implemented an alternative model of the conformal factor $\Phi(r)$ to analyze the compact star, which is given by
\begin{eqnarray}
    \Phi(r)=H+N\,r,
\end{eqnarray}
where $H,N$ are arbitrary constants.
This linear form of the conformal factor has been widely studied in the literature \cite{C3, C4}. Our work is well motivated by these articles. By imposing the above linear form of the conformal vector into the motion equations, we get the solution of the field equation as follows.
\begin{eqnarray}
    e^{\nu(r)} &=& H^2r^2,\nonumber \\
    e^{\lambda(r)} &=& \Big(\frac{I}{H+N\,r}\Big),\nonumber\\
    \rho^{\text{eff}} &=& -\frac{3 H^2 m+3 H N m r-2 B_g  I^2 r^2}{2 I^2 r^2},\nonumber\\
    p^{\text{eff}} &=& -\frac{H^2 m+H N m r+2 B_g  I^2 r^2}{2 I^2 r^2},\nonumber\\
    E^2 &=&\frac{1}{2I^2r^2}\big(5 H^2 m+11 H N m r-2 I^2 m+I^2 n r^2-2 B_g  I^2 r^2+6 N^2 m r^2\big).  
\end{eqnarray}
Upon examining the solution of the field equation, it becomes apparent that there are three constants: $H$, $I$, and $N$. To establish the additional requirement for model-II, we will employ the second fundamental condition of the continuity equation, which states that $ p(r=r_b)$ must equal zero, where $r_b$ corresponds to the stellar radius. The derived constants for model-II are given by
\begin{eqnarray}
H = \mp \frac{\sqrt{6 m M-3 M n r_{b}^2+4 M-2 r_{b}}}
{\sqrt{18 m M r_{b}^2-2 r_{b}^3}},
\qquad
H = \pm \frac{\sqrt{6 m M-3 M n r_{b}^2+4 M-2 r_{b}}}
{\sqrt{18 m M r_{b}^2-2 r_{b}^3}}, \nonumber \\[0.3cm]
I = -\frac{m \left(6 m M-3 M n r_{b}^2+4 M-2 r_{b}\right)}
{4 B_g r_{b}^3 (9 m M-r_{b})},
\qquad
I = +\frac{m \left(6 m M-3 M n r_{b}^2+4 M-2 r_{b}\right)}
{4 B_g r_{b}^3 (9 m M-r_{b})}, \nonumber \\[0.3cm]
N = \pm \frac{\sqrt{2 M \left(6 m-3 n r_{b}^2+4\right)-4 r_{b}}
\,\mathcal{F}}
{8 B_g r_{b} \left[-r_{b}^2 (r_{b}-9 m M)\right]^{3/2}},
\qquad
N = \mp \frac{\sqrt{2 M \left(6 m-3 n r_{b}^2+4\right)-4 r_{b}}
\,\mathcal{F}}
{8 B_g r_{b} \left[-r_{b}^2 (r_{b}-9 m M)\right]^{3/2}}.
\end{eqnarray}

where $\mathcal{F}=\big(6 m^2 M+m M \left(4-3 r_{b}^2 (n-12 B_g)\right) -2 m r_{b}-4 B_g  r_{b}^3\big)$.
In the next section, we analyze and compare the physical properties and stability of the two constructed models.
\begin{table}[t]
 \caption{The corresponding numerical values of the constants for different model parameters where we have taken the measured mass of the pulsar PSR J1614-2230.}\label{table1}
 \centering
  \begin{tabular}{@{}ccccccccccccc@{}}
            \hline\hline
             &  & Model-I & \\
            \hline
             $m$ & $n$ & H & $C$\\
             \hline
            2 & 0.02 & $0.189272$ & $0.0387991$\\
            3 & 0.05 & $0.701516$ & $0.0208653$ \\
            0.2 & 0.1 & $1427.6$ & $0.21629$\\
            -0.5 & 0.4 & $-802.576$ & $0.255184$\\
            \hline\hline
             &  & \,\,\,\,\,\,\,\, Model-II&  &\\ \hline\hline
            m & n & H & I & N \\
            \hline
            0.5 & -1 & $1.25289$ & $479.831$ & $-601.298$\\
            1.5 & -2 & $0.551493$ & $278.909$ & $-153.87$\\
            -0.5 & 1.5 & $0.532217 $ & $-86.584$ & $46.0298$\\
            -2 & 4 & $0.530841$ & $-344.547$ & $182.848 $\\
             \hline
        \end{tabular}
\end{table}
We have given the numerical values of the constants for model-I and model-II by varying the model parameter $m,n$ in table \ref{table1}. The corresponding numerical values of the constants were calculated by considering the measured mass of the pulsar PSR J1614-2230.

\section{Physical analysis}\label{seciv1}
To attain a well-behaved and feasible solution, the following conditions must be satisfied for a stellar configuration:
\begin{enumerate}
    \item \textbf{Metric potential:}
The metric potentials, which encode the geometric and causal structure of spacetime, must remain finite and bounded within the stellar interior ($0 \le r \le r_b$). From Fig.~\ref{metricconf}, we observe that model-I exhibits finite, continuous, and monotonically increasing metric functions throughout the star. In contrast, model-II develops a central singularity owing to the linear choice of the conformal factor, although the metric potentials remain well behaved away from the core. This indicates that the power-law conformal factor is more suitable for constructing physically viable compact star models.

\begin{figure*}[htbp]
    \includegraphics[width=8cm, height=5cm]{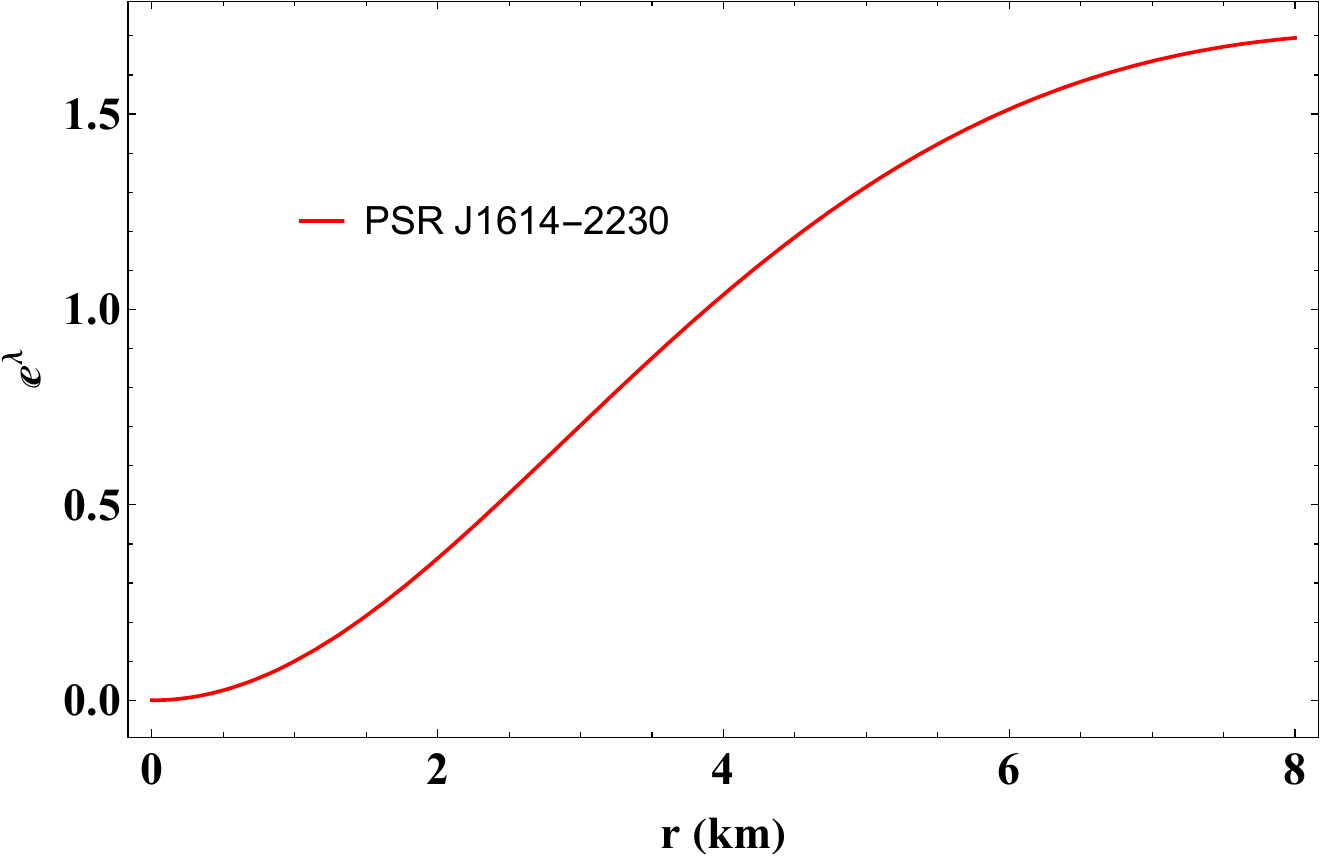}
    \includegraphics[width=8cm, height=5cm]{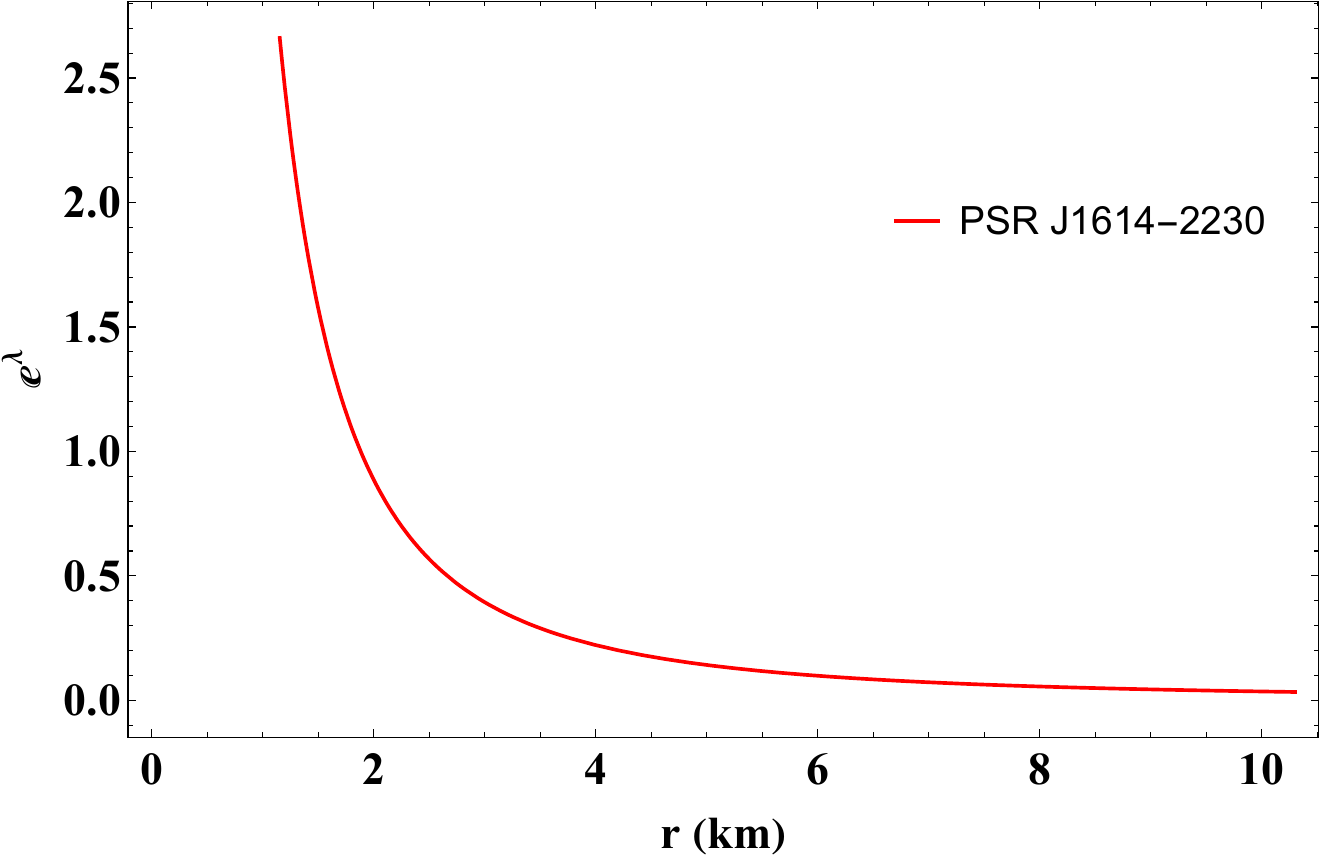}
      \caption{The metric coefficients for the model-I(left panel) and model-II (right panel) for measured mass of pulsar PSR J1614-2230. Here, we consider $m=-2,n=0.02$ for model-I and $m=2,n=-1$ for model-II.\label{metricconf}}
\end{figure*}

\item \textbf{Nature of physical quantities:}
For a physically viable compact star model, the effective pressure and energy density must satisfy standard regularity conditions. In particular, the surface pressure must vanish $p^{\text{eff}}(r_b)=0$, while both pressure and density should remain positive, attain their maximum values at the center, and decrease monotonically towards the stellar boundary:
\[
p^{\text{eff}}(0)>0,\quad \left.\frac{dp^{\text{eff}}}{dr}\right|_{0}=0,\quad
\left.\frac{d^{2}p^{\text{eff}}}{dr^{2}}\right|_{0}<0,
\]
\[
\rho^{\text{eff}}(0)>0,\quad \left.\frac{d\rho^{\text{eff}}}{dr}\right|_{0}=0,\quad
\left.\frac{d^{2}\rho^{\text{eff}}}{dr^{2}}\right|_{0}<0.
\]

Using observational mass–radius data for compact stars such as PSR J1614–2230, PSR J1903+327, Vela X-1, Cen X-3, and SMC X-1, we verify these conditions. As shown in Fig.~\ref{pressure2}, both the pressure and energy density attain their maximum values at the stellar center and decrease smoothly towards the surface, where the pressure vanishes. This behavior confirms the physical acceptability of the constructed models, with the concave nature of the profiles arising from the combined effects of conformal symmetry and electric charge.

\begin{figure*}[htbp]
    \includegraphics[width=8cm, height=5cm]{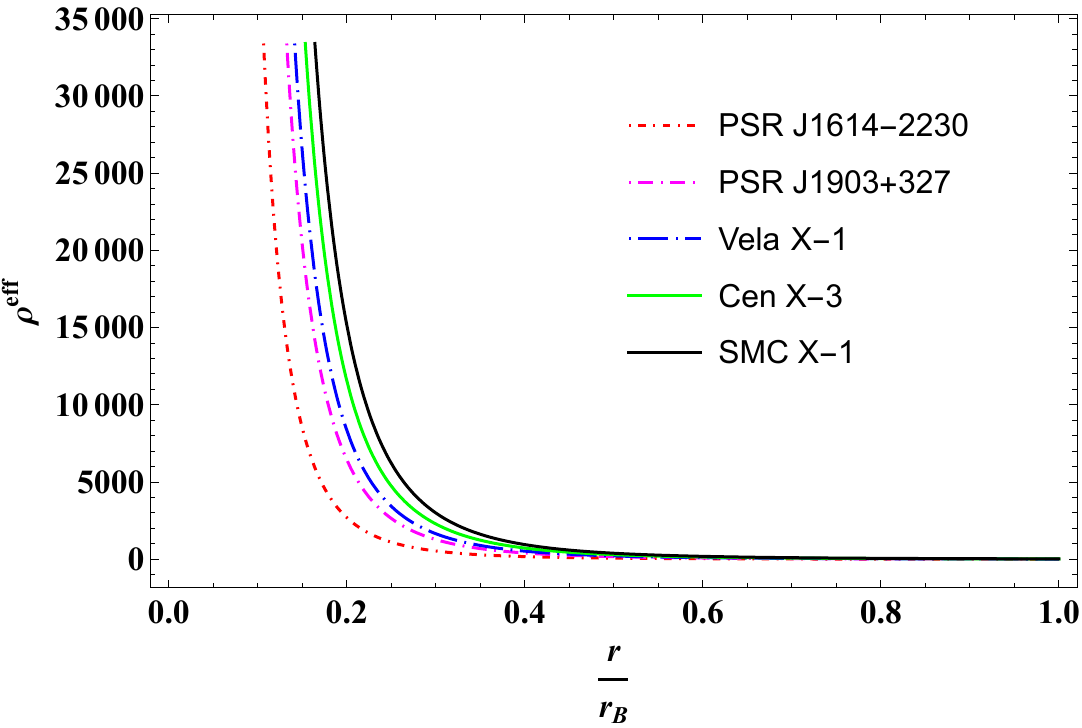}
    \includegraphics[width=8cm, height=5cm]{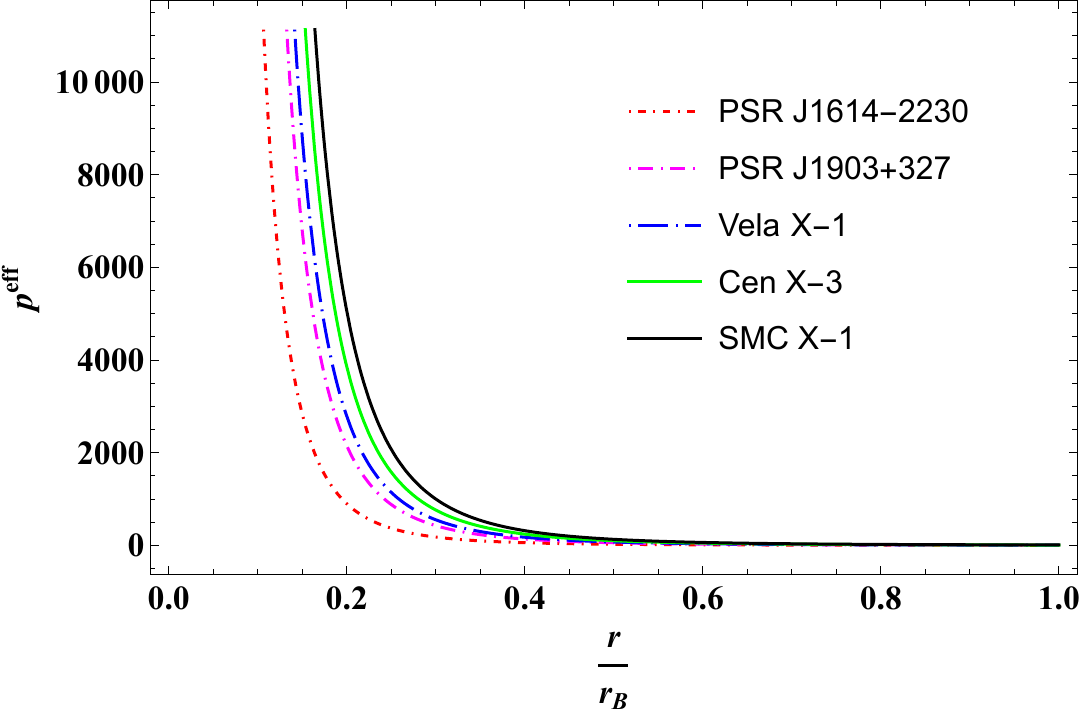}
    \includegraphics[width=8cm, height=5cm]{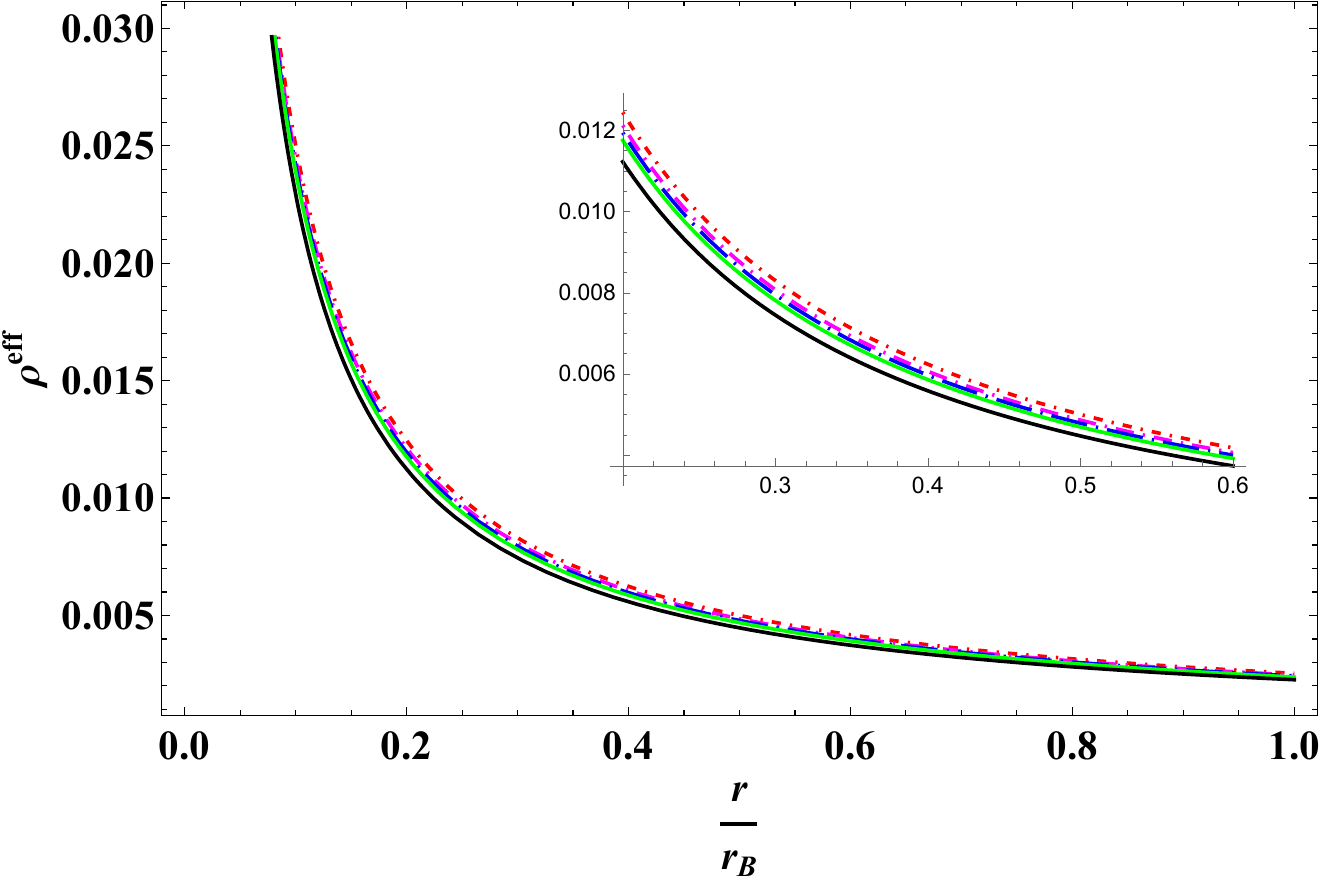}
    \includegraphics[width=8cm, height=5cm]{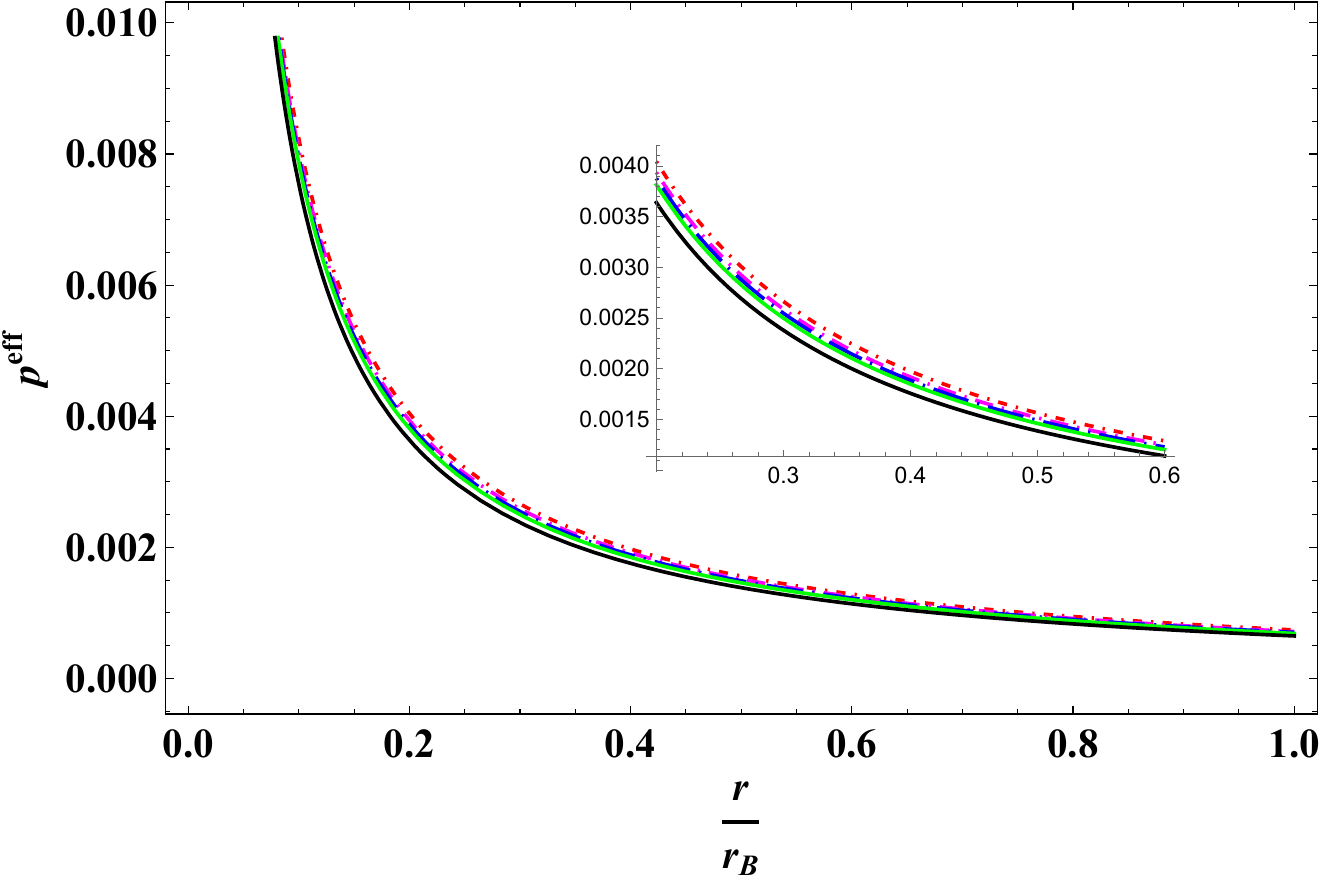}
      \caption{Behavior of the pressure and the matter density ($\text{km}^{-2}$) for model-I (upper panel) and model-II (lower panel). Here, we consider $m=-2,n=0.02$ for model-I and $m=2,n=-1$ for model-II.\label{pressure2}}
\end{figure*}
Moreover, the current investigation results in negative values for the derivatives of the energy density and pressure functions concerning the radial coordinate, denoted as $\frac{d\rho^{\text{eff}}}{dr}$ and $\frac{dp^{\text{eff}}}{dr}$, respectively. The presence of negative gradients in Fig.~\ref{dpdr} indicates that the solutions we have discovered meet the physical requirements and are physically acceptable for both models.
\begin{figure*}[htbp]
    \includegraphics[width=8cm, height=5cm]{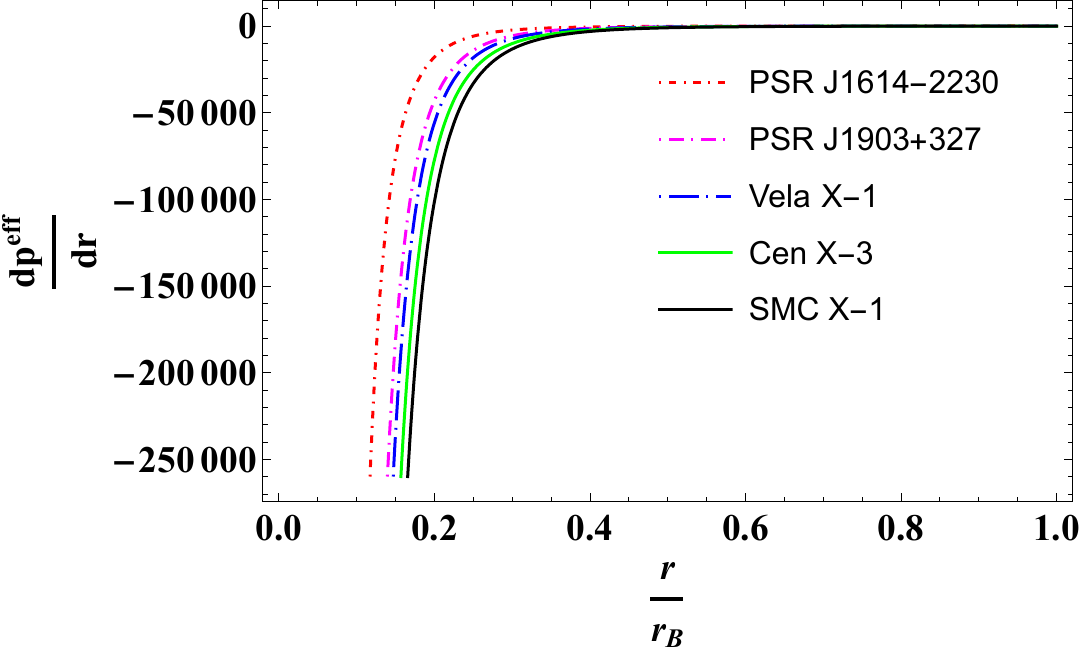}
    \includegraphics[width=8cm, height=5cm]{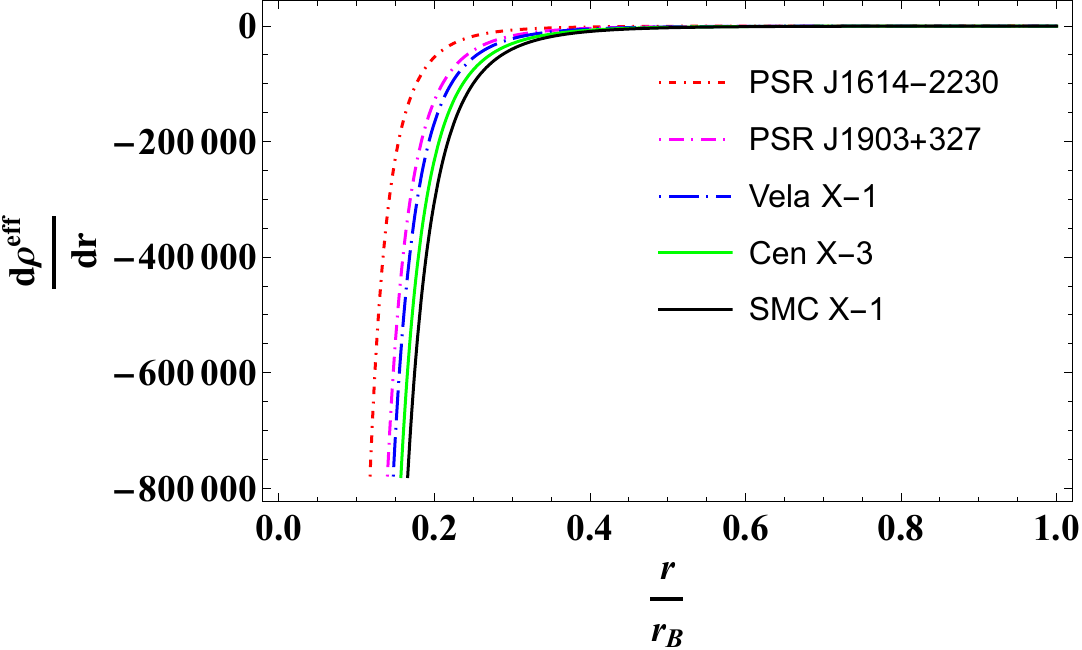}
    \includegraphics[width=8cm, height=5cm]{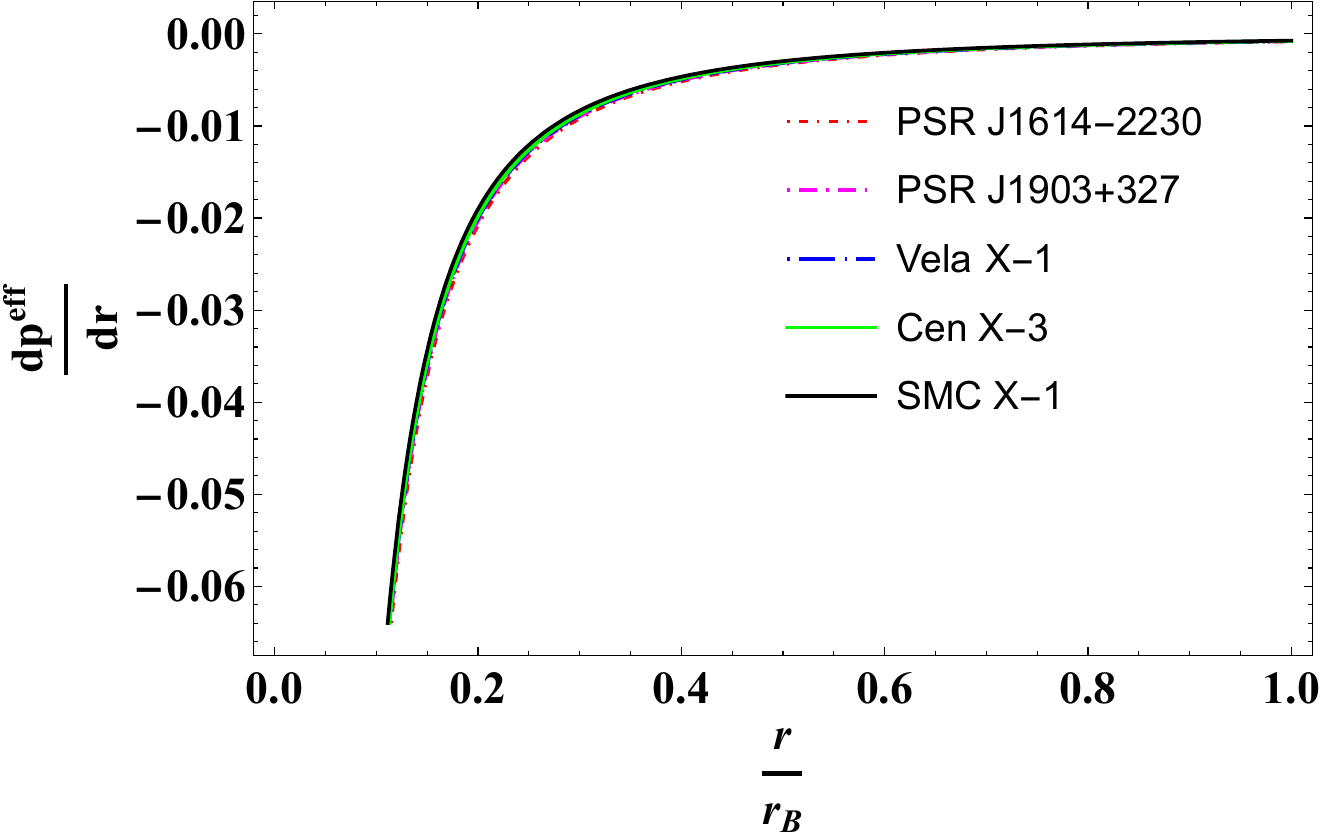}
    \includegraphics[width=8cm, height=5cm]{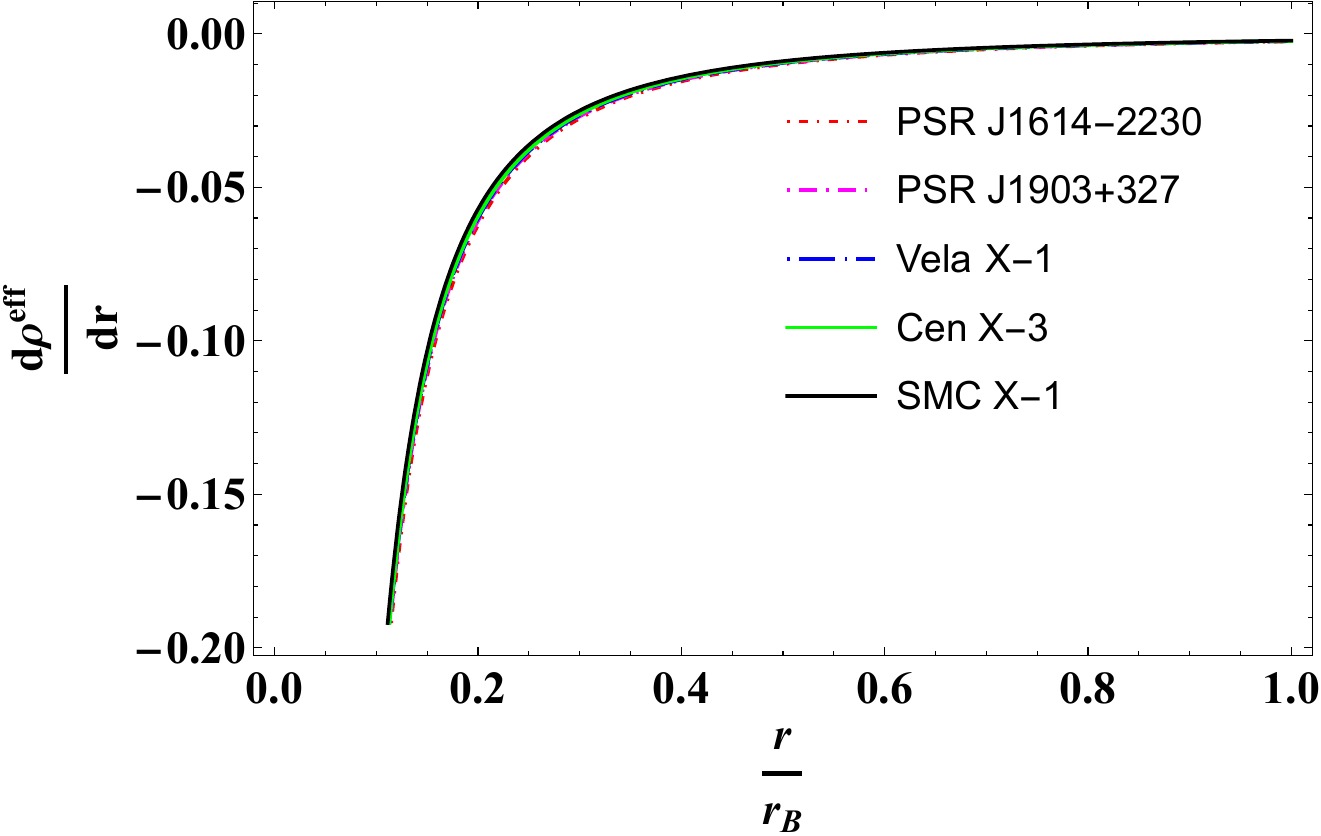}
      \caption{Behavior of pressure gradient and matter density gradient for model-I (upper panel) and model-II (lower panel). Here, we consider $m=-2,n=0.02$ for model-I and $m=2,n=-1$ for model-II.\label{dpdr}}
\end{figure*}
        \item \textbf{Energy conditions:}
Energy conditions provide essential consistency checks for physically realistic stellar models and have been extensively discussed earlier. In the present analysis, we verify the null, weak, strong, and dominant energy conditions using the effective density and pressure profiles. From Figs.~\ref{pressure2} and \ref{energy}, it is evident that all energy conditions are satisfied throughout the stellar interior for both constructed models, confirming the physical viability of charged compact stars under Bardeen spacetime. As expected from conformal symmetry, the physical quantities exhibit a central singularity, which remains an inherent limitation of the conformal motion approach and is consistent with earlier studies.

\begin{figure*}[htbp]
    \includegraphics[width=8cm, height=5cm]{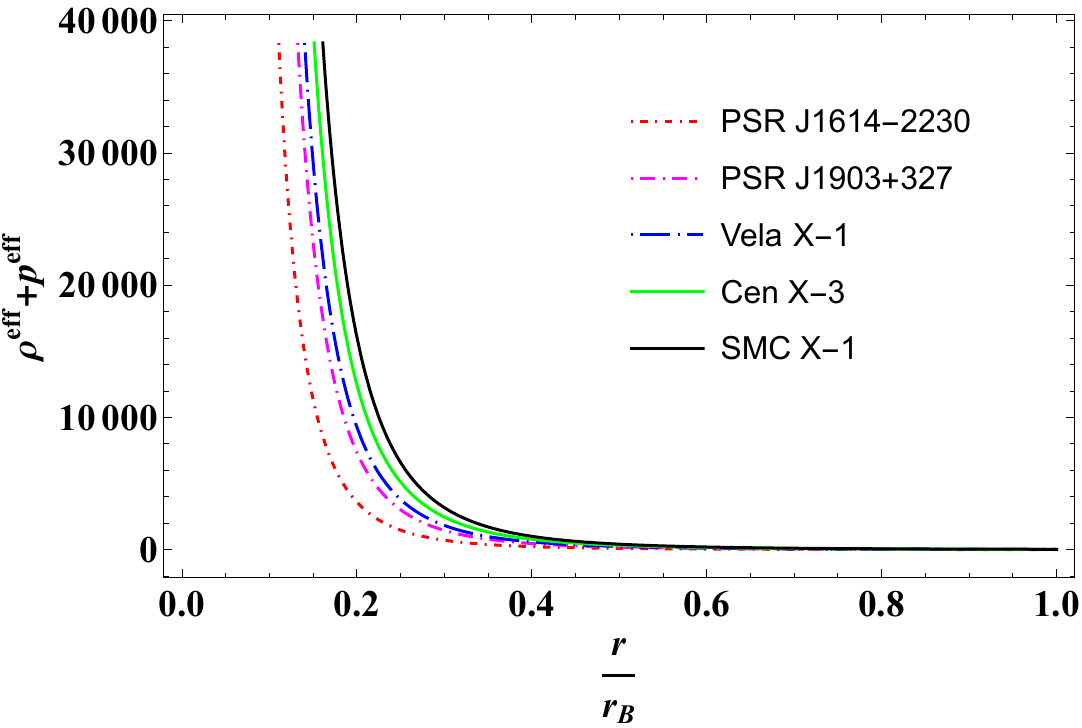}
    \includegraphics[width=8cm, height=5cm]{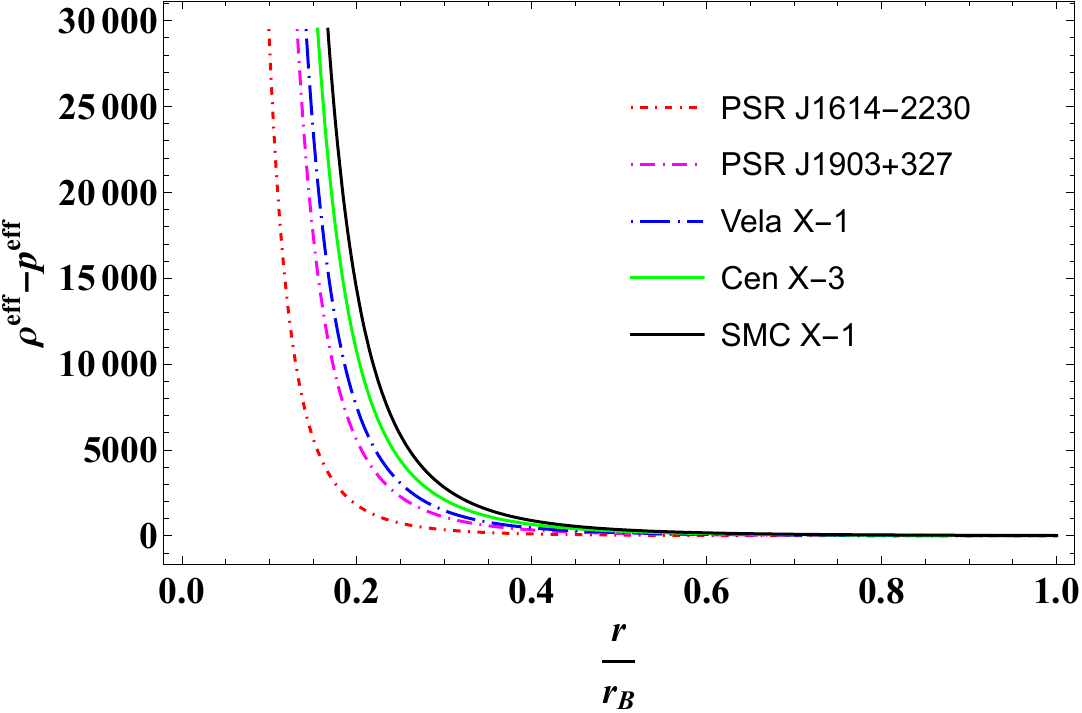}
    \includegraphics[width=8cm, height=5cm]{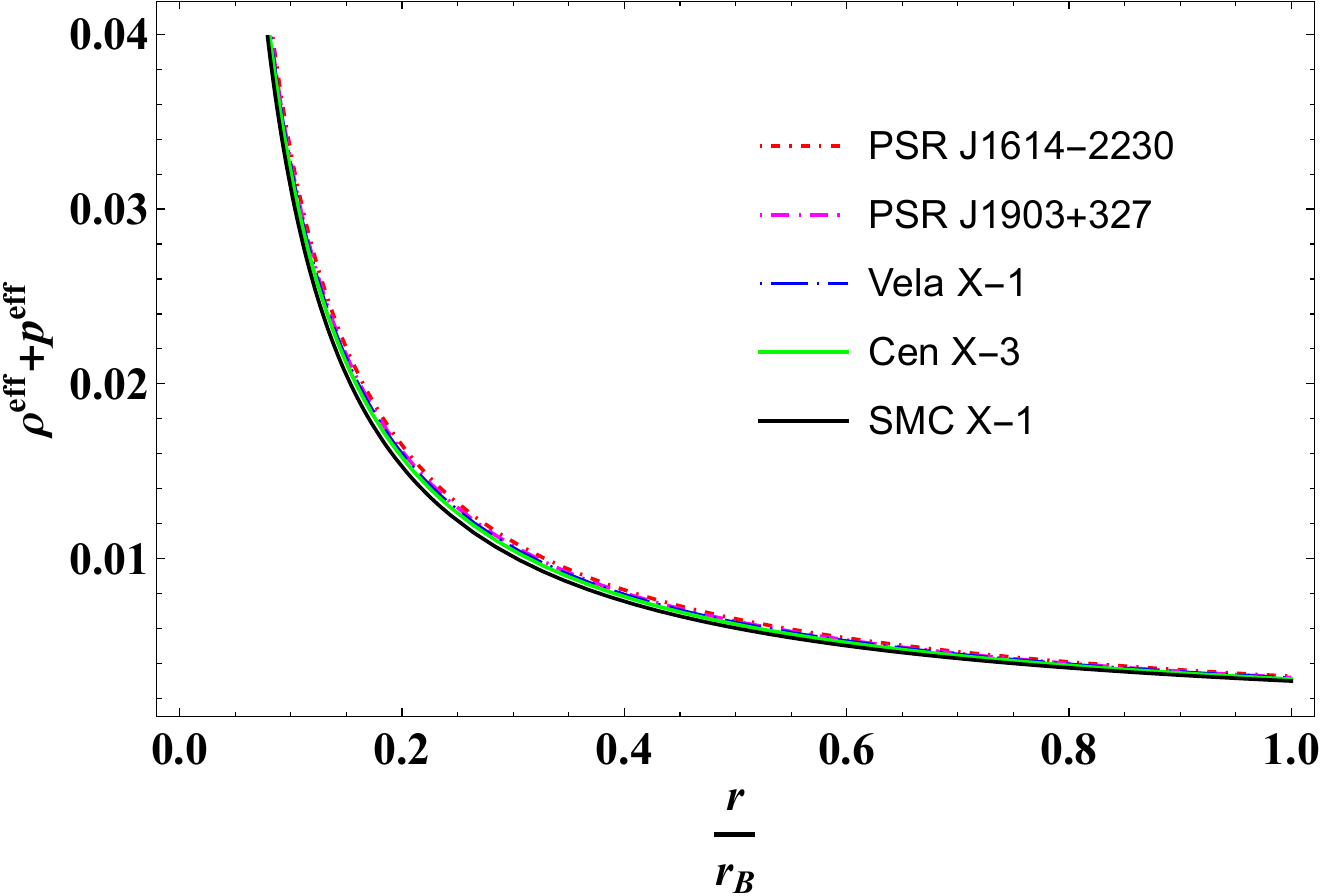}
    \includegraphics[width=8cm, height=5cm]{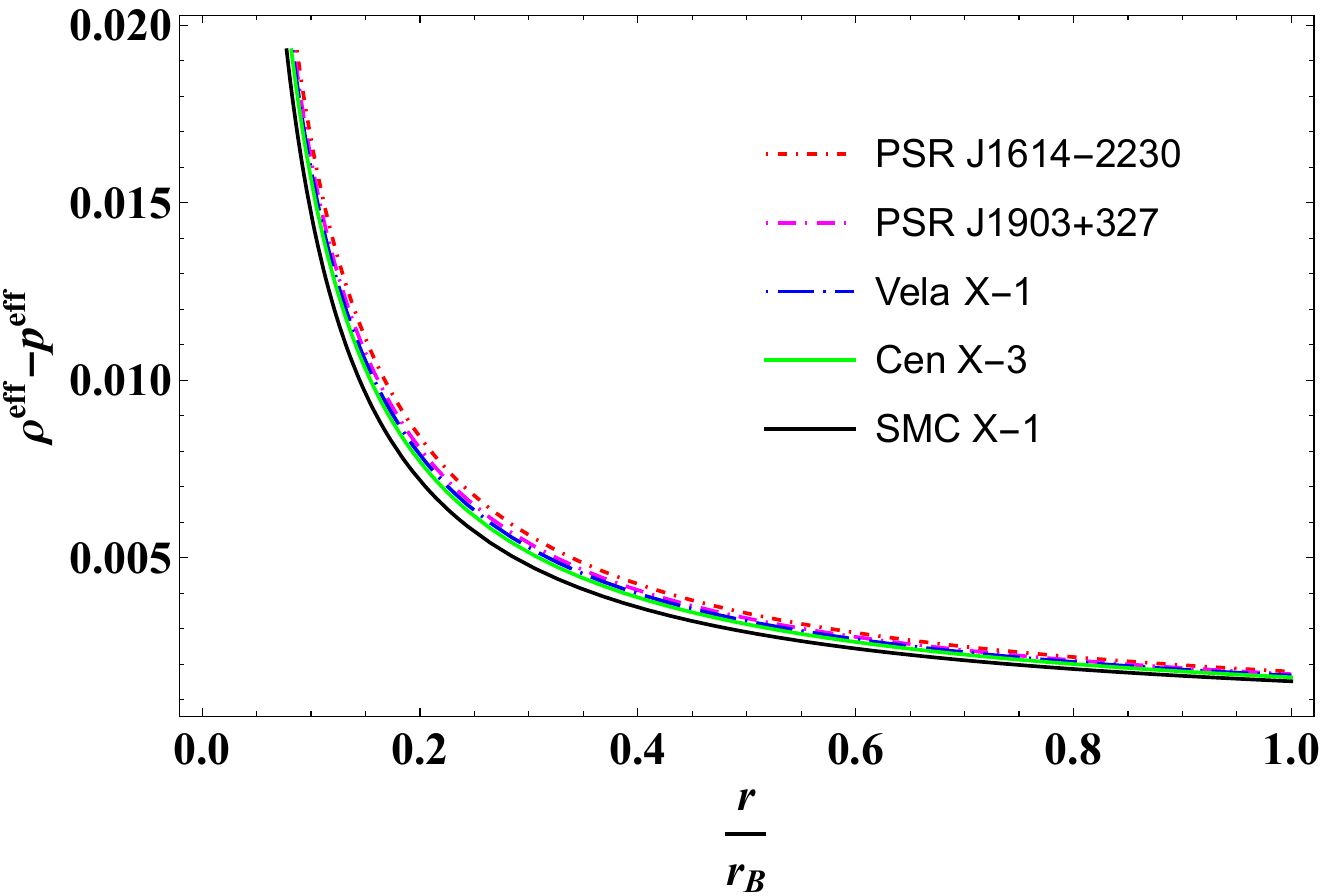}
      \caption{Behavior of energy conditions for model-I (upper panel) and model-II (lower panel). Here, we consider $m=-2,n=0.02$ for model-I and $m=2,n=-1$ for model-II.\label{energy}}
\end{figure*}

\end{enumerate}
\section{Equilibrium and stability analysis}\label{secv2}

\begin{enumerate}
    \item \textbf{Causality criterion:} The causality condition must be maintained, which says that the magnitude of the speed of sound must be lower than the speed of light. In other words, the inequality $0\leq v^2=\frac {d p^{\text{eff}}}{d\rho^{\text{eff}}}\leq 1$ must be satisfied, as discussed earlier. In this investigation, we determined that the square of the speed of sound $v^2=\frac{1}{3}$ maintains the stability criterion above for both model-I and model-II.
    \item \textbf{Relativistic adiabatic index:}
The relativistic adiabatic index provides an important criterion for assessing the dynamical stability of compact stars. As discussed earlier, stability requires $\Gamma> 4/3$. From Fig.~\ref{adbconf}, we observe that $\Gamma$ remains above this critical value throughout the stellar interior for all compact stars considered. This confirms that the charged compact star models admitting conformal motion and matched with Bardeen exterior geometry satisfy the stability condition based on the adiabatic index.

\begin{figure*}[htbp]
    \includegraphics[width=8cm, height=5cm]{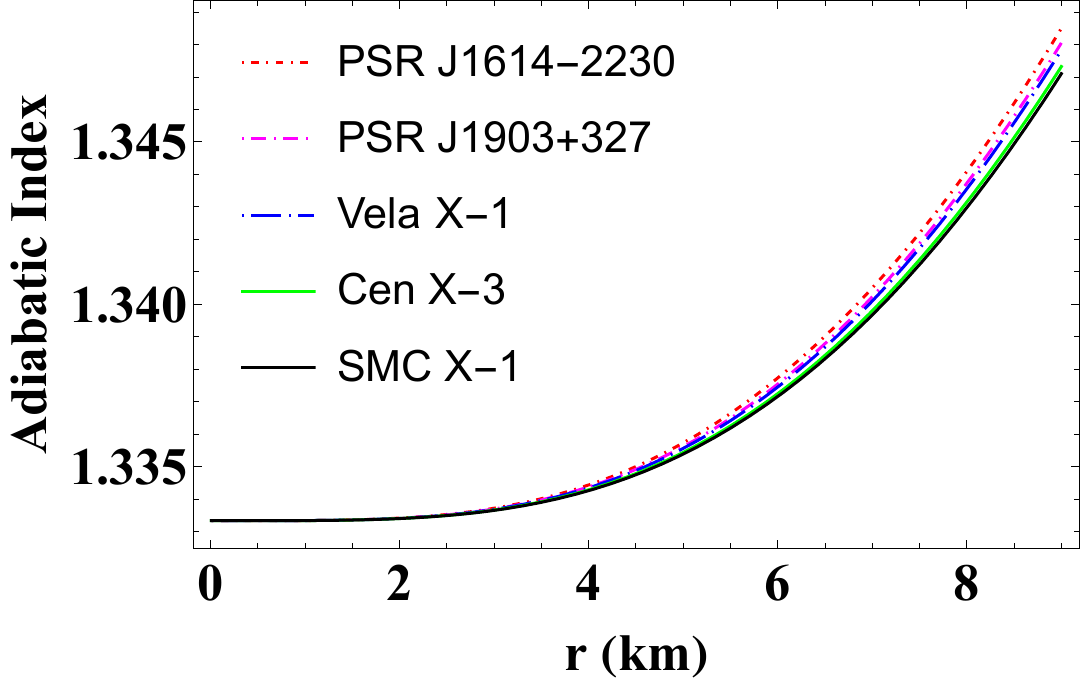}
    \includegraphics[width=8cm, height=5cm]{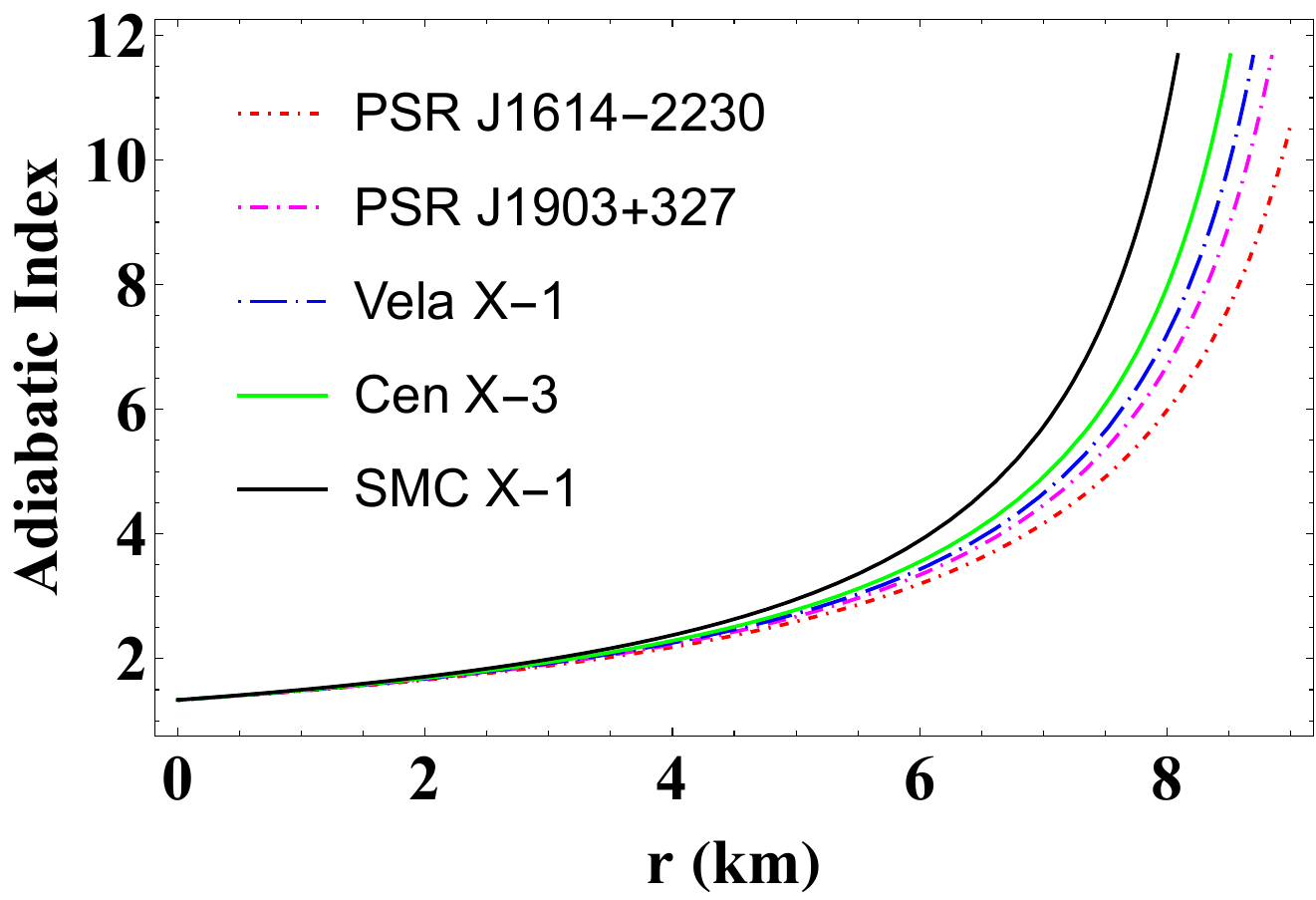}
      \caption{Behavior of adiabatic index for model-I (left panel) and model-II (right panel).  Here, we consider $m=-2,n=0.02$ for model-I and $m=2,n=-1$ for model-II. \label{adbconf}}
\end{figure*}
   \item \textbf{Equilibrium conditions:}
The equilibrium of a charged compact star is governed by the TOV equation, which ensures the balance between gravitational, hydrostatic, and electric forces. For a static and spherically symmetric configuration, these forces are given by
\[
F_g=-\frac{\nu'}{2}(\rho^{\text{eff}}+p^{\text{eff}}), \quad
F_h=-\frac{dp^{\text{eff}}}{dr}, \quad
F_e=\sigma E e^{\lambda/2},
\]
and satisfy the equilibrium condition
\[
F_g+F_h+F_e=0.
\]

The behavior of these forces for different compact stars is illustrated in Fig.~\ref{Forces}. For model~I, the gravitational and hydrostatic forces are found to balance each other throughout the stellar interior, while the electric force remains well controlled, indicating a stable equilibrium configuration. In contrast, although model-II shows partial force balance, the electric force exhibits behavior that can destabilize the system. This analysis clearly demonstrates that the power-law conformal factor leads to a more stable and physically acceptable charged compact star model than the linear conformal factor.

\begin{figure*}
\includegraphics[width=8cm, height=5cm]{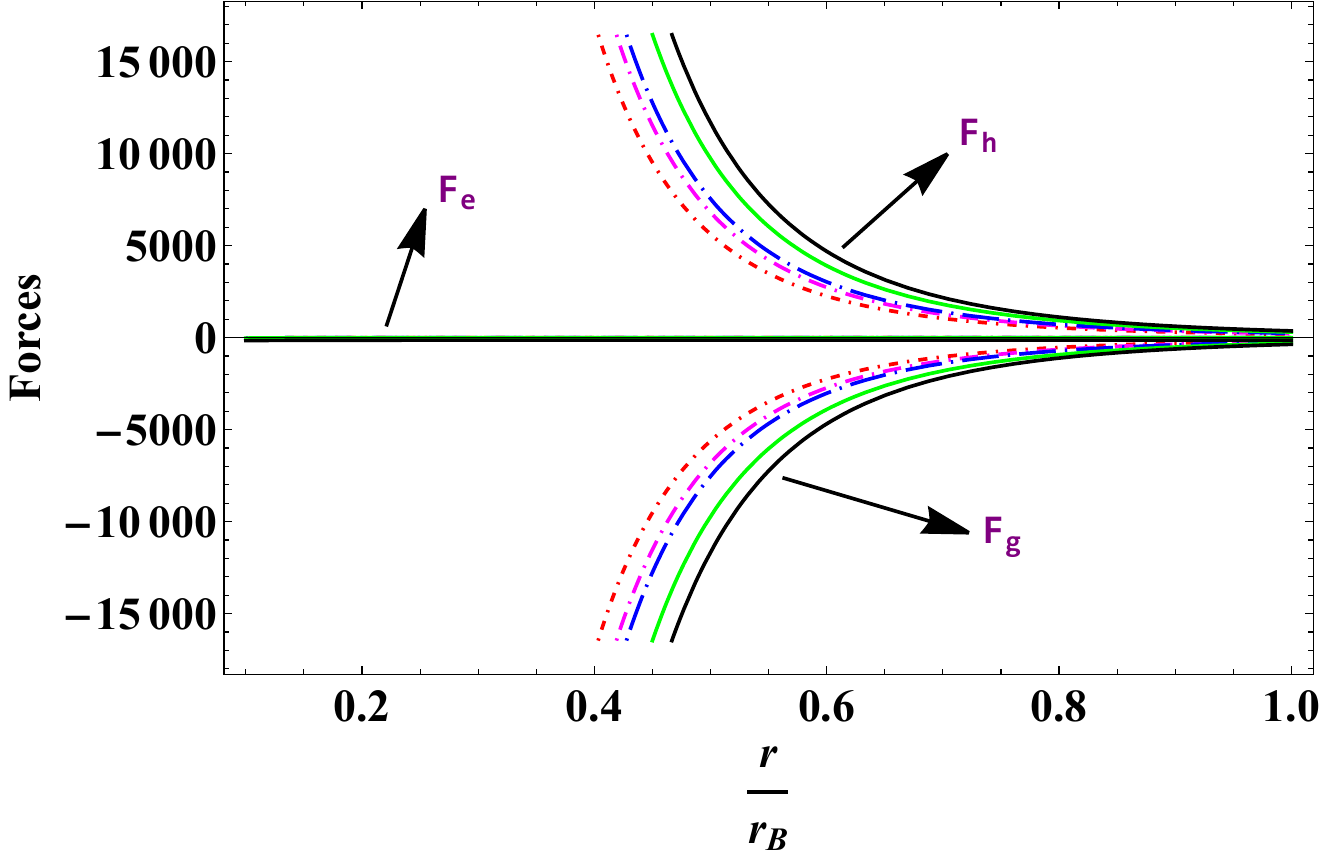} \includegraphics[width=8cm, height=5cm]{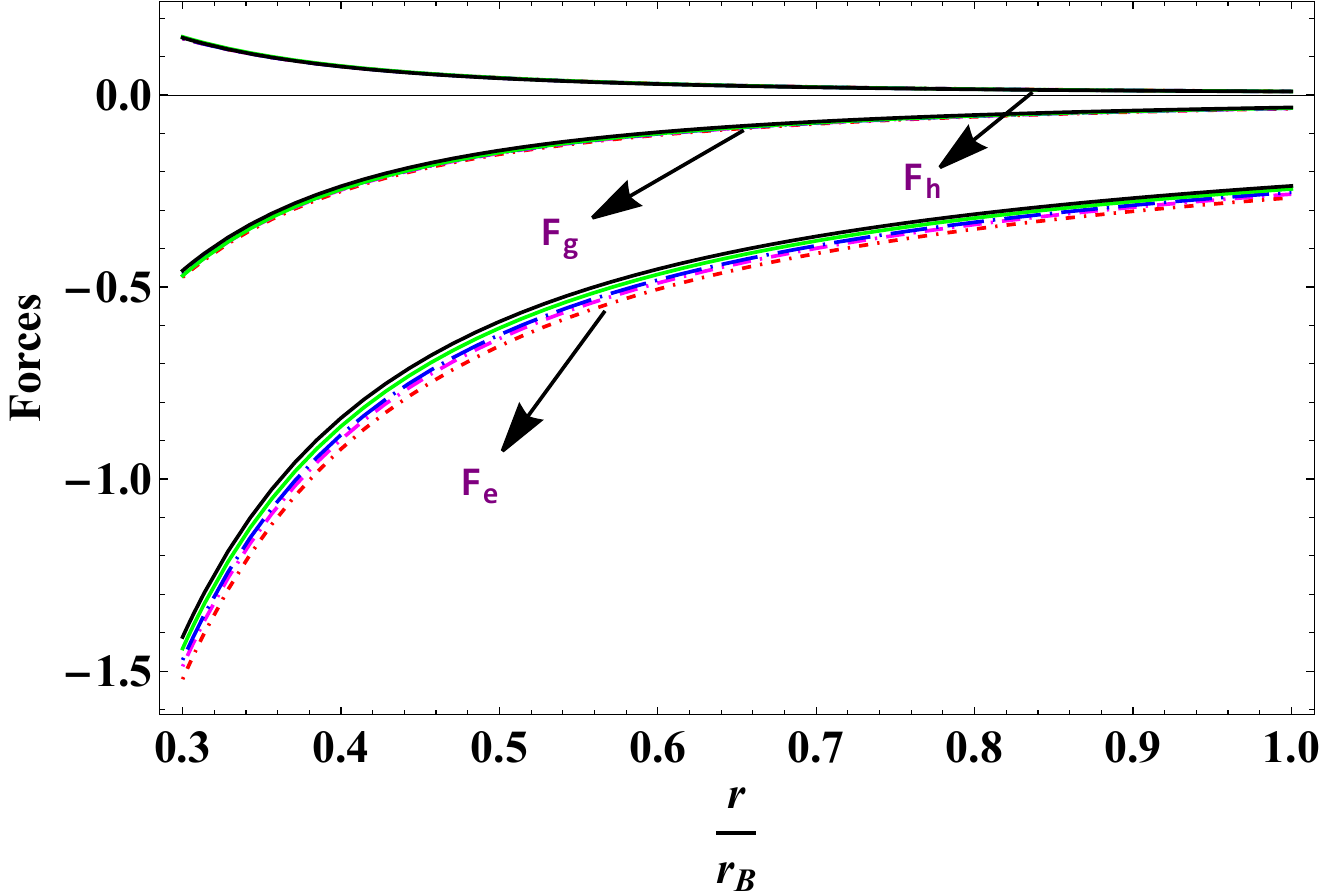} \caption{Behavior of different forces for model-I (left panel) and model-II (right panel). The different colors represents, PSR J1614-2230($\textcolor{red}\star$), PSR J1903+327 ($\textcolor{magenta}\star$), Vela X-1 ($\textcolor{blue}\star$), Cen X-3 ($\textcolor{green}\star$), and SMC X-1 ($\textcolor{black}\star$). Here, we consider $m=-2,n=0.02$ for model-I and $m=2,n=-1$ for model-II.\label{Forces}}
\end{figure*}

\item \textbf{Andr\'{e}asson's limit:} For an uncharged, static, spherically symmetric compact object, Buchdahl's inequality~\cite{buchdahl1959} states that gravitational equilibrium is possible only when the stellar radius satisfies $R > 9M/4$; otherwise, the star inevitably collapses to form a Schwarzschild black hole. In the presence of an electric charge $\mathcal{Q}$, Coulombic repulsion partially counterbalances gravitational attraction, allowing more compact configurations to remain in equilibrium. Andr\'{e}asson~\cite{y3} generalized Buchdahl's result to charged spherically symmetric objects, deriving the upper bound
\begin{equation}\label{andreasson}
    \sqrt{M} \leq \sqrt{\frac{R}{3}} + \sqrt{\frac{R}{9} + \frac{\mathcal{Q}^{2}}{3R}}\,,
\end{equation}
which any physically admissible charged compact configuration must obey.

Physically, the violation of Andr\'{e}asson's bound signals the onset of gravitational collapse to a charged (Reissner--Nordstr\"{o}m) black hole, whose horizons are located at $r_{\pm} = M \pm \sqrt{M^{2} - \mathcal{Q}^{2}}$. The bound therefore separates stable charged stellar configurations from configurations that must collapse. To illustrate this, consider two representative cases:

\textit{(i) Sub-extremal regime ($\mathcal{Q} < M$, e.g., $\mathcal{Q} = M/2$):} A genuine event horizon exists at $r_{+} = M + \sqrt{M^{2} - \mathcal{Q}^{2}} \approx 1.87 M$. If the matter distribution violates the Andr\'{e}asson bound, gravitational collapse proceeds, and the resulting configuration is a sub-extremal Reissner--Nordstr\"{o}m black hole with two distinct horizons. The critical radius $R_{c}$ at which a charged fluid configuration becomes unstable lies just outside $r_{+}$, consistent with cosmic censorship.

\textit{(ii) Super-extremal regime ($\mathcal{Q} > M$, e.g., $\mathcal{Q} = 2M$):} The horizon condition $M^{2} - \mathcal{Q}^{2} \geq 0$ is violated, and no event horizon forms. The strong electrostatic repulsion exceeds gravitational self-attraction, preventing gravitational collapse to a black hole. In this regime, Andr\'{e}asson's bound is automatically satisfied for sufficiently compact yet stable configurations, and the object remains a regular charged star rather than collapsing to a (forbidden) naked singularity, in accordance with the weak cosmic censorship conjecture.

In the extremal limit $\mathcal{Q} \to M$, the two horizons merge ($r_{-} = r_{+} = M$), and the Andr\'{e}asson bound recovers Buchdahl's result modified by the electric charge contribution. The values of $\sqrt{M}$ and the corresponding right-hand side of Eq.~\eqref{andreasson} for both models considered in this work are listed in Table~\ref{table2}, confirming that all configurations satisfy Andr\'{e}asson's limit.
\begin{table}
\caption{The estimated values of mass and Andreasson's limit. Here, we consider $m=-2,n=0.02$ for model-I and $m=2,n=-1$ for model-II.}\label{table2}
\centering
  \begin{tabular}{@{}ccccccccccccc@{}}
            \hline\hline
             &   Model-I & \\
            \hline
             Stars & $\sqrt{M}$ & $\frac{\sqrt{r_b}}{3}+\sqrt{\frac{r_b}{9}+\frac{q^2}{3r_b}}$\\
             \hline
            PSR J164-2230\cite{s1} & $1.40357$ & $3.64223$ \\
            PSR J1903+327\cite{s2} & $1.33041$ & $3.55699$ \\
            Vela X-1\cite{s3} & $1.29112$ & $3.50867$\\
            Cen X-3\cite{s3} & $1.22066$ & $3.46229$ \\
            SMC X-1\cite{s3} & $1.13578$ &$3.30119$\\
            \hline\hline
             &   Model-II & \\
            \hline
             Stars & $\sqrt{M}$ & $\frac{\sqrt{r_b}}{3}+\sqrt{\frac{r_b}{9}+\frac{q^2}{3r_b}}$\\
             \hline
            PSR J164-2230\cite{s1} & $1.40357$ & $13.8682 $ \\
            PSR J1903+327\cite{s2} & $1.33041$ & $13.6346 $ \\
            Vela X-1\cite{s3} & $1.29112$ & $13.5013$\\
            Cen X-3\cite{s3} & $1.22066$ & $13.4968 $ \\
            SMC X-1\cite{s3} & $1.13578$ &$12.9331$\\
            \hline
        \end{tabular}
\end{table}
\end{enumerate}

\section{Comparison with Reissner-Nordström spacetime}\label{vi}

\begin{figure*}[htbp]
    \includegraphics[width=8cm, height=5cm]{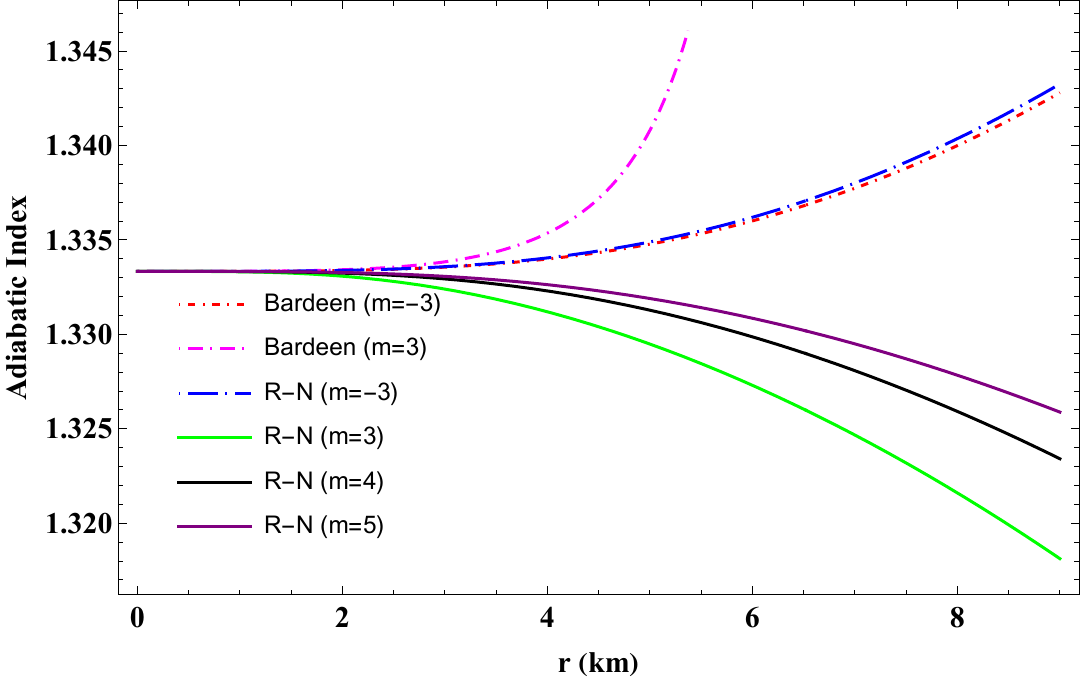}
    \includegraphics[width=8cm, height=5cm]{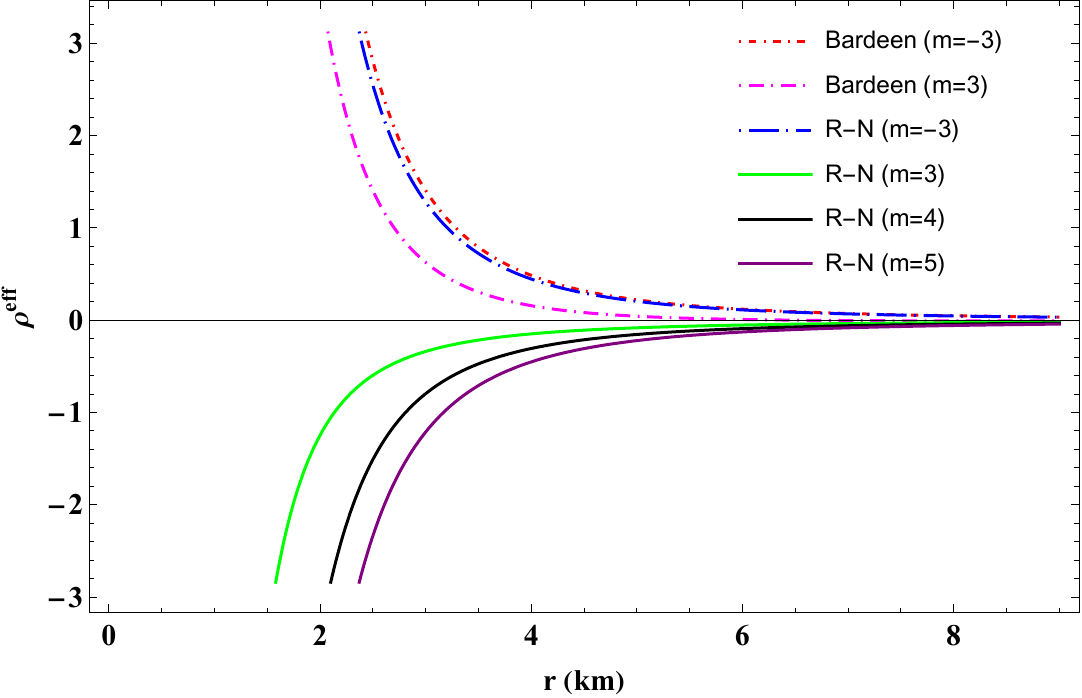}
      \caption{Comparison between the Bardeen and R-N spacetime. \label{compare}}
\end{figure*}

Here, we present a brief overview of the scenario with R-N spacetime as the external geometry for the matching condition. Additionally, this will allow us to make comparisons with our study. The R-N spacetime is defined by \cite{weitzenbock1923}
\begin{eqnarray}
ds^2 &=& -\Big(1-\frac{2 M}{r}+\frac{ \mathcal{Q}^{2}}{r^{2}}\Big) dt^2 +\Big(1-\frac{2 M}{r}+\frac{ \mathcal{Q}^{2}}{r^{2}}\Big)^{-1}+r^2 (d\theta^2+sin^2 \theta d\phi^2).
\end{eqnarray}
Similarly, as in the previous case, we implement the continuity equation to get the constant of our proposed model I, and we have shown those comparative studies in Fig. \ref{compare}.\\
In this comparative study, we examined the energy density and adiabatic index, which play important
roles in studying the physical behavior and stability of a stellar model. We have varied the model parameter in a wide range and observed that, for the Bardeen spacetime, the positive behavior of the energy density and its increasing behavior towards the star's core, but for the R-N model, as the model parameter $m$ increases, we get the negative energy density and it gradually decreases towards $-\infty$, which is not feasible in the regime of astrophysical object. In addition, from the second panel of Fig. \ref{compare}, readers can observe that our proposed model-I with the Bardeen spacetime gives the value $\Gamma>\frac{4}{3}$ for a wide range of $m$. But in the case of R-N spacetime, as $m$ increases, it maintains the adiabatic index limit $\Gamma>\frac{4}{3}$ for a certain range of radius, particularly in the core region, but as it approaches towards the outer surface region, it does not maintain the limit mentioned above, which does not support proving a viable model. However, our constructed model-II does not show any significant differences between the Bardeen and R-N spacetime.

\section{Conclusion}\label{vii}

In this chapter, we explored the structure of charged compact stars within the framework of $f(Q)$ gravity by employing CKVs. The stellar interior is modeled as a charged perfect fluid obeying the MIT Bag model EoS, whereas the spacetime geometry was determined through conformal motion. To examine the influence of conformal symmetry, we constructed two different stellar models by choosing a power-law and a linear form of the conformal factor.

Our analysis shows that the choice of the conformal factor plays a crucial role in determining the physical viability of the stellar configuration. The power-law model yields well-behaved metric potentials throughout the stellar interior, whereas the linear model develops an unavoidable central divergence. This clearly indicates that the power-law conformal factor provides a more reliable description of compact stars in $f(Q)$ gravity.

The physical behavior of the matter variables further supports this conclusion. In both models, the energy density and pressure remain positive inside the star and decrease smoothly towards the surface, where the pressure vanishes. All standard energy conditions are satisfied, which confirms the physical consistency of the models. The stability of the configurations is reinforced by the sub-luminal sound speed, the adiabatic index remaining above the critical value, and compliance with the charged mass bound. A detailed equilibrium analysis reveals that, in the power-law model, gravitational, hydrostatic, and electric forces balance each other effectively, while the linear model shows signs of instability due to the electric force.

We also compared the interior solutions matched with Bardeen and R-N exterior spacetimes. The Bardeen geometry consistently provides more realistic and stable results, particularly in terms of energy density behavior and stability criteria, highlighting its advantage as an exterior spacetime for charged compact stars.

In summary, our study demonstrates that charged compact stars admitting conformal motion can be successfully described within $f(Q)$ gravity when an appropriate conformal factor and exterior geometry are chosen. Although conformal symmetry inevitably introduces a central singularity, the resulting models remain well-behaved throughout the stellar envelope. Overall, the power-law conformal model matched with Bardeen geometry emerges as a stable and physically appealing framework for describing realistic compact stars.

In this chapter, we examined the formulation of 
$f(Q)$ gravity for a charged stellar configuration within the framework of conformal geometry. To check the effects of torsion based gravity and examine how different geometric foundations influence the structure and physical properties of compact stars, in the following chapter, we will investigate the properties of compact stars under the influence of 
$f(T)$ gravity.


\chapter{\fontsize{14}{16}\selectfont Gravitationally Deformed Polytropic Models in Extended Teleparallel Gravity and Influence of Decoupling Parameters on Constraining Mass-Radius Relation} 

\label{Chapter3} 


\definecolor{maroon}{RGB}{128, 0, 0}
\lhead{\textcolor{maroon}{\textit{\textbf{Chapter 3:}}} \emph{\textcolor{maroon}{Gravitationally deformed polytropic models in extended teleparallel gravity and \\ \hspace{2cm} influence of decoupling parameters on constraining mass-radius relation}}}

\blfootnote{*The work in this chapter is covered by the following publication:\\
\textit{Gravitationally Deformed
Polytropic Models in Extended Teleparallel Gravity and Influence of Decoupling Parameters on
Constraining Mass-Radius Relation}, Chinese Physics C, \textbf{49}(10), 105110 (2025).}

This chapter presents strange star (SS) configurations in the framework of modified $f(T)$ gravity using the gravitational decoupling (GD) method via minimal geometric deformation (MGD).
The main findings of this work are summarized below.
\begin{itemize}
    \item In this work, deformed stellar models are constructed by adopting the Buchdahl ansatz together with a quadratic polytropic EoS.

    \item We derive deformed SS models assuming that the energy-momentum components of the deformed
fluid obey $\rho =\Theta^0_0$ and $p_r=\Theta_1^1$. Here $\Theta^0_0$ and $\Theta^1_1$ are the additional matter sector that has been incorporated in this study. 

\item The influence of the model parameters is analyzed by comparing three scenarios: GR, $f(T)$ gravity, and
$f(T)$+ MGD.

\item Observational constraints from GW190814, PSR J1614-2230 and PSR J1903+327 are used to assess the physical viability and stability conditions of the models.

\item  The resulting mass–radius relations support the existence of massive SS within the modified $f(T)$ gravity framework.
 
\end{itemize}

\section{Introduction}\label{sec:1}

Modified theories of gravity have been extensively explored to address both the theoretical limitations and the observational challenges of GR. From a theoretical perspective, GR suffers from non-renormalizability, motivating the construction of extended gravitational frameworks \citep{Stelle:1976gc, Addazi:2021xuf}. Observationally, the accelerated expansion of the Universe during early and late-time epochs further supports the need for modifications beyond the Einstein–Hilbert action \citep{Capozziello:2011et}. Although many approaches extend GR through curvature-based modifications such as $f(R)$ gravity, an alternative and equally compelling route is provided by torsion-based formulations of gravity. $f(T)$ gravity offers a distinct geometric description of gravitation based on torsion rather than curvature \citep{Bengochea:2008gz, Linder:2010py, Cai:2015emx}. Unlike curvature-based theories, the field equations of $f(T)$ gravity remain second order, making the theory mathematically simpler and free from higher-derivative instabilities. Due to these features, $f(T)$ gravity has been widely studied in cosmology and has been shown to successfully describe inflationary and late-time acceleration scenarios \citep{Cai:2015emx}. However, its applications to compact stellar structures remain comparatively limited.

In GR, static and spherically symmetric compact stars are described through TOV equations supplemented by an appropriate EoS. These models provide well-established bounds on mass–radius ratios and stability criteria \citep{buchdahl1959, Andreasson:2007ck}. Extending this framework to modified gravity theories allows one to explore deviations in stellar structure, stability, and compactness. In this context $f(T)$ gravity is particularly appealing, as torsion-induced effects can influence the internal structure of highly dense objects such as NS and SS \citep{Boehmer:2011gw, Ilijic:2018ulf}.

To systematically construct anisotropic stellar configurations in modified gravity, the gravitational decoupling approach via MGD has proven to be a powerful and versatile method \citep{PhysRevD.95.104019, Ovalle_2019}. This technique allows one to generate new anisotropic solutions from known seed metrics and has been successfully applied to compact stars in various gravitational frameworks, including higher-dimensional and Gauss–Bonnet theories \citep{Maurya:2021tca}. Motivated by these developments, the present work employs GD through the MGD formalism within $f(T)$ gravity to construct a novel exact anisotropic stellar solution.
Specifically, we consider a polytropic fluid source and adopt the Buchdahl metric ansatz for the radial metric component. By combining these ingredients with the gravitational decoupling method, we obtain an exact and physically viable solution that describes a compact star in $f(T)$ gravity. This approach enables us to systematically investigate the influence of torsion, anisotropy, and decoupling effects on stellar structure and stability.

This chapter is organized as follows: First, section~\ref{sec2} outlines the basic mathematical framework of $f(T)$ gravity with an additional source. The MGD solution is developed in section~\ref{sec3}, while the model parameters are determined through smooth matching conditions at the stellar surface in section~\ref{sec4}. The deformed SS configurations and their astrophysical relevance are discussed in section~\ref{sec5}, followed by an analysis of the corresponding mass–radius relations in section~\ref{sec6}. The stability of the models is examined in section~\ref{7a} using the adiabatic index and the Harrison–Zeldovich–Novikov (HZN) stability criterion. Finally, the final remarks are presented in section~\ref{sec7}.

\section{Field equations in $f(T)$ gravity with an additional source}\label{sec2}

As outlined in the introduction, teleparallel gravity describes gravitation through the torsion scalar $T$ rather than the curvature of spacetime. In this framework $f(T)$ gravity arises by extending $T$ to a general function $f(T)$.
The E-H action can be expressed as \cite{aldrovandi2013}
\begin{eqnarray}\label{action}
    \mathcal{A}=\int \sqrt{-g} \Big[\frac{1}{16\pi }f(T)+\mathcal{L_{\mathrm{M}}}+\mathcal{L_{\mathrm{\Theta}}}\Big]d^4x.
\end{eqnarray}
The matter content is described by the Lagrangian density $\mathcal{L}_{\mathrm{M}}$, which is associated with the energy momentum tensor (EMT) $\mathcal{T}_{\mu\nu}$. In addition, $\mathcal{L}_{\Theta}$ represents an extra gravitational sector characterized by the source $\Theta_{\mu\nu}$. The presence of this sector modifies the matter distribution through effective EMT
$\mathcal{T}_{\mu\nu}^{\mathrm{eff}} = \mathcal{T}_{\mu\nu} + \alpha\,\Theta_{\mu\nu},$
where $\alpha$ denotes the coupling parameter governing the interaction between the matter fields and the $\Theta$ gravitational source. As a result, the stellar configuration can be modeled as an effective anisotropic fluid distribution
\begin{eqnarray}
\big\{\mathcal{T}_{\rho}^{~~\nu}\big\}^{\mathrm{eff}}=(\rho^{\mathrm{eff}}+p_t^{\mathrm{eff}}) u_{\mathrm{\nu}} u^{\mathrm{\rho}}-p_t^{\mathrm{eff}} \delta_{\nu}^{\rho}+
(p_r^{\mathrm{eff}}-p_t^{\mathrm{eff}}) v_{\mathrm{\nu}} v^{\mathrm{\rho}},
~~~~
\end{eqnarray}
where, $\rho^{\mathrm{eff}}$ denotes the effective energy density, while $p_r^{\mathrm{eff}}$ and $p_t^{\mathrm{eff}}$ represent the effective radial and tangential pressures, respectively. The time-like four-velocity is given by $u_\nu$, and $v_\nu$ denotes the unit space-like vector along the radial direction. Consequently, the distribution of matter within the star is modeled as an anisotropic fluid with effective EMT components
$\mathcal{T}^{\mu}_{\ \nu}{}^{\mathrm{eff}}=\mathrm{diag}\big(-\rho^{\mathrm{eff}},\,p_r^{\mathrm{eff}},\,p_t^{\mathrm{eff}},\,p_t^{\mathrm{eff}}\big)$ where
\begin{eqnarray}    \rho^{\text{eff}}=\rho+\alpha \Theta^{0}_{0}\, , \quad p_r^{\text{eff}}=p_r-\alpha \Theta^{1}_{1}\, , \quad p_t^{\text{eff}}=p_t-\alpha \Theta^{2}_{2}.~~~~
\end{eqnarray}
Regarding the effective term, the appropriate anisotropy factor is
\begin{eqnarray} \Delta^{\text{eff}}=p_t^{\text{eff}}-p_r^{\text{eff}}=(p_t-p_r)+\alpha(\Theta^{1}_{1}-\Theta^{2}_{2})
=\Delta_{\mathcal{T}}+\alpha \Delta_{\Theta}.~~~~
\end{eqnarray}
It is notable that within the current anisotropic compact stellar system, there exist two distinct forms of anisotropies $\Delta_{\mathcal{T}}$ and $\Delta_{\Theta}$ from $\mathcal{T}_{\mu\nu}$ and $\Theta_{\mu\nu}$ respectively. The additional form of anisotropy, $\Delta_{\Theta}$, becomes relevant due to GD, which plays a distinct role in transformation processes.
Now, for the standard static and spherically symmetric line element Eq. (\ref{metric1st}), the torsion scalar $T$ depends only on the radial coordinate $r$ and is given by
\begin{eqnarray}\label{torsion}
&&\hspace{0.5cm} T(r)=\frac{2 e^{-\lambda}}{r}\left[\nu^{\prime}+\frac{1}{r}\right].
\end{eqnarray}
For the metric (\ref{metric1st}), the tetrad matrix could be written as $e^n_{~i} =\Big(e^{\frac{\nu}{2}},e^{\frac{\lambda}{2}},r,rsin\theta\Big).$
Substituting the tetrad field and incorporating the torsion scalar into the field Eq.~(\ref{eq:TEGR_field_eq}), we can calculate the independent component of the field equation for an anisotropic fluid in $f(T)$ gravity as
\begin{small}
\begin{eqnarray}\label{fe1}
&&\hspace{0cm}8 \pi \rho^{\text{eff}} = \frac{f(T)}{2}+f_T\bigg[\frac{1}{r^2}+\frac{e^{-\lambda (r)}}{r} \big\{\lambda '(r)+\nu '(r)\big\}-T(r)\bigg],~~~~\\ 
\label{fe2}
 &&\hspace{0cm}8 \pi p_r^{\text{eff}} = -\frac{f(T)}{2}+f_T\Big[T(r)-\frac{1}{r^2}\Big], \\ \label{fe3}
 &&\hspace{0cm}8 \pi p_t^{\text{eff}} = -\frac{f(T)}{2}+f_T \Bigg[\frac{T(r)}{2}+e^{-\lambda (r)} \bigg\{\frac{\nu''(r)}{2}+\Big(\frac{\nu '(r)}{4}+\frac{1}{2 r}\Big)\big(\nu'(r)-\lambda'(r)\big)\bigg\}\Bigg]. 
\end{eqnarray}
\end{small}
The field equations discussed above in (\ref{fe1}-\ref{fe3}), directly produce the equivalent field equations in GR when considering the functional form of $f(T)=T$. However, in the context of $f(T)$ gravity, an additional non-diagonal quantity emerges as follows
\begin{eqnarray}\label{fe4}
    \frac{\cot\Theta}{2 r^2} T^{\prime}f_{TT} = 0 .
\end{eqnarray}
Where, $f_T=\partial f/\partial T$  and $f_{TT} =\partial^2f/\partial T^2$. Now, from Eq.~(\ref{fe4}) we get:\\
\textbf{\text{Case I}} : $T^{\prime}=0$ $\Rightarrow T=\text{constant}=T_0$ i.e., $T$ is independent of $r$ and therefore ${T}$, $f_{TT}$ remains constant.\\
   \textbf{{\text{Case II}}} : $f_{TT}=0$, which gives $f$ as a linear function of two model parameters $\zeta_1$ and $\zeta_2$ that is, $f(T)=\zeta_1 T+\zeta_2$.\\
 Our objective now is to address the solution of the $f(T)$ gravity field Eqs. (\ref{fe1}-\ref{fe3}) in the functional form $f(T)=\zeta_1 T+\zeta_2$. To achieve this goal, we aim to utilize the well-known methodology known as GD via the MGD approach, employing a polytropic fluid source.
Now, inserting Eq.~(\ref{torsion}) and the linear form of $f(T)$ into Eqs.~(\ref{fe1}-\ref{fe3}) we can obtain the equation of motion as follows
\begin{small}
\begin{eqnarray}\label{FE11}
&&\hspace{0cm}4 \pi \rho^{\text{eff}} = \frac{e^{-\lambda (r)}}{4r^2} \Big[-2 \zeta_1+e^{\lambda (r)} (2 \zeta_1+\zeta_2 r^2)+2 \zeta_1 r \lambda'(r)\Big],~~~
\\ \label{FE21}
&&\hspace{0cm}4\pi p_r^{\text{eff}} = \frac{e^{-\lambda (r)}}{4r^2} \Big[2 \zeta_1-e^{\lambda (r)} (2 \zeta_1+\zeta_2 r^2)+2 \zeta_1 r \nu '(r)\Big],~~~
\\ \label{FE31}
&&\hspace{0cm}4\pi p_t^{\text{eff}} = \frac{e^{-\lambda (r)}}{8r} \Big[-\zeta_1 (r \nu '(r)+2) (\lambda '(r)-\nu '(r))+2 \zeta_1 r \nu ''(r)-2 \zeta_2 r e^{\lambda (r)}\Big].
\end{eqnarray}
\end{small}
Furthermore, the conservation equation is given by, $\big(\nabla_{i}\big[\mathcal{T}_j^{i}\big]^{\text{eff}}=0\big)$, which gives
\begin{small}
\begin{eqnarray}\label{tov1}
    &&\hspace{-0.1cm}-\frac{\mathcal{\nu^{\prime}}}{2}\big(\rho^{\text{eff}}+p_r^{\text{eff}}\big)-\frac{dp_r^{\text{eff}}}{dr}+\frac{2}{r}\big(p_t^{\text{eff}}-p_r^{\text{eff}}\big)=0,\nonumber \\ \label{tov2}
    &&\hspace{-0.2cm}\rightarrow -\frac{dp_r}{dr}-\frac{\mathcal{\nu^{\prime}}}{2}(\rho+p_r)+\frac{2}{r}\big(p_t-p_r\big)+\alpha \frac{d \Theta_1^1}{dr}-\frac{\mathcal{\nu^{\prime}}}{2}\alpha (\Theta_0^0-\Theta_1^1)-\frac{2}{r}\alpha\big(\Theta_2^2-\Theta_1^1\big)=0.~~
\end{eqnarray}
\end{small}
Here, Eq.~(\ref{tov1}) represents the standard TOV equation in the decoupling framework \citep{ns3,Ovalle:2017wqi}. In the limit $\alpha \to 0$, the usual perfect-fluid field equations of $f(T)$ gravity are recovered. This motivates the use of gravitational decoupling via the MGD scheme to solve Eqs.~(\ref{FE11}–\ref{FE31}) for the compact star model, where the additional source $\Theta^{\mu}_{\ \nu}$ induces an effective quasi–$f(T)$ system through a suitable geometric deformation of the seed spacetime. The deform spacetime is
\begin{eqnarray}\label{mgd11}
\nu(r) \longrightarrow G(r)+\alpha\, \Phi(r),\\ \label{mgd21}
    \lambda(r) \longrightarrow-\log [H(r)+\alpha \, \psi (r)]. 
\end{eqnarray}
Here, $G(r)$ and $H(r)$ denote the temporal and inverse radial metric functions, respectively, of the seed spacetime corresponding to the pure perfect-fluid solution in $f(T)$ gravity, while $\Phi(r)$ and $\psi(r)$ represent the geometric distortions experienced by the temporal and radial metric components, respectively, induced by the additional source $\Theta^{\mu}_{\,\nu}$. For the MGD technique, we set $\Phi(r)\to 0$, therefore the metric in Eq.~(\ref{mgd11}) and (\ref{mgd21}) is minimally deformed by the underlying factor.
\begin{eqnarray}\label{MGD12}
  &&\hspace{0cm}   \nu(r)= G(r),\\ \label{MGD22}
  &&\hspace{0cm} \lambda(r)= -\log [H(r)+\alpha \, \psi (r)]. 
\end{eqnarray}
It should be noted that Eq.~(\ref{MGD22}) represents a linear deformation of the inverse radial metric component, consisting of a pure perfect-fluid contribution and an additional source by $\Theta_{\mu\nu}$. Substituting Eqs.~(\ref{MGD12}) and (\ref{MGD22}) into the field Eqs.~ (\ref{FE11}–\ref{FE31}), the system naturally decouples into two independent sets:
\begin{itemize}
    \item The normal field equations for perfect fluid $(\alpha=0)$, whose components are $\big\{\rho, p_r, p_t, G(r), H(r)\big\}$
    \begin{small}
 \begin{eqnarray}\label{em1}
&&\hspace{-0.6cm}\rho(r) = \frac{1}{16 \pi r^2}\Big[ 2 \zeta_1-2 \zeta_1 (r H'(r)+H(r))+\zeta_2 r^2\Big],~~
\\ \label{em2}
&&\hspace{-0.6cm}p_r = \frac{1}{16\pi r^2}\Big[-2 \zeta_1 H(r) (r G'(r)+1)+2 \zeta_1+\zeta_2 r^2\Big],~~~
\\ \label{em3}
&&\hspace{-0.6cm}p_t = \frac{1}{32\pi r}\Big[2 \zeta_1 r H(r) G''(r)+\zeta_1 \big(r G'(r)+2\big)\big(H(r) G'(r)+H'(r)\big)-2 \zeta_2 r\Big].
\end{eqnarray}
\end{small}
\item Another set of equations for $\Theta_{\mu\nu}$ source having components $\big\{\rho, p_r, p_t, G(r), \psi(r)\big\}$ 
\begin{small}
\begin{eqnarray}\label{t11}
&&\hspace{0cm}\Theta^0_0 = -\frac{\zeta_1}{8\pi r^2}\Big[ \big(r \psi '(r)+\psi (r)\big)\Big],
\\ \label{t21}
&&\hspace{0cm}\Theta_1^1 = \frac{-\zeta_1}{8\pi r^2}\Big[ \psi (r) \big(r G'(r)+1\big)\Big],
\\ \label{t31}
&&\hspace{0cm}\Theta^2_2 = \frac{-\zeta_1}{32\pi r}\Big[2 r \psi (r) G''(r)+(r G'(r)+2) \big(\psi (r) G'(r)+\psi'(r)\big)\Big].
\end{eqnarray}
\end{small}
\end{itemize}
Assuming that there is no exchange of energy--momentum between the perfect fluid $\mathcal{T}_{\mu\nu}$ and the additional source $\Theta_{\mu\nu}$, and that their interaction is purely gravitational, Eq.~(\ref{tov2}) reduces to
\begin{small}
\begin{eqnarray}
     -\frac{dp_r}{dr}-\frac{\mathcal{\nu^{\prime}}}{2}(p_r+\rho)-\frac{2}{r}\big(p_r-p_t\big)=0, \quad~~~
     \frac{d \Theta_1^1}{dr}-\frac{\mathcal{\nu^{\prime}}}{2} (\Theta_0^0-\Theta_1^1)-\frac{2}{r}\big(\Theta_2^2-\Theta_1^1\big)=0.
\end{eqnarray}
\end{small}
These are referred to as the improved TOV equations for pure $f(T)$ gravity and the $\Theta$ gravitational sector resulting from $\nabla_{\mu}\mathcal{T}^{\mu\nu}=0$ and $\nabla_{\mu}\Theta^{\mu\nu}=0$ respectively. Furthermore, the active gravitational mass function for two systems may be expressed as follows.
\begin{eqnarray}
    M_{\mathcal{T}}(r)=\int_0^r 4 \pi \tilde{r}^2 \rho(\tilde{r})\, d\tilde{r} ;\quad  M_{\Theta}(r)= \int_0^r 4 \pi \tilde{r}^2 \, \Theta_0^0(\tilde{r}) \, d \tilde{r}
\end{eqnarray}
 In the framework of $f(T)$ gravity, the pertinent mass functions for the sources $\mathcal{T}_{\mu\nu}$ and $\Theta_{\mu\nu}$ are denoted as $ M_{\mathcal{T}}(r)$ and $ M_{\Theta}(r)$, respectively. Subsequently, within the context of minimally deformed spacetime, the total interior mass function can be represented as
 \begin{eqnarray}
     M(r) = M_{\mathcal{T}}(r) -\frac{\zeta_1 \alpha }{2} r\, \psi(r).
 \end{eqnarray}

\section{Minimally gravitationally decoupled solution in $f(T)$ gravity}\label{sec3}
This section focuses on the two coupled systems of Eqs.~(\ref{em1}–\ref{em3}) and (\ref{t11}–\ref{t31}), corresponding to sources $\mathcal{T}_{\mu\nu}$ and $\Theta_{\mu\nu}$, respectively. The equations governing the $\Theta$ sector explicitly depend on the solutions of the seed system, which makes it natural to first solve Eqs.~(\ref{em1}–\ref{em3}). These field equations are highly non-linear and involve five unknown functions $\{\rho, p_r, p_t, G(r), H(r)\}$ but only three independent equations. To close the system, additional assumptions are required. A standard approach is to prescribe one metric potential and impose a physically motivated condition such as an EoS or an embedding constraint to determine the remaining variables. In the present work, we adopt a quadratic polytropic EoS
\begin{eqnarray}\label{eos}
    p_r = \gamma~ \rho^{1+\frac{1}{n}}+ \beta ~\rho +\chi ~.
\end{eqnarray}
A detailed discussion of this EoS is provided in the introduction section. Due to the high non-linearity in finding an exact solution, here we consider a polytropic index $n=1$. Now, substituting the expression of density and radial pressure from Eqs.~(\ref{em1}) and (\ref{em2}) into the EoS (\ref{eos}) we get
\begin{small}
\begin{eqnarray}\label{un}
 &&\hspace{0cm} 4 \zeta_1 r^2 H(r) (r G'(r)+1)+4 \beta  \zeta_1 r^2 (r H'(r)+H(r))-\gamma  (2 \zeta_1-2 \zeta_1 (r H'(r)+H(r))+\zeta_2 r^2)^2\nonumber\\ &&\hspace{0cm}-2 \beta  \zeta_2 r^4-2 \zeta_2 r^4-4 r^4 \chi -4 \beta  \zeta_1 r^2-4 \zeta_1 r^2 = 0.~~
\end{eqnarray}
\end{small}
There are two unknowns in Eq.~(\ref{un}) namely $G(r)$ and $H(r)$ correspond to $g_{tt}$ and $g_{rr}$ components, respectively. Consequently, there are two ways to find the exact solution, either by considering a suitable metric potential along the temporal components or a radial direction. For the present study, we consider a well-known ansatz Buchdahl metric for $H(r)$ as \citep{Prasad_2021,Kumar:2005kuy,Mkenyeleye:2014dwa,Gupta:2003guj}
\begin{eqnarray}\label{buch}
    H(r) = \frac{A+B\,r^2}{A(1+B\,r^2)}, \quad 0<A<1~,
\end{eqnarray}
where, $A$ is dimensionless and $B$ is the parameter with dimension $\text{km}^{-2}$. 
The metric function and its radial derivative exhibit non-singularity at the center of the stellar structure, fulfilling the necessary condition i.e.
\begin{eqnarray}
    H(0)=1 ~~~~ \text{and} ~~~~~~ \partial_r H(r)_{r=0}=0.
\end{eqnarray}
The most interesting characteristic of the Buchdahl solution is that assuming \( C = -A/R^2 \) allows retrieving the Vaidya and Tikekar solution \cite{Vaidya:1982vt}, and for \( A = -2 \), one obtains the Durgapal and Bannerji solution \cite{Durgapal}.
Now, for the known $H(r)$ we have a first-order non-linear differential equation in $G(r)$ (Eq.~\ref{un}) for which we can obtain the solution for another metric potential $G(r)$ as 
{\footnotesize
\begin{eqnarray}\label{metric}
 &&\hspace{-0.5cm}   G(r)= \frac{1}{4 A \zeta_1}\bigg[\frac{1}{2 B(1+Br^2)^2} \Big\{A^2 (B r^2+1)^3 (\zeta_2 (2 \beta +\gamma  \zeta_2+2)+4 \chi )-8 (A-1) B^2 \gamma  \zeta_1^2-16 (A-2) B^2 \gamma  \zeta_1^2 (B r^2+1)\Big\}\nonumber\\&&\hspace{0.6cm}+\frac{\text{log}(-A-Br^2)}{2B(A-1)}\times\Big\{A^4 (2 (\beta +1) \zeta_2+\gamma  \zeta_2^2+4 \chi )-2 A^3\big(2 (\beta +1) \zeta_2+2 B \zeta_1 (\beta +\gamma  \zeta_2+1)  +\gamma  \zeta_2^2+4 \chi \big)+A^2 \nonumber\\&&\hspace{0.6cm} \big(2(\beta +1) \zeta_2+4 B^2 \gamma  \zeta_1^2+8 B \zeta_1 (2 \beta +2 \gamma  \zeta_2+1)+\gamma  \zeta_2^2+4 \chi \big)-4 A B \zeta_1 (3 \beta +6 B \gamma  \zeta_1+3 \gamma  \zeta_2+1)+36 B^2 \gamma  \zeta_1^2\Big\}\nonumber\\&&\hspace{0.6cm}+\frac{1}{A-1}\Big\{2 \zeta_1 \log (B r^2+1) \times\big(A^2 (2 \beta +B \gamma  \zeta_1+2 \gamma  \zeta_2)-2 A (\beta +\gamma  (3 B \zeta_1+\zeta_2))+9 B \gamma  \zeta_1\big)\Big\}\bigg]+\mathcal{G}.
\end{eqnarray}
}
Here, $\mathcal{G}$ represents an arbitrary integration constant. Employing the expressions for $G(r)$ and $H(r)$, we derive the expressions for $\rho$, $p_r$, and $p_t$ as follows.
{\footnotesize
\begin{eqnarray}\label{s1}
     &&\hspace{-0.2cm}\rho = \frac{(A-1) B \zeta_1 (B r^2+3)}{A (B r^2+1)^2}+\frac{\zeta_1}{2}, \\ \label{s2}
  &&\hspace{-0.2cm} p_r=\frac{1}{4 A^2 (1+Br^2)^4}\bigg[4 B^2 (3 + B r^2)^2 \gamma \zeta_1^2 - 4 A B (3 + B r^2) \zeta_1  \big((1 + B r^2)^2 \beta + 2 B (3 + B r^2) \gamma \zeta_1 + (1 + B r^2)^2 \nonumber\\&&\hspace{0.6cm} \gamma \zeta_2\big) +A^2 \Big((2 B (3 + B r^2) \zeta_1 + (1 + B r^2)^2 \zeta_2) \big(2 (1 + B r^2)^2 \beta + 2 B (3 + B r^2) \gamma \zeta_1 + (1 + B r^2)^2 \gamma \zeta_2\big) \nonumber\\&&\hspace{0.6cm}+ 4 (1 + B r^2)^4 \chi\Big)\bigg], \\ \label{s3}
 &&\hspace{-0.2cm}    p_t = \frac{1}{8}\Bigg[ \frac{A+Br^2}{A^2(A-1)(1+Br^2)^5}\bigg\{-16 (A-1) B^2 (3 - 2 A + B r^2 (7 A-9)+ 3 (-2 + A) B^2 r^4) \gamma \zeta_1^2-4 B \nonumber\\&&\hspace{0.6cm} (-1 + B r^2)(1 + B r^2)^2 \zeta_1 \big(9 B \gamma \zeta_1 + A(2 (-1 + A) \beta + (A-6) B \gamma \zeta_1 + 2 (-1 + A) \gamma \zeta_2)\big)+A^2 (A-1) \nonumber\\&&\hspace{0.6cm} (1+Br^2)^4(\zeta_2 (2 + 2 \beta + \gamma \zeta_2) + 4 \chi)+P_{t1}(r)\Bigg].
\end{eqnarray}}
where,
{\footnotesize
\begin{eqnarray}
   &&\hspace{-0.2cm}  \mathcal{F}_{\mathrm{1}}(r)=\Big(A^2 (\zeta_2 (2 \beta +\gamma  \zeta_2+2)+B^4 r^4 (4 r^4 \chi +(2 \zeta_1+\zeta_2 r^2)  (2 \gamma  \zeta_1+r^2 (2 \beta +\gamma  \zeta_2+2)))+4 B^3 (r^6 (\zeta_2 (2 \beta +\gamma  \zeta_2+2)\nonumber\\&&\hspace{0.6cm}+4 \chi )+\zeta_1 r^4 (5 \beta  +5 \gamma  \zeta_2+1)+6 \gamma  \zeta_1^2 r^2)+2 B^2 (18 \gamma  \zeta_1^2+3 r^4 (\zeta_2 (2 \beta +\gamma  \zeta_2+2)+4 \chi ) +2 \zeta_1 r^2 (7 \beta +7 \gamma  \zeta_2-1))\nonumber\\&&\hspace{0.6cm}+4 B (\zeta_1 (3 \beta +3 \gamma  \zeta_2-1)+r^2 (\zeta_2 (2 \beta  +\gamma  \zeta_2+2)+4 \chi ))+4 \chi )+4 A B \zeta_1 \big(-(B r^2+1)^2 (3 \beta +(\beta +1) B r^2 -1)\nonumber\\&&\hspace{0.6cm}-2 B \gamma  \zeta_1  (B r^2+3)^2-\big\{\gamma  \zeta_2 (B r^2+3) (Br^2+1)^2\big\})\big)+4 B^2 \gamma  \zeta_1^2 (B r^2+3)^2\Big).
\end{eqnarray}
}

The given Eqs.~(\ref{s1}-\ref{s3}) provide the full spacetime geometry for the internal structure. However, to address the $\Theta$ sector, it is necessary to determine the solution for the second set of Eqs.~(\ref{t11}-\ref{t31}). To solve the second set of equations, we propose utilizing well-established techniques, specifically: (i) mimicking the density constraint, where $\rho=\Theta_0^0$, and (ii) mimicking the radial pressure constraint, where $p_r=\Theta_1^1$. These well-known techniques are physically motivated and thoroughly elaborated in the reference \citep{PhysRevD.95.104019}. 
\subsection{Mimicking the density constraints i.e. ($\rho=\Theta_0^0$)} \label{solution1}
To address the solution of the $\Theta$-sector, we replicate the seed density to $\Theta_0^0$, i.e., $\rho=\Theta^0_0$. From Eq.~(\ref{em1}) and (\ref{t11}) we get an ordinary differential equation in the deformation function $\psi(r)$ as
\begin{small}
\begin{eqnarray}
     \frac{d\psi(r)}{dr}+\frac{\psi(r)}{r} =\frac{1}{2r \zeta_1} \Big[2\zeta_1\big\{rH^{\prime}(r)+H(r) -1\big\}-r^2 \zeta_2\Big].~~~~
\end{eqnarray}
\end{small}
Using the known metric function $H(r)$ from Eq.~(\ref{buch}) we solved the above first order and first degree differential equation to get the deformation function as
\begin{small}
\begin{eqnarray}\label{deform1}
    \psi(r) =-\frac{r^2 \big(6 (A-1) B \zeta_1+A B \zeta_2 r^2+A \zeta_2\big)}{2 A \zeta_1 \big(3 B r^2+3\big)}+\frac{C_1}{r}.~~~~
\end{eqnarray}
\end{small}
where $C_1$ is the integrating constant which is set to zero for the sake of a non-singular solution.
Now, utilizing the above deformation function (\ref{deform1}), we obtain the solution of the $\Theta$ sector from Eqs.~(\ref{t21}-\ref{t31}) as
\begin{small}
\begin{eqnarray}
&&\hspace{-0.2cm}    \Theta^1_1 = \Big[4ABr^2(1 + Br^2)^3 \zeta_1+4(-1 + A)^2 B^2 r^2 (3 + Br^2)^2  \gamma \zeta_1^2
+4(-1 + A)ABr^2(1 + Br^2)^2 \zeta_1 (1 + 3\beta \nonumber\\&&\hspace{0.6cm}+ 3\gamma \zeta_2 + Br^2(1 + \beta + \gamma \zeta_2))
(A^2)((1 + Br^2)^3)
\big\{4\zeta_1+r^2 \big(B r^2+1\big) (\zeta_2 (2 \beta +\gamma  \zeta_2+2)+4 \chi )\big\}\Big]\nonumber\\&& \hspace{0.6cm} \times \frac{6 (A-1) B \zeta_1+A B \zeta_2 r^2+A \zeta_2}{24 A^2 \zeta_1 \big(B r^2+1\big)^4 \big(A+B r^2\big)},\\
 &&\hspace{-0.2cm}   \Theta_2^2= \frac{1}{288 (B \zeta_1 r^2+\zeta_1)^2}\Bigg[\frac{6 r^2}{A^2} (B \zeta_1 r^2+\zeta_1) (A B \zeta_2 r^2+6 (A-1) B \zeta_1+A \zeta_2)\bigg\{-\frac{64 (A-2) r^2 \gamma  \zeta_1^2 B^3}{(B r^2+1)^3}\nonumber\\&& \hspace{0.6cm}-\frac{96 (A-1) r^2 \gamma  \zeta_1^2 B^3}{(B r^2+1)^4}+\frac{16 (A-2) \gamma  \zeta_1^2 B^2}{(B r^2+1)^2}+\frac{16 (A-1) \gamma  \zeta_1^2 B^2}{(B r^2+1)^3}-\frac{(A-1)^{-1}}{ (B r^2+1)^2}\big[8 r^2 \zeta_1 (9 B \gamma  \zeta_1 \nonumber\\&& \hspace{0.6cm}+A (2 (A-1) \beta +(A-6) B \gamma  \zeta_1+2 (A-1) \gamma  \zeta_2)) B^2\big]+\frac{(A-1)^{-1}}{ (B r^2+1)}\big[4 \zeta_1 (9 B \gamma  \zeta_1+A (2 (A-1) \beta\nonumber\\&& \hspace{0.6cm} +(A-6) B \gamma  \zeta_1+2 (A-1) \gamma  \zeta_2)) B\big]+\Theta_{22}(r)\Bigg].
\end{eqnarray}
\end{small}
\subsection{Mimicking the pressure constraints i.e. ($p_r=\Theta_1^1$)} \label{solution2}
In this technique, next, we mimic the radial pressure $p_r$ to $\Theta_1^1$ i.e, $p_r=\Theta^1_1$. Therefore, from Eqs~(\ref{em2}), (\ref{t21}) and using Eq.~(\ref{metric}) we get the deformation function as
\footnotesize
\begin{eqnarray}\label{deform2}
    &&\hspace{-0.1cm}\psi(r)= \frac{1}{\mathcal{F}_3(r)}\bigg[2 \beta A^{-1}  r^2 (B r^2+1) (A+B r^2) \big(2 B \zeta_1(1-A) (B r^2+3)-A \zeta_2 (B r^2+1)^2\big)\nonumber\\&&\hspace{0.6cm}-\frac{1}{A^2(1+Br^2)}\big\{\gamma  (A+B r^2) \big(2 (A-1) B \zeta_1 r (B r^2+3)+A \zeta_2 r (B r^2+1)^2\big)^2\big\}\bigg].
\end{eqnarray}
where 
\begin{small}
    $\mathcal{F}_3(r)=4 A^{-1}(A-1)^2 B^2 \gamma  \zeta_1^2 r^2 \big(B r^2+3\big)^2+4 \zeta_1 \big(B r^2+1\big)^2 \big(A B r^2 \big(3 \beta +B r^2 (\beta +\gamma  \zeta_2+1)\\~~~~~+3 \gamma  \zeta_2+2\big)+A-B r^2 \big(B r^2+3\big) (\beta +\gamma  \zeta_2)\big)+A r^2 \big(B r^2+1\big)^4 (\zeta_2 (2 \beta +\gamma  \zeta_2+2)+4 \chi ).\nonumber$
\end{small}\\
Using the above deformation function Eq.~(\ref{deform2}) we have determined the other components of the $\Theta$ sector as:
\footnotesize
\begin{eqnarray}
   &&\hspace{-0.2cm} \Theta_0^0 = \frac{1}{\psi_{11}}\big[4 \zeta_1 (B r^2+1)^2\Big\{-2A^{-2}(1+Br^2)^{-4} (A-1) B^2 \zeta_1 r^2 (B r^2+5) (A+B r^2) \Big(A \big(\beta  (B r^2+1)^2+2 B\nonumber\\&&\hspace{0.6cm}  \gamma  \zeta_1 (B r^2+3)+\gamma  \zeta_2 (B r^2+1)^2\big)-2 B \gamma  \zeta_1 (B r^2+3)\Big)\Big\{\mathcal{L}+4 \zeta_1 (B r^2+1)^2 (A B r^2 (3 \beta +B r^2 (\beta +\gamma  \zeta_2+1)\nonumber\\&&\hspace{0.6cm} +3 \gamma  \zeta_2+2)+A-B r^2 (B r^2+3) (\beta +\gamma  \zeta_2))+\Theta_{00}(r)\big],\\
   &&\hspace{0cm} \Theta^2_{2}=\frac{-\zeta_1}{4r}\Bigg[\frac{1}{\psi_{33}}\bigg\{8 B \zeta_1 r^3 \big(B r^2+1\big)^3 \big(A+B r^2\big)(-\mathcal{N}\beta-\mathcal{N}^2\gamma-\chi) (1+Br^2)^{-4} \Big(-16 (A-2) B^2\nonumber\\
&&\hspace{0.6cm} \gamma  \zeta_1 r^2 \big(B r^2+1\big)-24 B^2 \gamma  \zeta_1 \times (A-1)r^2-2 B (A-1)^{-1} \big(B r^3+r\big)^2 (A (2 (A-1) \beta +(A-6) B \gamma  \zeta_1+2\nonumber\\
&&\hspace{0.6cm} (A-1) \gamma  \zeta_2)+9 B \gamma  \zeta_1)+(A-1)^{-1}\times\big(B r^2+1\big)^3  (A (2 (A-1) \beta +(A-6) B \gamma  \zeta_1+2 (A-1) \gamma  \zeta_2)+9 B \nonumber\\
&&\hspace{0.6cm}\gamma  \zeta_1)+4 (A-2) B \gamma  \zeta_1 \big(B r^2+1\big)^2+4 (A-1) B \gamma  \zeta_1 \times\big(B r^2+1\big)\Big)+\frac{A^4(A-1)^{-1}}{(Br^2+1)^4} \big(\zeta_2 (2 \beta +\gamma  \zeta_2+2)\nonumber\\
&&\hspace{0.6cm}+4 B \zeta_1 (\beta +\gamma  \zeta_2+1)+3 B r^2 (\zeta_2 (2 \beta +\gamma  \zeta_2+2)+4 \chi )+4 \chi \big)-A^3 \big(\zeta_2 (2 \beta +\gamma  \zeta_2+2)+B^2 \big(4 \gamma  \zeta_1^2\nonumber\\
&&\hspace{0.6cm}-\big(r^4 (\zeta_2 (2 \beta +\gamma  \zeta_2+2)+4 \chi )\big)+4 \zeta_1 r^2 (\beta +\gamma  \zeta_2+1)\big)+4 B \big(\zeta_1 (4 \beta +4 \gamma  \zeta_2+2)+r^2 (\zeta_2 (2 \beta +\gamma  \zeta_2\nonumber\\
&&\hspace{0.6cm}+2)+4 \chi )\big)+4 \chi \big)+A^2 B \big(4 \zeta_1 \big(3 \beta +2 B r^2 (2 \beta +2 \gamma  \zeta_2+1)+3 \gamma  \zeta_2+1\big)-\big(r^2 \big(B r^2-1\big) (\zeta_2 (2 \beta \nonumber\\
&&\hspace{0.6cm}+\gamma  \zeta_2+2)+4 \chi )\big)+4 B \gamma  \zeta_1^2 \big(B r^2+6\big)\big)-4 A B^2 \zeta_1\big(r^2 (3 \beta +6 B \gamma  \zeta_1+3 \gamma  \zeta_2+1)+9 \gamma  \zeta_1\big)+36 B^3 \gamma  \zeta_1^2 r^2\bigg\}\nonumber\\
&&\hspace{0.6cm}+\Theta^{22}(r)\Bigg].
\end{eqnarray}
For long expression, we have not given detailed expression of $P_{t1}(r)$, $\Theta_{22}(r)$, $\psi_{11}$, $\mathcal{L}$, $\Theta_{00}(r)$, $\psi_{33}$, $\mathcal{N}$ and $\Theta^{22}(r)$.

\section{Matching condition: exterior spacetime}\label{sec4}

An essential requirement in modeling stellar configurations is the continuity of the spacetime geometry across the stellar surface 
\( r=R \), separating the interior (\( r<R \)) and exterior 
(\( r>R \)) regions. In the context of \( f(T) \) gravity, the presence of the off-diagonal field 
Eq.~(\ref{fe4}) imposes strong constraints on admissible solutions. As shown by Boehmer \textit{et al.}~\cite{Boehmer:2011gw}, 
for a static spherically symmetric spacetime with a diagonal tetrad, physically viable solutions arise only when either 
\( T'=0 \) or \( f_{TT}=0 \), both of which have been discussed in the previous section. To determine the exterior spacetime, we consider the vacuum region where the matter energy-momentum tensor vanishes, i.e., $\rho=0,~ p_r=0, ~ p_t=0 .$ Under this condition, the field Eqs.~(\ref{FE11})--(\ref{FE31}) reduce to the following system
\begin{eqnarray}
&&\hspace{0cm}\nu' + \lambda' = 0, \label{r1}\\
&&\hspace{0cm}T =\frac{\zeta_2}{\zeta_1} + \frac{2}{r^2}, \label{r2}\\
&&\hspace{0cm}\frac{\zeta_2}{2} + \zeta_1 e^{-\lambda} 
\Bigg[\frac{\nu''}{2}
+ \left(\frac{\nu'}{4} + \frac{1}{2r}\right)
(\nu'-\lambda') \Bigg] = 0. \label{r3}
\end{eqnarray}
These equations govern the exterior vacuum geometry in \( f(T) \) gravity and form the basis for matching the interior 
stellar solution at the boundary \( r=R \).
Now, by integrating w.r.t $r$ from Eq~(\ref{r1}) one can get the form of the metric potential as
    $\nu(r)=-\lambda(r)+ b_0.$ For the Schwarzschild solution $\nu(r),\lambda(r)\to 0$ as $r\to \infty$ representing the flat spacetime, the constants $b_0$ must be zero. Now, for anti de-Sitter spacetime metric (AdS$_4$), $e^{\nu(r)}=\sqrt{1+\frac{r^2}{a^2}}$ and $e^{\lambda(r)}=\frac{1}{\sqrt{1+\frac{r^2}{a^2}}}$.
Therefore, $\nu(r)=\frac{1}{2}\text{ln}\big(1+\frac{r^2}{a^2}\big)$ and $\lambda(r)=-\frac{1}{2}\text{ln}\big(1+\frac{r^2}{a^2}\big)$, which imply $\nu(r)=\text{ln}(\frac{r}{a})$ and $\lambda(r)=-\text{ln}(\frac{r}{a})$ as $r \to \infty$, that is, in this limit, $\nu(r)+\lambda(r)=0$ which gives $b_0$=0, gives a result similar to the previous case as $\nu(r)=-\lambda(r)$.
Now, using Eq.~(\ref{r1}) and (\ref{r2}) we get
\begin{eqnarray}
    &&\hspace{-0.2cm}\frac{\zeta_2}{\zeta_1}+\frac{2}{r^2}=\frac{2e^{-\lambda(r)}}{r}\Big(\nu^{\prime}(r)+\frac{1}{r}\Big).\nonumber\\ \nonumber
    &&\hspace{-0.2cm} \rightarrow \frac{r^2\zeta_2}{2\zeta_1}+1=\frac{d}{dr}\Big(re^{-\lambda(r)}\Big),\\ \nonumber
   &&\hspace{-0.2cm} \rightarrow  e^{-\lambda(r)}=1+\frac{r^2\zeta_2}{6\zeta_1}+\frac{\text{const.}}{r},\\
    &&\hspace{-0.2cm} \rightarrow g_{rr}=\Big(1+\frac{r^2\zeta_2} {6\zeta_1}+\frac{\text{const.}}{r}\Big)^{-1}.
\end{eqnarray}

 In the limit of small values of $r$, the Newtonian approximation yields a constant value equal to $2M$, where $M$ represents the gravitational mass. Furthermore, one can note that the above spacetime solution will represent the Schwarzschild anti de Sitter solution if we consider the cosmological constant $\Lambda=-\frac{\zeta_2}{2\zeta_1}$ and const.$=-2M$ then
\begin{eqnarray}
    g_{tt}=(g_{rr})^{-1} = 1-\frac{2M}{r}-\frac{\Lambda r^2} {3}.
\end{eqnarray}

Considering the preceding discussion, we select the Schwarzschild anti-de Sitter (SAdS$_4$) metric within the framework of $f(T)$ gravity to describe the outer region of spacetime as
\begin{eqnarray}\label{+}
    ds_{+}^2 = -\Big(1-\frac{2M}{r}-\frac{\Lambda r^2} {3}\Big) dt^2 + \Big(1-\frac{2M}{r}-\frac{\Lambda r^2} {3}\Big)^{-1} dr^2  + ~r^2 (d\theta^2+sin^2 \theta d\phi^2).
\end{eqnarray}
Moreover, the following line element provides the interior metric encompassing the geometric distortion as
\begin{eqnarray}\label{-}
    ds_{-}^2 = -e^{G(r)} dt^2+[H(r)+\alpha \psi(r)]^{-1} dr^2 +r^2(d\theta^2+sin^2\theta d\phi^2).
\end{eqnarray}
According to the Israel–Darmois junction conditions \citep{darmois1927memorial,Israel:1966rt}, a physically acceptable stellar configuration requires a smooth match of the interior spacetime $ds_-^2$ with the exterior spacetime $ds_+^2$ across the boundary hypersurface $\Sigma$. This matching is ensured by the continuity of the first and second fundamental forms in $\Sigma$, which are obtained by appropriately identifying the metric functions of the interior and exterior geometries at the stellar surface
\begin{eqnarray}
g_{tt}^{-}|_{r=R}=g_{tt}^{+}|_{r=R} \quad \text{and} \quad 
g_{rr}^{-}|_{r=R}=g_{rr}^{+}|_{R}\,.
\end{eqnarray}
Taking into account Eq.~(\ref{-}) and Eq.~(\ref{+}), it takes an explicit form as follows
\begin{eqnarray}\label{C1}   H(R)+\alpha ~\psi(R) = \big(1-\frac{2 {M}}{R}-\frac{\Lambda R^2}{3}\big),\\ \label{C2}   e^{G(R)} = \big(1-\frac{2 M}{R}-\frac{\Lambda R^2}{3}\big).
\end{eqnarray}
Alternatively, the second fundamental form assumes the following expression
\begin{eqnarray}\label{C3}
    p_r^{\text{eff}}(r)|_{\Sigma}=\left[p(r)-\alpha \,\Theta_{r}^{r}(r)\right]_{\Sigma}=0.
\end{eqnarray}
To compute the numerical values of the constants, we utilize Eqs.~(\ref{C1}),(\ref{C2}) and (\ref{C3})  to determine the unspecified parameters, including the constant $\mathcal{G}$, the mass ($M$) and the polytropic constant $\chi$.

\section{Deformed SS models and their astrophysical relevance}\label{sec5}
In this section, we investigate the physical viability and astrophysical relevance of the deformed SS models by analyzing key thermodynamic quantities, namely, the effective energy density, effective radial and tangential pressures, and the effective anisotropy parameter. The analysis is carried out for both solutions: solution~\ref{solution1} ($\Theta^0_0=\rho$) and solution~\ref{solution2} ($\Theta^1_1=p_r$), and for three distinct scenarios: GR, $f(T)$ gravity, and $f(T)+\mathrm{MGD}$.

\begin{figure*}
    \includegraphics[width=4.2cm, height=4.5cm]{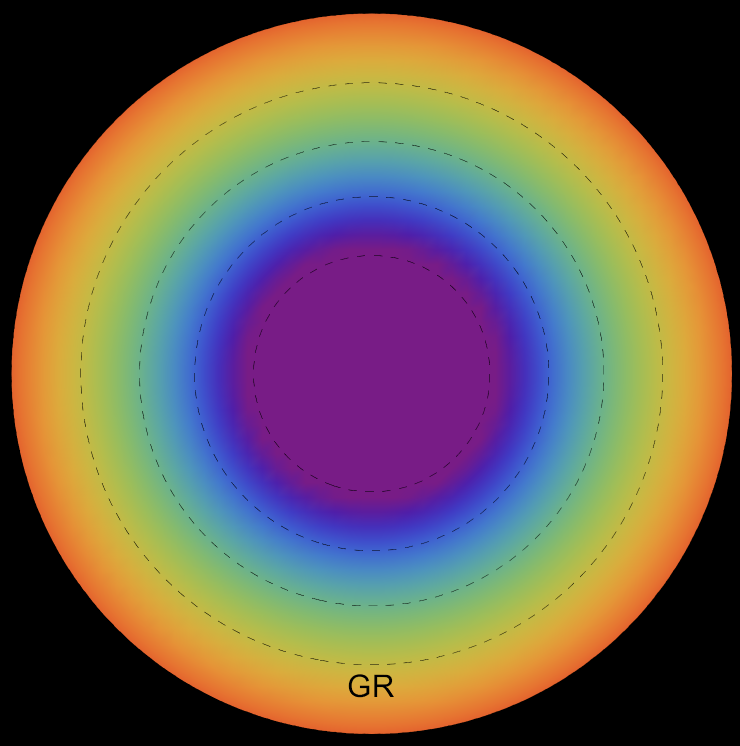}
     \includegraphics[width=0.9cm, height=4.5cm]{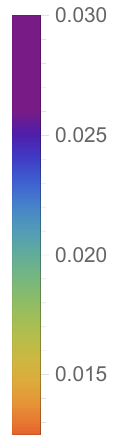}
      \includegraphics[width=4.2cm, height=4.5cm]{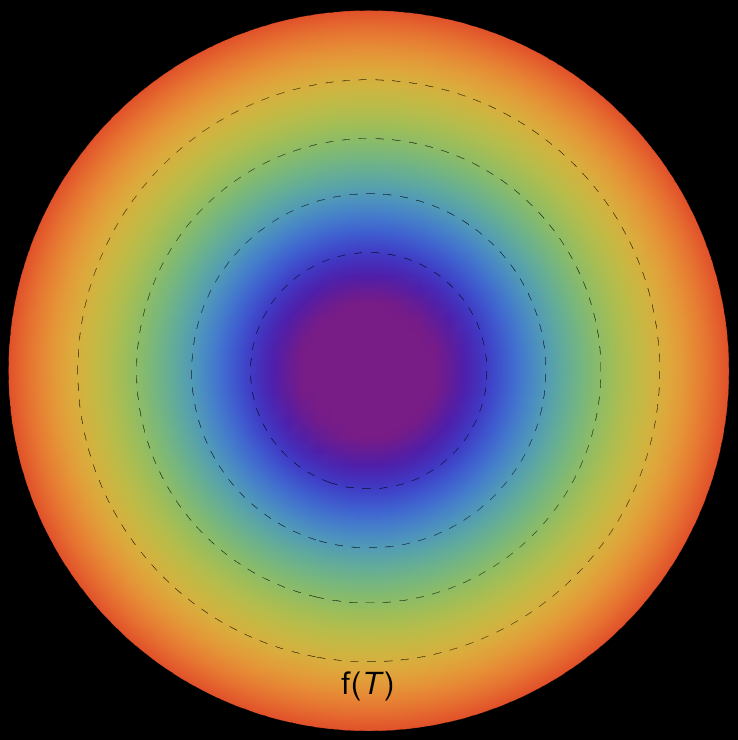}
      \includegraphics[width=0.9cm, height=4.5cm]{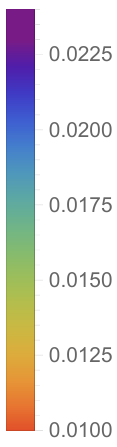}
      \includegraphics[width=4.2cm, height=4.5cm]{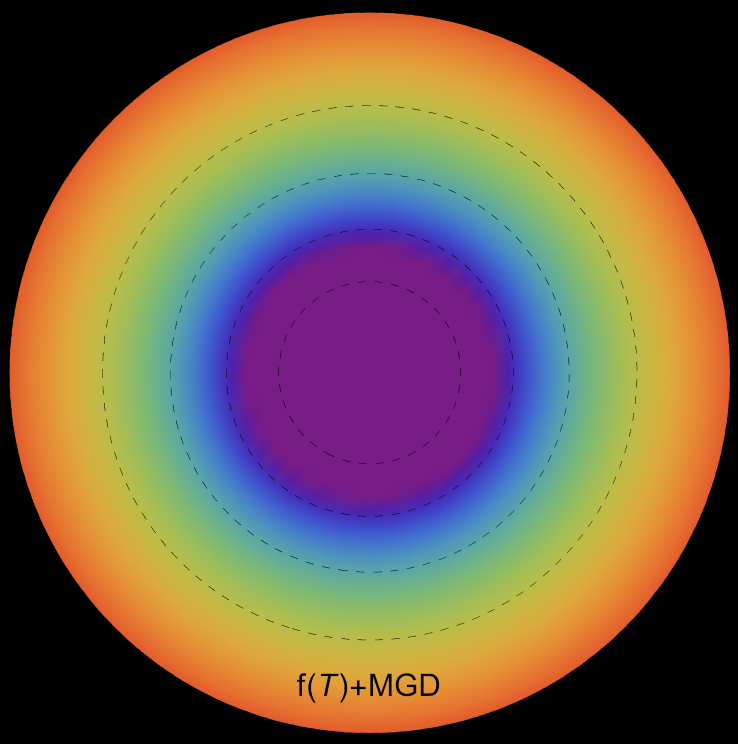}
      \includegraphics[width=0.9cm, height=4.5cm]{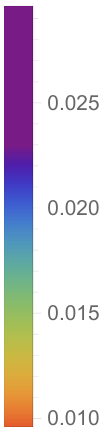}
      \caption{The density profile [$\rho^{\text{eff}}(r)$] in [$\text{km}^{-2}$] along to the radial distance $r$ of the stellar model for solution~\ref{solution1} [$\Theta^0_0=\rho$] in context of GR \big($\alpha=0.0$, $\zeta_1=1.0$, $\zeta_2=0.0$~$[\text{km}^{-2}]$\big)-left panel, $f(T)$ $\big(\alpha=0.0$, $\zeta_1=0.8$, $\zeta_2=2\times 10^{-6}$~$[\text{km}^{-2}]$\big)-middle panel, and $f(T)+\text{MGD}$ \big($\alpha=0.2$, $\zeta_1=0.8$, $\zeta_2=2\times 10^{-6}$~$[\text{km}^{-2}]$\big)-right panel. The fixed values of the constant parameters are:  $A = -1.5$, $B=0.006$, $\gamma=10$, $\beta=0.33$, and $R=11~\text{km}$.} \label{f1}  
\end{figure*}

\begin{figure*}
    \includegraphics[width=4.2cm, height=4.5cm]{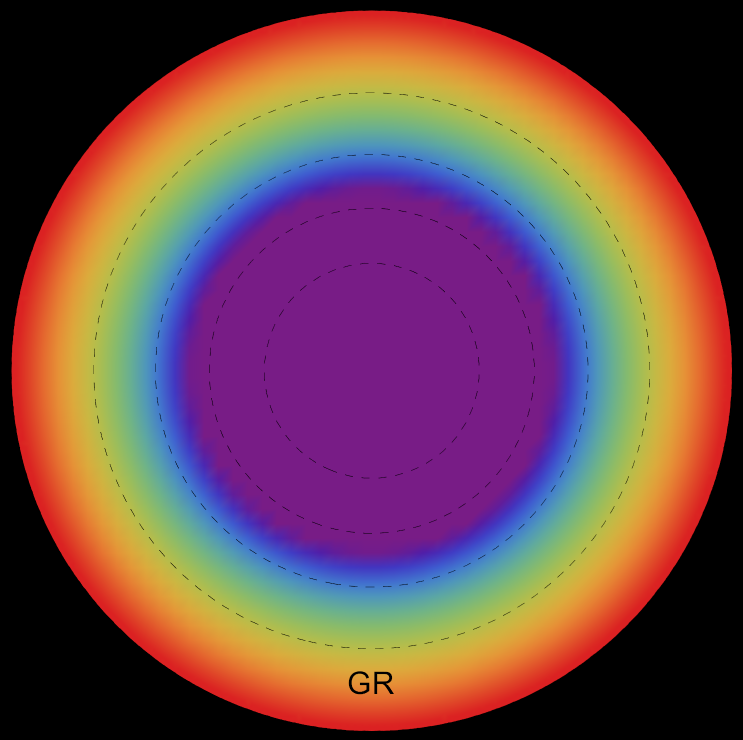}
     \includegraphics[width=0.9cm, height=4.5cm]{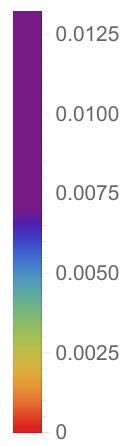}
      \includegraphics[width=4.2cm, height=4.5cm]{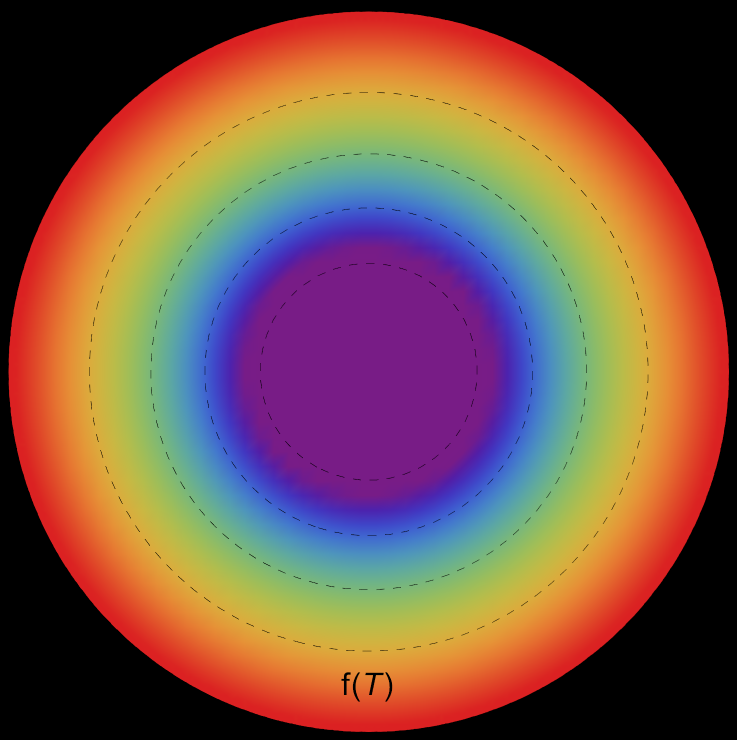}
      \includegraphics[width=0.9cm, height=4.5cm]{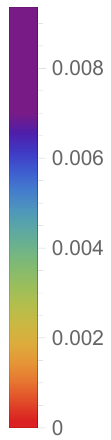}
      \includegraphics[width=4.2cm, height=4.5cm]{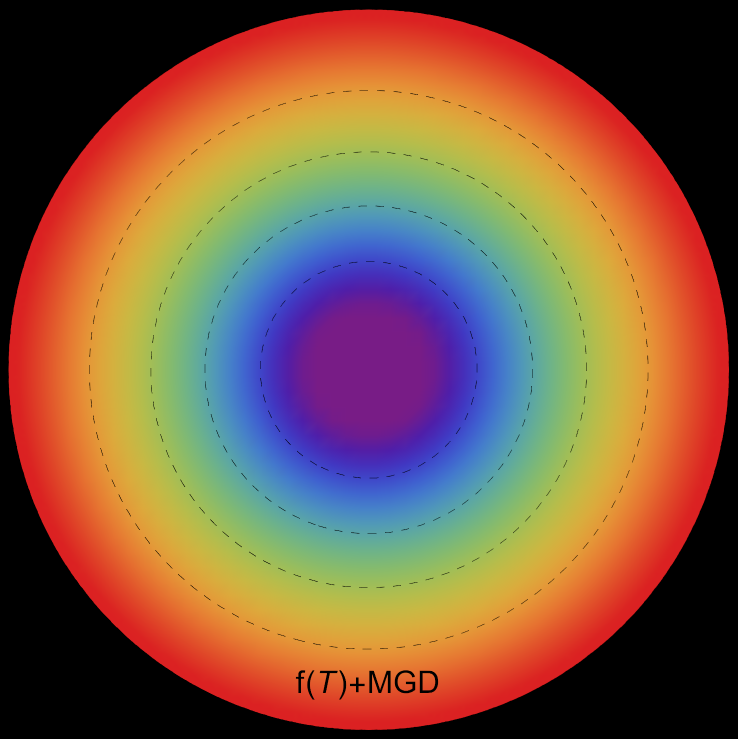}
      \includegraphics[width=0.9cm, height=4.5cm]{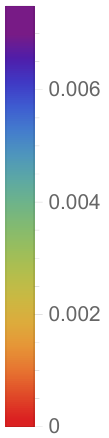}
        \caption{The radial pressure profile [$p^{\text{eff}}_r(r)$] in [$\text{km}^{-2}$] along to the radial distance $r$ of the stellar model for solution~\ref{solution1} [$\Theta^0_0=\rho$] in context of GR \big($\alpha=0.0$, $\zeta_1=1.0$, $\zeta_2=0.0$~$[\text{km}^{-2}]$\big)-left panel, $f(T)$ $\big(\alpha=0.0$, $\zeta_1=0.8$, $\zeta_2=2\times 10^{-6}$~$[\text{km}^{-2}]$\big)-middle panel, and $f(T)+\text{MGD}$ \big($\alpha=0.2$, $\zeta_1=0.8$, $\zeta_2=2\times 10^{-6}$~$[\text{km}^{-2}]$\big)-right panel.} \label{f2} 
\end{figure*}

\begin{figure*}
    \includegraphics[width=4.2cm, height=4.5cm]{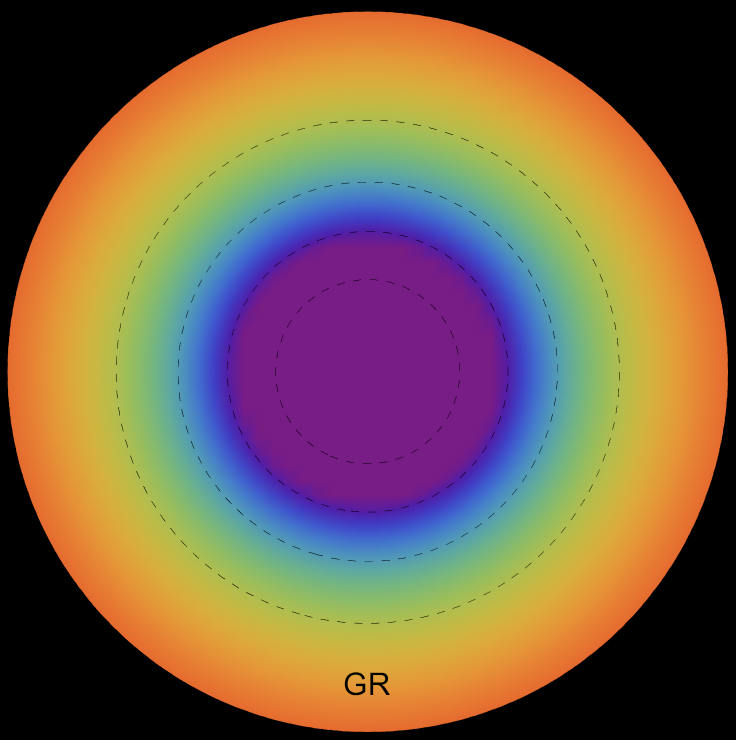}
     \includegraphics[width=0.9cm, height=4.5cm]{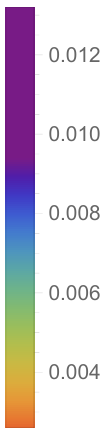}
      \includegraphics[width=4.2cm, height=4.5cm]{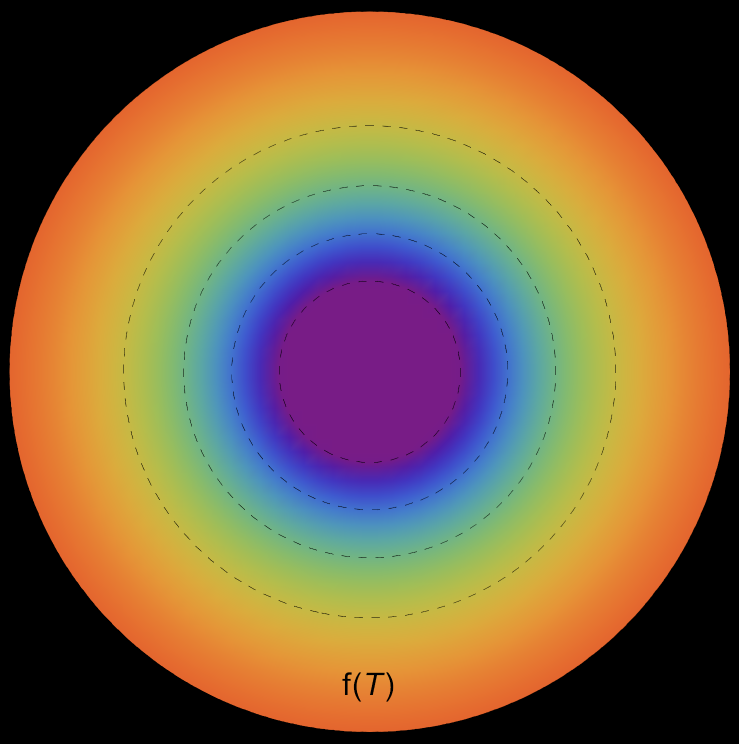}
      \includegraphics[width=0.9cm, height=4.5cm]{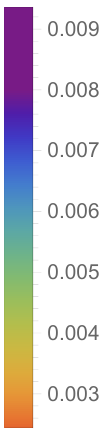}
      \includegraphics[width=4.2cm, height=4.5cm]{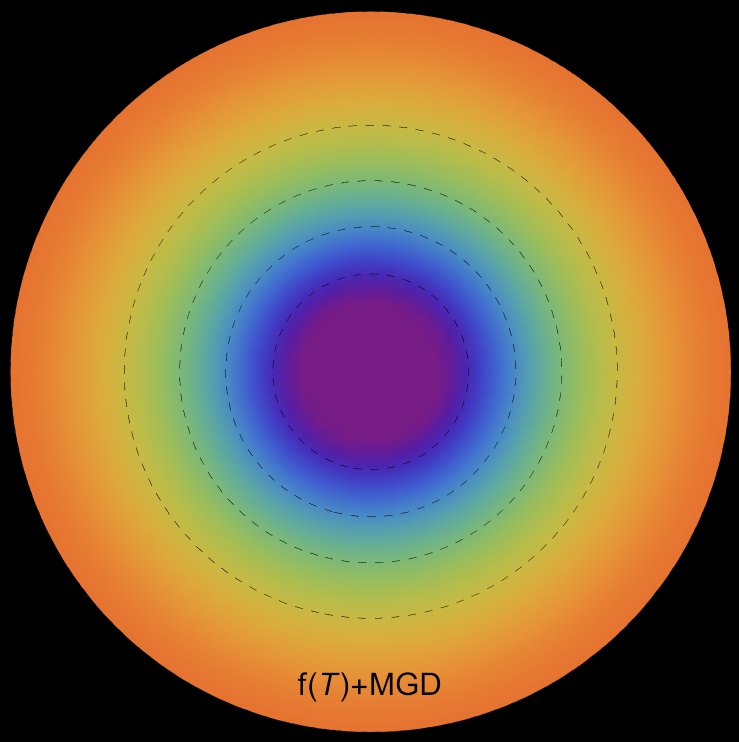}
      \includegraphics[width=0.9cm, height=4.5cm]{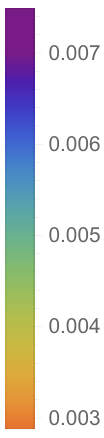}
       \caption{The effective tangential pressure profile [$p^{\text{eff}}_t(r)$] in [$\text{km}^{-2}$] along to the radial distance $r$ of the stellar model for solution ~\ref{solution1} [$\Theta^0_0=\rho$] in context of GR \big($\alpha=0.0$, $\zeta_1=1.0$, $\zeta_2=0.0$~$[\text{km}^{-2}]$\big)-left panel, $f(T)$ $\big(\alpha=0.0$, $\zeta_1=0.8$, $\zeta_2=2\times 10^{-6}$~$[\text{km}^{-2}]$\big)-middle panel, and $f(T)+\text{MGD}$ \big($\alpha=0.2$, $\zeta_1=0.8$, $\zeta_2=2\times 10^{-6}$~$[\text{km}^{-2}]$\big)-right panel.} \label{f3}
\end{figure*}

\begin{figure*}
    \includegraphics[width=4.2cm, height=4.5cm]{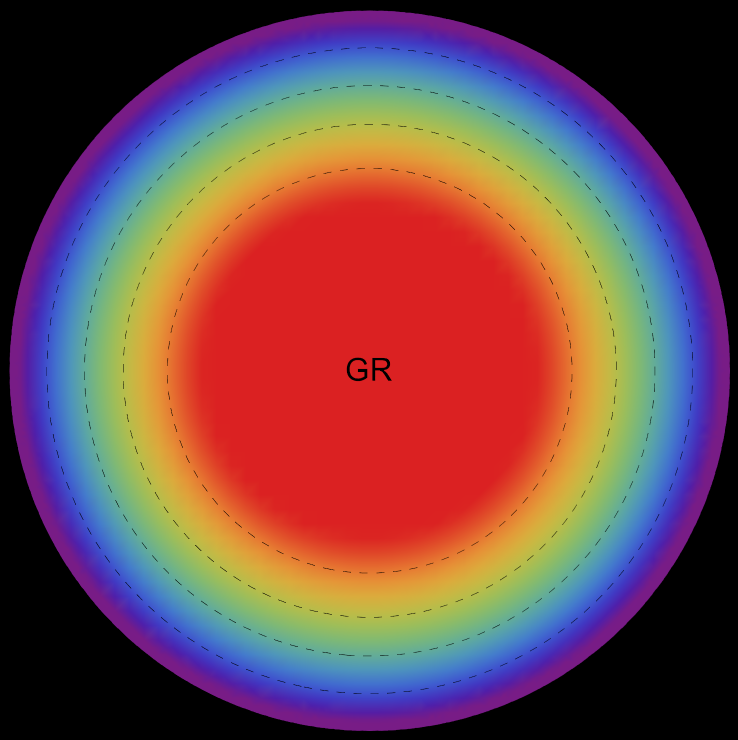}
     \includegraphics[width=0.9cm, height=4.5cm]{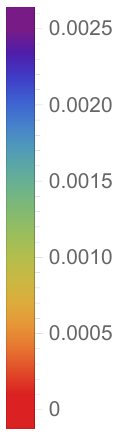}
      \includegraphics[width=4.2cm, height=4.5cm]{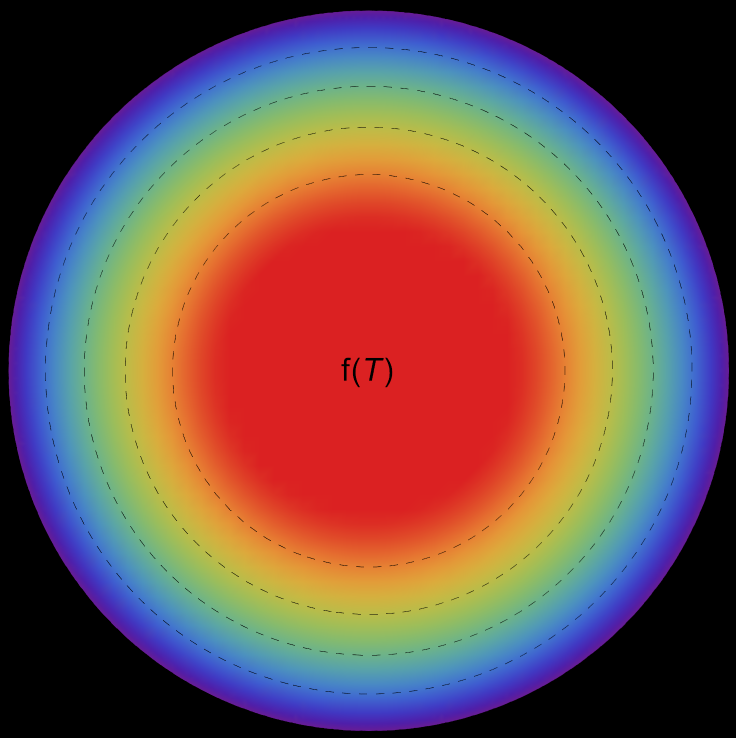}
      \includegraphics[width=0.9cm, height=4.5cm]{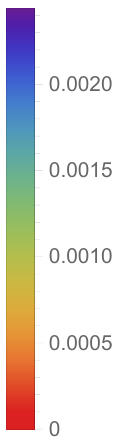}
      \includegraphics[width=4.2cm, height=4.5cm]{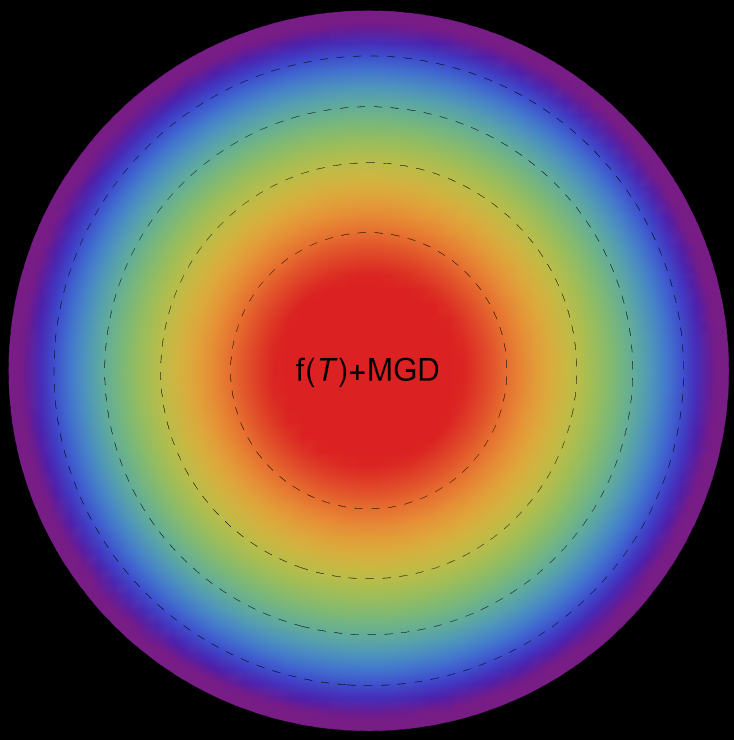}
      \includegraphics[width=0.9cm, height=4.5cm]{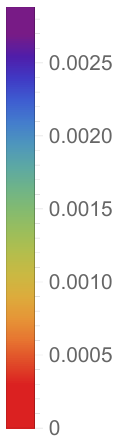}
       \caption{The effective anisotropy profile [$\Delta^{\text{eff}}(r)$] in [$\text{km}^{-2}$] along to the radial distance $r$ of the stellar model for solution ~\ref{solution1} [$\Theta^0_0=\rho$] in context of GR \big($\alpha=0.0$, $\zeta_1=1.0$, $\zeta_2=0.0$~$[\text{km}^{-2}]$\big)-left panel, $f(T)$ $\big(\alpha=0.0$, $\zeta_1=0.8$, $\zeta_2=2\times 10^{-6}$~$[\text{km}^{-2}]$\big)-middle panel, and $f(T)+\text{MGD}$ \big($\alpha=0.2$, $\zeta_1=0.8$, $\zeta_2=2\times 10^{-6}$~$[\text{km}^{-2}]$\big)-right panel.} \label{f4}
\end{figure*}

\begin{figure*}
    \includegraphics[width=4.2cm, height=4.5cm]{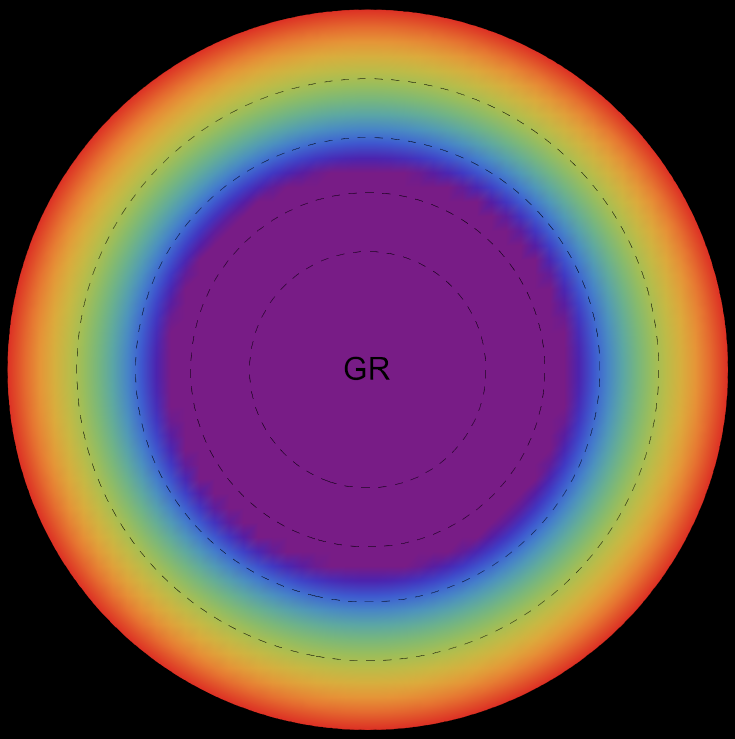}
     \includegraphics[width=0.9cm, height=4.5cm]{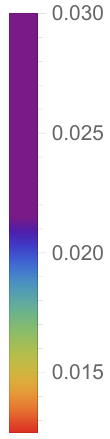}
      \includegraphics[width=4.2cm, height=4.5cm]{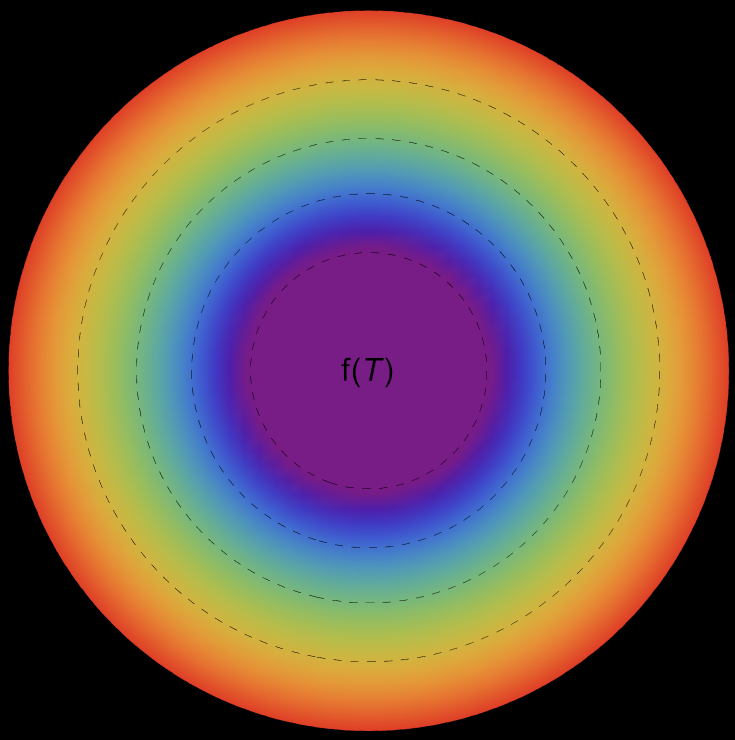}
      \includegraphics[width=0.9cm, height=4.5cm]{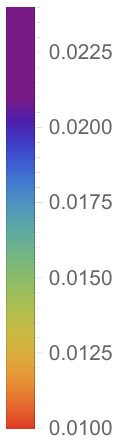}
      \includegraphics[width=4.2cm, height=4.5cm]{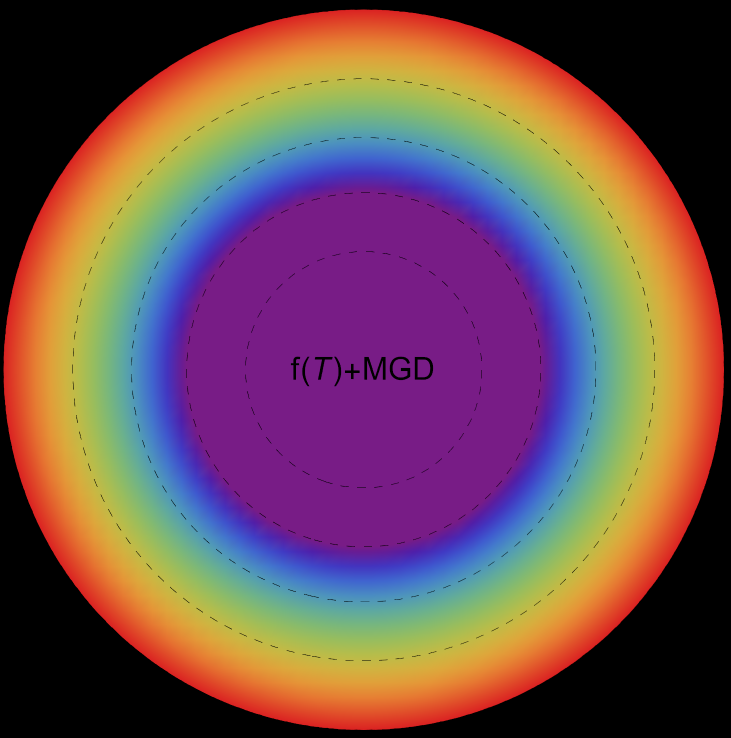}
      \includegraphics[width=0.9cm, height=4.5cm]{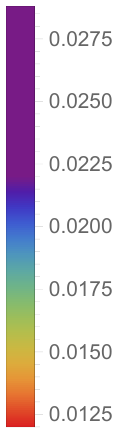}
       \caption{The effective density profile [$\rho^{\text{eff}}(r)$] in [$\text{km}^{-2}$] along to the radial distance $r$ of the stellar model for solution~\ref{solution2} [$\Theta^1_1=p_r$] in context of GR \big($\alpha=0.0$, $\zeta_1=1.0$, $\zeta_2=0.0$~$[\text{km}^{-2}]$\big)-left panel, $f(T)$ $\big(\alpha=0.0$, $\zeta_1=0.8$, $\zeta_2=2\times 10^{-6}$~$[\text{km}^{-2}]$\big)-middle panel, and $f(T)+\text{MGD}$ \big($\alpha=0.2$, $\zeta_1=0.8$, $\zeta_2=2\times 10^{-6}$~$[\text{km}^{-2}]$\big)-right panel.} \label{f5}
\end{figure*}

\begin{figure*}
    \includegraphics[width=4.2cm, height=4.5cm]{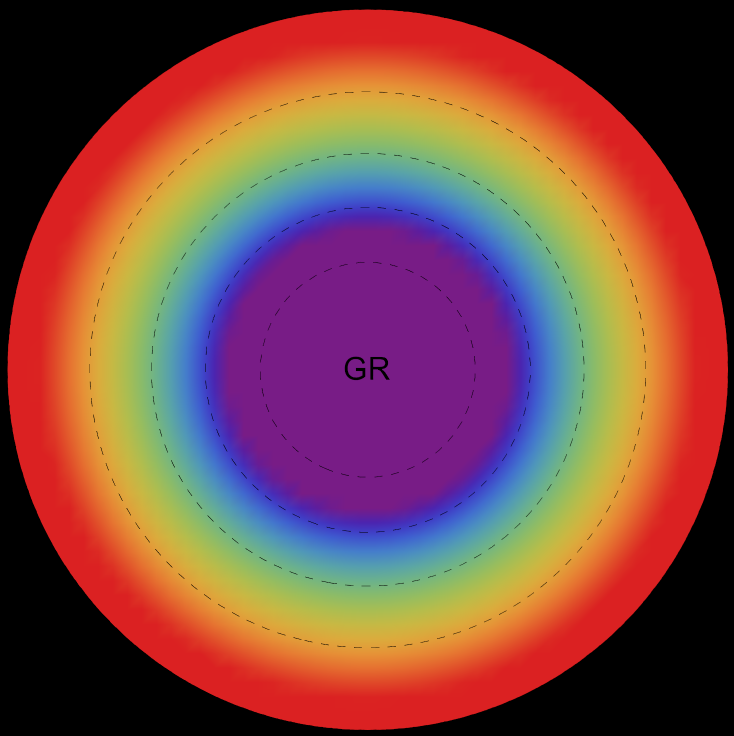}
     \includegraphics[width=0.9cm, height=4.5cm]{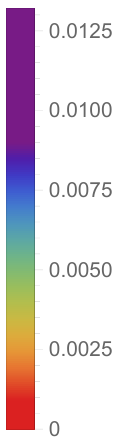}
      \includegraphics[width=4.2cm, height=4.5cm]{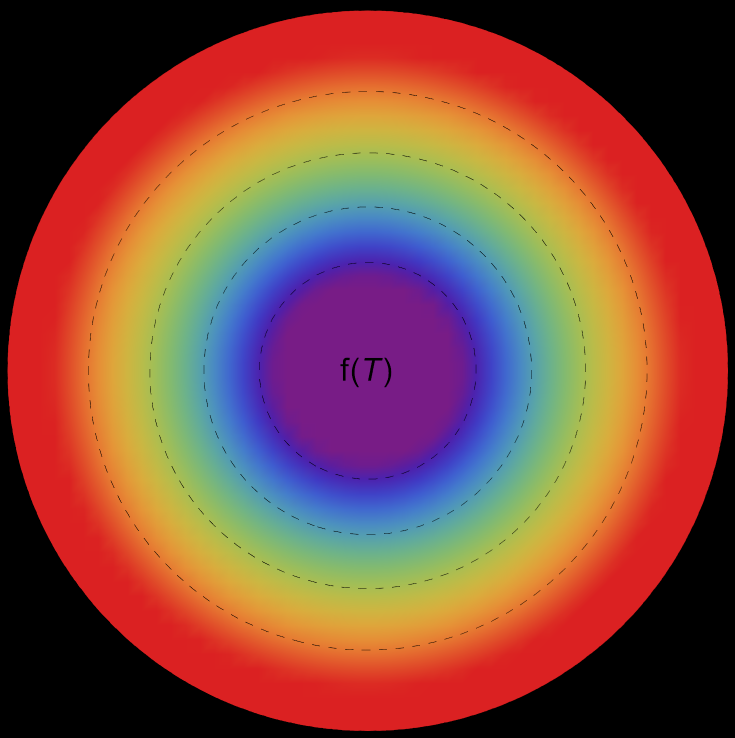}
      \includegraphics[width=0.9cm, height=4.5cm]{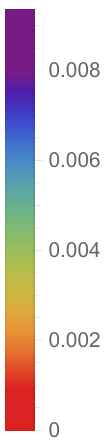}
      \includegraphics[width=4.2cm, height=4.5cm]{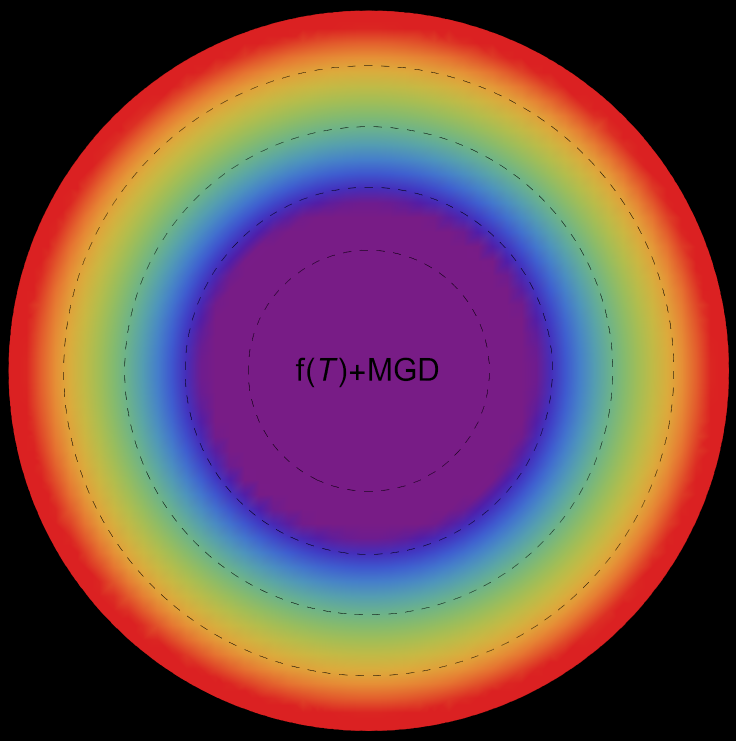}
      \includegraphics[width=0.9cm, height=4.5cm]{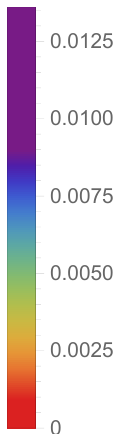}
       \caption{The effective radial pressure profile [$p^{\text{eff}}_r(r)$] in [$\text{km}^{-2}$] along to the radial distance $r$ of the stellar model for solution~\ref{solution2} [$\Theta^1_1=p_r$] in context of GR \big($\alpha=0.0$, $\zeta_1=1.0$, $\zeta_2=0.0$~$[\text{km}^{-2}]$\big)-left panel, $f(T)$ $\big(\alpha=0.0$, $\zeta_1=0.8$, $\zeta_2=2\times 10^{-6}$~$[\text{km}^{-2}]$\big)-middle panel, and $f(T)+\text{MGD}$ \big($\alpha=0.2$, $\zeta_1=0.8$, $\zeta_2=2\times 10^{-6}$~$[\text{km}^{-2}]$\big)-right panel.} \label{f6}
\end{figure*}

\begin{figure*}
    \includegraphics[width=4.2cm, height=4.5cm]{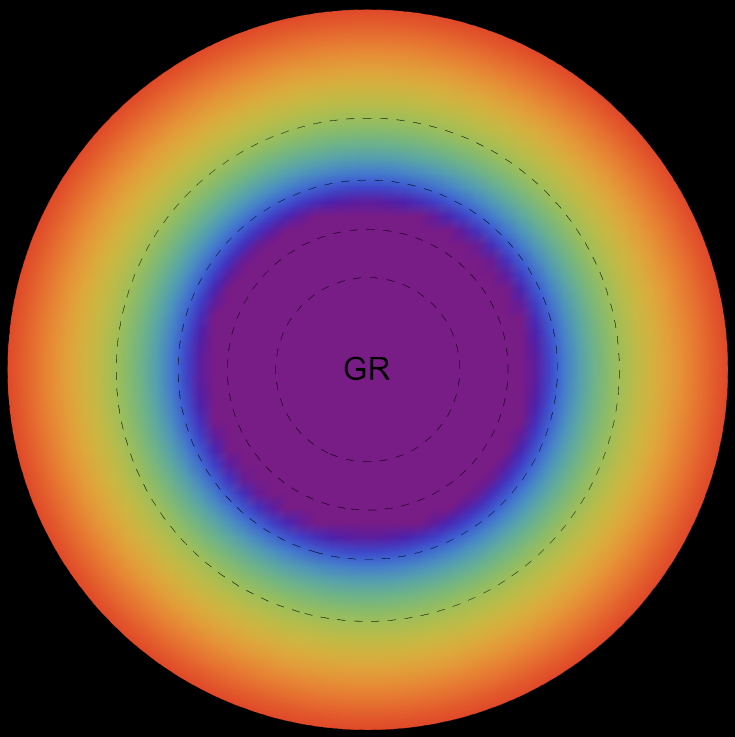}
     \includegraphics[width=0.9cm, height=4.5cm]{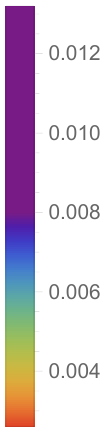}
      \includegraphics[width=4.2cm, height=4.5cm]{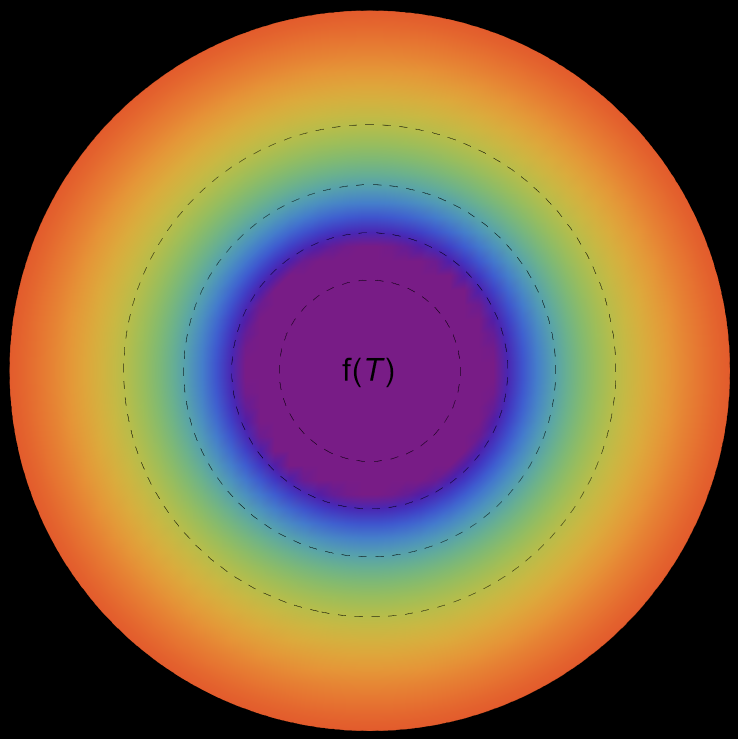}
      \includegraphics[width=0.9cm, height=4.5cm]{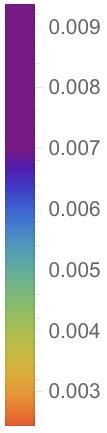}
      \includegraphics[width=4.2cm, height=4.5cm]{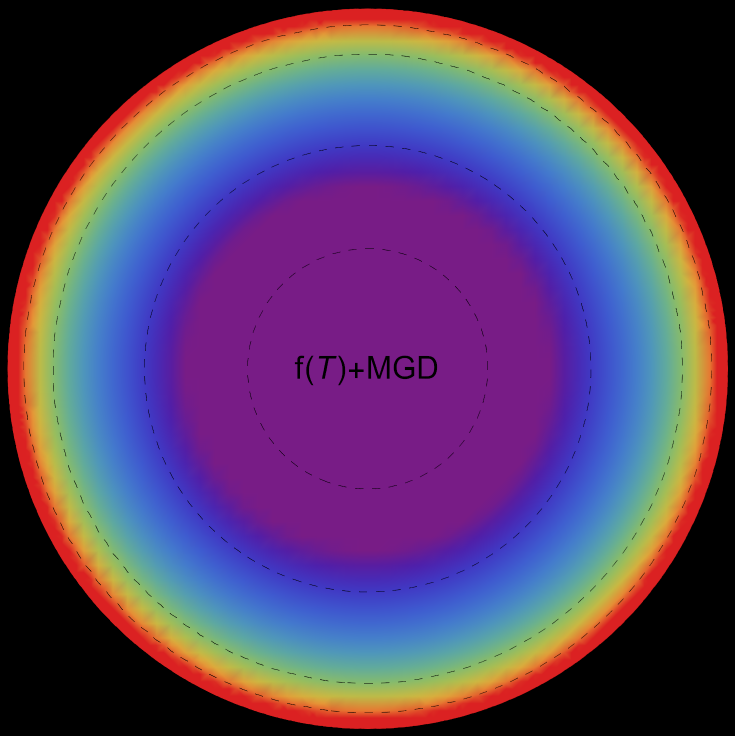}
      \includegraphics[width=0.9cm, height=4.5cm]{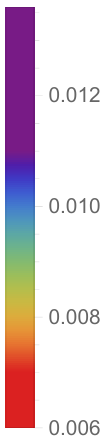}
       \caption{The effective tangential pressure profile [$p^{\text{eff}}_t(r)$] in [$\text{km}^{-2}$] along to the radial distance $r$ of the stellar model for solution~\ref{solution2} [$\Theta^1_1=p_r$] in context of GR \big($\alpha=0.0$, $\zeta_1=1.0$, $\zeta_2=0.0$~$[\text{km}^{-2}]$\big)-left panel, $f(T)$ $\big(\alpha=0.0$, $\zeta_1=0.8$, $\zeta_2=2\times 10^{-6}$~$[\text{km}^{-2}]$\big)-middle panel, and $f(T)+\text{MGD}$ \big($\alpha=0.2$, $\zeta_1=0.8$, $\zeta_2=2\times 10^{-6}$~$[\text{km}^{-2}]$\big)-right panel.} \label{f7} 
\end{figure*}

\begin{figure*}
    \includegraphics[width=4.2cm, height=4.5cm]{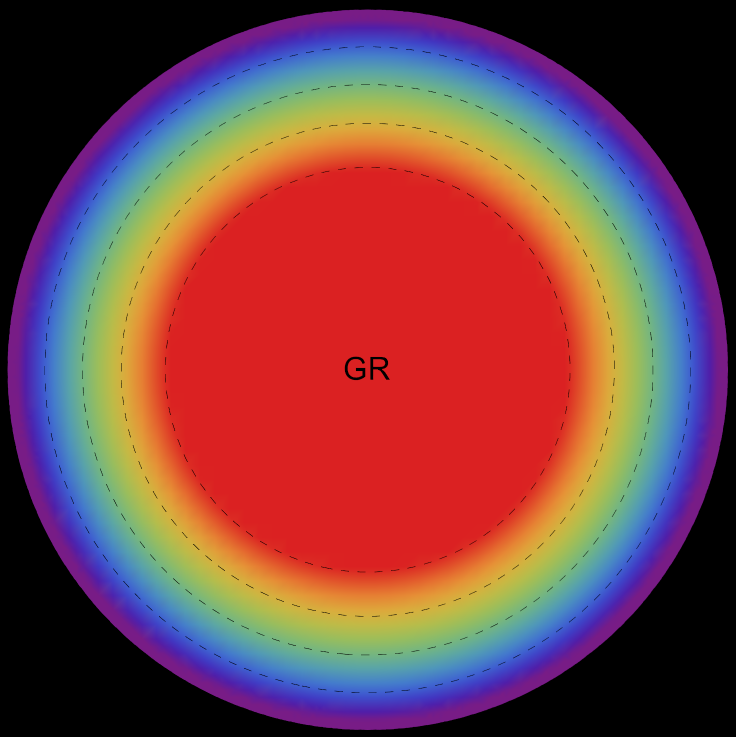}
     \includegraphics[width=0.9cm, height=4.5cm]{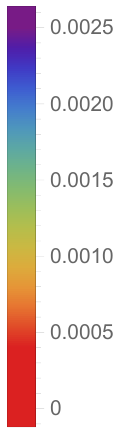}
      \includegraphics[width=4.2cm, height=4.5cm]{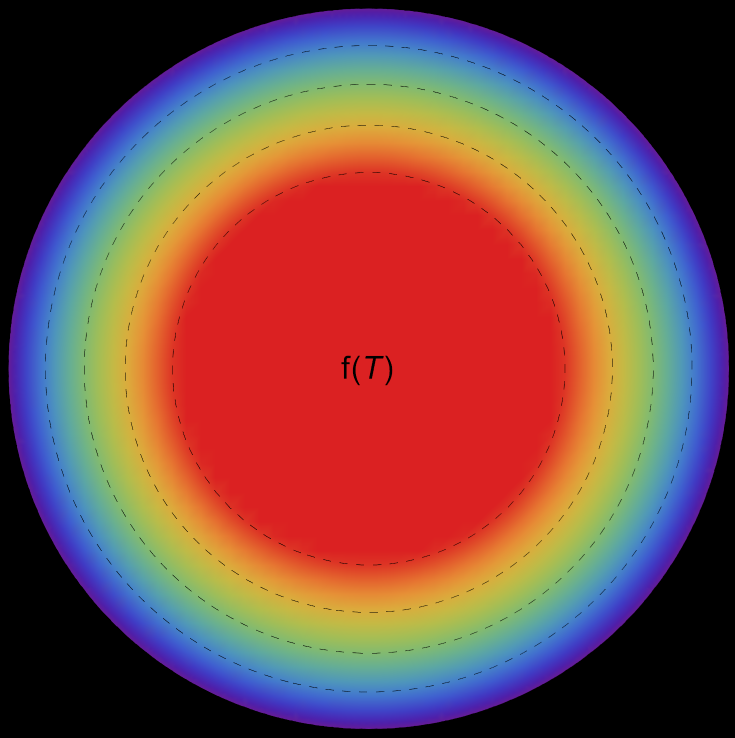}
      \includegraphics[width=0.9cm, height=4.5cm]{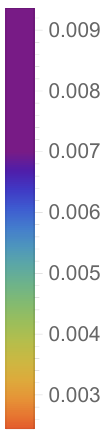}
      \includegraphics[width=4.2cm, height=4.5cm]{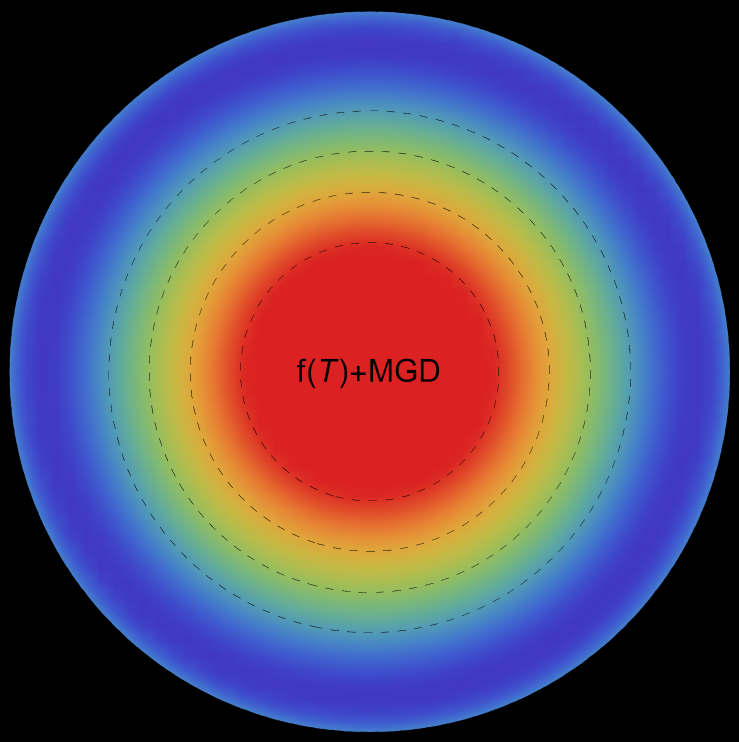}
      \includegraphics[width=0.9cm, height=4.5cm]{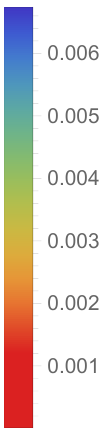}
       \caption{The effective anisotropy profile [$\Delta^{\text{eff}}(r)$] in [$\text{km}^{-2}$] along to the radial distance $r$ of the stellar model for solution~\ref{solution2} [$\Theta^1_1=p_r$] in context of GR \big($\alpha=0.0$, $\zeta_1=1.0$, $\zeta_2=0.0$~$[\text{km}^{-2}]$\big)-left panel, $f(T)$ $\big(\alpha=0.0$, $\zeta_1=0.8$, $\zeta_2=2\times 10^{-6}$~$[\text{km}^{-2}]$\big)-middle panel, and $f(T)+\text{MGD}$ \big($\alpha=0.2$, $\zeta_1=0.8$, $\zeta_2=2\times 10^{-6}$~$[\text{km}^{-2}]$\big)-right panel.} \label{f8} 
\end{figure*}

The effective energy density is found to be maximum at the stellar center and decreases monotonically towards the surface, remaining finite and regular throughout the interior for all scenarios (see Figs.~\ref{f1} and \ref{f5}). Compared to GR, the $f(T)$ model yields a lower central density due to the modified torsional contributions, while the inclusion of MGD enhances the central density, leading to a stronger concentration of matter in the inner regions. Notably, variations in the model parameters have a negligible influence near the stellar surface.
The radial and tangential pressure profiles, shown in Figs.~\ref{f2}, \ref{f3}, \ref{f6}, and \ref{f7}, are continuous and monotonically decreasing functions of the radial coordinate, with the effective radial pressure vanishing at the stellar boundary. For solution~\ref{solution1}, the MGD contribution lowers the central pressures, indicating a compacting effect on the fluid. In contrast, for the solution~\ref{solution2}, the inclusion of MGD increases the central pressures, suggesting a relaxing effect on stellar matter. Despite these differences, both solutions exhibit physically acceptable pressure behavior throughout the interior.
The effective anisotropy parameter $\Delta(r)$ is shown in Figs.~\ref{f4} and \ref{f8}. In all cases, $\Delta(r)$ starts from zero in the center and increases outward, reaching positive values near the surface. The presence of MGD significantly amplifies the anisotropy compared to GR and pure $f(T)$ gravity, effectively doubling its magnitude. This enhanced anisotropy generates an outward-directed repulsive force that counterbalances gravity, thereby improving the stability of the stellar configuration, particularly in the outer layers.

\section{Mass-radius relation for minimally deformed strange star models and their relevance to astrophysics}\label{sec6}

\begin{figure*}
\centering
\includegraphics[width=7cm,height=6.5cm]{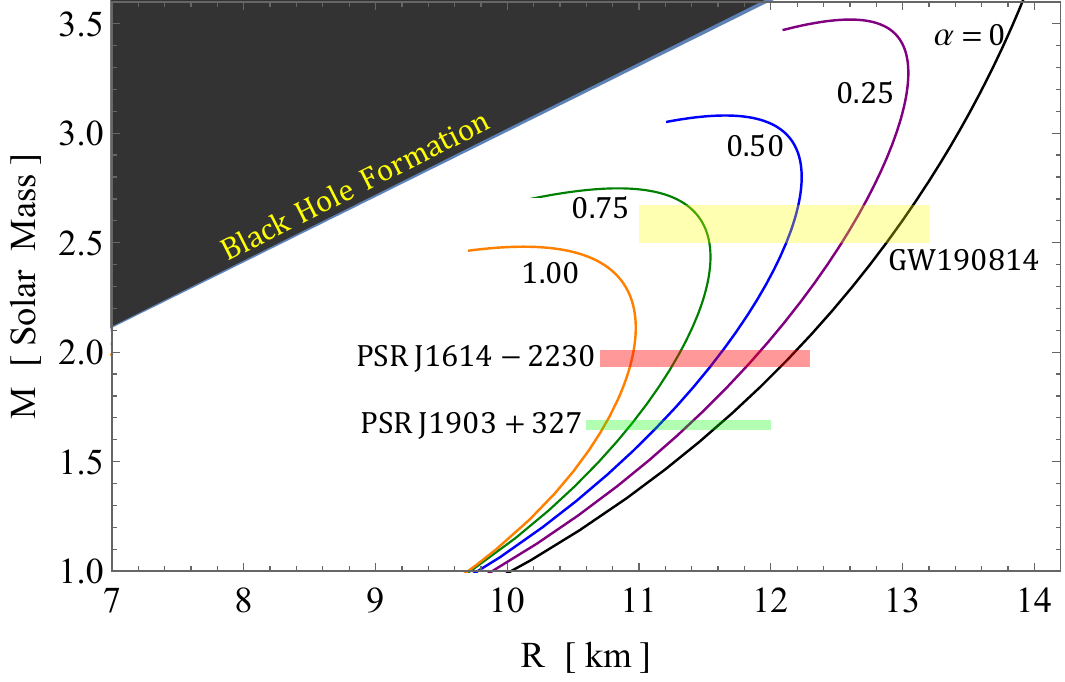}~~~
\includegraphics[width=7cm,height=6.5cm]{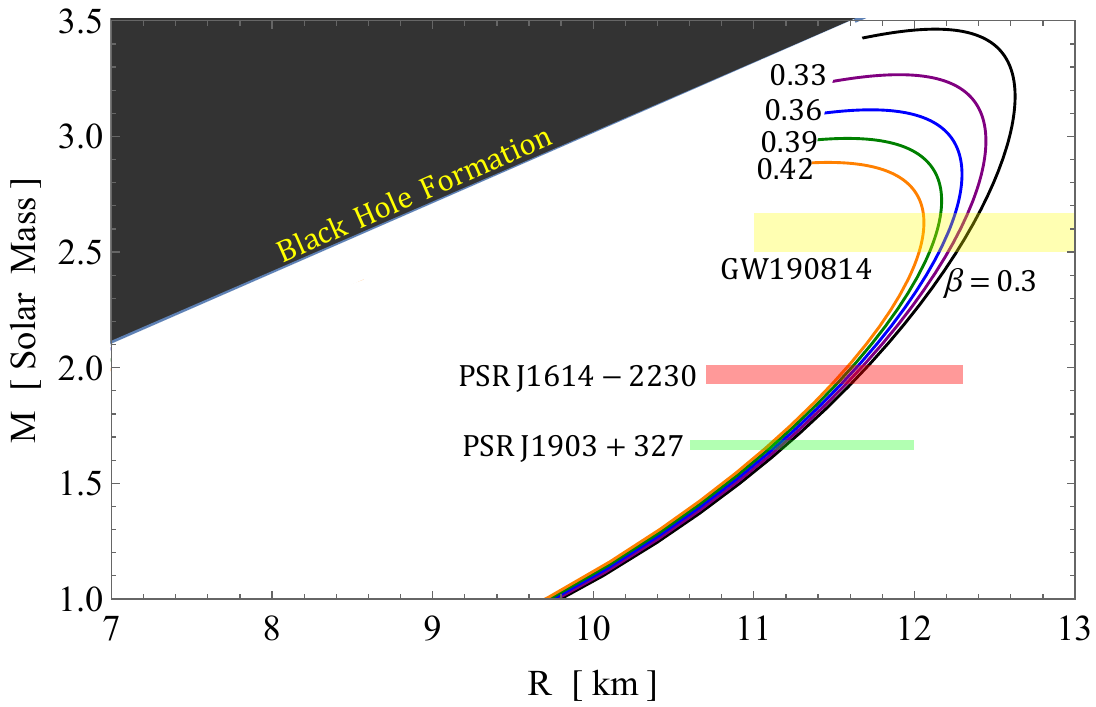}
\caption{The $M-R$ curves for different free parameters values $\alpha$-left panel with fixed $\beta=0.33$, $\gamma=20$,  $\zeta_1= 0.5$ and right figure shows the $M-R$ curves for different $\beta$ with fixed $\alpha=0.5$, $\gamma=15$, and $\zeta_1=0.5$. The both figures represent the mass-radius relation for solution~\ref{solution1} $\left[\Theta^0_0=\rho\right]$. }
\label{f9} 
\end{figure*}

\begin{figure*}
\centering
\includegraphics[width=7cm,height=6.5cm]{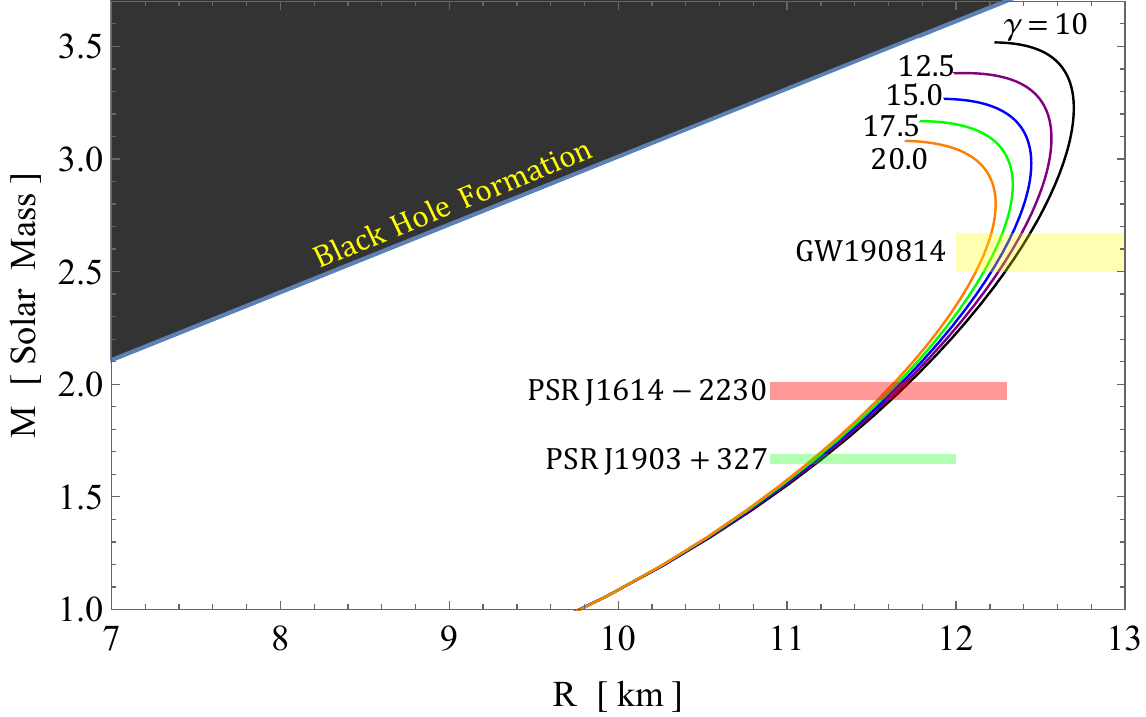}~~~
\includegraphics[width=7cm,height=6.5cm]{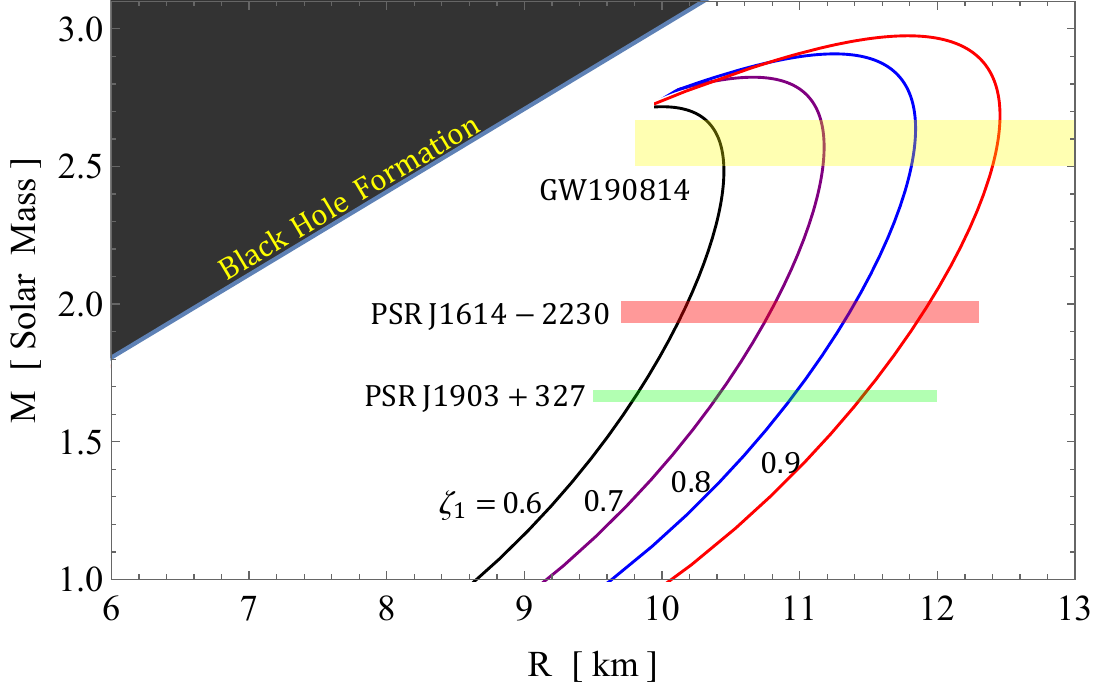}
\caption{The $M-R$ curves for different free parameters values $\gamma$-left panel with fixed $\beta=0.33$, $\alpha=0.5~\text{km}^2$,  $\zeta_1=0.5$ and right figure shows the $M-R$ curves for different $\zeta_1$ with fixed $\alpha=0.5$, $\gamma=15$, and $\beta=0.33$. The both figures represent the mass-radius relation for solution~\ref{solution1} $\left[\Theta^0_0=\rho\right]$.}
\label{f10} 
\end{figure*}

\begin{figure*}
\centering
\includegraphics[width=7cm,height=6.5cm]{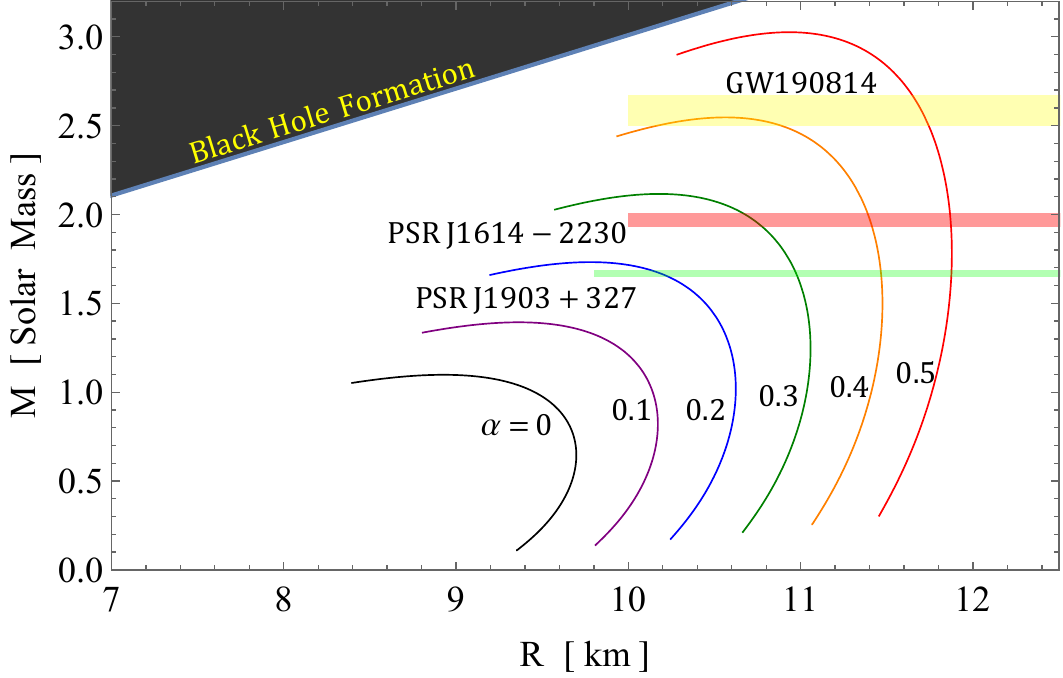}~~~~
\includegraphics[width=7cm,height=6.5cm]{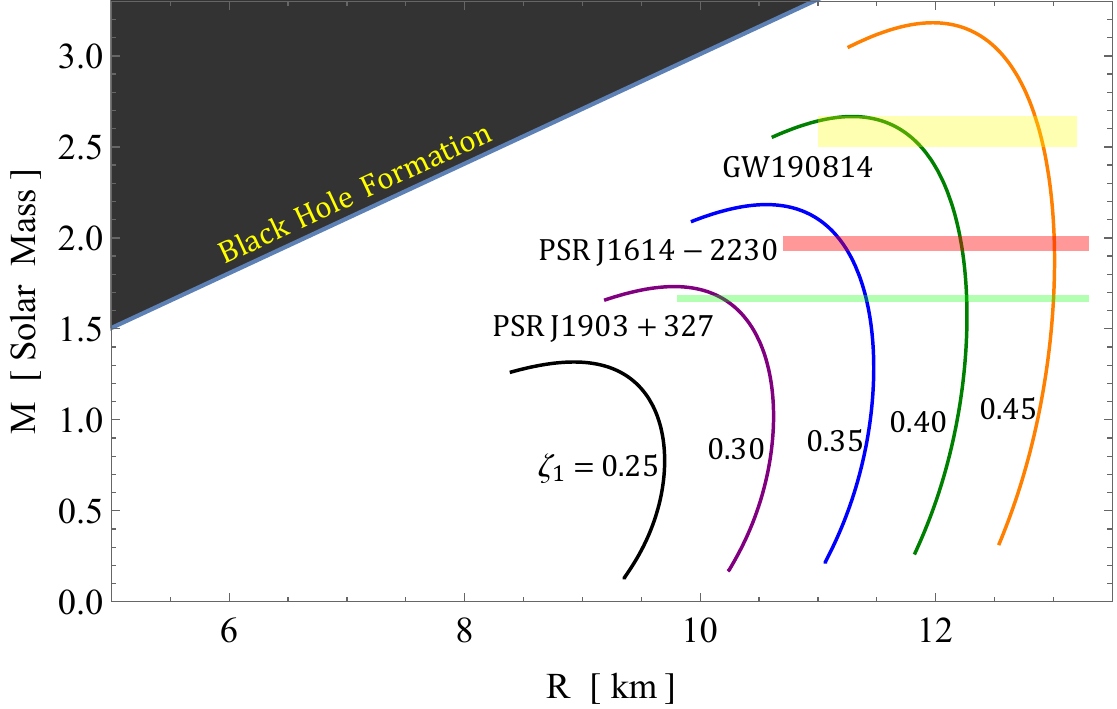}
\caption{The $M-R$ curves for different free parameters values $\alpha$-left panel with fixed $\beta=0.33$, $\gamma=20$,  $\zeta_1=0.3$ and right figure shows the $M-R$ curves for different $\zeta_1$ with fixed $\alpha=0.2$, $\gamma=20$, and $\beta=0.33$. The both figures represent the mass-radius relation for solution~\ref{solution2} $\left[\Theta^1_1=p_r\right]$.}
\label{f11} 
\end{figure*}

In this work, we construct SS models within $f(T)$ gravity using the Buchdahl metric and investigate the corresponding mass-radius ($M$--$R$) relations. The $M$-$R$ curves, shown in Figs.~\ref{f9}-\ref{f11}, are obtained by solving the TOV equation with a quadratic polytropic EoS for both solution~\ref{solution1} ($\Theta^0_0=\rho$) and solution~\ref{solution2} ($\Theta^1_1=p_r$), considering different values of the parameters $\alpha$, $\beta$, $\gamma$, and $\zeta_1$. Observational mass constraints of selected SS are indicated by horizontal bands. Curves intersecting unphysical regions, such as black hole formation limits, are excluded, allowing us to estimate viable stellar radii summarized in tables~\ref{Table1}--\ref{Table3}.

For solution~\ref{solution1}, the $M$--$R$ curves increase smoothly up to a maximum mass, followed by a rapid decrease and eventual flattening at larger radii. NSs with masses in the range $2.4$--$3.5\,M_{\odot}$ correspond to radii between $9.80^{+0.02}_{-0.01}$ and $13.01^{+0.01}_{-0.01}$ km, depending on the chosen parameters. Increasing $\alpha$, $\beta$, and $\gamma$ shifts the peak of the curves downward and toward smaller radii, indicating softer stellar configurations. In contrast, increasing $\zeta_1$ shifts the peak upward and to the right, corresponding to a stiffer EoS capable of supporting more massive and larger stars.
A similar qualitative behavior is observed for solution~\ref{solution2}. The $M$--$R$ curves again exhibit a well-defined maximum mass followed by a decline, with NS in the mass range $2.4$--$3.5\,M_{\odot}$ yielding comparable radii. In this case, the primary parameters that influence the stellar structure are $\alpha$ and $\zeta_1$, while variations in $\beta$ and $\gamma$ play a negligible role. Larger values of $\alpha$ and $\zeta_1$ consistently lead to higher maximum masses and larger radii, reinforcing the interpretation of a stiffer EoS. The shaded region in the upper-left corner of $M-R$ curves, labelled \textit{Black Hole Formation}, marks configurations for which $R \leq 2M$, corresponding to the Schwarzschild horizon condition. Although the present study is performed within $f(T)$ gravity, the exterior spacetime of a static, spherically symmetric star reduces asymptotically to the Schwarzschild solution, so the same horizon-based bound $R = 2M$ is adopted as in GR to identify the BH formation region. This choice also facilitates direct comparison with the standard GR mass--radius diagrams. All stable stellar configurations obtained in our analysis lie comfortably outside this forbidden region. Our results are consistent with earlier studies on modified gravity. For instance, Astashenok \emph{et al.} \citep{Astashenok:2021peo} demonstrated that $R^2$ gravity allows compact stars with masses approaching $3\,M_{\odot}$, remaining below the causal limit. Similar conclusions were drawn for SS in the context of GW190814. Further investigations in axion $R^2$ gravity \citep{Astashenok:2020qds} and inflationary models \citep{Odintsov:2023ypt} also reported compact stars with masses exceeding $2.5\,M_{\odot}$ while respecting causality bounds.


\begin{table*}
\centering \caption{Maximum masses and corresponding radii for different values of $\alpha$ and $\beta$ for $\rho=\Theta^0_0$.} \label{Table1}
\scalebox{0.65}{%
\begin{tabular}{| *{12}{c|}} \hline
\quad \multirow{3}{*}{Objects} \quad & \quad  \multirow{3}{*}{$M/M_\odot$} & \multicolumn{10}{c|}{Predicted $R$ [km]}\\
\cline{3-12}
&  & \multicolumn{5}{c|}{$\alpha$} & \multicolumn{5}{c|}{$\beta$}\\
\cline{3-7} \cline{8-12}
& & \quad $0$ &\quad $0.25$  & \quad $0.50$  & \quad $0.75$ & \quad $1.0$ & \quad $0.3$ &\quad $0.33$  & \quad $0.36$  & \quad $0.39$ & \quad $0.42$\\
\hline
PSR J1903+327 & $1.667\pm 0.021$ & $11.61^{+0.04}_{-0.03}$ & $11.36^{+0.04}_{-0.02}$  & $11.15^{+0.04}_{-0.02}$ & $10.94^{+0.02}_{-0.02}$ & $10.74^{+0.02}_{-0.01}$ & $11.21^{+0.03}_{-0.04}$ & $11.18^{+0.03}_{-0.04}$  & $11.15^{+0.03}_{-0.03}$ & $11.13^{+0.02}_{-0.03}$ & $11.10^{+0.03}_{-0.02}$\\ 
\hline
PSR J1614-2230  & $1.97\pm 0.04$  & $12.14^{+0.04}_{-0.06}$ & $11.87^{+0.06}_{-0.05}$  & $11.58^{+0.05}_{-0.04}$ & $11.28^{+0.04}_{-0.03}$ & $10.94^{+0.02}_{-0.01}$ & $11.66^{+0.05}_{-0.06}$ & $11.63^{+0.05}_{-0.05}$  & $11.60^{+0.05}_{-0.05}$ & $11.57^{+0.04}_{-0.03}$ & $11.54^{+0.04}_{-0.04}$\\ \hline
GW190814  & $2.5-2.67$  & $12.96^{+0.11}_{-0.06}$ & $12.60^{+0.11}_{-0.07}$  & $12.17^{+0.03}_{-0.05}$ & $11.50^{+0.03}_{-0.12}$ & - & $12.32^{+0.08}_{-0.06}$ & $12.26^{+0.07}_{-0.05}$  & $12.20^{+0.06}_{-0.05}$ & $12.13^{+0.03}_{-0.04}$ & $12.05^{+0.01}_{-0.02}$\\ 
\hline
\end{tabular}}
\end{table*}
\begin{table*}
\centering 
\caption{Maximum masses and corresponding radii for different values of $\gamma$ and $\zeta_1$ for $\rho=\Theta^0_0$.} \label{Table2}
\scalebox{0.68}{%
\begin{tabular}{| *{12}{c|}} \hline
\quad \multirow{3}{*}{Objects} \quad & \quad  \multirow{3}{*}{$M/M_\odot$} & \multicolumn{9}{c|}{Predicted $R$ [km]}\\
\cline{3-11}
&  & \multicolumn{5}{c|}{$\gamma$} & \multicolumn{4}{c|}{$\zeta_1$}\\
\cline{3-7} \cline{8-11}
& & \quad $10.0$ &\quad $12.5$  & \quad $15.0$  & \quad $17.5$ & \quad $20$ & \quad $0.6$ &\quad $0.7$  & \quad $0.8$  & \quad $0.9$ \\
\hline
PSR J1903+327 & $1.667\pm 0.021$ & $11.20^{+0.04}_{-0.03}$ & $11.195^{+0.04}_{-0.03}$  & $11.19^{+0.03}_{-0.02}$ & $11.186^{+0.02}_{-0.02}$ & $11.18^{+0.02}_{-0.03}$ & $9.80^{+0.02}_{-0.01}$ & $10.40^{+0.03}_{-0.02}$  & $10.94^{+0.03}_{-0.01}$ & $11.46^{+0.03}_{-0.02}$ \\ 
\hline
PSR J1614-2230  & $1.97\pm 0.04$  & $11.67^{+0.06}_{-0.05}$ & $11.66^{+0.06}_{-0.05}$  & $11.64^{+0.05}_{-0.04}$ & $11.61^{+0.05}_{-0.05}$ & $11.59^{+0.05}_{-0.04}$ & $10.15^{+0.04}_{-0.04}$ & $11.78^{+0.04}_{-0.04}$  & $11.35^{+0.06}_{-0.04}$ & $11.90^{+0.05}_{-0.05}$ \\
\hline
GW190814  & $2.5-2.67$  & $12.35^{+0.09}_{-0.06}$ & $12.31^{+0.08}_{-0.06}$  & $12.27^{+0.08}_{-0.05}$ & $12.22^{+0.05}_{-0.06}$ & $12.16^{+0.05}_{-0.05}$ & $10.43^{+0.02}_{-0.12}$ & $11.17^{+0.01}_{-0.01}$  & $11.83^{+0.02}_{-0.01}$ & $12.44^{+0.02}_{-0.04}$ \\ 
\hline
\end{tabular}}
\end{table*}
\begin{table*}
\centering \caption{$M-R$ curve and prediction of radii for different values of $\alpha$ and $\zeta_1$ for $p_r=\Theta^1_1$.} \label{Table3}
\scalebox{0.68}{%
\begin{tabular}{| *{12}{c|}} 
\hline
\quad \multirow{3}{*}{Objects} \quad & \quad  \multirow{3}{*}{$M/M_\odot$} & \multicolumn{9}{c|}{Predicted $R$ [km]}\\
\cline{3-11}
&  & \multicolumn{4}{c|}{$\alpha$} & \multicolumn{5}{c|}{$\zeta_1$}\\
\cline{3-6} \cline{7-11}
& & \quad $0.2$  & \quad $0.3$ & \quad $0.4$ & \quad $0.5$ &\quad $0.25$  & \quad $0.30$  & \quad $0.35$ & \quad $0.40$  & \quad $0.45$ \\
\hline
PSR J1903+327 & $1.667\pm 0.021$  & $10.18^{+0.06}_{-0.05}$ & $10.97^{+0.01}_{-0.01}$  & $11.47^{+0.01}_{-0.01}$ & $11.88^{+0.01}_{-0.01}$ & - & $10.20^{+0.05}_{-0.04}$ & $11.41^{+0.01}_{-0.01}$  & $12.26^{+0.01}_{-0.01}$ & $13.00^{+0.01}_{-0.01}$ \\ 
\hline
PSR J1614-2230  & $1.97\pm 0.04$  & - & $10.73^{+0.05}_{-0.06}$  & $11.41^{+0.01}_{-0.02}$ & $11.87^{+0.01}_{-0.01}$ & - & - & $11.20^{+0.05}_{-0.03}$  & $12.21^{+0.01}_{-0.02}$ & $13.01^{+0.01}_{-0.01}$ \\
\hline
GW190814  & $2.5-2.67$  & - & -  & - & $11.71^{+0.03}_{-0.06}$ & - & - & -  & $11.75^{+0.13}_{-0.29}$ & $12.89^{+0.03}_{-0.05}$ \\ 
\hline
\end{tabular}}
\end{table*}

\section{Stability analysis}\label{7a}
\subsection{Stability analysis via adiabatic index}
\begin{figure*}
\centering
\includegraphics[width=7.5cm,height=6.5cm]{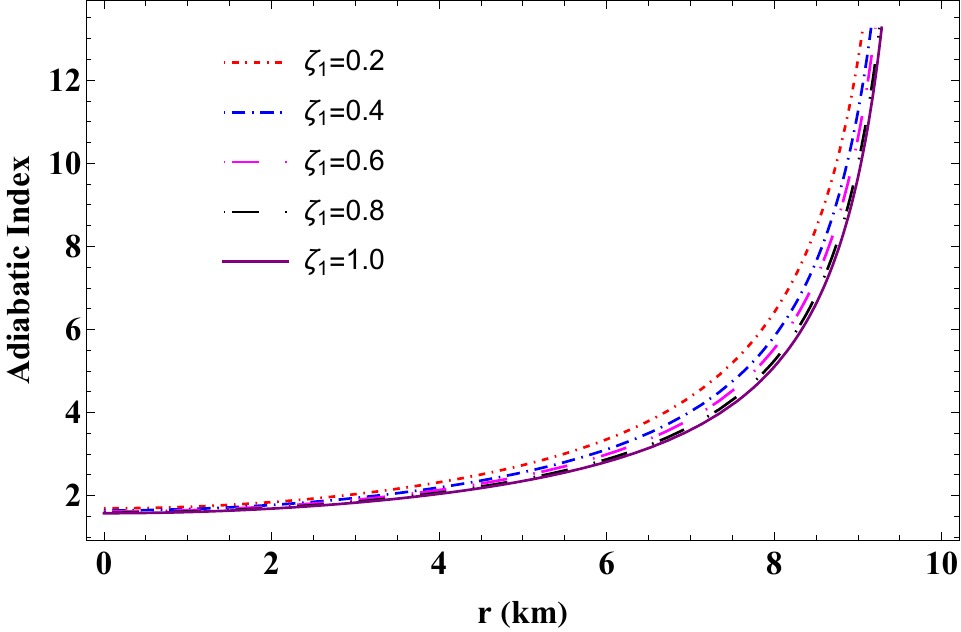}~~~
\includegraphics[width=7.5cm,height=6.5cm]{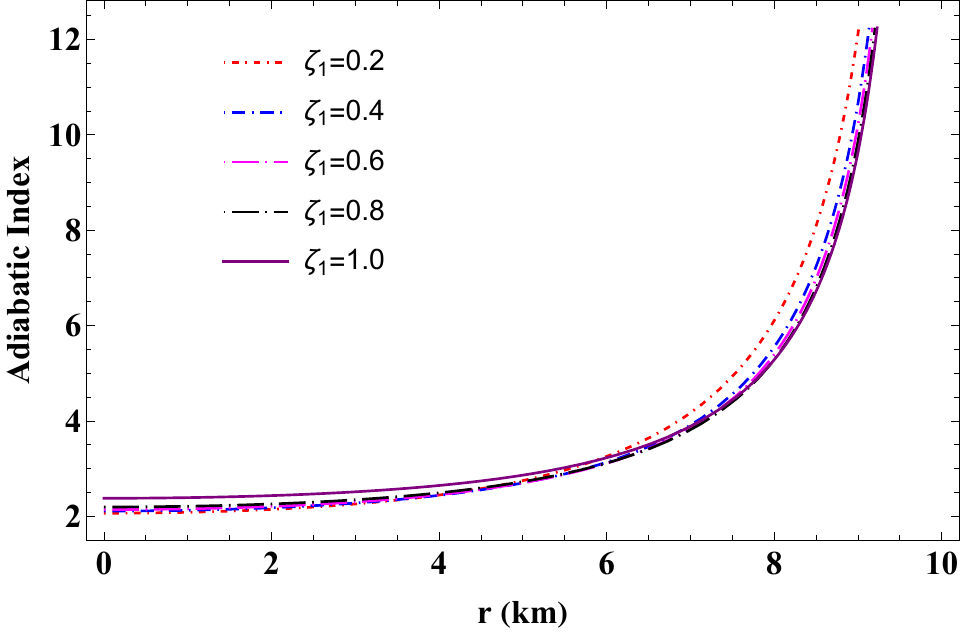}
\caption{Graphical analysis of adiabatic index for different values of the model parameter $\zeta_1$ for the solution $\rho=\Theta_0^0$ and  $p_r=\Theta^1_1$ respectively, where $\alpha=0.5~\text{km}^2$ and $\alpha=0.2~\text{km}^2$ respectively. }
\label{adb} 
\end{figure*}

The stability of the constructed anisotropic SS models is examined using the relativistic adiabatic index. Since the theoretical formulation and stability criteria have already been discussed in the introduction, we focus here only on the physical implications of our results.
From Fig.~\ref{adb}, it is evident that the adiabatic index $\Gamma$ remains well above the Chandrasekhar limit $\Gamma = 4/3$ throughout the stellar interior for all models considered. This confirms that both solutions satisfy the relativistic stability condition $\Gamma > \Gamma_{\text{crit}}$, indicating strong stability against radial perturbations.

For the solution $\Theta_0^0 = \rho$, an increase in the $f(T)$ coupling parameter $\zeta_1$ slightly reduces the degree of stability, as seen in the left panel of Fig.~\ref{adb}. In contrast, for the solution $\Theta_1^1 = p_r$, increasing $\zeta_1$ enhances the stability of the stellar core, as illustrated in the right panel of the same figure. This highlights the distinct role played by the additional gravitational source in determining the stability behavior.
A detailed numerical analysis, summarized in Table~\ref{adbtable}, further shows that increasing the MGD coupling parameter $\alpha$ tends to reduce overall stability for both solutions. Nevertheless, even for higher values of $\alpha$ and $\zeta_1$, the adiabatic index always remains above the critical threshold, ensuring that none of the models violates the stability condition.

\begin{table}
\caption{Numerical values of adiabatic index for model parameter $\zeta_1$ and MGD constant $\alpha$ at the core of the stellar region $(r=0.01~\text{km})$.}\label{adbtable}
\centering
  \begin{tabular}{@{}cc|cc@{}}
            \hline\hline
             &  Model-I ($\rho=\Theta_0^0$)\quad &    \quad Model-II ($p_r=\Theta_1^1$)\\
            \hline
              $\zeta_1$ & $\Gamma$&  $\zeta_1$ & $\Gamma$\\ \hline
            0.2 &2.224 &0.2 & 2.031 \\
            0.4 &2.104 & 0.4&2.049\\
            0.6 & 2.065 & 0.4 &2.055\\
            0.8 &2.058 &0.8 &2.057\\
            1.0&2.056&1.0&2.059\\
            \hline\hline
              $\alpha$ & $\Gamma$&  $\alpha$ & $\Gamma$\\ \hline
            0.05 &2.276&0.05&2.415\\
            0.10&2.038 &0.10& 2.279\\
            0.15&1.845 & 0.15& 2.149\\
            0.20& 1.694 &0.20&2.031\\
            0.25&1.567 &0.25&1.935\\
             \hline\hline
        \end{tabular}
\end{table}

\subsection{Stability analysis via Harrison-Zel\'dovich-Novikov criterion}
\begin{figure*}
\centering
\includegraphics[width=8cm,height=6.5cm]{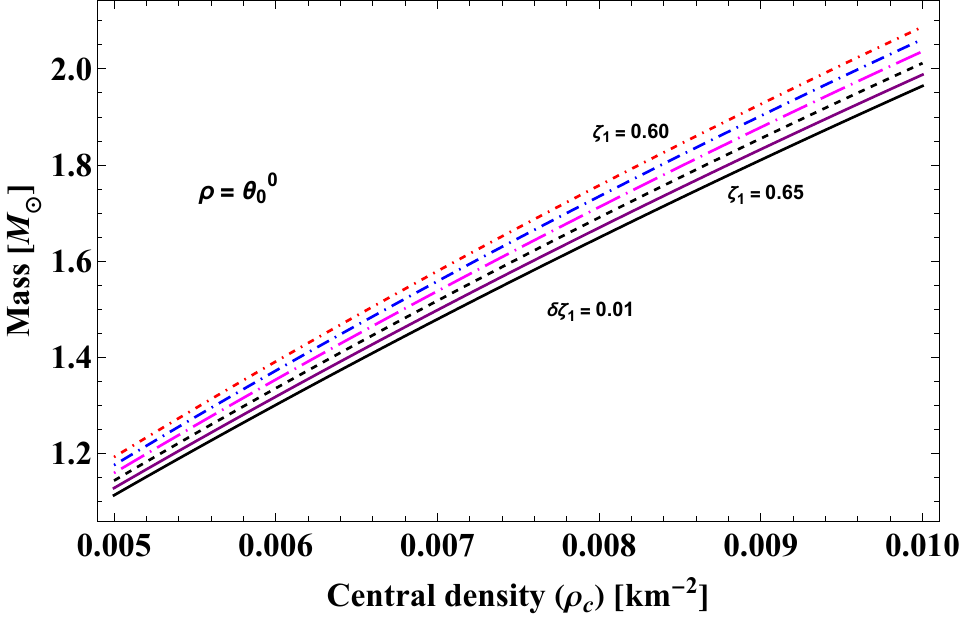}~~~
\includegraphics[width=8cm,height=6.5cm]{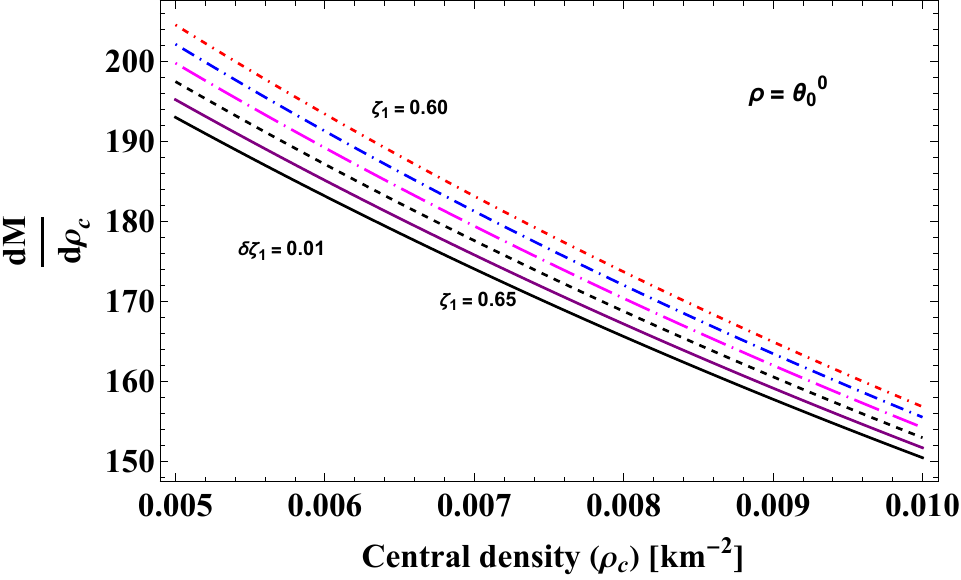}
\caption{Graphical analysis of mass and $\frac{dM}{d\rho_c}$ with respect to central density ($\rho_c$) for different values of the model parameter for the solution $\rho=\Theta^0_0$. }
\label{stab2} 
\end{figure*}

\begin{figure*}
\centering
\includegraphics[width=8cm,height=6.5cm]{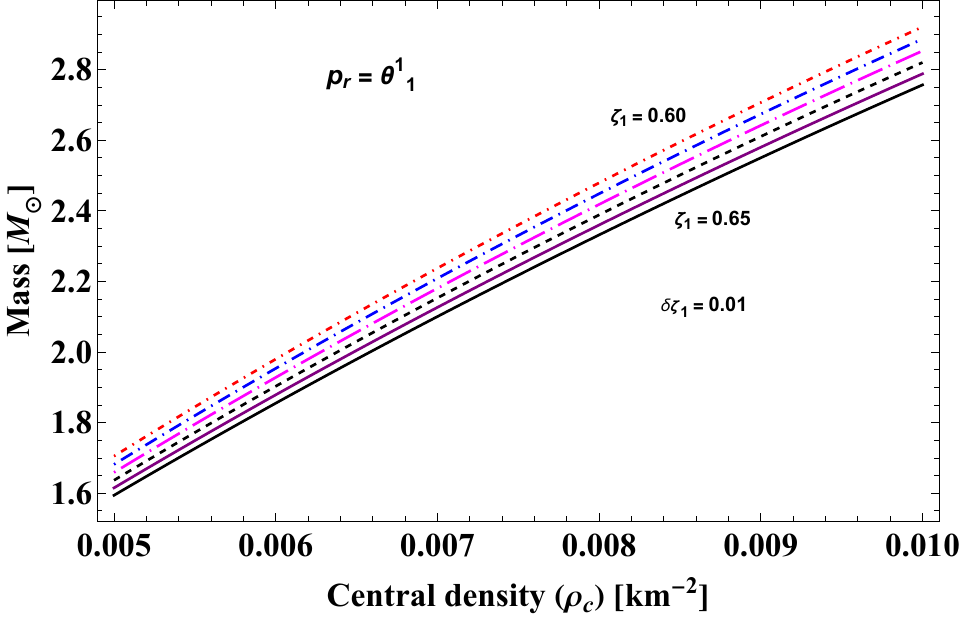}~~~
\includegraphics[width=8cm,height=6.5cm]{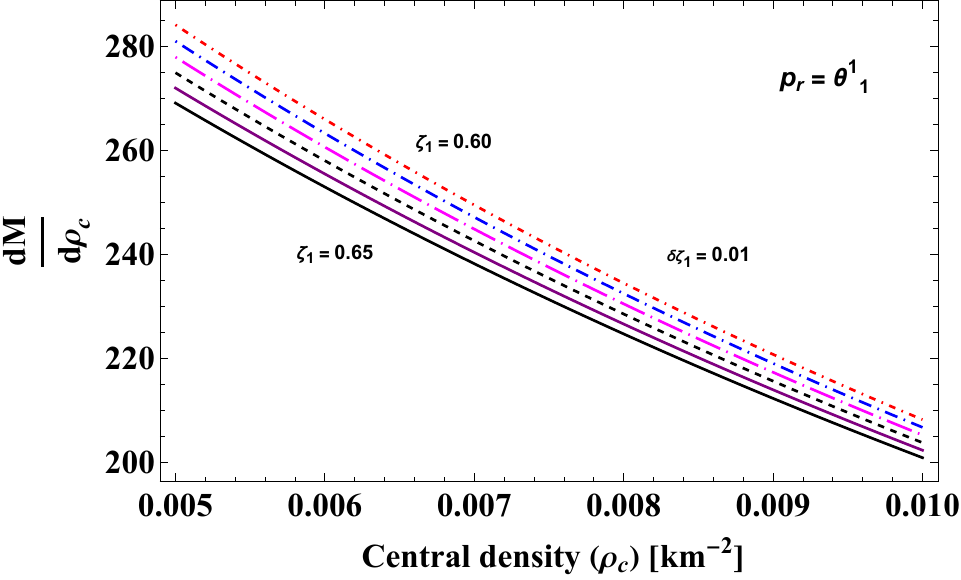}
\caption{Graphical analysis of mass and $\frac{dM}{d\rho_c}$ w.r.t. central density ($\rho_c$) for different values of the model parameter for the solution $p_r=\Theta^1_1$. }
\label{stab3} 
\end{figure*}

In the previous study, Chandrasekhar introduced a method to evaluate the stability of a stellar system when exposed to radial perturbations. However, in this approach, Harrison-Zeldovich-Novikov's (HZN) stability criterion involves examining the perturbation along the physical parameters such as the metric functions, pressure, and density. Based on the HZN stability criterion \citep{stab1,stab2}, the following constraints are established
\begin{eqnarray}
    \frac{dM}{d\rho_c}>0~~\Longrightarrow \text{Stable configuration}\nonumber\\
    \frac{dM}{d\rho_c}<0~~\Longrightarrow \text{Unstable configuration}\nonumber
\end{eqnarray}
In this context, we analyze the stability of the solution $\rho=\Theta^0_0$  and $p_r=\Theta_1^1$ in Fig.~\ref{stab2} and Fig.~\ref{stab3}, respectively. 
In the left panel of the figure, we present the mass profile as a function of the central density $\rho_c$. 
It is evident that the mass increases monotonically as the central density grows. Moreover, For a given central density of the anisotropic star, lower values of the model parameter $\zeta_1$ lead to an increase in the total mass $M$ of the system. Next, varying the model parameter $\zeta_1$ for both models, we examine the rate of changes in total mass with respect to $\rho_c$ in the right panel of Figs.~\ref{stab2} and \ref{stab3}. The results show that the rate of change of mass with central density is positive and exhibits a linear trend across the entire stellar region. Therefore, the current anisotropic star models meet the stability condition.

\section{Conclusion}\label{sec7}

In this work, we investigated the configurations of compact anisotropic stars within the framework of $f(T)$ gravity employing the gravitational decoupling method through the MGD approach. Starting from a well-defined seed system, the application of MGD naturally led to two decoupled sectors: the isotropic seed solution and an additional $\Theta$-sector, allowing us to explore the role of anisotropy in a controlled manner.
Assuming a quadratic polytropic EoS that reduces to the MIT bag model in the appropriate limit and adopting the Buchdahl metric ansatz, we obtained physically acceptable interior solutions for strange star models. Two distinct configurations, namely $\Theta_0^0=\rho$ and $\Theta_1^1=p_r$, were analyzed, each leading to different but viable stellar structures.

Our results show that the effective energy density and pressures are regular, finite, and monotonically decreasing from the stellar center to the surface for all the scenarios considered: GR, $f(T)$, and $f(T)+\text{MGD}$. The inclusion of MGD significantly modifies the central behavior of matter variables and enhances anisotropy, which plays a crucial role in counterbalancing gravitational collapse and stabilizing the stellar configuration.

The mass-radius analysis, constrained by observational data from massive pulsars and the GW190814 event, demonstrates that the proposed models can support strange stars with masses up to $\sim 3.5\,M_{\odot}$ and radii in the range $9.8$--$13.0$ km. The dependence of the $M$--$R$ relation on the model parameters indicates that larger values of $\zeta_1$ and appropriate choices of $\alpha$, $\beta$, and $\gamma$ correspond to a stiffer EoS, favoring the existence of massive compact stars within $f(T)$ gravity.
Stability analyzes based on the generalized Chandrasekhar adiabatic index and HZN analysis confirm that all obtained configurations remain stable against radial perturbations throughout the stellar interior. 

Therefore, this study highlights the effectiveness of GD through the MGD framework in $f(T)$ gravity for constructing realistic, stable, and observationally consistent models of massive anisotropic compact stars. It provides a deeper understanding of the role of torsion and anisotropy in relativistic astrophysics. In the subsequent chapter, we extend this work by employing the complete gravitational decoupling method, allowing for nonzero deformations in both the temporal and radial directions.


\chapter{\fontsize{14}{16}\selectfont The Stability of Anisotropic Compact Stars Influenced by Dark Matter under Teleparallel Gravity: An Extended Gravitational Deformation Approach} 

\label{Chapter4} 

\definecolor{maroon}{RGB}{128, 0, 0}
\lhead{\textcolor{maroon}{\textit{\textbf{Chapter 4:}}} \emph{\textcolor{maroon}{The Stability of Anisotropic Compact Stars Influenced by Dark Matter under \\ \hspace{1.9cm} Teleparallel Gravity: An Extended Gravitational Deformation Approach}}} 

\blfootnote{*The work in this chapter is covered by the following publication:\\
\textit{The Stability of Anisotropic
Compact Stars Influenced by Dark Matter under Teleparallel Gravity: An Extended Gravitational
Deformation Approach, The European Physical Journal C}, \textbf{85}(2), 127 (2025).}

In this chapter, we develop geometrically deformed SS models within the framework of teleparallel gravity by employing the gravitational decoupling approach through the complete gravitational deformation (CGD) technique. The main aspects of this chapter are summarized below:
\begin{itemize}
    \item We construct geometrically deformed SS configurations using the CGD method and obtain exact solutions by imposing the vanishing complexity factor condition.

    \item To account for dark matter (DM) effects, we introduce spacetime deformations in the metric potentials, where the decoupling parameter $\alpha$ controls the strength of the DM contribution.

    \item We support our theoretical predictions with observational constraints from GW190814 and compact stars such as EXO 1785--248, 4U 1608--52, and PSR J0952--0607.

    \item We examine the physical viability and stability of the models by analyzing the causality conditions, Herrera’s cracking method, the adiabatic index and the HZN criterion.

    \item We demonstrate that the proposed model can successfully describe a wide range of recently observed pulsars, with predicted masses and radii presented for different values of the parameters $\alpha$ and $\zeta_1$.
\end{itemize}

\section{Introduction}

In our previous chapter, we investigated anisotropic compact stars in the framework of teleparallel $f(T)$ gravity by employing the gravitational decoupling technique through the MGD approach \citep{PhysRevD.95.104019,Ovalle:2017wqi}. Although MGD has proven to be an efficient method for generating anisotropic solutions from isotropic seed configurations, it restricts the deformation to the radial metric component alone. This limitation motivates the exploration of more general decoupling schemes capable of capturing richer gravitational interactions.
In this chapter, we extend our earlier analysis by adopting the CGD approach, which allows simultaneous deformations of both temporal and radial metric components \citep{Ovalle_2019}. Recent studies have demonstrated that CGD plays a crucial role in the modification of the internal structure, complexity, and stability of self-gravitating systems, particularly in static and symmetric spherical configurations \citep{gd6,gd7,gd8,gd9}. The CGD formalism has been successfully applied to a variety of gravitational settings, including stellar interiors, cosmological solutions, and systems with vanishing complexity \citep{gd6,gd8}. In contrast to MGD, the CGD approach relaxes the restrictive assumptions on the seed fluid and allows the additional gravitational source to influence the full spacetime geometry. This makes CGD particularly suitable for modeling realistic compact stars where anisotropy, multiple matter components, and non-trivial energy exchange are expected to coexist \citep{sd2,sd13}.

Motivated by these developments, we employ the CGD method within teleparallel gravity to examine anisotropic compact stars influenced by dark matter. Dark matter is expected to accumulate inside compact stars through gravitational capture, forming a dense core or halo that modifies the effective EoS and enhances the overall compactness of the system \citep{dn1,dn2}. The presence of dark matter not only alters the pressure balance, but also plays a significant role in stabilizing massive stellar configurations by effectively stiffening the matter sector \citep{dn3,dn4}.
The primary aim of this work is to investigate how the combined effects of CGD and dark matter influence the stability of anisotropic compact stars. We perform a detailed stability analysis using established criteria, including Herrera’s cracking condition \citep{sd19,sd20}, the relativistic adiabatic index, and the HZN criterion \citep{stab1,stab2}. This study therefore represents a natural and systematic extension of our previous work, providing deeper insight into the role of CGD and dark matter in shaping the equilibrium and stability of compact objects within teleparallel gravity.

The outline of the chapter is as follows: The first
section~\ref{sec22} presents the field equations in $f(T)$ gravity using the CGD framework. In section~\ref{seciii}, we derive an exact anisotropic dark star solution based on the vanishing complexity condition for both the seed and $\Theta$ sectors. The physical properties and stability of the model are analyzed in sections~\ref{seciv} and~\ref{secv} respectively. Sections~\ref{secvi} and~\ref{secvii} examine the maximum mass and radius of compact objects through $M$–$R$ relations and equi-mass diagrams, respectively. The role of dark matter–induced energy exchange is discussed in section~\ref{secviii}. Finally, section~\ref{secx} summarizes the main conclusions.

\section{Field equations in $f(T)$ gravity via CGD}\label{sec22}

Since the mathematical and geometrical formulation of $f(T)$ gravity and the corresponding field equations for a static, spherically symmetric anisotropic configuration have already been discussed in the previous chapter, we now proceed directly to the implementation of the CGD methodology.
Using the torsion scalar given in Eq.~(\ref{torsion}) together with the linear form of the function $f(T)$, the three independent components of the field equations in teleparallel gravity take the form
\begin{eqnarray}\label{FE1}
&&8 \pi \rho^{\text{tot}} = \frac{e^{-\lambda (r)}}{2r^2} \Big[e^{\lambda (r)} (2 \zeta_1+\zeta_2 r^2)-2 \zeta_1+2 \zeta_1 r \lambda'(r)\Big],\\
\label{FE2}
&&8\pi p_r^{\text{tot}} = \frac{e^{-\lambda (r)}}{2r^2} \Big[2 \zeta_1-e^{\lambda (r)} (2 \zeta_1+\zeta_2 r^2)+2 \zeta_1 r \nu '(r)\Big],\\
\label{FE3}
&&8\pi p_t^{\text{tot}} = \frac{e^{-\lambda (r)}}{4r} \Big[-\zeta_1 (r \nu '(r)+2) (\lambda '(r)-\nu '(r))
+2 \zeta_1 r \nu ''(r)-2 \zeta_2 r e^{\lambda (r)}\Big].
\end{eqnarray}
To examine the influence of the additional source $\Theta_{\mu\nu}$ on the distribution of seed matter, we introduce the complete geometric deformation of both metric potentials as \citep{PhysRevD.95.104019}
\begin{eqnarray}\label{mgd1}
\nu(r) \longrightarrow G(r)+\alpha\, \Phi(r), \\
\label{mgd2}
e^{-\lambda(r)} \longrightarrow H(r)+\alpha\, \psi(r).
\end{eqnarray}
Here, $\Phi(r)$ and $\psi(r)$ denote the temporal and radial deformations, respectively.  
Since both deformations are non-vanishing, the procedure corresponds to the CGD method. We remember in the MGD method, we set $\Phi(r)=0$. Substituting Eqs.~(\ref{mgd1}) and (\ref{mgd2}) into the field equations leads to a natural splitting of the system.

\begin{itemize}
    \item The first is that the seed system corresponds to a perfect fluid in teleparallel gravity with $\alpha=0$, having components: $\{\rho, p_r, p_t, G(r), H(r)\}$.
 \begin{eqnarray}\label{sfe1}
&&\hspace{-0.6cm}\rho(r) = \frac{1}{16 \pi r^2}\Big[ 2 \zeta_1-2 \zeta_1 (r H'(r)+H(r))+\zeta_2 r^2\Big],~~
\\ \label{sfe2}
&&\hspace{-0.6cm}p_r = \frac{1}{16\pi r^2}\Big[-2 \zeta_1 H(r) (r G'(r)+1)+2 \zeta_1+\zeta_2 r^2\Big],~~~
\\ \label{sfe3}
&&\hspace{-0.6cm}p_t = \frac{1}{32\pi r}\Big[2 \zeta_1 r H(r) G''(r)+\zeta_1 \big(r G'(r)+2\big)\big(H(r) G'(r)+H'(r)\big)-2 \zeta_2 r\Big].
\end{eqnarray}
\end{itemize}
Moreover, in this reference, the interior space-time metric can be re-defined as 
\begin{eqnarray}
    ds_{-}^2=-e^{G(r)}dt^2+H^{-1}(r)dr^2+r^2(d\theta^2+sin^2\theta~d\phi^{2}).~~~~~~
\end{eqnarray}
\begin{itemize}
\item The second set of equations corresponding to the source $\Theta_{\mu\nu}$, which possesses components $\{\Theta_0^0,~ \Theta_1^1,~ \Theta_2^2,\\~ G(r), H(r), \Phi(r), \psi(r)\}$, is given by
\begin{eqnarray}\label{t1}
&&\hspace{-2.0cm}\Theta^0_0 = -\frac{\zeta_1}{8\pi r^2}\Big[ \big(r \psi '(r)+\psi (r)\big)\Big],
\\ \label{t2}
&&\hspace{-2.0cm}\Theta_1^1 = \frac{-\zeta_1}{8\pi r^2}\Big[ \psi (r) (r G'(r)+1)+r \Phi '(r) (H(r)+\alpha  \psi (r))\Big],~~~~~~
\\ \label{t3}
&&\hspace{-2.0cm}\Theta^2_2 = \frac{-\zeta_1}{32\pi r\alpha}\Big[-\big(\left(r G'(r)+2\right) (H(r) G'(r)+H'(r))\big)+2 r (H(r)+\alpha  \psi (r)) \left(G''(r)+\alpha  \Phi ''(r)\right)\nonumber\\&&\hspace{0.5cm}-2 r H(r)G''(r)+\left(r G'(r)+\alpha  r \Phi '(r)+2\right) \big((H(r)+ \alpha  \psi (r))\left(G'(r)+\alpha  \Phi '(r)\right)\nonumber\\&&\hspace{0.5cm}+H'(r)+ \alpha  \psi '(r)\big)\Big].
\end{eqnarray}
\end{itemize}

In the next section, we derive solutions for the proposed compact star model by enforcing the condition of vanishing complexity in the presence of a dark matter distribution.
\section{Complexity-free anisotropic dark star model via gravitational decoupling}\label{seciii}

In this section, we construct a complexity-free anisotropic dark star model by solving the two decoupled systems corresponding to the seed sector $\mathcal{T}_{\mu\nu}$ and the additional source $\Theta_{\mu\nu}$. Since the solution of the $\Theta$ sector explicitly depends on the geometry of the seeds, we first determine the solution of the seed system. The field Eqs.~(\ref{sfe1}–\ref{sfe3}) constitute a highly non-linear second-order system with five unknowns $\{\rho, p_r, p_t, G(r), H(r)\}$. To close the system, we adopt the well-known Tolman–Kuchowicz metric ansatz \cite{tolman,kucho}, given by
\begin{eqnarray}
dS^2=-e^{Cr^2+2\ln D}\,dt^2+\left(1+Ar^2+Br^4\right)^{-1}dr^2+r^2(d\theta^2+\sin^2\theta\,d\phi^2),
\end{eqnarray}
where $A$, $B$, and $C$ are constants of appropriate dimensions and $D$ is a dimensionless constant. This choice ensures regularity and freedom from singularities and has been widely employed in compact star modeling \cite{pb3}.
Substituting this ansatz into Eqs.~(\ref{sfe1}–\ref{sfe3}), we obtain the seed energy density and pressures as
\begin{eqnarray}\label{rhoeq}
\rho &=& -3A\zeta_1-5B\zeta_1 r^2+\frac{\zeta_2}{2},\\ \label{preq}
p_r &=& \zeta_1\!\left[r^2(2AC+B)+A+2BCr^4+2C\right]-\frac{\zeta_2}{2},\\ \label{pteq}
p_t &=& \zeta_1\!\left[Cr^4(AC+4B)+r^2(C(3A+C)+2B)+A+BC^2r^6+2C\right]-\frac{\zeta_2}{2}.
\end{eqnarray}

We now turn to the $\Theta$ sector, which captures the dark matter contribution. To construct a complexity-free configuration, we employ Herrera’s definition of the complexity factor \cite{gd10}. In teleparallel gravity, the total complexity factor reads
\begin{eqnarray}
Z^{T}_{TF}=8\pi(p_r^{\rm tot}-p_t^{\rm tot})-\frac{4\pi}{2r^3}\int_0^r y^3(\rho^{\rm tot})'(y)\,dy.
\end{eqnarray}
Imposing the vanishing complexity condition $Z^{T}_{TF}=0$ yields a differential relation between the metric potentials, which integrates with
\begin{eqnarray}
&&\hspace{0cm}\nu'(r)e^{\nu/2}=\mathcal{C}_1\, r\, e^{\lambda/2},\\
&&\hspace{-1cm}\implies\nu(r)=2\ln\!\left[\mathcal{C}_2+\mathcal{C}_1\int \frac{r}{\sqrt{H(r)+\alpha\psi(r)}}\,dr\right],
\end{eqnarray}
where $\mathcal{C}_1$ and $\mathcal{C}_2$ are constants fixed through boundary conditions. This relation closely parallels the corresponding result in GR \cite{cf1}. Next,
to model the dark matter distribution, we adopt the pseudo-isothermal density profile \cite{dm1,dm2}
\begin{eqnarray}
\rho_{\tiny{DM}}=\rho_0\left[1+\left(\frac{r}{r_0}\right)^2\right]^{-1},
\end{eqnarray}
which is regular and monotonically decreasing. Identifying $\Theta_0^0=\rho_{DM}$ and integrating Eq.~(\ref{t1}), we obtain the radial deformation function
\begin{eqnarray}
\psi(r)=\frac{L r^2(3Nr^2-5)}{15\zeta_1},
\end{eqnarray}
after enforcing the regularity in the center. However, in the present investigation, we treated the constants $L=\rho_0$ and $N=1/r_0^2$ as free parameters. The temporal deformation function $\Phi(r)$ then follows directly from the vanishing complexity condition and is obtained analytically in terms of seed potentials and $\psi(r)$. Substituting $\Phi(r)$ and $\psi(r)$ into Eqs.~(\ref{t1}–\ref{t3}) we can fully determine the components of the sector $\Theta$.
Consequently, the completely deformed metric functions take the form
\begin{eqnarray}
e^{-\lambda}(r)&=&(1+Ar^2+Br^4)+\alpha\,\frac{L r^2(3Nr^2-5)}{15\zeta_1},\\
\nu(r)&=&2\ln\!\left[\mathcal{C}_2+\mathcal{C}_1\int \frac{r}{\sqrt{H(r)+\alpha\psi(r)}}\,dr\right],
\end{eqnarray}
due to the long expression of $\nu(r)$ we are not giving the full expression here and the effective energy density and pressures are given by
\begin{eqnarray}
\rho^{\rm tot}&=&-3A\zeta_1-5B\zeta_1 r^2+\frac{\zeta_2}{2}+\frac{\alpha L}{Nr^2+1},\\
p_r^{\rm tot}&=&p_r+\alpha\,\Theta_1^1,\qquad
p_t^{\rm tot}=p_t+\alpha\,\Theta_2^2,
\end{eqnarray}
with $\Theta_1^1$ and $\Theta_2^2$ obtained from Eqs.~(\ref{t2}) and (\ref{t3}) respectively. The expressions $p_r$ and $p_t$ can be taken from Eq.~(\ref{preq}) and Eq.~(\ref{pteq}) respectively.
This completes the construction of a complexity-free anisotropic dark star model within the CGD framework of teleparallel gravity. 

The construction of a physically acceptable dark star model within the CGD framework requires a smooth match between the interior anisotropic spacetime and an exterior vacuum geometry at the stellar boundary \( r = R \). To this end, we impose the standard junction conditions by matching the CGD-modified interior solution with the Schwarzschild exterior spacetime.
From a methodological perspective, the continuity of the first fundamental form guaranties the smooth behavior of the metric potentials across the boundary, while the continuity of the second fundamental form enforces the physically necessary condition that the total radial pressure vanishes at the stellar surface
\begin{equation}
p_r^{\text{tot}}(R) = 0 .
\end{equation}
These boundary conditions are sufficient to uniquely determine the integration constants \( \mathcal{C}_1 \) and \( \mathcal{C}_2 \) appearing in the deformed temporal metric potential.
The primary objective of this matching procedure is therefore twofold: to ensure the absence of surface discontinuities in the spacetime geometry and to fix the free parameters of the CGD solution in terms of stellar mass and radius. Once these constants are determined, the interior solution becomes fully specified and suitable for subsequent physical and stability analyzes, which are presented in the following section.

\section{Physical analysis of the stellar model}\label{seciv}
In this section, we shall examine various physical properties of the dark star model to present the best model through our analysis. Let us start with the important and well-known fundamental energy profiles of the dark star in the following subsections.

\begin{figure*}
    \centering
    \includegraphics[height=6.2cm,width=7.8cm]{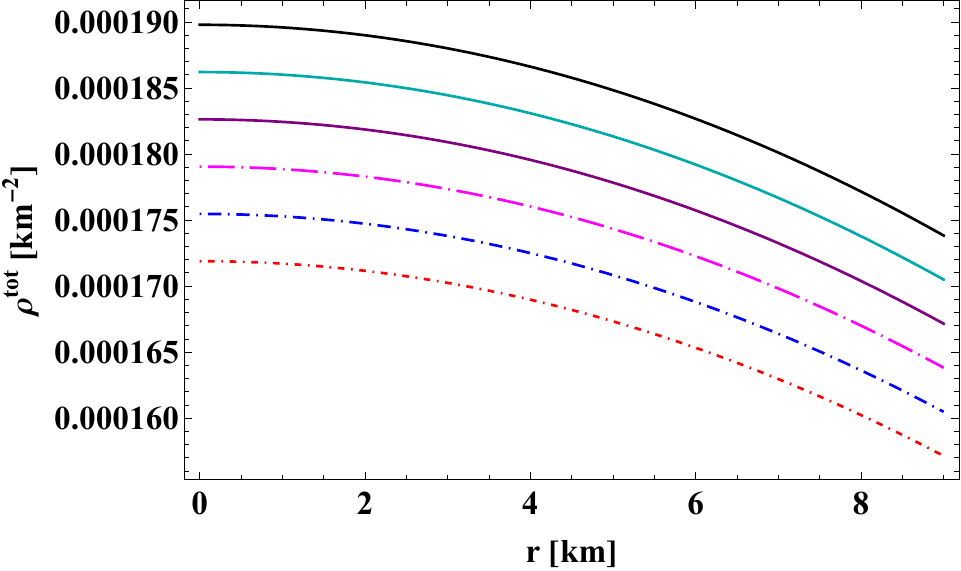}~~~
    \includegraphics[height=6.2cm,width=8cm]{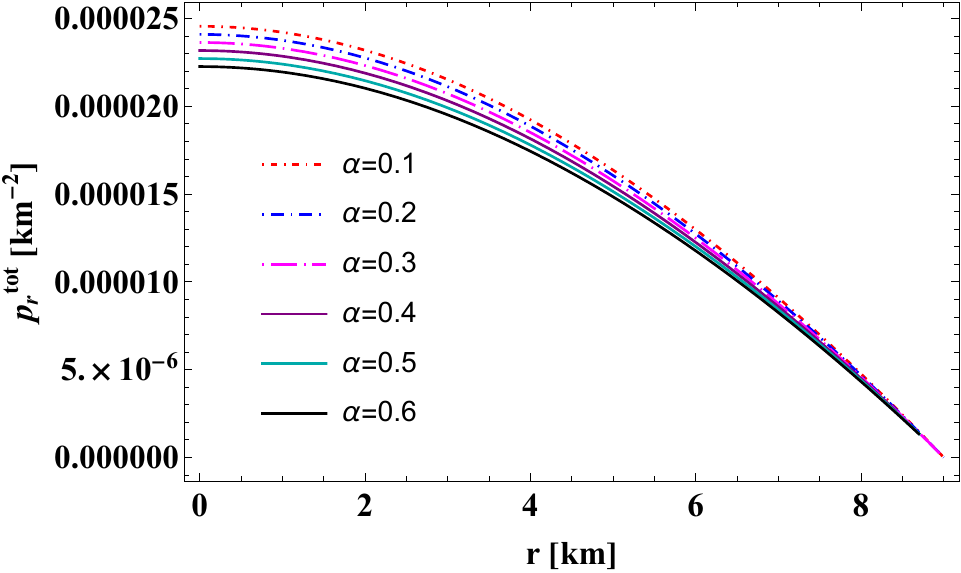}
    \caption{Graphical analysis of energy density (left) [$\alpha=0.1(\textcolor{red}\star),\alpha=0.2(\textcolor{blue}\star), \alpha=0.3(\textcolor{magenta}\star), \alpha=0.4(\textcolor{purple}\star), \alpha=0.5(\textcolor{green}\star),\alpha=0.6(\textcolor{black}\star) $] and radial pressure (right) for $C = 0.288~ \text{km}^{-2}; D = 0.1; A = 0.009~ \text{km}^{-2}; B = 0.000009~\text{km}^{-4}; L = 0.0009 ~\text{km}^{-2}; N = 0.0009~ \text{km}^{-2}$.}
    \label{fig11}
\end{figure*}

\begin{figure*}
    \centering
    \includegraphics[height=6.2cm,width=8cm]{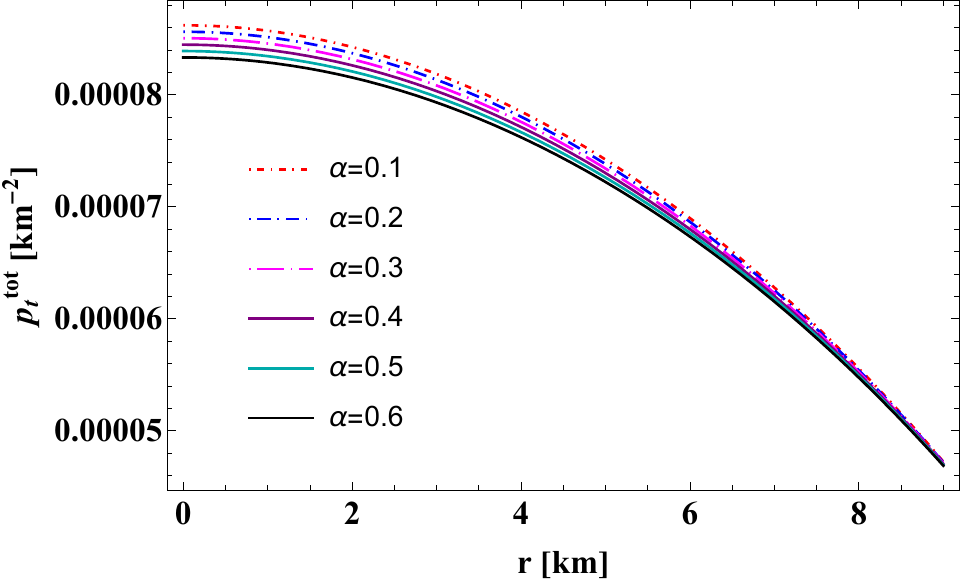}~~~
    \includegraphics[height=6.2cm,width=8cm]{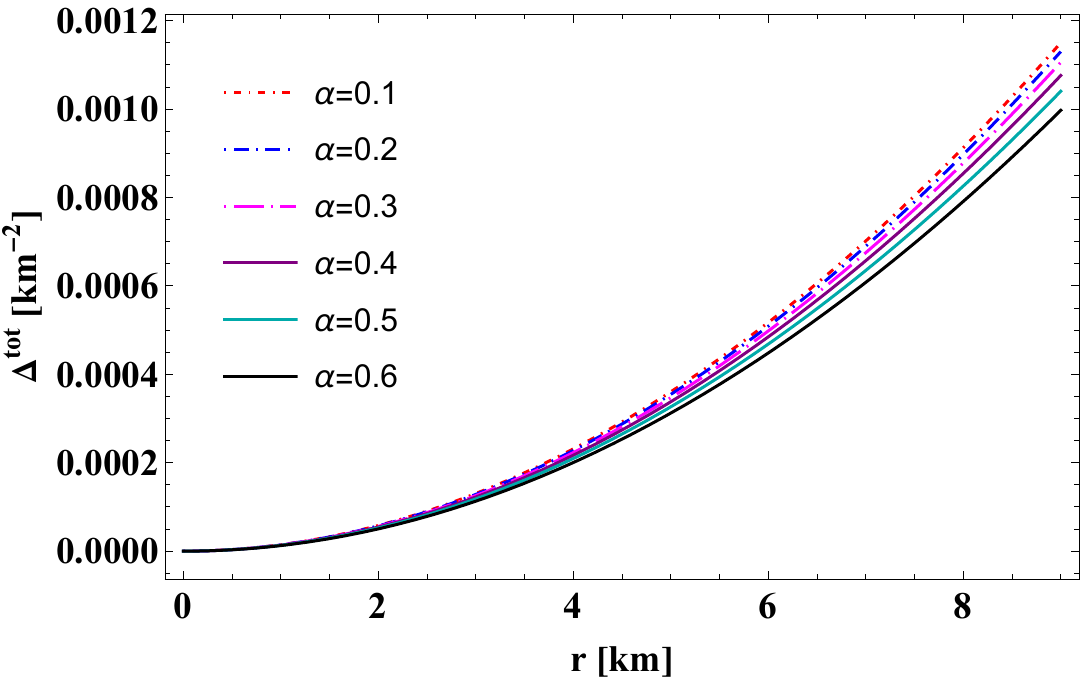}
    \caption{Graphical analysis of tangential pressure (left) and anisotropy (right) for $C = 0.288~ \text{km}^{-2}; D = 0.1; A = 0.009~ \text{km}^{-2}; B = 0.000009~\text{km}^{-4}; L = 0.0009 ~\text{km}^{-2}; N = 0.0009~ \text{km}^{-2}$.}
    \label{fig21}
\end{figure*}

\subsection{Density, pressure, and anisotropy}
This subsection will explore various physical properties for the above-analyzed star model configured with the perfect fluid and dark matter components. Quantities such as energy density ($\rho^{tot}$), radial pressure ($p_r^{tot}$), and transverse pressure ($p_t^{tot}$) are enough to help us discuss the energy conditions.
The graphical representation of $\rho^{tot}$, $p_r^{tot}$, $p_t^{tot}$ and anisotropy ($\Delta^{tot}$) in Figs. \ref{fig11} and \ref{fig21} provides a comprehensive view of the energy distribution within the stellar model. Our analysis reveals that $\rho^{tot}$, $p_r^{tot}$, and $p_t^{tot}$ satisfy the essential criteria for a valid stellar model. Specifically, these quantities are positive and finite throughout the stellar region, peaking at the center and decreasing as the radial distance increases. As depicted in Fig. \ref{fig11}, the total energy density ($\rho^{\text{tot}}$) and the radial pressure $p_r^{\text{tot}}$ reach zero at the stellar surface ($r = R$). Moreover, Fig. \ref{fig21} shows that anisotropy ($\Delta^{tot}$), defined as the difference between tangential and radial pressures ($p_t^{tot} - p_r^{tot}$), is positive and increases with radial distance. This positive anisotropy, where $p_r^{tot} < p_t^{tot}$, is crucial as it contributes to the stability of the stellar system by supporting hydrostatic equilibrium.

Let us delve more deeply into the behavior of the physical quantities $\rho^{tot}$, $p_r^{tot}$, and $p_t^{tot}$ with the changes of the decoupling parameter $\alpha$, as depicted in Figs. \ref{fig11}-\ref{fig21}. These figures reveal a fascinating trend: with each increment in $\alpha$, the energy density throughout the anisotropic dark star rises. It is suggested that by utilizing the new source term $\Theta_{\mu\nu}$, our model's dark matter component $\alpha$ leads to the formation of denser stellar objects. Examining Fig. \ref{fig11}, we see an opposite behavior with the total energy density in the radial pressure. In the central region, the radial pressure decreases with higher $\alpha$, then converges near the surface and eventually drops to zero around 9 km. A similar pattern is observed in Fig. \ref{fig21} for the tangential pressure, which converges to different finite non-zero values on the surface for each $\alpha$. This convergence indicates that the surface pressures, both radial and tangential, remain unaffected by changes in $\alpha$, producing effects similar to those seen in isotropic systems. One can see, Fig. \ref{fig21} provides further insight, showing that the anisotropy introduced by the dark matter component intensifies near the surface and converges towards the center as $\alpha$ increases. This demonstrates a monotonically increasing effect of $\alpha$ on the star's anisotropy from the center to the surface. Additionally, for positive values of $\alpha$, the dark-matter contribution constrains the central and surface values of both the energy density and the pressures to lie within the same order of magnitude, $\rho(0),\, \rho(R),\, p_r^{\text{tot}}(0),\, p_t(R) \sim \mathcal{O}(10^{14})\,\text{g/cm}^{3}$, where $\rho_c \equiv \rho(0)$ denotes the central density.

\subsection{Energy conditions and matter gradients}\label{seciva}

In this subsection, we examine the physical viability of the anisotropic dark star configuration by verifying the standard energy conditions, namely the null, weak, dominant, and strong energy conditions, using the total effective quantities \(\rho^{\text{tot}},\, p_r^{\text{tot}}\), and \(p_t^{\text{tot}}\). Since the formal expressions of these conditions have already been presented in Chapter-\ref{Chapter1}, we focus here only on their physical implications and results.

From Figs.~\ref{fig11} and \ref{fig21}, it is evident that all energy conditions are satisfied throughout the stellar interior. The effective energy density remains positive and finite, while both radial and tangential pressures respect the required inequalities associated with NEC, WEC, DEC, and SEC. In particular, the fulfillment of DEC ensures that the effective energy flow remains causal, whereas the validity of SEC confirms the attractive nature of gravity within the stellar configuration. We further observe that the decoupling parameter \(\alpha\), which encodes the influence of the dark matter sector, enhances the positivity of these inequalities. This indicates that a stronger dark matter contribution improves the physical admissibility of the model.

\begin{figure*}
    \centering   \includegraphics[height=4.8cm,width=5.2cm]{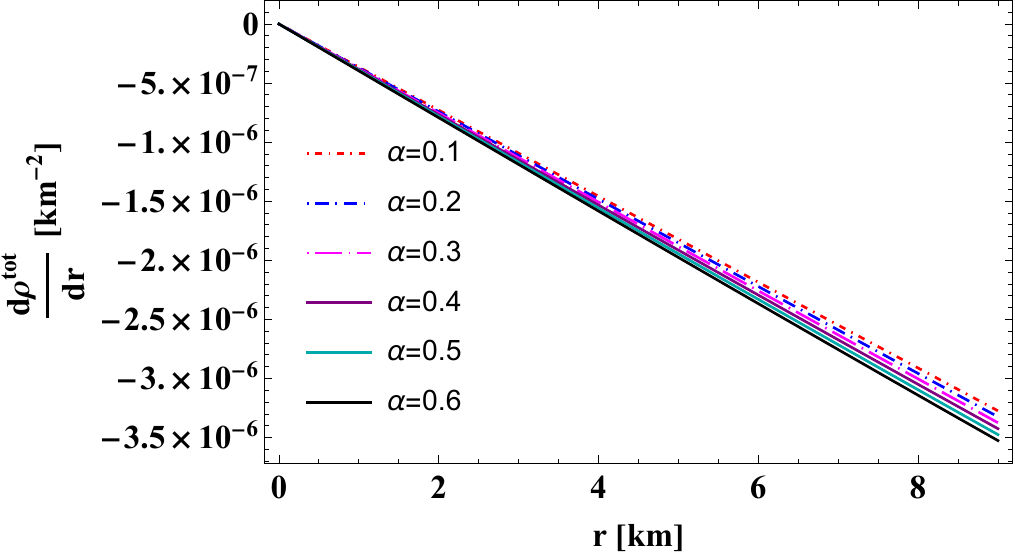}
    \includegraphics[height=4.8cm,width=5.2cm]{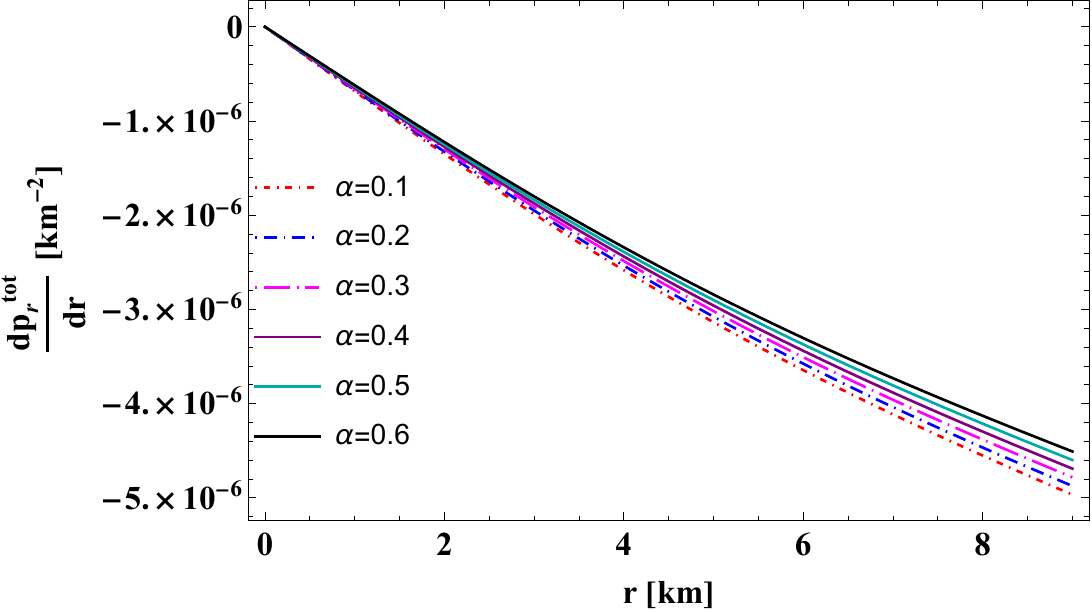}
    \includegraphics[height=4.8cm,width=5.2cm]{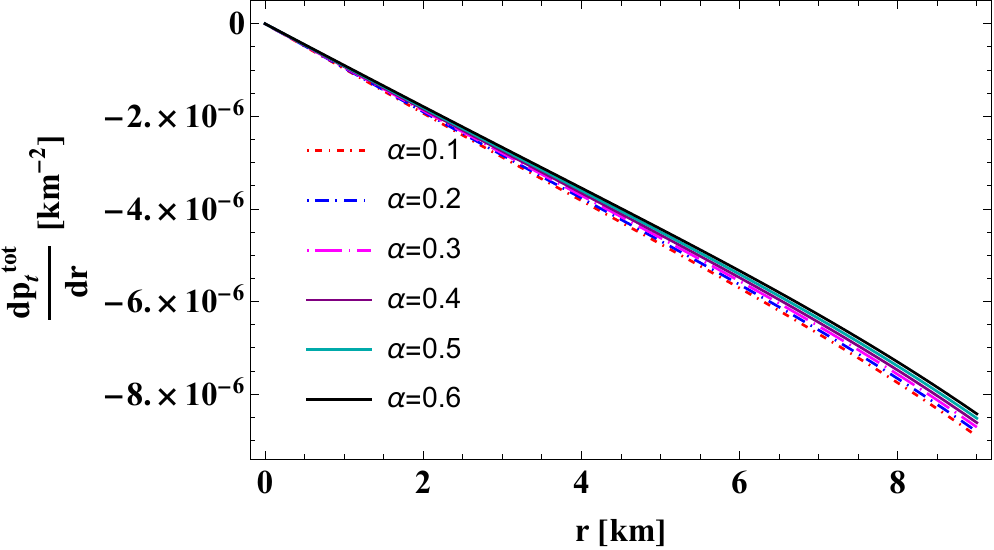}
    \caption{Graphical analysis of the density gradient (left), the radial pressure gradient (middle) and the tangential pressure gradient (right) with respect to '$r$,' for $C = 0.288~ \text{km}^{-2}; D = 0.1; A = 0.009~ \text{km}^{-2}; B = 0.000009~\text{km}^{-4}; L = 0.0009 ~\text{km}^{-2}; N = 0.0009~ \text{km}^{-2}$.}
    \label{fig41}
\end{figure*}

In addition to the energy conditions, we analyze the radial behavior of the gradients of the effective energy density and pressures to ensure a realistic stellar profile. As shown in Fig.~\ref{fig41}, the gradients of \(\rho^{\text{tot}},\, p_r^{\text{tot}}\) and \(p_t^{\text{tot}}\) are negative throughout the interior and vanish at the center, implying that all thermodynamic quantities reach their maximum values at \(r=0\) and decrease monotonically towards the boundary. This behavior confirms the regularity and stability of the matter distribution and further supports the physical consistency of the anisotropic dark star model constructed within the CGD framework.

\section{Stability analysis}\label{secv}

The anisotropic fluid distributions of the dark stellar structure can verify the stability of the stellar structure. These distributions can be examined through various stability analyzes, and some of these, such as adiabatic index, speed of sound through the dark matter fluid, and analysis of central density, shall be discussed in the following subsections.

\subsection{Stability analysis via adiabatic index}\label{secva}

\begin{figure}
    \centering
    \includegraphics[height=6.2cm,width=7.8cm]{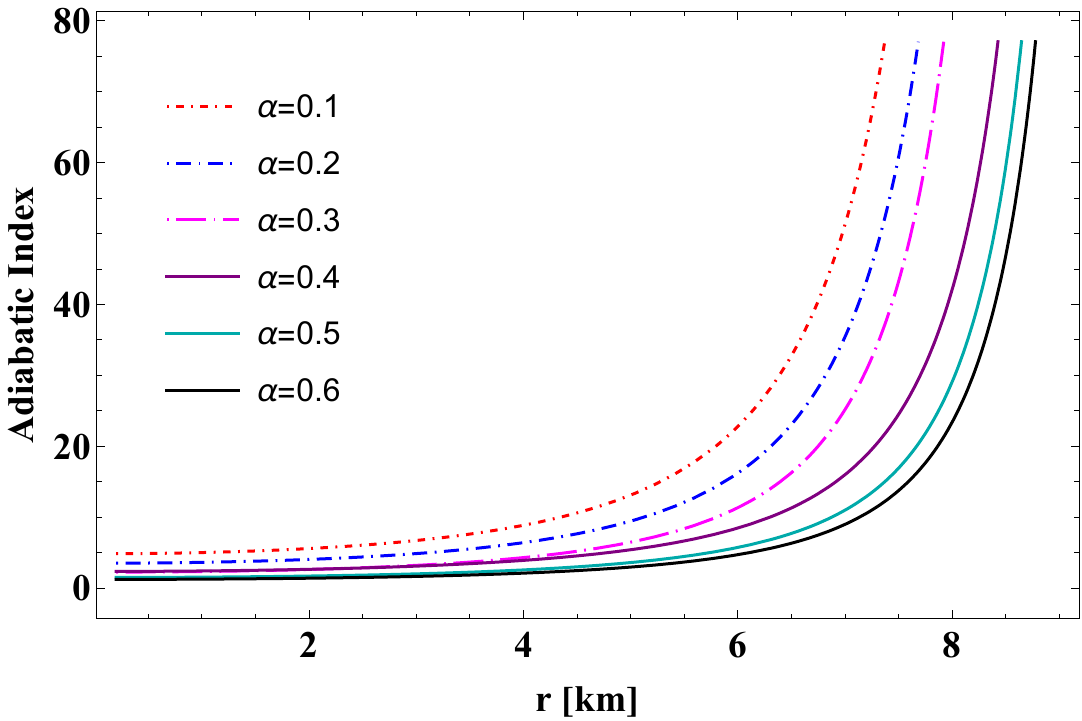}~~~
    \caption{Stability analysis via adiabatic index ($\Gamma$) for $C = 0.288~ \text{km}^{-2}; D = 0.1; A = 0.009~ \text{km}^{-2}; B = 0.000009~\text{km}^{-4}; L = 0.0009 ~\text{km}^{-2}; N = 0.0009~ \text{km}^{-2}$.}
    \label{fig31}
\end{figure}

We analyze the hydrostatic stability of the anisotropic dark star through the behavior of the relativistic adiabatic index \(\Gamma^{\text{tot}}\), focusing exclusively on the physical outcomes since its formal definition has already been discussed earlier.
Fig.~\ref{fig31} shows that \(\Gamma^{\text{tot}}\) remains everywhere greater than the critical value \(4/3\) for all values considered of the decoupling parameter \(\alpha\), confirming that the stellar configuration is stable against radial perturbations. Positive anisotropy further strengthens this stability.

A comparison between the anisotropy factor (Fig.~\ref{fig21}) and the adiabatic index (Fig.~\ref{fig31}) reveals that variations in \(\alpha\) have little influence on anisotropy near the stellar center, while they significantly affect \(\Gamma^{\text{tot}}\) in the inner regions. Smaller values of \(\alpha\) are sufficient to maintain core stability, whereas near the stellar surface the influence of \(\alpha\) becomes negligible, as the stability condition is satisfied throughout.

\subsection{Causality criterion \& Herrera's cracking method}\label{secvb}

\begin{figure*}
    \centering
    \includegraphics[height=5.2cm,width=5.2cm]{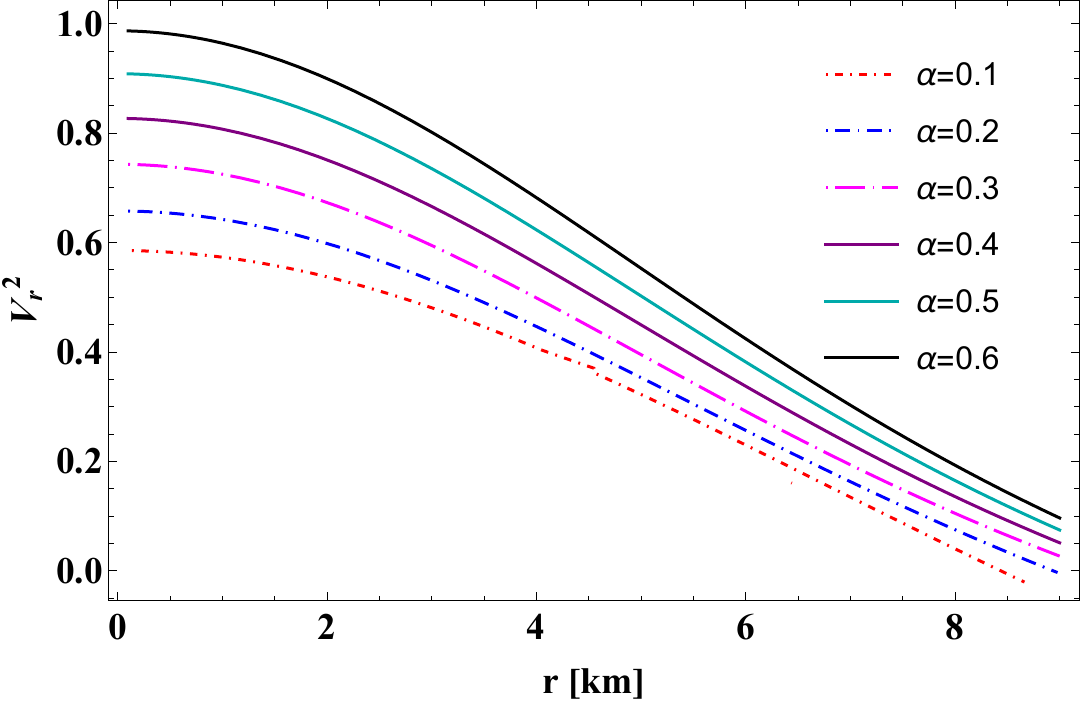}
    \includegraphics[height=5.2cm,width=5.2cm]{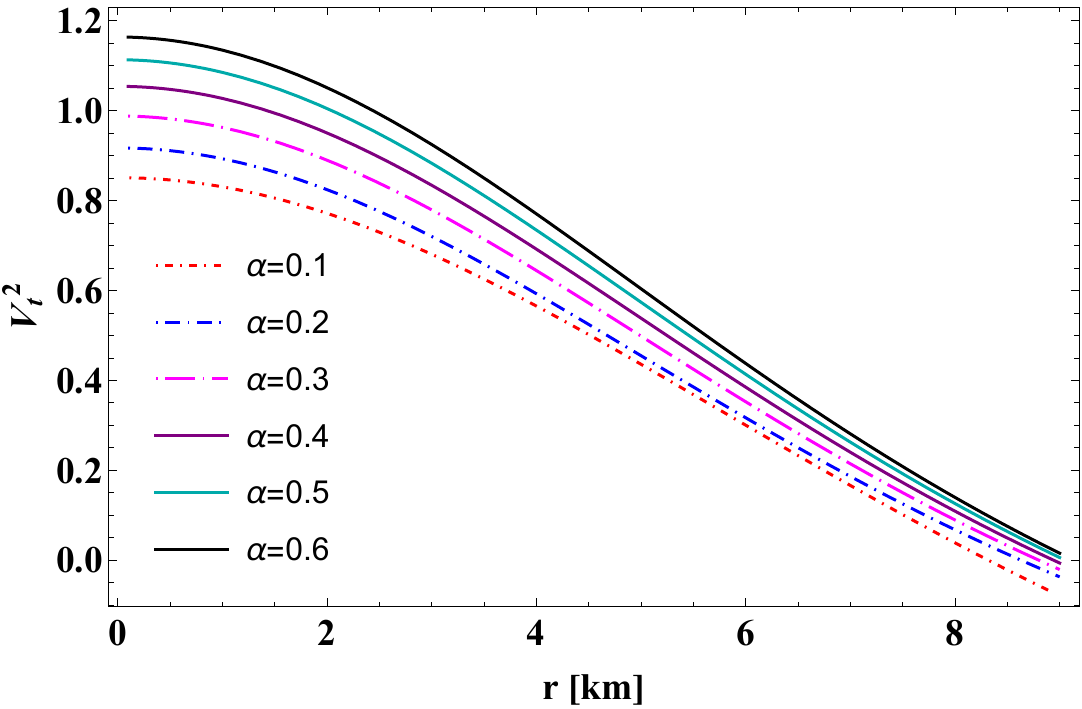}     \includegraphics[height=5.2cm,width=5.2cm]{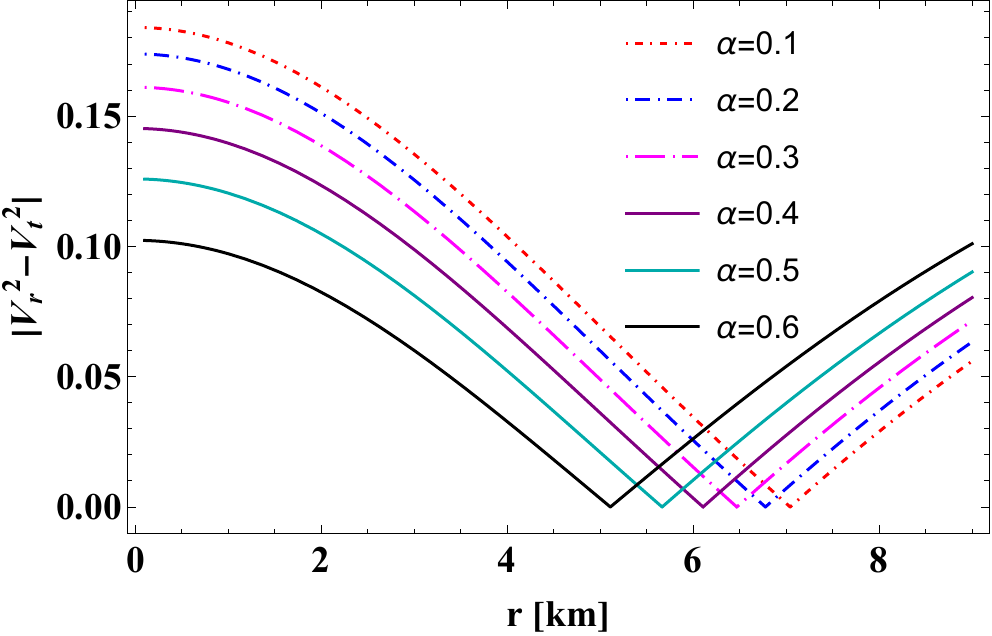}
    \caption{Stability analysis via speed of sound w.r.t. 'r' for $C = 0.288~ \text{km}^{-2}; D = 0.1; A = 0.009~ \text{km}^{-2}; B = 0.000009~\text{km}^{-4}; L = 0.0009 ~\text{km}^{-2}; N = 0.0009~ \text{km}^{-2}$.}
    \label{fig51}
\end{figure*}

The stability of the anisotropic dark star model is further examined using the causality criterion and Herrera’s cracking method, both of which rely on the behavior of the radial and transverse sound speeds. For a physically viable stellar configuration, these velocities must remain non-negative and subluminal.
As shown in Fig.~\ref{fig51}, both radial and tangential sound speeds satisfy the causality requirement $0\leq V_r^2 \leq 1$ and $0\leq V_t^2 \leq 1$ throughout the stellar interior. In addition, the difference between the squared sound speeds remains within the allowed range, thus fulfilling Herrera’s cracking condition $|V_r^2-V_t^2|\leq 1$ and confirming the absence of gravitational cracking \cite{sd19,
abr/2007}.

An important trend is observed with respect to the decoupling parameter \(\alpha\). As \(\alpha\) increases, both \(V_r^2\) and \(V_t^2\) approach their limiting values, indicating that excessive decoupling weakens the stability of the configuration. Consequently, moderate values of \(\alpha\) are favored for maintaining a stable anisotropic dark star.

\subsection{Harrison--Zeldovich--Novikov criterion}\label{secvc}

The stability of the anisotropic dark star model is further examined using the HZN criterion, which is based on the response of the stellar mass to variations in the central density. 
According to the HZN stability condition, a stellar configuration remains stable against radial perturbations if the total mass increases monotonically with the central density, i.e. $dM/d\rho_c > 0$. Physically, this implies that small compressions of the stellar core lead to an increase in mass, thereby preventing runaway collapse.

\begin{figure}
     \centering
\includegraphics[height=7cm,width=9.8cm]{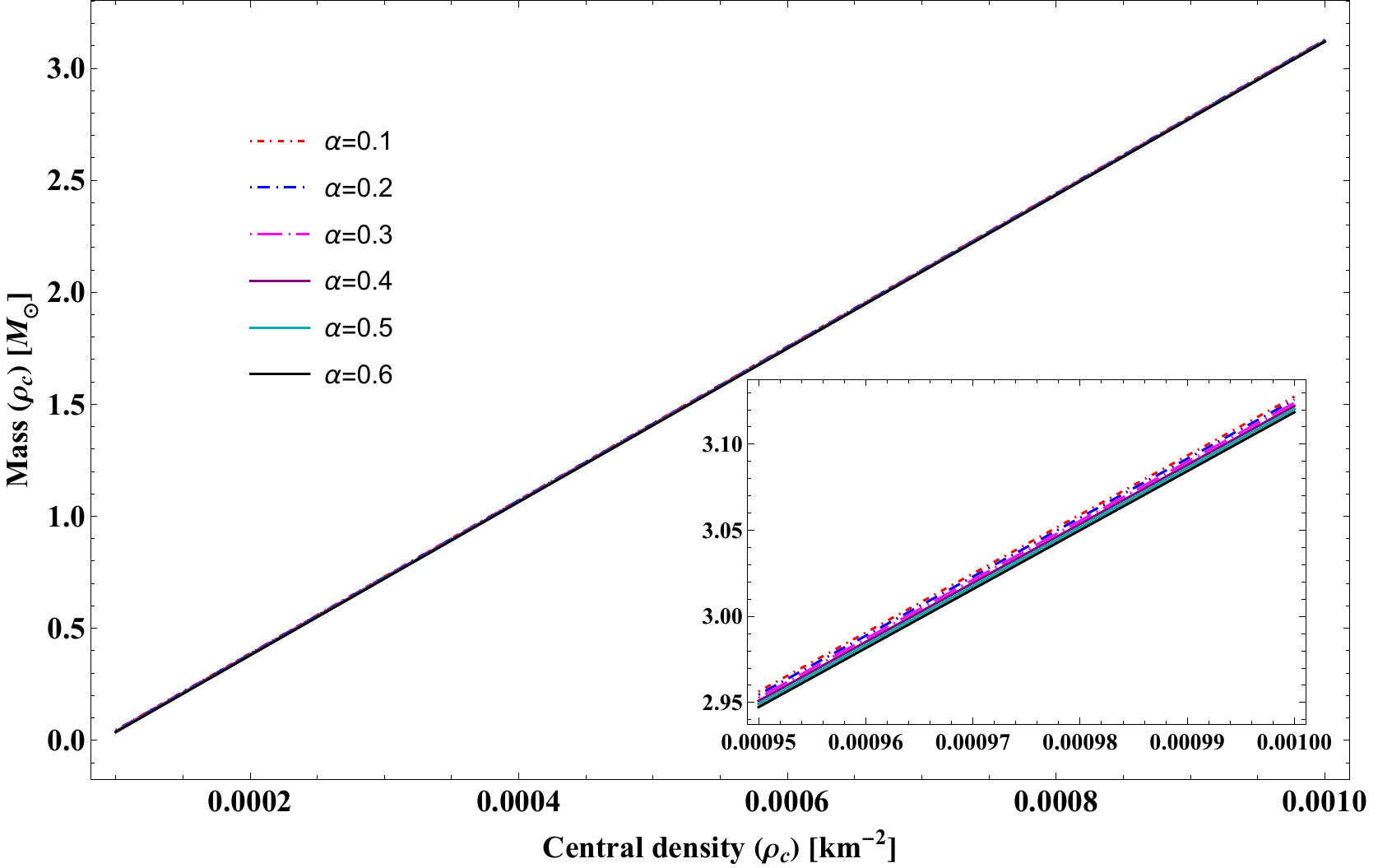}
    \caption{Graphical analysis of mass ($M_{\odot}$) with respect to central density $\rho_c$ for $C = 0.288~ \text{km}^{-2}; D = 0.1; A = 0.009~ \text{km}^{-2}; B = 0.000009~\text{km}^{-4}; L = 0.0009 ~\text{km}^{-2}; N = 0.0009~ \text{km}^{-2}$.}
    \label{fig61}
\end{figure}

From our analysis, we find that the derivative of the total mass with respect to the central density remains strictly positive throughout the parameter space considered. This condition is satisfied because of the positivity of the metric parameters involved, ensuring that the stellar mass grows linearly with increasing central density. The corresponding mass--central density profile, depicted in Fig.~\ref{fig61}, confirms this monotonic behavior, with the mass remaining finite and positive throughout the range.

Furthermore, it is observed that the decoupling parameter associated with the dark matter sector has only a marginal influence on the mass--density relation, indicating that the global stability of the configuration is primarily governed by the core matter distribution rather than the strength of the gravitational decoupling.
Hence, the present anisotropic dark star model fully satisfies the HZN criterion, reinforcing its stability against radial perturbations and confirming its physical viability.

\section{Measurement of maximum mass and corresponding radii via $M$--$R$ curves}\label{secvi}

\begin{figure*} \centering \includegraphics[height=6.2cm,width=7.6cm]{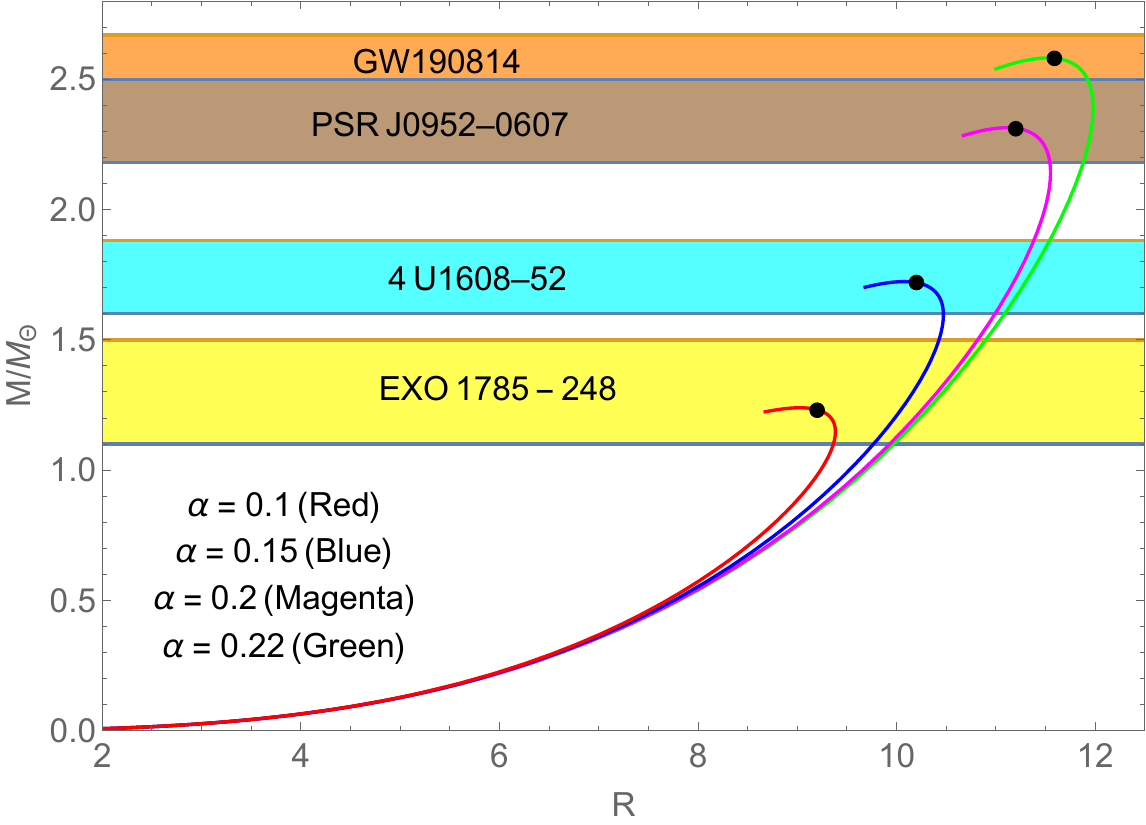}~~ \includegraphics[height=6.2cm,width=7.6cm]{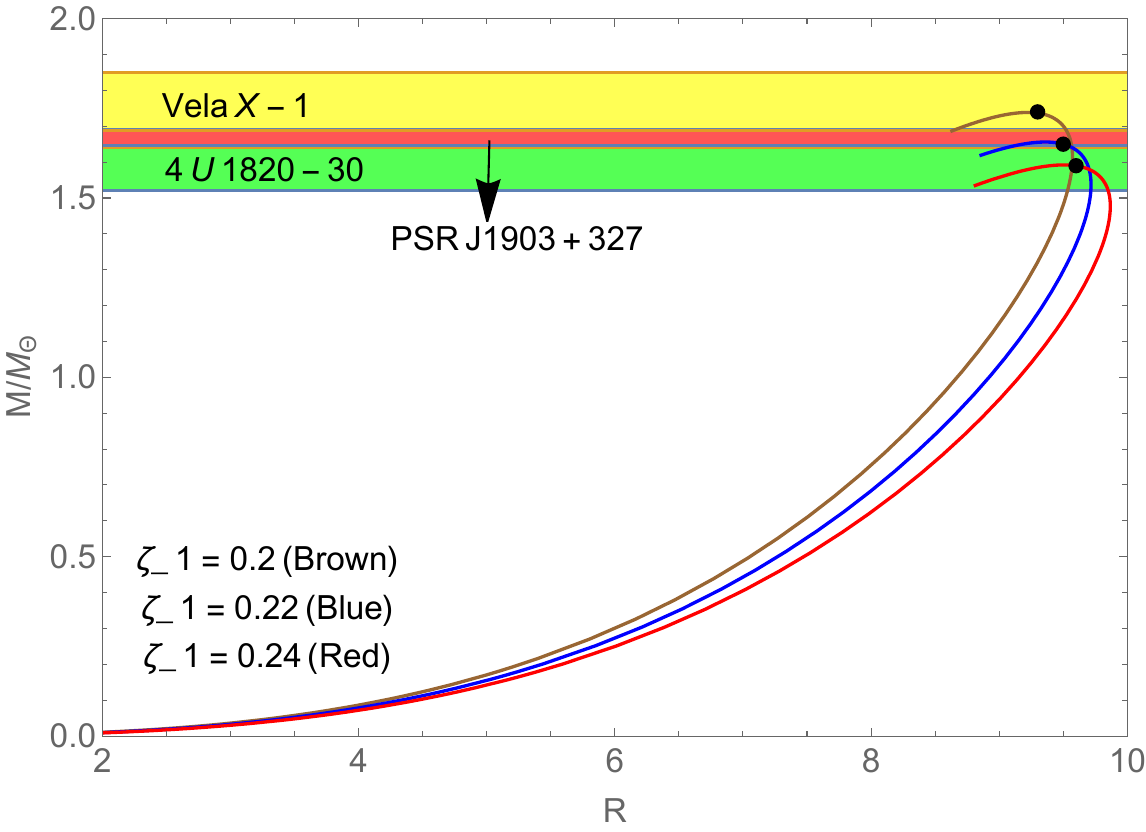} \caption{Prediction of radii of some well-known compact objects from our model for different values of the decoupling parameter $\alpha$ (left) and model parameter $\zeta_1$ (right). \label{mr}}
\end{figure*}

To examine the astrophysical viability of our model, we construct $M$--$R$ relations by varying the decoupling parameter $\alpha$ and the model parameter $\zeta_1$, as shown in Fig.~\ref{mr}. These curves help us to predict the radii of several observed compact objects and assess the maximum mass supported by the system.

In the left panel of Fig.~\ref{mr}, the $M$--$R$ relation is obtained for different values of the decoupling parameter $\alpha$. The predicted stellar radii lie in the range $9.2$--$11.59$ km, whereas the corresponding maximum masses span $(1.23$--$2.58)\,M_{\odot}$, depending on the chosen model parameters $\zeta_1$, $\zeta_2$, $A$, $B$, $N$, and $L$. Notably, increasing $\alpha$ shifts the curves toward higher masses and larger radii, indicating that the presence of dark matter stiffens the effective EoS and enables the star to sustain higher masses. For $\alpha=0.10$, $0.15$, $0.20$, and $0.22$, the model successfully accommodates the observed masses of EXO~1745$-$248, 4U~1608$-$52, PSR~J0952$-$0607, and the lighter component of the GW190814 event, respectively. The maximum mass attained in our model is $2.58\,M_{\odot}$ with a corresponding radius of $11.59$ km for $\alpha=0.22$, placing it within the lower edge of the observed mass gap and making it a viable candidate for objects similar to GW190814. A similar enhancement of maximum mass due to additional gravitational sources has also been reported in other modified gravity contexts \cite{Bhar:2023xku}.

The right panel of Fig.~\ref{mr} illustrates the influence of the model parameter $\zeta_1$ on the $M$--$R$ relation. For increasing values of $\zeta_1$, the maximum supported mass decreases while the stellar radius increases, indicating a less compact configuration. In this case, the curves encompass compact stars such as Vela~X$-$1, 4U~1820$-$30, and PSR~J1903$+$0327 for $\zeta_1=0.20$, $0.22$, and $0.24$, respectively. The predicted radii corresponding to these sources are summarized in table~\ref{tb2}.

Overall, the $M$--$R$ analysis demonstrates that the CGD-based teleparallel dark star model is capable of reproducing the observed properties of both canonical and massive compact stars. The decoupling parameter $\alpha$ plays a central role in regulating the dark matter contribution, thereby controlling the maximum mass and compactness of the stellar configuration.

\begin{table*}[t]
\centering
\caption{The prediction of radii for some well-known compact stars for different values of $\alpha$. }\label{tb1}
\begin{tabular}{@{}ccccccccccccc@{}}
\hline
$\alpha$& Observed mass && Predicted radius  && Matched with the mass of \\
& $M(M_{\odot})$ && (in km.) \\
\hline
0.10& 1.23 &&       9.2 && EXO 1745-248 \cite{Ozel:2008kb} \\
0.15& 1.72 &&       10.2 && 4U 1608-52 \cite{Rawls:2011jw} \\
0.20& 2.31 &&     11.2 && PSR J0952-0607 \cite{Romani:2022jhd}\\
0.22& 2.58 &&     11.59 && lighter component of \\
&&&&&GW 190814 event \cite{abott2}\\
\hline
\end{tabular}
\end{table*}

\begin{table*}[t]
\centering
\caption{The prediction of radii for some well-known compact stars for different values of $\zeta_1$. }\label{tb2}
\begin{tabular}{@{}ccccccccccccc@{}}
\hline
$\zeta_1$& Observed mass && Predicted radius  && Matched with the mass of \\
& $M(M_{\odot})$ && (in km.) \\
\hline
0.20& 1.74 &&       9.3 && Vela X-1 \cite{Rawls:2011jw} \\
0.22& 1.65 &&       9.5 && 4U 1820-30 \cite{Guver:2010td} \\
0.24& 1.59 &&     9.6 && PSR J1903+327 \cite{Freire:2010tf}\\
\hline
\end{tabular}
\end{table*}

\section{Mass measurement via equi-mass planes}\label{secvii}
\begin{figure*}
    \centering
    \includegraphics[height=5.2cm,width=5.2cm]{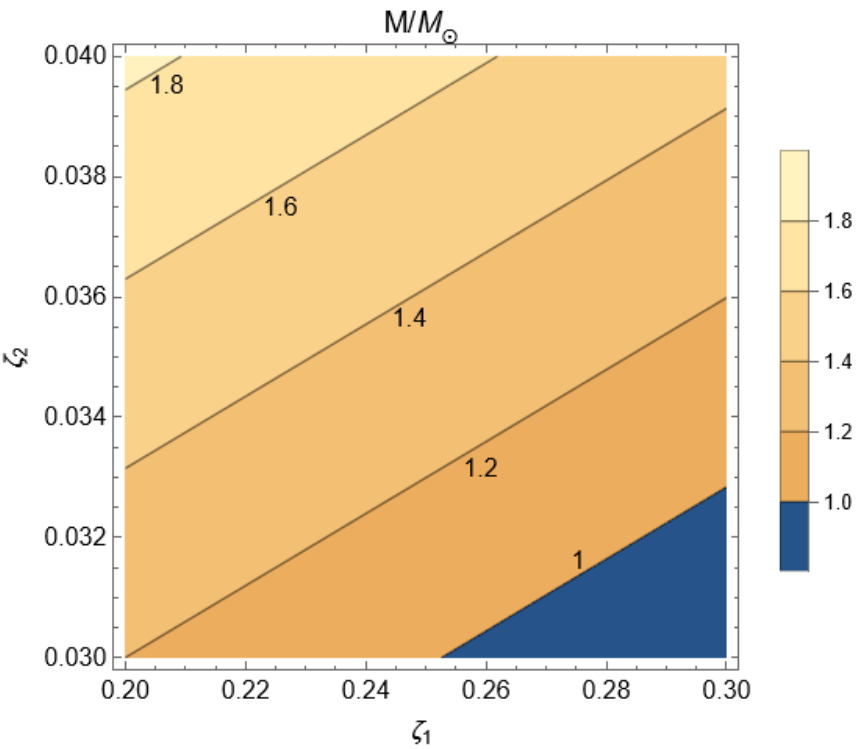}
    \includegraphics[height=5.2cm,width=5.2cm]{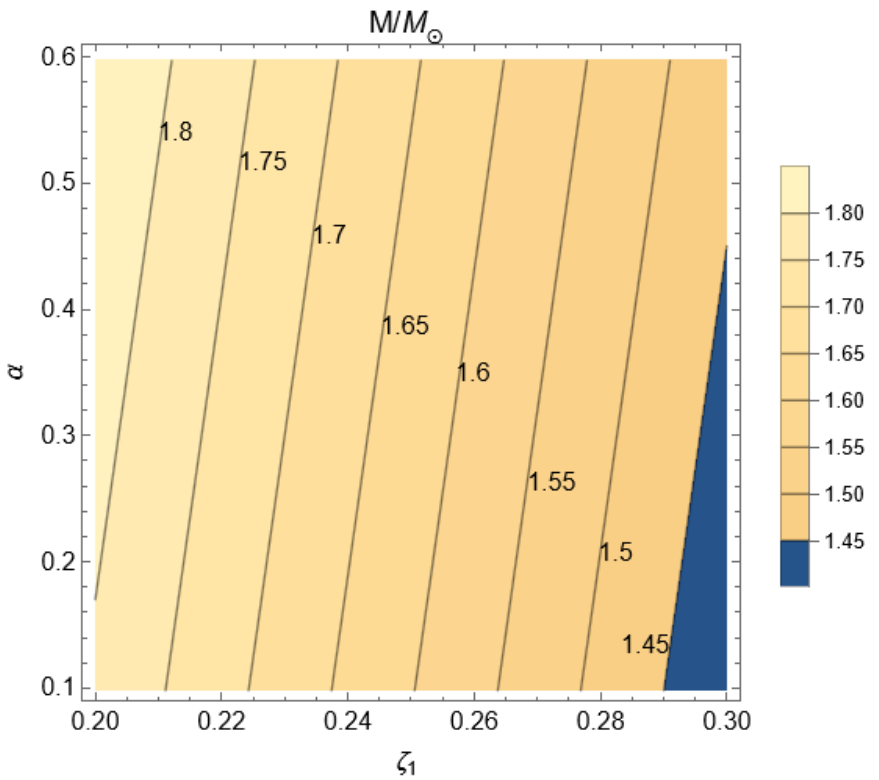}
    \includegraphics[height=5.2cm,width=5.2cm]{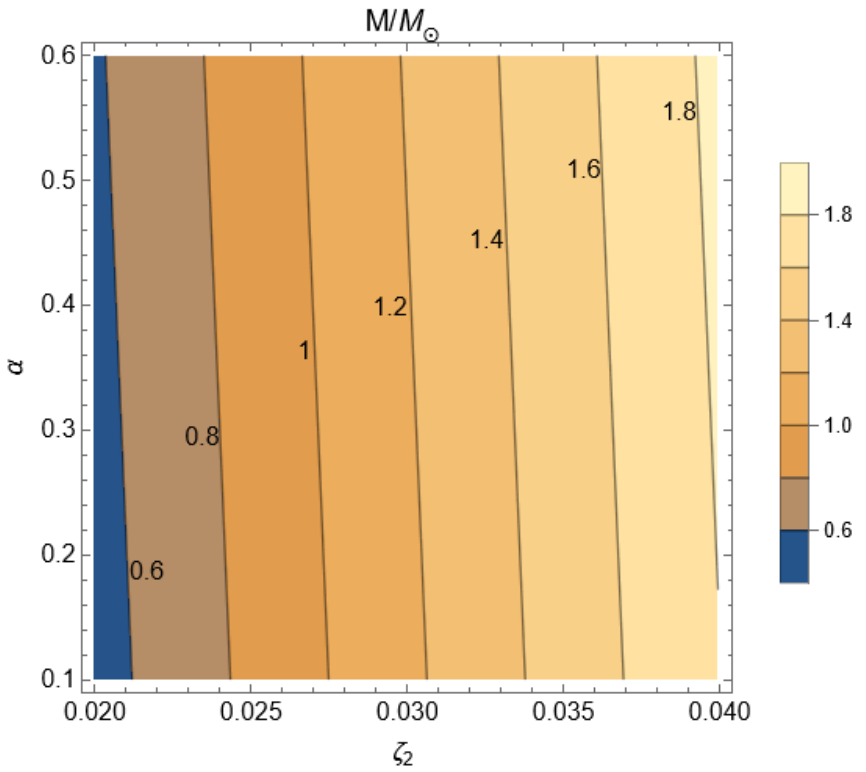}
    \caption{(Left panel) Equi-mass contour in $\zeta_1-\zeta_2$ plane, (middle panel) in $\zeta_1-\alpha$ plane, and (right panel) in $\zeta_2-\alpha$ plane are shown. }
    \label{con}
\end{figure*}

In this section, a couple of contour plots are drawn for in-depth analysis of the mass of our current model. 
We have displayed the equi-mass contours in the $\zeta_1-\zeta_2$ and $\zeta_1-\alpha$ planes in the left and middle panels, respectively, of Fig.~\ref{con}. The left panel of Fig.~\ref{con} illustrates this point: the maximum mass of the star decreases with increasing $\zeta_1$, whereas maximum mass grows with increasing $\zeta_2$.
On the other hand, one can notice an increase in the mass of the compact star with the growing value of $\alpha$ in the $\zeta_1-\alpha$ plane, as shown in the middle panel of Fig.~\ref{con}. In contrast, this profile indicates that the mass reduces as $\zeta_1$ increases.\\
The equi-mass contour on the $\zeta_2-\alpha$ plane is shown in the right panel of Fig.~\ref{con}. The profile shows that the mass increases as $\zeta_2$ and the decoupling parameter $\alpha$ increases. So, a higher values of mass are achievable with higher values of $\zeta_2$ and a high decoupling parameter $\alpha$ as confirmed by the graphical analysis.

\section{Energy exchange between fluid distributions under dark matter}\label{secviii}

Understanding the interaction between multiple matter sources is crucial for modeling realistic compact stellar interiors. In the present work, the stellar configuration consists of two gravitationally coupled sources: the ordinary anisotropic fluid described by $\mathcal{T}_{\mu\nu}$ and an additional dark matter sector represented by $\Theta_{\mu\nu}$. Although these sources together form the effective energy-momentum tensor $\mathcal{T}_{\mu\nu}^{\text{tot}}$, their mutual interaction can be explicitly examined through the contracted Bianchi identities,
$\nabla_{\mu}\mathcal{T}^{\mu\nu}+\nabla_{\mu}\Theta^{\mu\nu}=0.$
Two physically relevant scenarios arise from this relation. In the first case, both sources are independently conserved, i.e.
$\nabla_{\mu}\mathcal{T}^{\mu\nu}=0, ~ \nabla_{\mu}\Theta^{\mu\nu}=0,$
implying no energy transfer between them. The second and more realistic scenario allows for energy exchange between the two sectors, which is mathematically expressed as
$\nabla_{\mu}\mathcal{T}^{\mu\nu}=-\nabla_{\mu}\Theta^{\mu\nu}.$
Following earlier studies \cite{enexc1,cf1}, the energy exchanged between the two sources is quantified by the scalar $\delta E$, defined as
\begin{equation}
\delta E=\frac{\Phi'(r)}{2}\,(\rho+p_r),
\end{equation}
where $\Phi(r)$ is the temporal deformation function. Since the energy density $\rho$ and the radial pressure $p_r$ are positive throughout the stellar interior, the sign of $\delta E$ is entirely governed by $\Phi'(r)$. A positive $\Phi'(r)$ corresponds to $\delta E>0$, indicating a transfer of energy from the dark matter sector to the stellar fluid, while $\Phi'(r)<0$ implies the reverse process.
 Using seed solutions and the explicit form of $\Phi(r)$, the energy exchange term for our model can be written as
\begin{eqnarray}
\delta E = \frac{\Phi'}{2}\Big[\zeta_1\big(r^2(2AC+B)+A+2BCr^4+2C\big)-3A\zeta_1-5B\zeta_1 r^2\Big].
\end{eqnarray}

\begin{figure*} \centering \includegraphics[height=6.2cm,width=7.5cm]{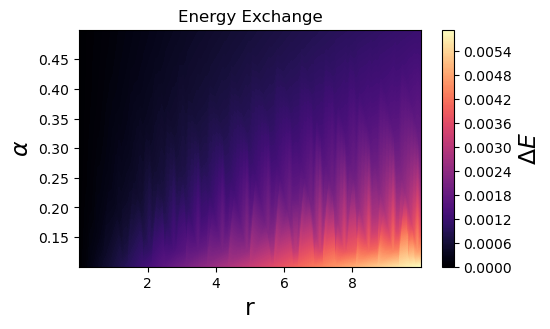}~~ \includegraphics[height=6.2cm,width=7.5cm]{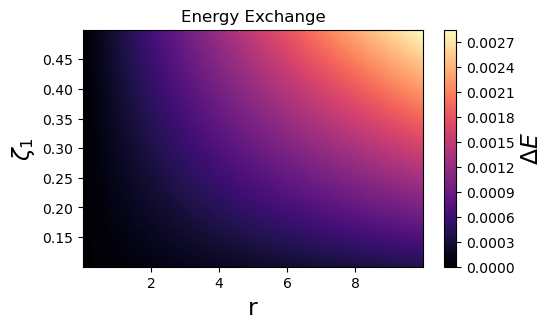} \caption{Energy exchange between the fluid and dark matter in the $r-\alpha$ plane (left) and $r-\zeta_1$ plane (right) for $C = 0.288~ \text{km}^{-2}; D = 0.1; A = 0.009~ \text{km}^{-2}; B = 0.000009~\text{km}^{-4}; L = 0.0009 ~\text{km}^{-2}; N = 0.0009~ \text{km}^{-2}$.} \label{enexpic} \end{figure*}

The radial behavior of $\delta E$ is shown in Fig.~\ref{enexpic} for the $r$--$\alpha$ and $r$--$\zeta_1$ planes. For fixed values of the decoupling parameter $\alpha$, the energy exchange remains positive across the stellar interior but exhibits a non-monotonic, oscillatory profile, reflecting the complex interplay between the fluid and dark matter components. In contrast, for a fixed $\zeta_1$, the energy exchange increases monotonically with radius. Although $\zeta_1$ has a negligible influence near the stellar center, its effect becomes increasingly significant toward the surface. In both cases, the energy transfer attains its maximum value near the stellar boundary, indicating a stronger interaction between the two sectors in the outer layers of the star.

\section{Conclusion}\label{secx}

In this chapter, we constructed an anisotropic strange star model within the framework of teleparallel gravity by employing the gravitational decoupling technique for the first time in its complete form. By adopting the CGD approach under the condition of a vanishing complexity factor, we derived an exact and physically viable solution that describes a strange deformed star influenced by dark matter. The seed configuration was obtained using the Tolman–Kuchowicz metric potentials, while the spacetime deformation was driven by an additional gravitational source associated with a cold dark matter halo. The decoupling parameter $\alpha$ controls the strength of this deformation and effectively encodes the dark matter contribution.

Our analysis shows that the inclusion of the $\Theta$-sector through CGD leads to denser and more massive stellar configurations without violating physical acceptability conditions. The energy density and pressures remain finite, positive, and monotonically decreasing from the center to the surface, whereas all standard energy conditions (NEC, WEC, DEC, and SEC) are satisfied throughout the stellar interior. The gradients of the physical variables disappear in the center, ensuring the regularity of the solution.

The stability of the constructed dark star model was thoroughly examined using multiple independent criteria. The relativistic adiabatic index remains above the critical value $4/3$ everywhere inside the star, confirming the hydrostatic equilibrium. Herrera’s cracking condition is satisfied for moderate values of the decoupling parameter, indicating stability against local anisotropic perturbations, while very large values of $\alpha$ may induce instability. Furthermore, the HZN criterion is fulfilled as the total mass increases monotonically with the central density, ensuring global stability under radial perturbations.

From an astrophysical perspective, the predicted mass–radius relations demonstrate that the model successfully accommodates a wide range of observed compact objects, including massive neutron stars such as PSR J0952$-$0607 and the lighter component of GW190814. An increase in the decoupling parameter $\alpha$ enhances the maximum mass, indicating that dark matter can play a crucial role in supporting heavier compact stars. The analysis of energy exchange between the perfect fluid and dark matter reveals a non-trivial interaction that becomes increasingly significant toward the stellar surface.


Overall, the present study highlights the effectiveness of the CGD framework combined with vanishing complexity in teleparallel gravity for modeling anisotropic dark stars. Our results provide deeper insight into the interplay between anisotropy, dark matter, and gravitational geometry, and establish a robust theoretical foundation for exploring compact objects beyond GR.

So far, we have illustrated how the mass and other physical properties of compact stars can vary significantly with arbitrary choices of model parameters. While this demonstrates the flexibility of the framework, such unconstrained tuning lacks predictive power and physical reliability. To establish the robustness and observational viability of the model, it is therefore crucial to perform a systematic statistical analysis that constrains the parameters using current astrophysical data. The next chapter is devoted to this comprehensive investigation.

\chapter{\fontsize{14}{16}\selectfont Bayesian Inference of Neutron Star Properties in $f(Q)$ Gravity Using NICER Observations} 

\label{Chapter5} 

\definecolor{maroon}{RGB}{128, 0, 0}
\lhead{\textcolor{maroon}{\textit{\textbf{Chapter 5:}}} \emph{\textcolor{maroon}{Bayesian Inference of Neutron Star Properties in $f(Q)$ Gravity Using NICER Observations}}} 

\blfootnote{*The work in this chapter is covered by the following submission:\\
\textit{Bayesian Inference of Neutron Star
Properties in $f(Q)$ Gravity Using NICER Observations}, arXiv:2601.16227.}


In this chapter, we investigate neutron stars (NSs) in the strong-field regime within the framework of $f(Q)$ gravity. Our aim is to explore how modified gravity affects neutron star structure using a Bayesian approach.  
The main aspects of this work are summarized below:
\begin{itemize}

    \item We study three representative $f(Q)$ gravity models, namely linear, logarithmic, and exponential forms, focusing on their impact on NS properties.

    \item To isolate the effects of modified gravity, we fix the dense matter EoS to the DDME2 model.

    \item We perform a Bayesian inference analysis by investigating theoretical mass--radius predictions with NICER observations of PSR J0030+0451, PSR J0740+6620, PSR J0437+4715 and PSR J0614+3329.

    \item Then, we compare competing gravity models using Bayes factor analysis and find that the exponential $f(Q)$ model is statistically preferred over linear and logarithmic cases.

    \item For the favored exponential model, we obtain well-constrained neutron star properties, including radius and tidal deformability at $1.4\,M_\odot$, consistent with current observational bounds.

    \item Our results demonstrate that NSs provide powerful probes of symmetric teleparallel gravity in the strong-field regime.

\end{itemize}

\section{Introduction}\label{sec:1}


From a theoretical perspective, a wide variety of modified gravity theories have been proposed and investigated in the context of compact stars. These include curvature-based extensions such as $f(R)$ gravity \cite{defelice2010,sotiriou2010,Capozziello:2011et,Nojiri:2011}, torsion-based formulations such as $f(T)$ gravity \cite{Vilhena_2023}, and other generalized frameworks. Among these, STEGR offers a geometrically distinct formulation in which both curvature and torsion vanish, and gravity is entirely encoded in spacetime non-metricity \cite{Aldrovandi:2013,Zhao:2021zab}. 
Several recent studies have explored compact stars in $f(Q)$ gravity and related non-Riemannian theories, demonstrating that non-metricity-induced corrections can substantially affect stellar equilibrium, maximum mass, radius, and stability properties \cite{wang/2022,log,gaurav}. These works indicate that departures from GR may either stiffen or soften stellar configurations depending on the functional form of $f(Q)$ and the choice of model parameters. However, in most existing analyzes, the parameters of the gravity sector are selected phenomenologically, often without direct comparison with observational data, leaving the astrophysical viability of such models largely unconstrained.

On the observational front, the last decade has witnessed remarkable progress in neutron-star astrophysics. Precise mass measurements of heavy pulsars with masses close to or exceeding $2M_{\odot}$ have imposed stringent constraints on the EoS \cite{demorest2010,antoniadis2013,Cromartie:2019,Fonseca:2021}. In parallel, the \textit{Neutron Star Interior Composition Explorer} (NICER) has provided the first simultaneous mass--radius measurements of millisecond pulsars through relativistic X-ray pulse-profile modeling, notably for PSR J0030+0451 and PSR J0740+6620 \cite{Gendreau:2017,Miller:2021,Riley:2021,Karan_2025}. Complementary constraints arise from gravitational-wave observations of binary neutron-star mergers, particularly GW170817, which place strong upper bounds on tidal deformability and neutron-star radii \cite{Abbott:2018}. Together, these multi-messenger observations now offer unprecedented opportunities to test gravity theories in the strong-field regime.

Motivated by these advances, there has been growing interest in investigating modified gravity models with compact-star observations using statistically robust methods. Bayesian inference has emerged as a powerful framework for combining theoretical models with observational data, allowing simultaneous constraints on dense-matter and gravity-sector parameters \cite{Raaijmakers:2021,Essick:2020,Jiang_2023,PhysRevLett.121.062701,article}. Although such analyzes have been successfully applied to curvature-based theories, including $f(R)$ gravity \cite{2023PhRvD.107l4045N,PhysRevD.109.064048}, observationally driven constraints on symmetric teleparallel gravity remain relatively unexplored.

The primary aim of this chapter is to address this gap by systematically investigating neutron stars within the framework of $f(Q)$ gravity from both theoretical and observational perspectives. We consider three representative models such as linear, logarithmic, and exponential forms of $f(Q)$ that capture a broad class of deviations from GR \cite{wang/2022,log,gaurav}. To minimize uncertainties associated with dense-matter physics, we adopt the realistic DDME2 EoS, which is consistent with nuclear experiments and astrophysical constraints \cite{Xia:2022,Lalazissis:2005de}.
Most importantly, for the first time in the literature on $f(Q)$ gravity, we employ a Bayesian inference framework to constrain the free parameters of the theory using neutron-star observations from NICER. By investigating theoretical mass--radius relations and internal stellar properties with observational data, we aim to assess the astrophysical viability of different $f(Q)$ models in the strong-field regime. This approach moves beyond phenomenological parameter choices and provides statistically meaningful bounds on deviations from GR, highlighting the role of compact stars as sensitive probes of symmetric teleparallel gravity beyond cosmological scales.

The organization of this chapter is as follows: Modified TOV equations governing the neutron-star structure in $f(Q)$ gravity are derived in section~\ref{sec:3}. In section~\ref{sec:4}, we introduce the three representative $f(Q)$ models considered in this study. The numerical methodology for solving the modified TOV equations and the Bayesian inference framework used for parameter estimation are presented in sections.~\ref{sec:5} and \ref{sec:6}, respectively. Our results, including a comparative analysis of the different models, are discussed in section~\ref{sec:7}. Finally, we summarize our findings and present concluding remarks in section~\ref{sec:9}.
\section{Modified TOV in $f(Q)$ gravity}\label{sec:3}
The modified field equations corresponding to the isotropic fluid with components of the energy momentum tensor ($\mathcal T_{\mu \nu}$) as ($\varepsilon$, $p$, $p$, $p$) and the static spherically symmetric metric, the independent components of the field equations can be derived as
\begin{eqnarray}
\label{Eq: eom-dens}
&&\hspace{-0.7cm}\varepsilon = \frac{f}{2} - f_Q\left( Q+\frac{1}{r^2}+\frac{e^{-\lambda}}{r}(\nu^{\prime}+\lambda^{\prime})\right)
\,, \\
\label{Eq: eom-presr}
&&\hspace{-0.7cm}p =- \frac{f}{2} + f_Q\left(Q+\frac{1}{r^2}\right)
\, , \\
\label{Eq: eom-prest}
&&\hspace{-0.7cm}p = -\frac{f}{2} + f_Q\Big\{ \frac{Q}{2}-e^{-\lambda}\big[\frac{\nu^{\prime\prime}}{2}+\big(\frac{\nu^{\prime}}{4}+\frac{1}{2r}\big) (\nu^{\prime}-\lambda^{\prime})\big]\Big\},~~~~~~~ \\
\label{Eq: eom-impos}
&&\hspace{-0.6cm}\frac{\cot\theta}{2}Q^{\prime}f_{QQ}=0
\,.
\end{eqnarray}
Using the static spherically symmetric line element in the context of symmetric teleparallel gravity, the non-metricity scalar \( Q \) becomes a function of the radial coordinate and is given by  
\begin{align}
\label{Eq: spherical-nms}
Q(r) = -\frac{2e^{-\lambda}}{r}\left(\nu^{\prime}+\frac{1}{r}\right)
\,,
\end{align}
where the prime denotes a derivative with respect to \( r \). Upon substituting the expression for \( Q \) from Eq.~(\ref{Eq: spherical-nms}) into the equations of motion Eq.~(\ref{Eq: eom-dens}) and Eq.~(\ref{Eq: eom-presr}), the modified version of the TOV equation in \( f(Q) \) gravity takes the form of
\begin{eqnarray}
&& p^{\prime}+\frac{\nu^{\prime}}{2}\left(\varepsilon+p\right) = f_{Q}\frac{e^{-\lambda}}{2r} \left[ \frac{2}{r}\left(\nu^{\prime}+\lambda^{\prime}\right) - \nu^{\prime}\left(\nu^{\prime}-\lambda^{\prime}\right) \right. \left. + \frac{4}{r^{2}}\left(1-e^{\lambda}\right) - 2\nu^{\prime\prime} \right] = 0.
\end{eqnarray}
By solving Eqs.~(\ref{Eq: eom-dens})--(\ref{Eq: eom-prest}) for a specific functional form of
$f(Q)$, together with appropriate boundary conditions, the metric functions
$\nu(r)$ and $\lambda(r)$ can be determined. As an illustrative case, for vacuum
solutions with $\mathcal{T}_{\mu\nu}=0$, Eqs.~(\ref{Eq: eom-dens}) and (\ref{Eq: eom-presr})
imply the relation
\begin{equation}
\nu'(r) + \lambda'(r) = 0 .
\end{equation}
In general, Eqs.~(\ref{Eq: eom-dens})--(\ref{Eq: eom-prest}) can be reformulated as
a system of modified TOV equations in the framework of $f(Q)$ gravity. These equations describe the internal structure of neutron
stars and are supplemented by the continuity equation arising from the
conservation of the energy-momentum tensor $\mathcal{T}_{\mu\nu}$, given by
\begin{eqnarray}
&&\hspace{0cm}p' + \frac{\nu'}{2}\,(\varepsilon + p) = 0, \\
&&\hspace{0cm}\lambda' = \frac{-k r (\rho + p)e^{\lambda} - \nu'}{f(Q)}, \\
&&\hspace{0cm}\nu'' = \frac{1}{2r f_Q}
\Big[
2rQf_Q e^{\lambda}
- (r\nu' + 2)(\nu' - \lambda')f_Q 
- 2\bigl(2kp + f(Q)\bigr) r e^{\lambda}
\Big].
\end{eqnarray}
By specifying an EoS and providing suitable initial conditions
for the metric functions $\nu(r)$ and $\lambda(r)$, as well as for the energy
density $\varepsilon$ and pressure $p$, the stellar structure of the NSs can be fully determined in
$f(Q)$ gravity. In the particular limit $f(Q)=-Q$,
Eqs.~(\ref{Eq: eom-dens})--(\ref{Eq: eom-prest}) reduce to the standard field
equations of GR. Therefore, these equations allow for the computation of NS
configurations for various $f(Q)$ gravity models.
\section{$f(Q)$ models}\label{sec:4}
In this work, we investigate the internal structure of NSs in the context of 
$f(Q)$ gravity by considering three representative functional forms, namely linear, logarithmic, and exponential models. These formulations extend the standard gravitational action through both linear and nonlinear dependencies on the non-metricity scalar 
$Q$, leading to modified field equations that govern stellar equilibrium. A systematic examination of these models allows us to assess how deviations from GR influence the macroscopic characteristics of NSs.
\subsection{Linear model}
To investigate the role of different functional forms of $f(Q)$, we first focus on the linear model \citep{wang/2022},
\begin{eqnarray}
f(Q)=Q_0 \bigg[\alpha+\beta \Big( \frac{Q}{Q_0} \Big)  \bigg],
\end{eqnarray}
where, $\alpha,\beta,Q_0$ are the model parameters. The above equat emerges directly from Eq.~(\ref{Eq: eom-impos}) upon imposing the condition $f_{QQ}=0$. This constraint restricts the gravitational action to depend linearly on the nonmetricity scalar $Q$. The choice of the linear form of the model is primarily motivated by the dimensional consistency and its theoretical analogy with GR. In the Einstein-Hilbert action, the gravitational Lagrangian density is proportional to the Ricci scalar $R$. In STEGR, the Ricci scalar is replaced by a function of the non-metricity scalar, $f(Q)$. Therefore, in order to maintain the dimensional consistency of the gravitational action, the function $f(Q)$ must carry the same physical dimension as $R$, namely $[\mathrm{length}]^{-2}$.
In a static spherically symmetric spacetime, the non-metricity scalar $Q$ also has dimension $[\mathrm{length}]^{-2}$. To construct generalized functional forms such as logarithmic and exponential corrections, we introduce a constant scale $Q_0$ with the same dimension as $Q$. This guaranties that the ratio $Q/Q_0$ is dimensionless, ensuring that functions such as $\ln(Q/Q_0)$ or $\exp(Q/Q_0)$ are mathematically well defined.
Consequently, the parameters $\alpha$ and $\beta$ are dimensionless constants that quantify deviations from GR, while $Q_0$ sets the characteristic non-metricity scale of the theory. The adopted linear form is therefore not arbitrary; rather, it is motivated by dimensional consistency, structural coherence with the Einstein-Hilbert framework, and uniformity with the logarithmic and exponential models considered in this work.

The linear $f(Q)$ model constitutes the minimal extension of STG and is exactly reduced to GR for $\alpha=0$ and $\beta = -1$. Owing to the absence of higher-order derivatives, the resulting field equations remain second order, ensuring mathematical consistency and freedom from additional dynamical degrees of freedom. Consequently, this model provides a well-controlled theoretical baseline for isolating leading-order gravitational modifications and serves as a natural reference against which nonlinear $f(Q)$ models can be systematically assessed in the context of NS structure.

\subsection{Logarithmic model}

We now turn to a logarithmic form of the $f(Q)$ function, which introduces scale-dependent corrections to the gravitational action and has been widely studied in \cite{log}. The model is specified as
\begin{equation}
f(Q) = -Q + \alpha \ln\!\left[ \beta \left( \frac{Q}{Q_0} \right) \right],
\end{equation}
where $\alpha$ and $\beta$ are constant parameters controlling the magnitude and scale of the logarithmic correction, and $Q_0$ is a constant with dimensions of $[\mathrm{length}]^{-2}$ introduced to ensure a dimensionless argument of the logarithm. In the limit $\alpha \rightarrow 0$, the logarithmic contribution vanishes and the theory is reduced to the symmetric teleparallel formulation equivalent to GR.

The logarithmic $f(Q)$ model is motivated by its ability to introduce controlled scale-dependent deviations from GR. Logarithmic corrections naturally emerge in effective field theories and quantum-inspired gravity \cite{eft}, encoding running gravitational effects that remain negligible in the weak-field limit while becoming relevant in intermediate and strong-field regimes, such as those encountered in compact objects.
From a cosmological standpoint, logarithmic $f(Q)$ models have been widely applied to study late-time cosmic acceleration, dynamical dark energy, and phase-space stability \cite{log1,gaurav}. These properties provide a compelling motivation to investigate their implications for NS interiors, where strong gravity may amplify scale-dependent effects.

\subsection{Exponential model}

Finally, we consider an exponential form of the $f(Q)$ function, which introduces nonlinear corrections to the gravitational action and has been widely explored in the literature \cite{simu}. The model is defined as
\begin{equation}
f(Q) = -Q + \alpha Q_0 \left[ 1 - \exp\!\left(-\beta \sqrt{\frac{Q}{Q_0}} \right) \right],
\end{equation}
where $\alpha$ and $\beta$ are dimensionless parameters characterizing the strength and scale of deviations from GR, and $Q_0$ is a constant with dimensions of $[\mathrm{length}]^{-2}$ introduced to render the argument of exponential dimensionless. In the limit $\beta \rightarrow 0$, the exponential correction vanishes and the theory is reduced to the symmetric teleparallel formulation equivalent to GR.

The exponential model is physically motivated by its ability to generate smooth and controlled deviations from GR while remaining well behaved in both the weak and strong field regimes. The exponential suppression ensures that nonlinear corrections become significant only above a characteristic scale set by $Q_0$, allowing the theory to recover GR at low curvatures while permitting appreciable modifications in high-density environments. Such properties make this model particularly attractive for studying compact objects, where strong gravitational fields can amplify deviations from standard gravity.

Beyond astrophysical applications, exponential $f(Q)$ models have been extensively employed in cosmological contexts, including inflationary dynamics, constraints from big bang nucleosynthesis, and phase-space analyzes of cosmic evolution \cite{bbn}. These successes motivate their application to NS interiors, where they provide a theoretically consistent framework to probe the impact of nonlinear non-metricity effects on stellar structure.

\section{Boundary conditions and equation of state}\label{sec:5}

For each $f(Q)$ model, we computed the interior and exterior structure of neutron stars by numerically integrating the modified TOV equations. The system consists of coupled ordinary differential equations for the enclosed mass $m(r)$, the metric functions $\lambda(r)$ and $\nu(r)$, and the pressure $p(r)$. These equations are solved using a fourth-order Runge-Kutta (RK4) scheme.
The integration is initiated at the stellar center, where regularity imposes the boundary conditions
\begin{eqnarray}
m(0) = 0, \qquad \lambda(0) = 0, \qquad \nu'(0) = 0 .
\end{eqnarray}
The central pressure $p_c$ is determined from a realistic tabulated EoS.In this work, we adopt the density-dependent relativistic mean-field DDME2 equation of state (EoS)~\cite{Lalazissis:2005de}, which provides a consistent relation between pressure and energy density across the full density range relevant to neutron-star interiors. Among the various EoSs introduced in the introduction, the DDME2 model is preferred here for several reasons: (i) it is derived from a microscopic, nuclear-physics-based framework in which nucleons interact through density-dependent meson exchanges, with parameters calibrated to a wide range of nuclear data, in contrast to phenomenological EoSs (polytropic, MIT Bag, linear); (ii) it is consistent with current multi-messenger observational constraints, including the maximum-mass bound $M_{\text{max}} \gtrsim 2.07\,M_{\odot}$ from massive pulsars~\cite{Fonseca:2021}. These features make DDME2 a robust and physically well-motivated choice for studying neutron-star structure within the $f(Q)$ gravity framework. The central energy density $\varepsilon_c$ is varied in the range $100$--$1700~\mathrm{MeV\,fm^{-3}}$, corresponding to several times the nuclear saturation density.
For a given $\varepsilon_c$, the modified TOV equations are integrated radially outward with a fixed step size $\Delta r = 0.01~\mathrm{km}$ until the pressure vanishes $p(R)=0$, which defines the stellar surface at radius $R$. The total gravitational mass is then obtained as $M = m(R)$. To ensure a physically consistent global solution, the central value of the metric function $\nu(r_0)$ is determined using a shooting method: the trial value of $\nu(r_0)$ is iteratively adjusted until the interior solution aligns smoothly with the exterior Schwarzschild spacetime on the surface, satisfying $e^{\nu(R)} = e^{-\lambda(R)}$. Repeating this procedure over the full range of central energy densities yields the mass--radius relation for each $f(Q)$ model. This allows us to systematically assess the impact of non-metricity-induced corrections on neutron-star structure and to compare the resulting configurations with observational constraints.

\section{Bayesian estimation}\label{sec:6}
A Bayesian approach allows one to perform a detailed statistical analysis of the parameters of a model for a given set of data. This approach is based on \textit{Bayes' theorem}. Combining prior knowledge with fit data produces the \textit{posterior probability} distribution for model parameters. This posterior distribution effectively quantifies the probability of every hypothesis of a possible parameter. From this we can study individual parameter distributions and examine the correlations between them. For a given dataset $\mathcal{D}$ and a hypothesis $H(\theta)$, \textit{Bayes' theorem} is expressed as
\[
P(\theta | \mathcal{D}, H) = \frac{{\mathcal{L}(\mathcal{D} | \theta, H) P(\theta | H)}}{P(\mathcal{D} | H)}.
\]
The components of \textit{Bayes' theorem} are defined as follows:
\begin{itemize}
    \item \textbf{Posterior probability \( P(\theta \mid \mathcal{D}, H) \): }The updated probability of the parameters \(\theta\) after observing the data \(\mathcal{D}\) within the model \(H\).
    \item \textbf{Likelihood \( \mathcal{L}(\mathcal{D} \mid \theta, H) \):} The probability of observing the data \(\mathcal{D}\) given the values of the specific parameters \(\theta\) according to hypothesis \(H\).
    \item \textbf{Prior probability \( P(\theta \mid H)\):} The initial probability assigned to \(\theta\) before seeing the data, based on existing knowledge or assumptions.
    \item \textbf{Marginal likelihood (evidence) \( P(\mathcal{D} \mid H) \):} A normalizing constant representing the probability of the data \(\mathcal{D}\) under \(H\), averaged over all possible values of \(\theta\).
\end{itemize}
\subsection{Likelihood} 
In Bayesian analysis, the likelihood function quantifies the probability of observed data for a given set of model parameters  $\theta$. It can be computed for a given posterior distribution from astrophysical observations. observations. We explain this method below with
the help of an example.

\textbf{X-ray observations (NICER):} 
X-ray observations of NS hot spots by the NICER mission have given some of the strongest constraints on NS masses and radii. The X-ray measurements provide the mass and radius of individual pulsars, both isolated and in binaries. These measurements offer a direct way to study the dense-matter EoS.

The main NICER results come from two key pulsars. For PSR J0030+0451 (at a canonical mass), the mass-radius posteriors were independently published by two groups: Riley { et al.~\cite{riley12} ~~ \href{https://zenodo.org/records/8239000}{https://zenodo.org/records/8239000} and Miller et al.} ~\cite{miller12}~\footnote{~\href{https://zenodo.org/record/3473466}{https://zenodo.org/record/3473466}}. For PSR J0740+6620 (near the maximum mass), similar posteriors were released by Riley {\sl et al.}~\cite{Riley:2021}~\footnote{~\href{https://zenodo.org/records/4697625}{https://zenodo.org/records/4697625}} and Miller {\sl et al.}~\cite{Miller:2021}~\footnote{~\href{https://zenodo.org/records/4670689}{https://zenodo.org/records/4670689}}.  Recently, NICER has provided additional mass–radius measurements for other pulsars such as PSR J0437+4715 by Choudhury {\sl et al.}~\cite{choudhuri}~\footnote{~\href{https://zenodo.org/records/13766753}{https://zenodo.org/records/13766753}} and PSR J0614+3329 by Mauviard {\sl et al.}~\cite{mau}~\footnote{~\href{https://zenodo.org/records/15603406}{https://zenodo.org/records/15603406}}. 

For the NICER mass-radius observations, the likelihood is formally expressed by marginalizing the stellar mass
\begin{eqnarray}
&&\hspace{0cm}\mathcal{L}^{\rm NICER}(\boldsymbol{\theta}_{\rm f(Q)})
\equiv P\!\left(d_{\rm X\text{-}ray}\mid \boldsymbol{\theta}_{\rm f(Q)}\right) \nonumber \\
&&\hspace{0cm}= \int_{M_l}^{M_u} dm \;
P\!\left(m \mid \boldsymbol{\theta}_{\rm f(Q)}\right)
P\!\left(d_{\rm X\text{-}ray} \mid m, R\!\left(m, \boldsymbol{\theta}_{\rm f(Q)}\right)\right).~~~~~~~
\end{eqnarray}

Here, $M_l$ denotes the lower mass bound, taken to be $1\,M_\odot$, and
$M_u(\boldsymbol{\theta}_{\rm f(Q)})$ represents the maximum NS
mass obtained by solving the modified TOV
equations for a given set of $f(Q)$ model parameters
$\boldsymbol{\theta}_{\rm f(Q)}$, assuming the DDME2 EoS.
Suppose there are three statistically independent datasets $A$, $B$, and $C$, with corresponding Bayesian likelihoods
$\mathcal{L}^A(\boldsymbol{\theta})$,
$\mathcal{L}^B(\boldsymbol{\theta})$, and
$\mathcal{L}^C(\boldsymbol{\theta})$,
where $\boldsymbol{\theta}$ denotes the model parameters.
Under the assumption of statistical independence, the joint likelihood
is given by
\begin{equation}
\mathcal{L}(\boldsymbol{\theta})
= \prod_{i=A,B,C} \mathcal{L}^i(\boldsymbol{\theta}) .
\end{equation}

\subsection{Bayes factor}\label{bayesfactor}

The \textit{Bayes factor} (BF) is a Bayesian model-comparison statistic that quantifies the strength of evidence provided by observational data in favor of one hypothesis over another. Given two competing hypotheses, $H_1$ and $H_2$, the Bayes factor is defined as
\begin{equation}
\text{BF}_{12} = \frac{P(D \mid H_1)}{P(D \mid H_2)},
\end{equation}
where $P(D \mid H_1)$ and $P(D \mid H_2)$ denote the marginal likelihoods (evidence) of the data $D$ under hypotheses $H_1$ and $H_2$, respectively. A value $\text{BF}_{12} > 1$ indicates that the data favor $H_1$, while $\text{BF}_{12} < 1$ favors $H_2$.

\subsubsection{Logarithmic Bayes factor}

For practical applications, it is customary to work with the logarithm of the Bayes factor, which provides a more convenient and interpretable scale. The logarithmic Bayes factor is given by
\begin{equation}
\ln(\text{BF}_{12}) = \ln\!\left(\frac{P(D \mid H_1)}{P(D \mid H_2)}\right).
\label{logBayes}
\end{equation}
This form is particularly useful in numerical analyzes and allows for a straightforward assessment of the relative support for competing hypotheses.

\subsubsection{Interpretation of the Bayes factor}\label{IntpBF}

The strength of evidence associated with a given value of the logarithmic Bayes factor is commonly interpreted using established empirical scales \cite{Kass01061995,Jarosz2014}. In particular,
\begin{itemize}
    \item $\ln(\text{BF}) > 1$: strong evidence in favor of hypothesis $H_1$,
    \item $0 < \ln(\text{BF}) < 1$: weak to moderate evidence in favor of $H_1$,
    \item $\ln(\text{BF}) = 0$: no preference between the two hypotheses,
    \item $-1 < \ln(\text{BF}) < 0$: weak to moderate evidence in favor of $H_2$,
    \item $\ln(\text{BF}) < -1$: strong evidence in favor of hypothesis $H_2$.
\end{itemize}
This criterion provides a statistically meaningful framework for model comparison, enabling a quantitative assessment of competing theoretical descriptions based on observational data.
\section{Results}\label{sec:7}
\subsection{Priors}
\begin{figure*}[!ht]
    \centering
    \includegraphics[height=5.8cm,width=\linewidth]{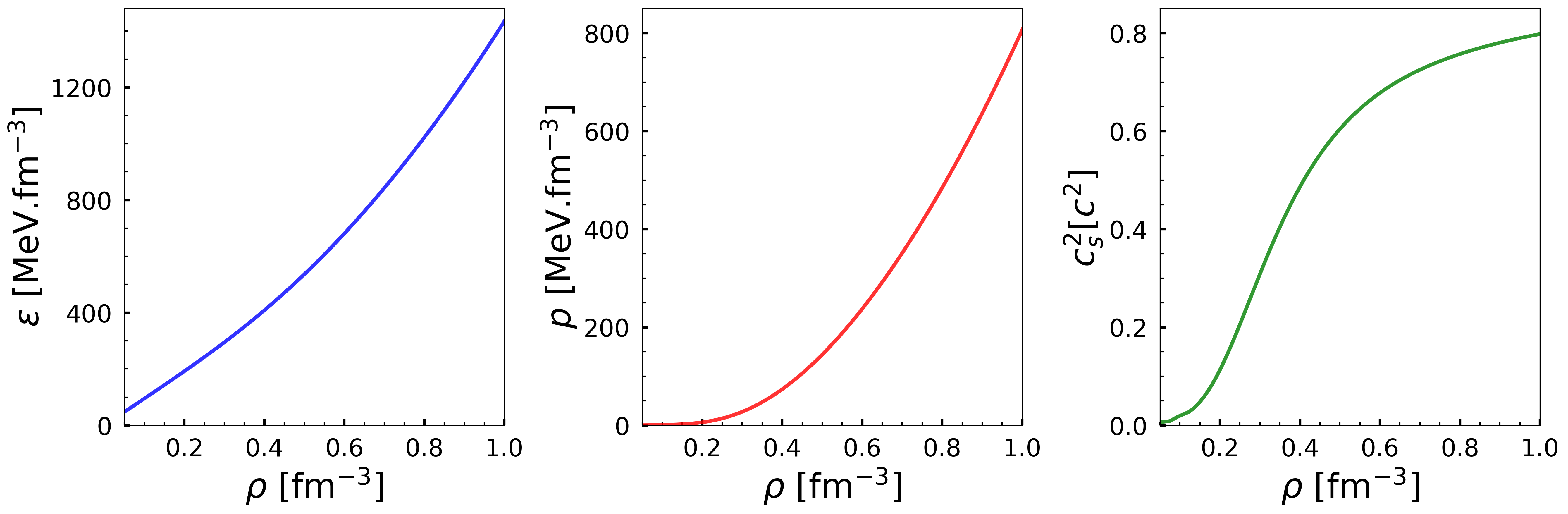}
    \caption{The energy density ($\varepsilon$) (in ${\rm MeV ~
fm^{-3}}$)(left panel), pressure ($p$) (in MeV.fm$^{-3}$) (middle panel), and speed of sound ($c_s^2$) (in $c^2$) (right panel) as a function of baryon density ($\rho$) (in fm$^{-3}$) for DDME2.}
    \label{fig1}
\end{figure*}

We investigate the impact of observational data on NS properties mass, radius, tidal deformability, and compactness using a set of $f(Q)$ models: the linear, logarithmic, and exponential forms detailed in Sec. \ref{sec:4}. These models are constrained within a Bayesian framework using NICER mass–radius posteriors. To assess the flexibility of each $f(Q)$ model, we adopt the DDME2 hadronic EoS as a baseline. In Fig.~\ref{fig1}, we plot the energy density ($\varepsilon$)(left), pressure ($p$) (middle), and speed of sound ($c_s^2$) (right) as a function of baryon density $\rho$ for DDME2. The prior distributions for the model parameters are listed in table \ref{tab1}. Since these specific models have not been explored previously in the context of astrophysical objects, we have adopted sufficiently broad prior ranges for the parameters $\alpha,\beta,Q_0$. The chosen intervals are as wide as possible while still ensuring the existence of physically stable solutions.

\begin{table}
\caption{\label{tab1}Uniform priors on the $f(Q)$ model parameters.}
\centering
\begin{tabular}{lc}
\hline\hline
Parameter & Prior \\
\midrule
$\alpha$ & $[-15,15]$ \\
$\beta$  & $[-3,5]$ \\
$Q_0$    & $[-1,1]$ \\
\hline
\hline
\end{tabular}
\end{table}

\subsection{Posterior distribution of $f(Q)$ model parameters and NS properties}

\begin{figure*}
    \centering
    \includegraphics[height=13.6cm,width=12.0cm]{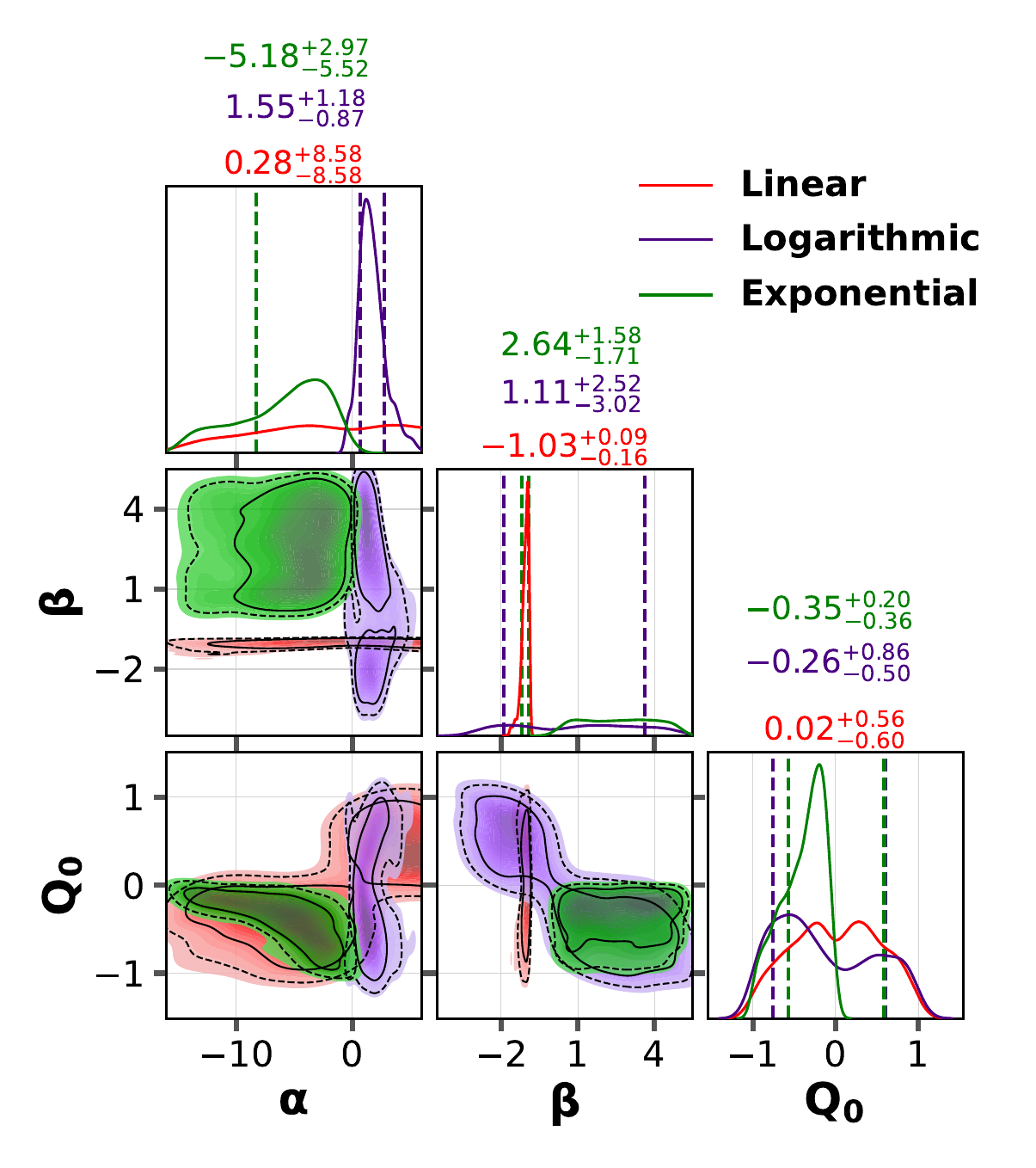}
    \caption{The marginalized posterior distributions of the
$f(Q)$ model parameters, obtained through Bayesian inference, for linear  (red), logarithmic  (purple), and exponential  (green) models. The vertical lines indicate
the 68\% confidence interval of the parameters. The confidence ellipses
for two-dimensional posterior distributions are plotted with 1$\sigma$, 2$\sigma$ and 3$\sigma$ confidence intervals.}
    \label{fig2}
\end{figure*}

The corner plots for the marginalized posterior distributions for $f(Q)$ model parameters of linear (red), logarithmic (purple), and exponential (green) models are shown in Fig.~\ref{fig2}. The marginalized one-dimensional posterior distributions for the parameters are displayed along the diagonals of the corner plots. The median values of the parameters and their uncertainties $2\sigma$, drawn from the marginalized posterior distributions, are shown. The vertical lines indicate the 68\% confidence interval of the parameters. We also plot the 2D confidence ellipses along with the off-diagonal corner plots, corresponding to the credible intervals $1\sigma$, $2\sigma$, and $3\sigma$. The shapes and orientations of these ellipses indicate the presence or absence of correlations between the parameters.

\begin{table}
\caption{\label{tab2}The median values and associated 68\%(90\%) uncertainties for the parameters from their marginalized posterior
distributions. The results are obtained for linear, logarithmic and exponential models. } 
  \centering
\renewcommand{\arraystretch}{1.4}
  
  \begin{tabular}{cccc}
  \hline\hline
\multirow{2}{*}{Models} & \multicolumn{3}{c}{Parameters} \\[1.0ex]
\cline{2-4}
                       & $\alpha$ & $\beta$ & $Q_0$ \\ \hline
 \hline
 Linear & $0.28^{+8.58(12.45)}_{-8.58(-12.84)}$ & $-1.03^{0.09(0.13)}_{-0.16(-0.34)}$ & $0.022^{+0.56(0.81)}_{-0.60(-0.88)}$ \\ [1.3ex]
 Logarithmic & $1.55^{+1.18(2.41)}_{-0.87(1.33)}$ & $1.11^{+2.52(3.45)}_{-3.02(3.77)}$ & $-0.26^{+0.86(1.12)}_{-0.50(0.66)}$\\ [1.3ex]
 Exponential & $-5.18^{+2.97(3.89)}_{-5.52(8.31)}$ & $2.64^{+1.58(2.15)}_{-1.71(2.14)}$ & $-0.35^{+0.20(0.28)}_{-0.36(0.54)}$\\ [1.3ex]
 \hline \hline
  \end{tabular}
   
\end{table}

\begin{table*}
\caption{\label{tab3}The median values and associated 68\%(90\%) uncertainties for the NS properties, namely the tidal deformability ($\Lambda_{1.4}$), radii ($R_{1.4}$ and $R_{2.07}$ and maximum
mass ($M_{max}$) from their posterior distribution.} 
  \centering

  \begin{tabular}{ccccc}
  \hline\hline
\multirow{2}{*}{Models} & \multicolumn{4}{c}{NS Properties} \\[1.0ex]
\cline{2-5}
                       & $\Lambda_{1.4}$& $R_{1.4}$ (km) & $R_{2.07}$ (km)& $M_{max}$($M_{\odot}$)\\ \hline
 \hline
 Linear & $143.99^{+86.30(166.57)}_{-38.62(-53.07)}$ & $11.19^{+0.53(0.94)}_{-0.38(-0.57)}$ & $11.68^{+0.39(0.73)}_{-0.40(-0.93)}$ &$2.30^{+0.34(0.54)}_{-0.20(-0.29)}$\\ [1.3ex]
 Logarithmic & $135.82^{+64.59(179.65)}_{-23.33(32.21)}$ & $11.09^{+0.48(1.05)}_{-0.22(0.32)}$ & $12.08^{+0.36(0.65)}_{-0.25(0.42)}$ &$2.36^{+0.28(0.42)}_{-0.23(0.32)}$\\ [1.3ex]
 Exponential & $156.95^{+84.02(177.68)}_{-41.73(55.77)}$ & $11.27^{+0.53(0.97)}_{-0.36(0.52)}$ & $11.76^{+0.50(0.83)}_{-0.36(0.50)}$ &$2.29^{+0.11(0.16)}_{-0.13(0.21)}$\\ [1.3ex] \hline \hline
  \end{tabular}
 
\end{table*}
The correlation between $\alpha$ and $Q_0$ is moderate in the linear case ($r \sim 0.64$) and in the exponential case ($r \sim -0.64$) and negligible in the logarithmic case ($r \sim -0.14$), where $r$ is the Pearson's correlation
coefficient. A strong correlation between $\beta$ and $Q_0$ is found only for the logarithmic model ($r \sim -0.79$). The median values of the model parameters and the corresponding confidence intervals of 68\% (90\%) obtained from the
marginalized posterior distributions are listed in Table~\ref{tab2}. The Bayesian analysis indicates that the constraining power of the data depends sensitively on how each parameter modifies the effective gravitational sector in the three $f(Q)$ parameterizations. In the linear model, the parameter 
$\beta = -1.03^{+0.09}_{-0.16}$ 
 is tightly constrained, reflecting its direct role in determining the effective linear coupling to $Q$. Since the GR limit corresponds to a specific linear scaling, the narrow credible interval indicates that the data strongly restrict deviations from this behavior. 
For the logarithmic model, the amplitude parameter 
$\alpha = 1.55^{+1.18}_{-0.87}$ 
is comparatively well constrained, indicating that the strength of the logarithmic correction is observationally relevant. However, 
$\beta = 1.11^{+2.52}_{-3.02}$ 
exhibits large asymmetric uncertainties, signaling degeneracy between the amplitude and scale of the modification, and allowing partial compensation that weakens individual constraints. The scale parameter 
$Q_0 = -0.26^{+0.86}_{-0.50}$ 
is only moderately constrained, suggesting a limited sensitivity of the data to the precise normalization inside the logarithmic term.
In the exponential model, the strongest constraint is obtained for 
$Q_0 = -0.35^{+0.20}_{-0.36}$, 
indicating that the characteristic scale governing the exponential suppression of deviations from GR is well determined. The parameter 
$\beta = 2.64^{+1.58}_{-1.71}$, 
 which controls the rate at which the model departs from the GR limit, is moderately constrained. 
Overall, parameters that directly control the effective linear coupling or the characteristic transition scale away from the GR limit are more tightly constrained, whereas additive or amplitude like parameters remain weakly determined due to degeneracies in the background expansion behavior.

\begin{figure*}
    \centering    \includegraphics[height=12cm,width=12cm]{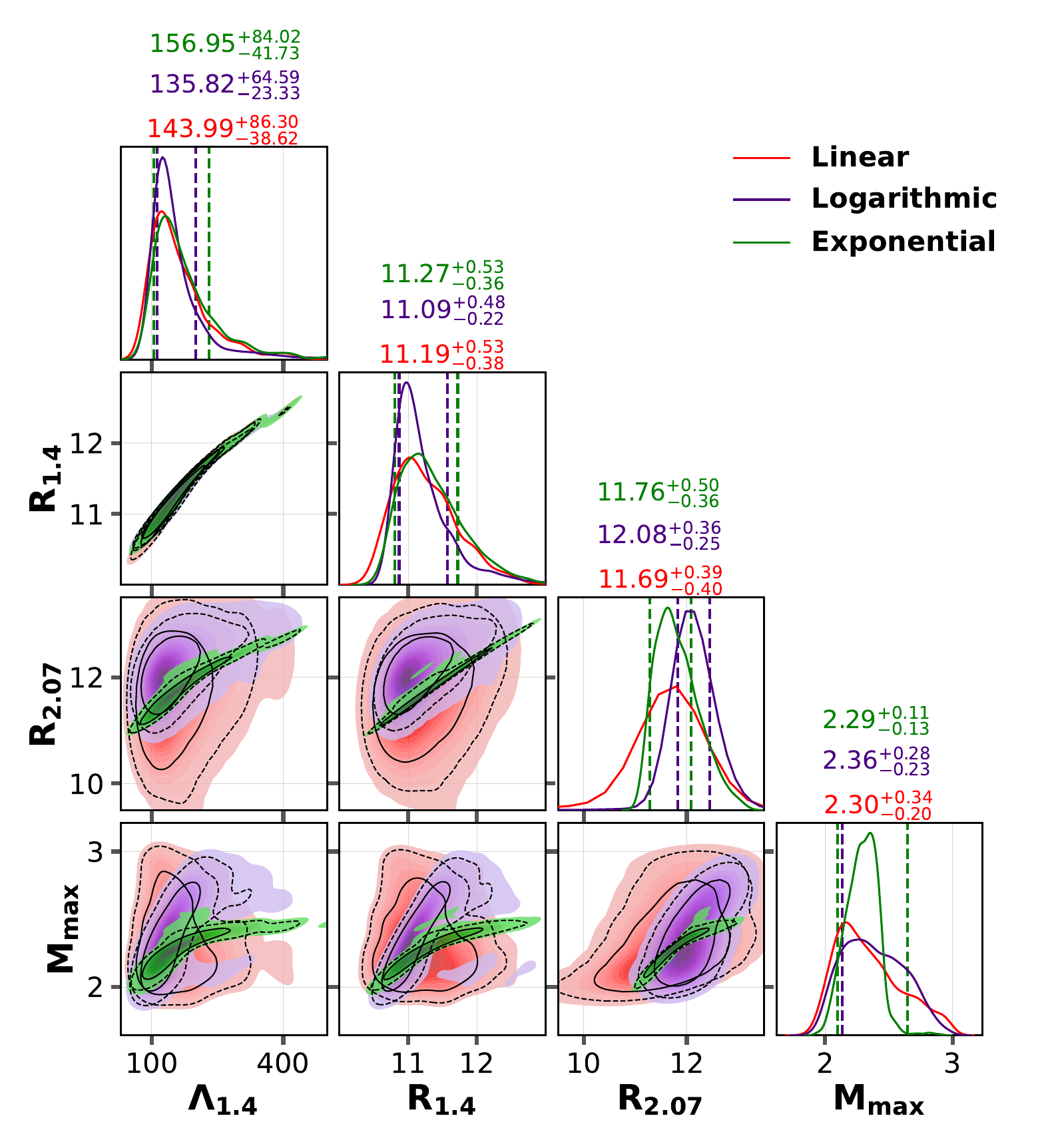}
    \caption{Corner plots for the marginalized posterior distributions
of the tidal deformability $\Lambda_{1.4}$, radii $R_{1.4}$ (km) and $R_{2.07}$ (km) and the maximum mass $M_{max}$ ($M_\odot$) for linear (red), logarithmic (purple), and exponential (green). }
    \label{fig3}
\end{figure*}

We have obtained distributions of NS properties such as tidal deformability $\Lambda_{1.4}$, radii $R_{1.4}$ (km) and $R_{2.07}$ (km) and the maximum mass $M_{max}$ ($M_\odot$)  using the posterior distributions for the parameters corresponding to the linear, logarithmic and exponential models as listed in table~\ref{tab2}. The corner plots for these NS properties are shown in Fig.~\ref{fig3}. The three observables reported in table~\ref{tab3} are: (i) the \textit{dimensionless tidal deformability} of a canonical neutron star, $\Lambda_{1.4} \equiv \Lambda(M = 1.4\,M_{\odot})$, constrained by GW170817 (ii) the radius of a $1.4\,M_{\odot}$ neutron star, $R_{1.4}$, the standard reference radius in the literature; and (iii) the radius of a $2.07\,M_{\odot}$ neutron star, $R_{2.07}$, corresponding to the mass of the heavy pulsar PSR~J0740+6620. Including $R_{2.07}$ provides a direct probe of the high-mass end of the mass--radius diagram and tests the ability of each $f(Q)$ model to support the most massive observed neutron stars. The same as Fig.\ref{fig2}, the vertical lines, median values with $2\sigma$ uncertainties, and 2D confidence ellipses along the off-diagonal line are shown. It is clear from the off-diagonal graphs that $\Lambda_{1.4}$ is strongly correlated with $R_{1.4}$ for all models ($r\sim 0.96$). In the linear model, the maximum mass $M_{\max}$ is found to be statistically uncorrelated with the other neutron star observables, as indicated by nearly circular and non-tilted confidence contours. This absence of correlation implies that variations in the maximum mass occur independently of changes in other stellar properties, suggesting that the linear modification does not introduce significant structural coupling in the equilibrium configuration.
In contrast, the logarithmic model exhibits a mild but noticeable correlation between $M_{\max}$ and other neutron star properties, reflected by moderately tilted confidence contours. This indicates that the logarithmic correction induces a partial coupling between the maximum mass and the internal stellar structure, allowing variations in one quantity to influence the other within the allowed parameter space.
However, the exponential model demonstrates a strong and pronounced correlation between $M_{\max}$ and the remaining properties of neutron stars, characterized by highly elongated and tilted contours. This behavior signifies a substantial structural interdependence introduced by the exponential modification, where changes in the gravitational parameters directly impact the global stellar configuration. Consequently, the maximum mass in this case is highly sensitive to the underlying modified gravity parameters, revealing a stronger departure from the effective GR-like behavior compared to the linear and logarithmic scenarios.
  We have summarized the median values of the properties of NS together with 68\% (90\%) confidence intervals in table~\ref{tab3}. It shows how different models for $f(Q)$ theory affect these astrophysical quantities.
The logarithmic model predicts the highest maximum mass ($2.36_{-0.23}^{+0.28}$) while the linear and exponential models show the widest uncertainty ranges for most properties, particularly for $\Lambda_{1.4}$, indicating different constraint sensitivities. The radii predictions for $R_{1.4}$ and $R_{2.07}$ are remarkably similar for all models , differing by less than $\sim$ 0.4 km in the median values. This suggests that within these $f(Q)$ models, the predicted radius does not change significantly with mass in this range.

\begin{figure*}
    \centering   \includegraphics[width=\linewidth]{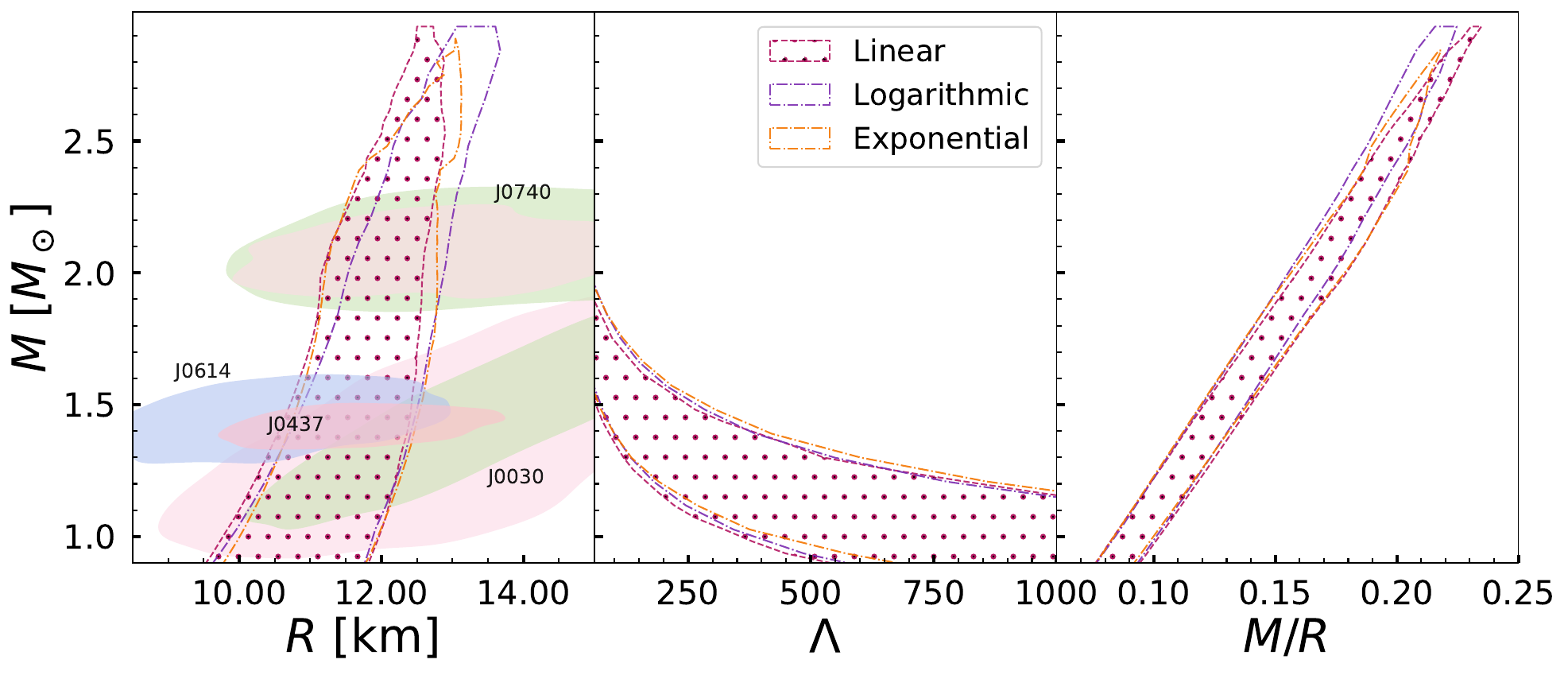}
    \caption{The 95\% confidence interval distributions for the radius $R$ (km) (left panel), tidal deformability $\Lambda$ (middle panel) and compactness $M/R$ (right panel) as a function of NS mass $M$ ($M_\odot$).}
    \label{fig4}
\end{figure*}

\begin{figure*}
    \centering
    \includegraphics[width=\linewidth]{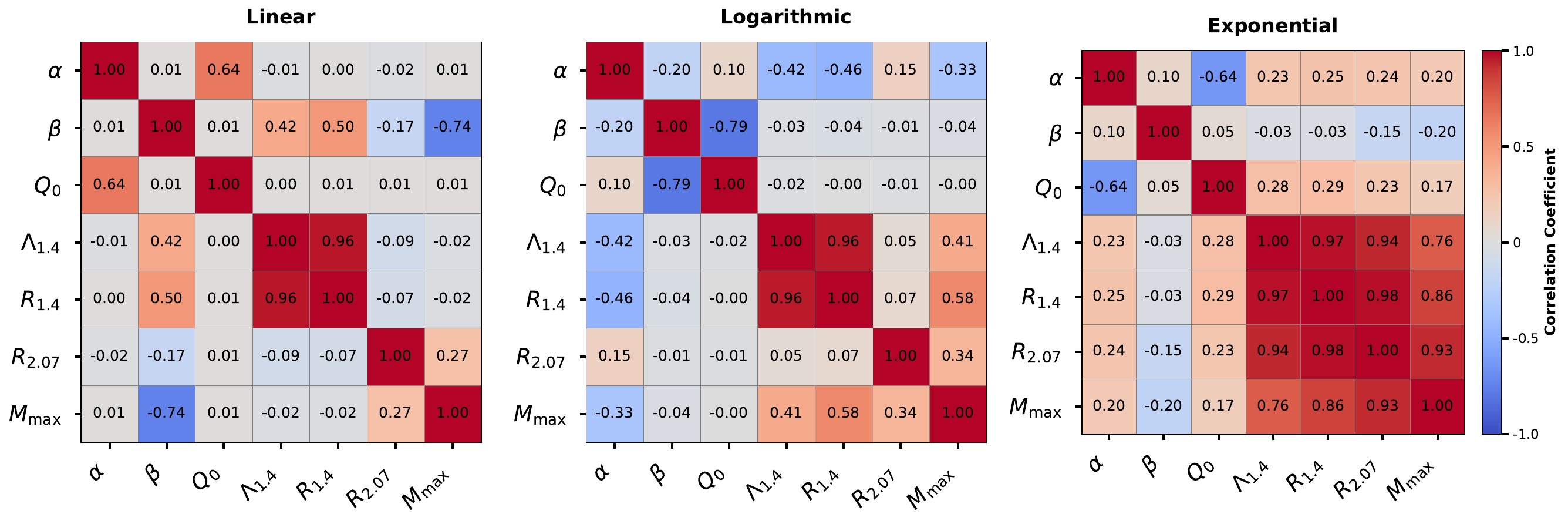}
    \caption{The Pearson's correlation coefficients among parameters and selected NS properties for linear (left panel), logarithmic (middle panel), and exponential(right panel).}
    \label{fig5}
\end{figure*}

Figure~\ref{fig4} displays the 95\% confidence bands for the $M$–$R$ (left panel), $M$–$\Lambda$ (middle panel), and $M$–$M/R$ (right panel) relations for the linear (magenta), logarithmic (purple), and exponential (orange) models. All NICER observational constraints incorporated in the Bayesian framework are also shown.
The $M-R$ curves for all models pass through the centers of the observational bands, indicating that the Bayesian inference successfully reproduces the observed data. The exponential model exhibits the widest confidence band, particularly in the 1–2 $M_\odot$ mass range, consistent with its larger uncertainties of the parameters  (see Table~\ref{tab3}). In contrast, the linear model shows the narrowest band, reflecting its tighter parameter constraints. The logarithmic model yields broader distributions above $2 M_\odot$, suggesting a greater capacity to produce stiffer high-mass $M$–$R$ configurations. 
A key result of our analysis is that all three constrained $f(Q)$ models predict a maximum neutron star mass exceeding $2.5\,M_{\odot}$. This places the stable configurations within the lower mass gap region ($\sim 2.5$--$5\,M_{\odot}$), where the nature of compact objects remains uncertain. The $95\%$ confidence bands of the mass--radius relation consistently extend beyond $2.5\,M_{\odot}$ and reach nearly about $3~M_{\odot}$, indicating that the Bayesian-constrained $f(Q)$ gravity parameters allow significantly more massive stable stars than typically obtained in standard GR scenarios. The appearance of mass-gap configurations therefore emerges naturally from the constrained parameter space and constitutes a clear, testable prediction of the present $f(Q)$ framework.
For the $M-\Lambda$ distributions, all three models follow similar trends. In terms of compactness, the linear model reaches higher $M/R$ values than the others at a given mass. The different compactness trends between models highlight how each $f(Q)$ functional form alters the mass–radius relation, particularly in the high-density regime.\\
The Pearson's correlation coefficients among all \( f(Q) \) model parameters (\( \alpha \), \( \beta \), and \( Q_0 \)) and the NS properties (\( \Lambda_{1.4} \), \( R_{1.4} \), \( R_{2.07} \), and \( M_{\text{max}} \)) are shown in Fig.~\ref{fig5}. The left, middle, and right panels correspond to the linear, logarithmic, and exponential models, respectively. In the linear model, \( \beta \) is strongly correlated with the $M_{\text{max}}$, radius and tidal deformability at the canonical mass, with correlation coefficients of \( -0.74\), \(-0.79 \) and \( -0.70 \), respectively. This indicates that \( \beta \) plays a key role in the definition of the canonical NS properties in the linear case. In the logarithmic model, \( \alpha \) is moderately correlated with these canonical properties. In contrast, for the exponential model, no strong correlation is found between individual parameters and NS properties, suggesting that the parameters collectively relate to NS properties.

\subsection{Model comparison}

\begin{figure*}
    \centering
    \includegraphics[width=\linewidth]{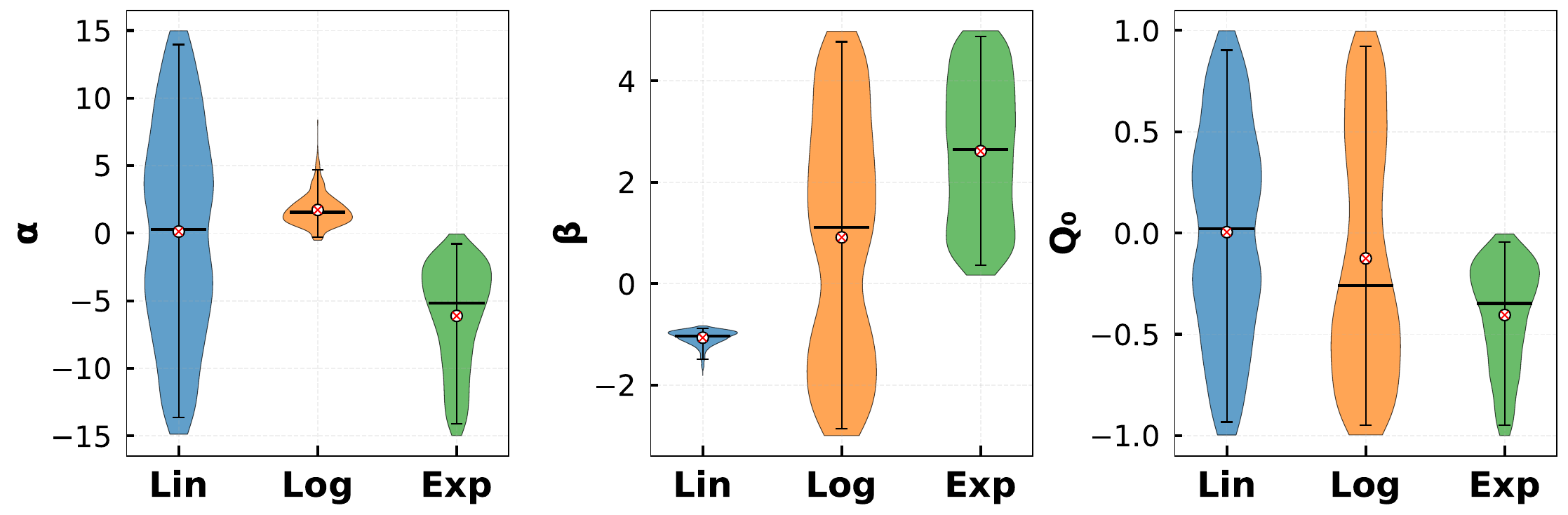}
    \caption{Violin plots of the posterior distributions for the \( f(Q) \) model parameters \( \alpha \) (left), \( \beta \) (middle), and \( Q_0 \) (right) for the linear (Lin), logarithmic (Log), and exponential (Exp) models.}
    \label{fig6}
\end{figure*}

\begin{table}
\caption{The logarithmic Bayes factor for all $f(Q)$ models considered for this analysis. Here we consider exponential as \( H_2 \) hypothesis always and other models as \( H_1 \) to compute \(\ln(\text{BF}_{12})\) using Eq. (\ref{logBayes}).\label{tab4}}
\centering
\begin{tabular}{cc}
\hline\hline
\textbf{Model} & \textbf{\(\ln(\text{BF}_{12})\) } \\
\hline
Linear & -1.81 \\
Logarithmic & -0.43\\
\hline \hline
\end{tabular}
\end{table}

Since we use a large number of $f(Q)$ models for this analysis, it would be good to compare these models statistically using the
Bayes factor as discussed in Sec.~\ref{bayesfactor}. In Table~\ref{tab4}, we listed the logarithmic values of the Bayes factor computed using Eq. (\ref{logBayes}) for all models. The interpretation of the Bayes factor is discussed in Sec. \ref{IntpBF}. We consider the exponential model as hypothesis \( H_2 \) (because of large negative values of log evidence) and other models as \( H_1 \) to compute \(\ln(\mathrm{BF}_{12}) \). From the large negative values of $\ln(\mathrm{BF}_{12})$, it is clear that the exponential is definitely the best model in Bayesian inference.  The Bayes factors computed for all models consistently rank the models in the same order: exponential as the best, followed by logarithmic, and then linear.

Violin plots provide a detailed view of posterior distributions by merging the summary statistics of a box plot with a smooth density curve. This allows for easy visual comparison of features such as skewness, multimodality, and the overall spread of model parameter constraints.
Figure~\ref{fig6} shows the violin plots of the posterior distributions for the \( f(Q) \) model parameters \( \alpha \) (left), \( \beta \) (middle), and \( Q_0 \) (right) for linear (Lin), logarithmic (Log), and exponential (Exp) models. In the {exponential} case, all parameter distributions are nearly Gaussian and unimodal, indicating a single, well-defined solution favored by the observed data. For the {logarithmic} model, the posteriors for \( \beta \) and \( Q_0 \) deviate from a perfect Gaussian shape (bimodal), reflecting greater uncertainty and potential degeneracies. The {linear} model shows the most pronounced issues: its \( \alpha \) and \( Q_0 \) distribution is distinctly bimodal, suggesting two separate solutions and its \( \alpha \) posterior is very broad. This bimodality and the associated parameter degeneracies lower the model's Bayesian evidence relative to the others. This visualization highlights why the exponential model is statistically favored, followed by the logarithmic model and then the linear model.

\section{Summary and outlook}\label{sec:9}

We employed Bayesian inference with three modified gravity $f(Q)$ models: linear, logarithmic, and exponential by integrating astrophysical constraints from NICER mass-radius measurements. NICER data includes PSR J0030+0451, PSR J0740+6620, PSR J0437+4715, and PSR J0614+3329. For the first time in the literature on $f(Q)$ gravity, this analysis systematically investigates the theory with neutron star observables in the strong field regime.

The exponential model appeared as the statistically preferred choice, supported by multiple lines of evidence: its \( Q_0 \) parameter is the most tightly constrained ($-0.35^{+0.20}_{-0.36}$), its posterior distributions of all parameters are nearly Gaussian and unimodal.   In contrast, the logarithmic model shows broader parameter uncertainties and mild degeneracies, particularly for \( \beta \) ($1.11^{+2.52}_{-3.02}$). However, it predicts the highest maximum neutron-star mass (\( 2.36\,M_\odot \)). The linear model exhibits pronounced bimodality in its \( \alpha \) and \( Q_0 \) posterior and the broadest  \( \alpha \) distribution, reflecting strong parameter degeneracies that reduce its statistical credibility. The deformability parameter $\Lambda_{1.4}$ and radius $R_{1.4}$ for the exponential model were found to be $156.95^{+84.02}_{-41.73}$ and $11.27^{+0.53}_{-0.36}$ km, respectively. One of the most significant findings of this work is that all three observationally constrained $f(Q)$ models robustly predict maximum neutron star masses exceeding $2.5\,M_{\odot}$. The corresponding $95\%$ confidence regions of the mass--radius relation consistently enter the lower mass gap ($\sim 2.5$--$5\,M_{\odot}$), a regime where the nature of compact objects remains unresolved. This result is not imposed by construction, but emerges directly from the Bayesian-constrained parameter space, demonstrating that the modified gravity effects enhance the high-density gravitational support and allow substantially heavier stable configurations than typically expected within standard GR frameworks. 
The presence of stable neutron stars in the mass-gap region therefore constitutes a testable prediction of the present $f(Q)$ models. Future precise mass measurements in this regime could provide a powerful discriminator between GR and extended gravity scenarios.

Model comparisons based on the computed Bayes factors confirm this hierarchy. The logarithmic Bayes factors, \( \ln(\text{BF}) \), decisively favor the exponential model over the logarithmic and linear alternatives, with values of \(-0.35\) and \(-1.28\), respectively. According to the established scale, this indicates strong evidence for the exponential model. Therefore, the exponential \( f(Q) \) model emerges as the best constrained and statistically supported description of the neutron star structure among the three, highlighting its viability as a candidate for modified gravity in the strong-field regime.

In the present analysis, we adopted a single baseline dense-matter description (DDME2) to reduce nuclear-physics systematics and focus on the impact of the $f(Q)$ sector, with the NICER likelihood and inferred maximum mass explicitly conditioned on this choice. A natural next step is to quantify the robustness of the inferred gravity parameters against EoS uncertainties by performing Bayesian inference for alternative representative nuclear matter EoSs (RMF-based or hybrid), and performing Bayesian model comparison across EoSs in analogy with our Bayes-factor ranking of the $f(Q)$ models. More generally, one can carry out a joint (hierarchical) Bayesian inference in which the $f(Q)$ parameters and a flexible EoS parameterization (e.g., piecewise-polytropie or spectral representations) are sampled simultaneously, thereby marginalizing over the EoS and directly addressing the known degeneracy between strong-field gravity and microphysics; such joint gravity-EoS strategies have proven informative in other modified-gravity contexts. Extending the data evidence beyond NICER (e.g. incorporating experimental nuclear data, empirical nuclear inputs, and theoretical predictions ) would further help break down degeneracies and tighten the constraints on both sectors.


\chapter{Concluding Remarks \& Future Perspective} 

\label{Chapter7} 

\lhead{Chapter 6. \emph{Concluding Remarks and future Perspectives}} 

The primary objective of this thesis has been to investigate compact astrophysical objects within the framework of modified theories of gravity. Across chapters \ref{Chapter2}-\ref{Chapter5}, four concrete and complementary studies have been carried out, each addressing different theoretical formulations and physical aspects of compact stars. These investigations collectively examine stellar structure, stability, and observational viability under deviations from GR. The results demonstrate how modified gravity effects can significantly influence the internal properties and global characteristics of compact objects. Overall, this thesis provides a systematic and physically consistent assessment of compact stars as probes of strong-field gravity.\\
Chapter~\ref{Chapter1} provides the theoretical foundation of the thesis by reviewing compact objects, the relativistic stellar structure, and the limitations of GR. Introduces the geometric trinity of gravity such as curvature, torsion, and non-metricity motivating modified gravity theories such as 
$f(T)$ and $f(Q)$. The chapter also summarizes the various types of EoSs and key physical and stability criteria relevant to compact stars.\\
In Chapter \ref{Chapter2}, we showed that charged compact stars in
$f(Q)$ gravity can be consistently modeled using conformal Killing vectors and the MIT bag EoS. Among the two configurations studied, the power-law conformal factor yields regular metric functions, stable equilibrium, and satisfies all physical and stability conditions, whereas the linear model develops central pathologies and instability. A comparison of exterior space-times reveals that matching with the Bardeen geometry leads to more realistic and stable stellar configurations than the R-N case. Overall, the power-law conformal model with Bardeen exterior emerges as the most physically viable description of charged compact stars in 
$f(Q)$ gravity.\\
Chapter~\ref{Chapter3} focuses on anisotropic compact stars in $f(T)$ gravity using the MGD approach, starting from a well-defined isotropic seed configuration. By adopting the Buchdahl ansatz with a quadratic polytropic EoS, two viable anisotropic models were obtained, yielding regular and physically acceptable stellar interiors in the GR, $f(T)$, and $f(T)$+MGD scenarios. The inclusion of anisotropy through MGD enhances stability and supports more massive configurations by counteracting gravitational collapse. The mass-radius relations constrained by GW190814 and massive pulsars show that models can sustain compact stars with masses up to 3.5 $M_{\odot}$. Stability analyzes based on the adiabatic index and the HZN criterion confirm the robustness of all configurations, demonstrating the effectiveness of the GD–MGD framework in 
$f(T)$ gravity.\\
In Chapter~\ref{Chapter4}, we constructed an anisotropic strange star model in teleparallel gravity using the CGD approach under the vanishing complexity condition, incorporating dark matter through an additional gravitational source. The resulting exact solution is physically viable, with finite and monotonic matter variables satisfying all energy conditions and regularity at the center. Stability analyzes using the adiabatic index, Herrera’s cracking condition, and the HZN criterion confirm both local and global stability for moderate dark matter coupling. The model successfully reproduces observed massive compact stars, including PSR J0952$-$0607 and GW190814, with increasing decoupling strength enhancing the maximum mass. These results demonstrate that CGD in teleparallel gravity provides a robust framework for modeling realistic dark matter–admixed compact stars beyond GR.\\
In Chapter~\ref{Chapter5}, we performed a Bayesian analysis of neutron stars in $f(Q)$ gravity using NICER mass–radius data for PSR J0030+0451, PSR J0740+6620, PSR J0437+4715, and PSR J0614+3329, considering linear, logarithmic, and exponential models. For the first time in $f(Q)$ gravity, this study systematically explores strong-field neutron-star observables within the framework of modified gravity. The exponential model emerges as statistically preferred, exhibiting tightly constrained and nearly Gaussian posteriors. In contrast, the logarithmic and linear models show broader uncertainties and parameter degeneracies. Comparisons of Bayes-factors decisively favor the exponential model, establishing it as the most viable $f(Q)$ description of the neutron-star structure. 

The present work opens several promising directions for future investigations in compact-star physics within modified gravity frameworks. A natural extension of this study would involve incorporating multiple realistic nuclear EoSs to examine model dependence in neutron star observables using a Bayesian framework. Machine learning techniques, particularly deep neural networks, can be employed to capture nonlinear correlations between gravity-sector parameters, nuclear matter properties, and observational data. The inclusion of ultra-strong magnetic fields would allow for a more realistic modeling of magnetars and their impact on stellar structure and stability in modified gravity. Furthermore, studying neutron star oscillation modes and their coupling to modified gravity effects can provide additional observational signatures through gravitational waves. Exploring phase transitions in hybrid stars within the 
$f(Q)$ or $f(T)$ gravity context also remains an important aspect. Finally, extending the present analysis to rapidly rotating and binary systems would strengthen the connection between theoretical predictions and multi-messenger astrophysical observations.






\addtocontents{toc}{\vspace{2em}} 

\backmatter


\label{References}
\lhead{\emph{References}}

\cleardoublepage
\pagestyle{fancy}

\label{Publications}
\lhead{\emph{List of Publications}}

\chapter{List of Publications}
\section*{Thesis Publications}
\begin{enumerate}

\item \textbf{S. Pradhan}, P.K. Sahoo, 
\textit{A comprehensive study of massive compact star admitting conformal motion under Bardeen geometry}, 
\textcolor{blue}{Nuclear Physics B} \textbf{1002}, 116523 (2024).

\item \textbf{S. Pradhan}, S.K. Maurya, A. Errehymy, G. Mustafa, P.K. Sahoo, 
\textit{Gravitationally deformed polytropic models in extended teleparallel gravity and influence of decoupling parameters on constraining mass-radius relation}, 
\textcolor{blue}{Chinese Physics C} \textbf{49}(10), 105110 (2025).

\item \textbf{S. Pradhan}, P. Bhar, S. Mandal, P.K. Sahoo, K. Bamba, 
\textit{The stability of anisotropic compact stars influenced by dark matter under teleparallel gravity: an extended gravitational deformation approach}, 
\textcolor{blue}{The European Physical Journal C} \textbf{85}(2), 127 (2025).

\item \textbf{Sneha Pradhan}, N. K. Patra, Kai Zhou, P.K. Sahoo, \textit{Bayesian inference of neutron star properties in $f(Q)$ gravity using NICER observations}, \textcolor{blue}{	arXiv:2601.16227}.

\end{enumerate}
\section*{Other Publications}

\begin{enumerate}

\item P. Balani, C. S. Medisetti, G. Mustafa, \textbf{S. Pradhan}, P. K. Sahoo,
\textit{Traversable wormholes with Van der Waals equation of state in $f(R,L,T)$ gravity},
\textcolor{blue}{Annalen der Physik} \textbf{538}(1), e00408 (2025).

\item P. H. R. S. Moraes, R. V. Lobato, \textbf{S. Pradhan} P. K. Sahoo,
\textit{White dwarfs in $f(Q)$ gravity},
\textcolor{blue}{General Relativity and Gravitation} \textbf{57}(10), 149 (2025).

\item \textbf{S. Pradhan}, G.N. Gadibail, S. Mandal, P.K. Sahoo, K. Bamba, 
\textit{Dynamics of interacting dark energy and dark matter model in a curved FLRW scalar field cosmology}, 
\textcolor{blue}{Monthly Notices of the Royal Astronomical Society} \textbf{543}(2), 1273–1287 (2025).

\item S. Ghosh, \textbf{S. Pradhan}, P.K. Sahoo, 
\textit{Maximum mass limit of generalized MIT compact star: A theoretical study of thin shell dynamics and gravitational lensing}, 
\textcolor{blue}{Fortschritte der Physik} \textbf{73}(8), e70020 (2025).

\item \textbf{S. Pradhan}, Zinnat Hassan, P.K. Sahoo, 
\textit{Wormhole geometries supported by strange quark matter and phantom-like generalized Chaplygin gas within $f(Q)$ gravity}, 
\textcolor{blue}{Physics of the Dark Universe} \textbf{46}, 101620 (2024).

\item \textbf{S. Pradhan}, R. Solanki, P.K. Sahoo, 
\textit{Cosmological constraints on $f(Q)$ gravity models in the non-coincident formalism}, 
\textcolor{blue}{Journal of High Energy Astrophysics} \textbf{43}, 258 (2024).

\item \textbf{S. Pradhan}, S.K. Maurya, P.K. Sahoo, G. Mustafa, 
\textit{Geometrically deformed charged anisotropic models in $f(Q,T)$ gravity}, 
\textcolor{blue}{Fortschritte der Physik} \textbf{72}(9), 2400092 (2024).

\item O. Sokoliuik, \textbf{S. Pradhan}, A. Baransky, P.K. Sahoo, 
\textit{AdS Black Hole Thermodynamics and micro-structures from $f(Q)$ gravitation}, 
\textcolor{blue}{Fortschritte der Physik} \textbf{72}(1), 2300043 (2024).

\item S. Mandal, \textbf{S. Pradhan}, P.K. Sahoo, T. Harko, 
\textit{Cosmological observational constraints on the power law $f(Q)$ type modified gravity theory}, 
\textcolor{blue}{The European Physical Journal C} \textbf{83}, 1141 (2023).

\item P. Bhar, \textbf{S. Pradhan}, A. Malik, P.K. Sahoo, 
\textit{Physical characteristics and maximum allowable mass of hybrid star in the context of $f(Q)$ gravity}, 
\textcolor{blue}{The European Physical Journal C} \textbf{83}(7), 646 (2023).

\item \textbf{S. Pradhan}, D. Mohanty, P.K. Sahoo, 
\textit{Thin-Shell gravastar model in $f(Q,T)$ gravity}, 
\textcolor{blue}{Chinese Physics C} \textbf{47}(9), 095104 (2023).

\item \textbf{S. Pradhan}, S. Mandal, P.K. Sahoo, 
\textit{Gravastar in the framework of symmetric teleparallel gravity}, 
\textcolor{blue}{Chinese Physics C} \textbf{47}(5), 055103 (2023).

\item O. Sokoliuik, \textbf{S. Pradhan}, P.K. Sahoo, A. Baransky, 
\textit{Buchdahl quark stars within $f(Q)$ theory}, 
\textcolor{blue}{The European Physical Journal Plus} \textbf{137}(9), 1–15 (2022).

\end{enumerate}
\cleardoublepage
\pagestyle{fancy}

\label{Conferences}
\lhead{\emph{Conferences/Workshops Attended}}

\chapter{Conferences/Workshops Attended}
\begin{enumerate}

\item Participated in a one-week academic visit to the \textbf{\textit{Chennai Mathematical Institute} (CMI), India} and presented research work in the CMI Physics Seminar.

\item Presented a paper at the \textbf{\textit{Young Astronomers’ Meet (YAM--2025)}}, held at \textbf{\textit{Indian Institute of Technology, Hyderabad}~(IIT-H)}, India.

\item Presented a paper in the \textcolor{blue}{\textbf{\textit{International Conference}}} ``\textit{High Energy Astrophysics in Southern Africa 2025 (HEASA 2025)}'' hosted by the \textbf{Centre for Astro-Particle Physics (CAPP)}, \textbf{\textit{University of Johannesburg}}, South Africa. 

\item Participated and presented a paper in the \textbf{``\textit{Brainstorming Workshop on Neutron Stars}''} held at the \textbf{\textit{Institute of Mathematical Sciences} (IMSc)}, Chennai, India.

\item Presented a poster at the \textbf{``\textit{Summer School on Gravitational-Wave Astronomy}''} organized by the \textbf{\textit{International Center for Theoretical Sciences} (ICTS-TIFR)}, Bengaluru.

\item Participated and gave an oral presentation in the seminar sponsored by \textbf{IUCAA, Pune} , on \textit{\textbf{Advances in Cosmology}} organized by \textbf{\textit{Christ University}}, Bangalore.

\item Presented a paper at the \textbf{\textit{International Meeting IMRACEA-2025} (ICARD)} organized by the \textbf{\textit{University of North Bengal} (NBU)}, Darjeeling, India.

\item Attended the \textbf{\textit{Workshop on Mathematics for Machine Learning}} organized by the \textbf{Department of Mathematics}, \textbf{\textit{BITS Pilani Hyderabad Campus}}.

\item Presented a paper in the \textcolor{blue}{\textbf{\textit{International Conference}}} ``\textit{Quantum Extreme Universe: Matter, Information and Gravity}'' held at the \textbf{\textit{Okinawa Institute of Science and Technology} (OIST)}, Okinawa, Japan.

\item Presented a paper in the IUCAA-sponsored national conference \textbf{\textit{Classical and Quantum Gravity}} organized by the \textbf{\textit{Cochin University of Science and Technology} (CUSAT)}, Kochi, Kerala.

\item Attended the \textcolor{blue}{\textbf{\textit{International Summer School}}} ``\textit{Indo-French Astronomy School} (IFAS-09)'' jointly organized by the \textbf{\textit{University of Lyon}} and \textbf{\textit{IUCAA}}, France.

\item Presented a paper at the ``\textit{International Mathematical Society Conference (IMS-2023)}'' conducted by the \textbf{\textit{Department of Mathematics, BITS Pilani Hyderabad Campus} (BITS-H)}.

\item Presented a paper at the ``\textit{International Conference on Differential Geometry and Relativity (ICDGR)}'' organized by the \textbf{Tensor Society} in the \textbf{\textit{Department of Mathematics, SSJ University}}, Almora, Uttarakhand.

\item Participated in the \textbf{\textit{International Workshop on Astronomy Data Analysis with Python} (ADAP-23)} organized by the \textbf{\textit{Maulana Azad National Urdu University} (MANUU)}, Hyderabad, and sponsored by \textbf{IUCAA, Pune}.

\item Presented the final project work under the \textbf{\textit{S.N. Bhatt Summer Internship Programme}} at the \textbf{\textit{International Center for Theoretical Sciences} (ICTS-TIFR)}, Bengaluru.

\item Attended a two-day open workshop on \textbf{\textit{Gravitational Wave Data Analysis}} at the \textbf{\textit{International Center for Theoretical Sciences} (ICTS-TIFR)}, Bengaluru.


\end{enumerate}
\cleardoublepage
\pagestyle{fancy}
\lhead{\emph{Biography}}

\chapter{Biography}

\section*{Brief Biography of the Candidate:}
Ms. \textbf{Sneha Pradhan} obtained her Bachelor’s degree in Mathematics from Jogamaya Devi College, affiliated with the University of Calcutta, in 2018, where she secured the first position in the first class. She completed her Master’s degree in Applied Mathematics from the University of Calcutta in 2020. Subsequently, she joined the Ph.D. program under the NBHM Project Fellowship Scheme in 2022.

Over the past four years of her research career, Ms. Pradhan has published 16 research papers in high impact factor international journals and has presented her work at several national and international conferences.

\section*{Brief Biography of the Supervisor:}
\textbf{Prof. Pradyumn Kumar Sahoo} has over 24 years of immense research experience in Applied Mathematics, Cosmology, Astrophysical Objects, General Theory of Relativity, and Modified Theories of Gravity. He obtained his Ph.D. from Sambalpur University, Odisha, India, in 2004. In 2009, he joined the Department of Mathematics at BITS Pilani, Hyderabad Campus, as an Assistant Professor and is currently a Professor. He is also an Associate Member of IUCAA, Pune. In 2022, he received the ``Prof. S. Venkateswaran Faculty Excellence Award" from BITS Pilani. He has been awarded a visiting professor fellowship at Transilvania University of Brașov, Romania. According to a survey by researchers from Stanford University, he has been ranked among the top 2\% of scientists worldwide in the field of Nuclear and Particle Physics in the last six years. Throughout his career, he has published more than 270 research articles in various renowned national and international journals. As a visiting scientist, he had the opportunity to visit the European Organization for Nuclear Research (CERN) in Geneva, Switzerland, a renowned center for scientific research. He has participated in numerous national and international conferences, often presenting his work as an invited speaker. Prof. Sahoo has engaged in various research collaborations at both the national and international levels. He has contributed to BITS through five sponsored research projects: University Grants Commission (UGC 2012-2014), DAAD Research Internships in Science and Engineering (RISE) Worldwide (2018, 2019, 2023, 2024 and 2025), Council of Scientific and Industrial Research (CSIR 2019-2022), National Board for Higher Mathematics (NBHM 2022-2025), and Anusandhan National Research Foundation (ANRF), Department of Science and Technology (DST 2023-2026). He also serves as an expert reviewer for Physical Science Projects for ANRF, DST (Government of India), and UGC research schemes. Additionally, he is an editorial board member for various reputable journals, contributing to the research community.

\end{document}